\pdfoutput=1
\documentclass[11pt]{article}

\usepackage{jheppub}

\usepackage[english]{babel}

\usepackage{subfigure}
\usepackage{placeins}

\usepackage{braket}   

\newcommand{\sect}[1]{section~#1}
\newcommand{\sects}[1]{sections~#1}
\newcommand{\fig}[1]{figure~#1}
\newcommand{\figs}[1]{figures~#1}
\newcommand{\tab}[1]{table~#1}
\newcommand{\eqn}[1]{equation~#1}

\newcommand{\half}{{\textstyle\frac{1}{2}}}

\newcommand{\gev}{\operatorname{GeV}}
\newcommand{\mev}{\operatorname{MeV}}

\newcommand{\fm}{\operatorname{fm}}

\newcommand{\ms}{\mskip 1.5mu}
\newcommand{\bs}{\mskip -1.5mu}

\newcommand{\tvec}[1]{\boldsymbol{#1}}

\newcommand{\mvec}[1]{\vec{\mskip 0.5mu #1}\mskip 1.5mu}

\newcommand{\etac}{\eta_{\textit{\tiny C}}}
\newcommand{\etapt}{\eta_{\textit{\tiny PT}}}

\graphicspath{{plots/},{graphs/}}

\newcommand{\rev}[1]{#1}

\allowdisplaybreaks[2]

\title{Double parton distributions in the pion from lattice QCD}

\abstract{
We perform a lattice study of double parton distributions in the pion, using the relationship between their Mellin moments and pion matrix elements of two local currents.  A good statistical signal is obtained for almost all relevant Wick contractions.  We investigate correlations in the spatial distribution of two partons in the pion, as well as correlations involving the parton polarisation.  The patterns we observe depend significantly on the quark mass.  We investigate the assumption that double parton distributions approximately factorise into a convolution of single parton distributions.
}

\preprint{\vbox{
\hbox{DESY 20-098, CERN-TH-2020-086}}}

\author[a,b]{Gunnar S. Bali}
\author[a]{Luca Castagnini}
\author[c,d]{Markus Diehl}
\author[e]{Jonathan R. Gaunt}
\author[f]{Benjamin Gl{\"a}{\ss}le}
\author[a]{Andreas Sch{\"a}fer}
\author[a]{and Christian Zimmermann}

\affiliation[a]{Institute for Theoretical Physics, University of Regensburg, 93040 Regensburg, Germany}
\affiliation[b]{Department of Theoretical Physics, Tata Institute of Fundamental Research, Homi Bhabha Road, Mumbai 400005, India}
\affiliation[c]{Fachbereich Physik, University of Hamburg, 22761 Hamburg, Germany}
\affiliation[d]{Deutsches Elektronen-Synchroton DESY, 22607 Hamburg, Germany}
\affiliation[e]{CERN Theory Division, 1211 Geneva 23, Switzerland}
\affiliation[f]{Zentrum f\"ur Datenverarbeitung, Universit\"at T\"ubingen, W\"achterstr.\ 76, 72074 T\"ubingen, Germany}

\begin{document}

\maketitle

\section{Introduction}
\label{sec:intro}

Matrix elements of currents in a hadron offer a variety of ways to quantify and study hadron structure.  In particular, information about \emph{correlations} inside the hadron can be obtained from the matrix elements of two currents that are separated by a space-like distance.  Such matrix elements can be calculated in lattice QCD, and there has been considerable activity in this area over the years \cite{Barad:1984px,Barad:1985qd,Wilcox:1986ge,Wilcox:1986dk,Wilcox:1990zc,Chu:1990ps,Lissia:1991gv,Burkardt:1994pw,Alexandrou:2002nn,Alexandrou:2003qt,Alexandrou:2008ru}.  These studies address a broad range of physics questions, such as confinement \cite{Barad:1984px,Barad:1985qd}, the size of hadrons \cite{Wilcox:1986ge,Wilcox:1986dk,Wilcox:1990zc,Burkardt:1994pw}, density correlations \cite{Chu:1990ps}, comparison with quark models \cite{Lissia:1991gv}, or the non-spherical shape of hadrons with spin $1$ or larger \cite{Alexandrou:2002nn,Alexandrou:2003qt,Alexandrou:2008ru}.

We continued this line of investigation in a recent paper \cite{Bali:2018nde}. We performed a lattice computation of the matrix elements of two scalar, pseudoscalar, vector, or axial vector currents in the pion and compared our results with predictions of chiral perturbation theory.  For the first time, we computed all Wick contractions that contribute to these matrix elements, whilst earlier work had focused on the case in which the two currents are inserted on different quark lines between the hadron source and sink operators (see graph $C_1$ in \fig{\ref{fig:contractions}}).  We obtained signals with a good statistical accuracy for almost all contractions and were thus able to study their relative importance.  Our results were compared with different models in~\cite{Rinaldi:2020ybv,Courtoy:2020tkd}.

Extending our work in \cite{Bali:2018nde}, we will in the present paper use two-current matrix elements from the lattice to obtain information about double parton distributions (DPDs).  DPDs describe the correlated distribution of two partons inside a hadron and appear in the cross sections for double parton scattering, which occurs when there are two separate hard-scattering processes in a single hadron-hadron collision.  The study of this mechanism has a long history in collider physics, from early theoretical papers such as \cite{Landshoff:1978fq,Kirschner:1979im,Politzer:1980me,Paver:1982yp,Shelest:1982dg,Mekhfi:1983az,Sjostrand:1986ep} to the detailed investigation of QCD dynamics and factorisation that started about ten years ago \cite{Blok:2010ge,Diehl:2011tt,Gaunt:2011xd,Ryskin:2011kk, Blok:2011bu,Diehl:2011yj,Manohar:2012jr, Manohar:2012pe,Ryskin:2012qx,Gaunt:2012dd,Blok:2013bpa,Diehl:2017kgu}.  After early experimental studies \cite{Akesson:1986iv,Alitti:1991rd}, a multitude of double parton scattering processes has been measured at the Tevatron and the LHC, see \cite{Abe:1997xk,Abazov:2015nnn,Aaij:2016bqq,Aaboud:2018tiq,Sirunyan:2019zox} and references therein.  Some final states produced by double parton scattering are of particular interest because they are a background to search channels for new physics.  A prominent example are like-sign gauge boson pairs $W^+ W^+$ and $W^- W^-$ \cite{Kulesza:1999zh,Gaunt:2010pi,Sirunyan:2019zox,Ceccopieri:2017oqe, Cotogno:2018mfv,Cotogno:2020iio}, the decay of which can yield like-sign lepton pairs.  A wealth of further information about double parton scattering can be found in the monograph~\cite{Bartalini:2017jkk}.

Double parton distributions remain poorly known, and their extraction from experimental data is considerably more difficult than the extraction of single parton distributions (PDFs).  It is therefore important to have as much theoretical guidance  as possible about the properties and behaviour of DPDs.  Apart from approaches that focus on fulfilling theoretical constraints \cite{Gaunt:2009re,Golec-Biernat:2014bva,Golec-Biernat:2015aza,Diehl:2020xyg}, there exists a large number of model calculations for the DPDs of the nucleon \cite{Chang:2012nw,Rinaldi:2013vpa,Broniowski:2013xba, Rinaldi:2014ddl,Broniowski:2016trx,Kasemets:2016nio,Rinaldi:2016jvu, Rinaldi:2016mlk} and a smaller number for those of the pion \cite{Rinaldi:2018zng,Courtoy:2019cxq,Broniowski:2019rmu,Broniowski:2020jwk}.

A relation between the Mellin moments of DPDs and two-current matrix elements that can be computed on the lattice was written down in \cite{Diehl:2011tt,Diehl:2011yj}.  This generalises the relation between matrix elements of one current and the Mellin moments of PDFs, which has been extensively exploited in lattice studies, as reviewed for instance in \cite{Hagler:2009ni,Lin:2017snn,Lin:2020rut}.  Whilst knowledge of a few Mellin moments is insufficient for reconstructing the full DPDs, it allows one to investigate crucial features of these functions, such as their dependence on the distance between the two partons and on the parton polarisation.  In the present paper, we pursue this idea for the DPDs of the pion, focusing on their lowest Mellin moments.  We use the same lattice data as in our study \cite{Bali:2018nde}.  Corresponding work on the DPDs of the nucleon is in progress, and preliminary results have been presented in~\cite{Zimmermann:2019quf}.

This paper is organised as follows.  In \sect{\ref{sec:theory}}, we recapitulate some basics about DPDs and then elaborate on the relation between their Mellin moments and the two-current matrix elements we compute on the lattice.  This will in particular lead us to introduce the concept of skewed DPDs.  In \sect{\ref{sec:lattice}}, we describe the main elements of our lattice simulations (a full account is given in \cite{Bali:2018nde}) and investigate several lattice artefacts that are present in our data.  Our results for zero pion momentum are presented and discussed in \sect{\ref{sec:results}}.  In
\sect{\ref{sec:py-not-zero}}, we develop a parametrisation of the data for both zero and nonzero pion momenta, which will allow us to reconstruct the Mellin moments of pion DPDs, albeit in a model-dependent fashion.  Our main findings are summarised in \sect{\ref{sec:summary}}.

\section{Theory}
\label{sec:theory}

\subsection{Double parton distributions}
\label{sec:dps}

To begin with, we recall some basics about double parton distributions.  An extended introduction to the subject can be found in \cite{Diehl:2017wew}.

Factorisation for a double parton scattering process means that its cross section is given in terms of hard-scattering cross sections at parton level and double parton distributions for each of the colliding hadrons.  For pair production of colourless particles, such as $Z$, $W$ or Higgs bosons, this factorisation can be proven rigorously.  A DPD gives the joint probability for finding in a hadron two partons with longitudinal momentum fractions $x_1$ and $x_2$ at a transverse distance $\tvec{y}$ from each other.  The distributions for quarks and antiquarks are defined by operator matrix elements as
\begin{align}
\label{eq:dpd-def}
F_{a_1 a_2}(x_1,x_2,\tvec{y})
= 2p^+ \int dy^- \int \frac{dz^-_1}{2\pi}\, \frac{dz^-_2}{2\pi}\,
&      e^{i\ms ( x_1^{} z_1^- + x_2^{} z_2^-)\ms p^+}
\nonumber \\
& \times
  \bra{h(p)} \mathcal{O}_{a_1}(y,z_1)\, \mathcal{O}_{a_2}(0,z_2) \ket{h(p)}
\,.
\end{align}
We use light-cone coordinates $v^{\pm} = (v^0 \pm v^3) /\sqrt{2}$ and boldface letters for the transverse part $\tvec{v} = (v^1, v^2)$ for any four-vector $v^\mu$.  The definition \eqref{eq:dpd-def} refers to a reference frame in which the transverse hadron momentum is zero, $\tvec{p} = \tvec{0}$.  In a frame where the hadron moves fast in the positive $z$ direction, $x_1$ and $x_2$ can be interpreted as longitudinal momentum fractions.  \rev{The vectors $z_1^\mu$ and $z_2^\mu$  are lightlike with only $z_1^-$ and $z_2^-$ nonzero, whereas $y^\mu$ is spacelike with $y^+ = 0$.}
The hadron state is denoted by ${h(p)}$, and it is understood that an average over its polarisation is taken on the r.h.s.\ of \eqref{eq:dpd-def} if the hadron has nonzero spin.  Unless specified otherwise, the expressions of the present section hold both for a pion and for the nucleon (and in fact for any unpolarised hadron or nucleus).

The matrix element in \eqref{eq:dpd-def} involves the same twist-two operators that appear in the definition of ordinary PDFs.  For quarks, one has
\begin{align}
\label{eq:quark-ops}
\mathcal{O}_{a}(y,z)
&= \bar{q}\bigl( y - \half z \bigr)\, \Gamma_{a} \, q\bigl( y + \half z \bigr)
   \Big|_{z^+ = y^+_{} = 0,\, \tvec{z} = \tvec{0}}
\end{align}
with spin projections
\begin{align}
  \label{eq:quark-proj}
\Gamma_q & = \half \gamma^+ \,, &
\Gamma_{\Delta q} &= \half \gamma^+\gamma_5 \,, &
\Gamma_{\delta q}^j = \half i \sigma^{j+}_{} \gamma_5  \quad (j=1,2) \,.
\end{align}
The analogous expressions for antiquarks can e.g.\ be found in \cite[\sect{2.2}]{Diehl:2011yj}.  The form \eqref{eq:quark-ops} holds in light-cone gauge $A^+ = 0$, whereas in other gauges a Wilson line is to be inserted between the fields.  Since the two fields in \eqref{eq:quark-ops} have light-like separation from each other, their product requires renormalisation.  This results in a scale dependence of the two operators and of the DPD in \eqref{eq:dpd-def}, which we do not indicate for the sake of brevity.

Lorentz invariance implies that one can write
\begin{align} \label{eq:invar-dpds}
F_{q_1 q_2}(x_1,x_2, \tvec{y}) &= f_{q_1 q_2}(x_1,x_2, y^2) \,,
\nonumber \\
F_{\Delta q_1 \Delta q_2}(x_1,x_2, \tvec{y})
&= f_{\Delta q_1 \Delta q_2}(x_1,x_2, y^2) \,,
\nonumber \\
F_{\delta q_1 q_2}^{j_1}(x_1,x_2, \tvec{y})
&= \epsilon^{j_1 k} \tvec{y}^k\, m f_{\delta q_1 q_2}(x_1,x_2, y^2) \,,
\nonumber \\
F_{q_1 \delta q_2}^{j_2}(x_1,x_2, \tvec{y})
&= \epsilon^{j_2 k} \tvec{y}^k\, m f_{q_1 \delta q_2}(x_1,x_2, y^2) \,,
\nonumber \\
F_{\delta q_1 \delta q_2}^{j_1 j_2}(x_1,x_2, \tvec{y})
&= \delta^{j_1 j_2} f^{}_{\delta q_1 \delta q_2}(x_1,x_2, y^2)
   + \bigl( 2 \tvec{y}^{j_1} \tvec{y}^{j_2}
         - \delta^{j_1 j_2} \tvec{y}^2 \bigr)\ms
   m^2 f^{\ms t}_{\delta q_1 \delta q_2}(x_1,x_2, y^2)
\end{align}
with $y^2 = y^\mu y_\mu = - \tvec{y}^2$.  Due to parity invariance, one has $F_{q_1 \Delta q_2} = F_{\Delta q_1 q_2} = 0$, and time reversal invariance implies $F_{\delta q_1 \Delta q_2} = F_{\Delta q_1 \delta q_2} = 0$.
The hadron mass $m$ has been introduced on the r.h.s.\ of \eqref{eq:invar-dpds} so that all scalar functions $f$ have the dimension of an inverse area.
The operator $\mathcal{O}^j_{\delta q}$ is a vector whose direction gives the transverse quark spin direction, and $\epsilon^{j k}$ is the two-dimensional antisymmetric tensor with $\epsilon^{1 2} = +1$.  The density interpretation of the different distributions in \eqref{eq:invar-dpds} is then as follows:
\begin{itemize}
\item The unpolarised distribution $f_{q_1 q_2}$ gives the probability density to find two quarks with momentum fractions $x_1$ and $x_2$ at a transverse distance $\tvec{y}$, regardless of their polarisation.
\item $f_{\Delta q_1 \Delta q_2}$ is the density for finding two quarks with their longitudinal polarisations aligned minus the density for finding them with their longitudinal polarisations anti-aligned.
\item $f_{\delta q_1 \delta q_2}$ is the analogue of $f_{\Delta q_1 \Delta q_2}$ for transverse quark polarisations.
\item $f_{\delta q_1 \ms q_2}$ describes a correlation between the transverse polarisation of the quark $q_1$ and the distance $\tvec{y}$ of that quark from the unpolarised quark $q_2$.  In $f_{q_1 \delta q_2}$, the first quark is unpolarised and the second quark has transverse polarisation.
\item $f^t_{\delta q_1 \delta q_2}$ describes a correlation between the transverse polarisations of the two quarks and their transverse distance $\tvec{y}$.
\end{itemize}
Decompositions of the same form as \eqref{eq:invar-dpds} can be given for the cases where one replaces one or both of the quarks by an antiquark, with the same physical interpretation as given above for two quarks.

Note that the polarisation dependence of DPDs is not only interesting from the point of view of hadron structure, but can have measurable implications on double parton scattering, as was for instance shown in~\cite{Diehl:2011yj,Kasemets:2012pr,Cotogno:2018mfv,Cotogno:2020iio}.  Lattice calculations can give information about the strength of the different spin correlations we just discussed.

We note that cross sections for double parton scattering involve the product of two DPDs integrated over the interparton distance,
\begin{equation}
\int d^2\tvec{y} \;
    F_{a_1 a_2}(x_1, x_2, \tvec{y}) \, F_{b_1 b_2}(x_1', x_2', \tvec{y}) \,.
\end{equation}
The dependence of DPDs on $\tvec{y}$ can hence not be directly inferred from experimental observables.  If $\tvec{y}$ is small, one can use perturbation theory to compute $F_{a_1 a_2}$ in terms of PDFs and splitting functions \cite{Diehl:2011yj,Diehl:2019rdh}.  By contrast, for large distances the $\tvec{y}$ dependence is fully non-perturbative.  Lattice studies can give information about this dependence, whose knowledge is crucial for computing double parton scattering cross sections.

Both unpolarised and polarised DPDs can exhibit correlations in their dependence on $x_1$, $x_2$ and $\tvec{y}$.  We cannot address this aspect in our present study, because the  matrix elements we compute are related to the lowest Mellin moments of DPDs, i.e.\ their integrals over both $x_1$ and $x_2$.  In principle, one could investigate higher Mellin moments, i.e.\ integrals weighted with powers of $x_1$ and $x_2$.  This would require extending the set of currents in \eqref{eq:local-ops} to currents that involve covariant derivatives and is beyond the scope of the present work.

Phenomenological analyses often make the assumption that in unpolarised DPDs the two partons are independent of each other.  This gives the relation
\begin{align}
   \label{eq:dpd-fact}
F_{a_1 a_2}(x_1, x_2, \tvec{y})
& \overset{?}{=} \int d^2\tvec{b}\; f_{a_1}(x_1, \tvec{b} + \tvec{y})\,
                      f_{a_2}(x_2, \tvec{b}) \,,
\end{align}
where $f_{a}(x, \tvec{b})$ is an unpolarised impact parameter dependent single parton distribution.  The question mark above the equal sign in \eqref{eq:dpd-fact} indicates that this is a hypothesis.  Our lattice study allows us to test this indirectly in two different ways, as discussed in \sects{\ref{sec:fact-theory}}, \ref{sec:fact-A} and \ref{sec:fact-mellin}.

A related but different simplifying assumption is that unpolarised DPDs can be written as
\begin{align}
   \label{eq:dpd-pocket}
F_{a_1 a_2}(x_1, x_2, \tvec{y})
& \overset{?}{=} f_{a_1}(x_1)\, f_{a_2}(x_2) \, G(\tvec{y}) \,,
\end{align}
where $f_a(x)$ denotes a standard PDF and $G(\tvec{y})$ is a factor describing the dependence on the transverse parton distance.  This assumption leads to the so-called ``pocket formula'', which expresses double parton scattering cross sections in terms of the cross sections for each single scattering and a universal factor $\sigma_{\text{eff}}^{-1} = \int d^2 \tvec{y} \; [\ms G(\tvec{y}) \ms]^2$.  Whilst our study cannot address the factorisation between the $x_1$, $x_2$ and $\tvec{y}$ dependence assumed in \eqref{eq:dpd-pocket}, we can investigate the assumption that the $\tvec{y}$ dependence is the same for all parton combinations $(a_1, a_2)$ in a given hadron.  We will do this in \sect{\ref{sec:physical_combinations}}.


\subsection{Matrix elements of local currents}
\label{sec:tensor}

The matrix element \eqref{eq:dpd-def} involves fields at light-like distances and is hence not suitable for direct evaluation on a Euclidean lattice.  What we can study in Euclidean space-time are the matrix elements
\begin{align}
  \label{eq:mat-els}
M^{\mu_1 \cdots \mu_2 \cdots}_{q_1 q_2, i_1 i_2}(p,y)
&= \bra{h(p)} J^{\mu_1 \cdots}_{q_1, i_1}(y)\,
              J^{\mu_2 \cdots}_{q_2, i_2}(0) \ket{h(p)} \,,
\end{align}
where as in \eqref{eq:dpd-def} a polarisation average is understood if the hadron $h$ carries spin.  The local currents $J_{q, i}^{\mu \cdots}$ we will consider here are
\begin{align}
  \label{eq:local-ops}
J_{q, V}^\mu(y) &= \bar{q}(y) \ms \gamma^\mu\ms q(y) \,,
&
J_{q, A}^\mu(y) &= \bar{q}(y) \ms \gamma^\mu \gamma_5\, q(y) \,,
&
J_{q, T}^{\mu\nu}(y) &= \bar{q}(y) \ms \sigma^{\mu\nu} \ms q(y) \,.
\end{align}
For spacelike distances $y$, which we assume throughout this work, the two currents in \eqref{eq:mat-els} commute, so that one has
\begin{align}
   \label{eq:exchange-J}
M^{\mu_1 \cdots \mu_2 \cdots}_{q_1 q_2, i_1 i_2}(p,y)
   &= M^{\mu_2 \cdots \mu_1 \cdots}_{q_2 q_1, i_2 i_1}(p,-y) \,.
\end{align}
Together with the fact that the currents in \eqref{eq:local-ops} are Hermitian, it follows that the matrix elements \eqref{eq:mat-els} are real valued.

The currents transform under charge conjugation ($C$) and under the
combination of parity and time reversal ($PT$) as
\begin{align}
J_{q, i}^{\mu \cdots}(y)
 & \underset{C}{\to} \etac^i\, J_{q, i}^{\mu \cdots}(y) \,,
&
J_{q, i}^{\mu \cdots}(y)
 & \underset{PT}{\to} \etapt^i\, J_{q, i}^{\mu \cdots}(-y)
\end{align}
with sign factors
\begin{align}
  \label{eq:CPT-parity}
\etac^{i}
  &= +1  ~~~\text{for}~ i = A \,,
&
\etac^{i}
  &= -1  ~~~\text{for}~ i = V, T
\intertext{and}
\etapt^{i}
  &= +1  ~~~\text{for}~ i = V \,,
&
\etapt^{i}
  &= -1  ~~~\text{for}~ i = A, T \,.
\end{align}
The combination of a parity and time reversal transformation gives
\begin{align}
  \label{eq:time-reversal-constr}
M^{\mu_1 \cdots \mu_2 \cdots}_{q_1 q_2, i_1 i_2}(p,y)
  &= \etapt^{i_1}\, \etapt^{i_2}\,
     M^{\mu_1 \cdots \mu_2 \cdots}_{q_1 q_2, i_1 i_2}(p,-y)
\end{align}
and thus relates the matrix elements for $y$ and $-y$.


\paragraph{Symmetry relations for pion matrix elements.}
For pion matrix elements, one has additional relations due to charge conjugation and isospin invariance.  For $\etac^{i_1} \, \etac^{i_2} = 1$, which is the case for all current combinations considered in our work, one has
\begin{align}
M_{q_1 q_2}(y,p) \ms\big|_{\pi^+}
&= M_{q_1 q_2}(y,p) \ms\big|_{\pi^-} \,,
\end{align}
where we indicated for which hadron the matrix element is taken but for brevity omitted the Lorentz indices and the labels $i_1, i_2$ specifying the currents.  Still for $\etac^{i_1} \, \etac^{i_2} = 1$, one finds
\begin{align}
   \label{eq:pion-symm}
M_{u u}(y,p) \ms\big|_{\pi^c}
&= M_{d d}(y,p)  \ms\big|_{\pi^c} \,,
&
M_{u d}(y,p) \ms\big|_{\pi^c}
&= M_{d u}(y,p)  \ms\big|_{\pi^c}
\end{align}
for $c = +, -, 0$, as well as
\begin{align}
   \label{eq:val-sea-sum}
M_{u d}(y,p) \ms\big|_{\pi^+}
 + M_{u u}(y,p) \ms\big|_{\pi^+}
&= M_{u d}(y,p) \ms\big|_{\pi^0}
 + M_{u u}(y,p) \ms\big|_{\pi^0} \,.
\end{align}
A derivation of these relations can be found in \cite[\sect{2.1}]{Bali:2018nde}.


\paragraph{Tensor decomposition and extraction of twist-two functions.}

The matrix elements in \eqref{eq:mat-els} are related to the lowest Mellin moments of DPDs as
\begin{align}
  \label{eq:t2-mat-els}
\int_{-\infty}^{\infty} dy^-\, M_{q_1 q_2, V V}^{++}(p,y)
   \, \Big|_{y^+ = 0,\, \tvec{p} = \tvec{0}}
  &= 2 p^+\ms I^{}_{q_1 q_2}(y^2) \,,
\nonumber \\
\int_{-\infty}^{\infty} dy^-\ms M_{q_1 q_2, A A}^{++}(p,y)
   \, \Big|_{y^+ = 0,\, \tvec{p} = \tvec{0}}
  &= 2 p^+\ms I^{}_{\Delta q_1 \Delta q_2}(y^2) \,,
\nonumber \\
\int_{-\infty}^{\infty} dy^-\ms M_{q_1 q_2, T V}^{k_1 ++}(p,y)
   \, \Big|_{y^+ = 0,\, \tvec{p} = \tvec{0}}
  &= 2 p^+\ms \tvec{y}^{k_1} m I^{}_{\delta q_1 q_2}(y^2) \,,
\nonumber \\
\int_{-\infty}^{\infty} dy^-\ms M_{q_1 q_2, V T}^{+ k_2 +}(p,y)
   \, \Big|_{y^+ = 0,\, \tvec{p} = \tvec{0}}
  &= 2 p^+\ms \tvec{y}^{k_2} m I^{}_{q_1 \delta q_2}(y^2) \,,
\nonumber \\
\int_{-\infty}^{\infty} dy^-\ms M_{q_1 q_2, T T}^{k_1 + k_2 +}(p,y)
   \, \Big|_{y^+ = 0,\, \tvec{p} = \tvec{0}}
  &= 2 p^+\ms \bigl[\ms
      \delta^{k_1 k_2}\, I^{}_{\delta q_1 \delta q_2}(y^2)
\nonumber \\
  & \qquad \quad
     - \bigl( 2 \tvec{y}^{k_1} \tvec{y}^{k_2} - \delta^{k_1 k_2} \tvec{y}^2 \bigr)
     m^2 \ms I^t_{\delta q_1 \delta q_2}(y^2) \bigr]
\end{align}
with the Mellin moments given by
\begin{align}
  \label{eq:mellin-mom-def}
I_{a_1 a_2}(y^2)
&= \int_{-1}^{1} dx_1^{} \int_{-1}^{1} dx_2^{} \; f_{a_1 a_2}(x_1,x_2, y^2)
\nonumber \\
&= \int_0^1 dx_1^{} \int_0^1 dx_2^{} \, \Bigl[ f_{a_1 a_2}(x_1,x_2, y^2)
  + \etac^{i_1} f_{\bar{a}_1 a_2}(x_1,x_2, y^2)
\nonumber \\
& \qquad \qquad \qquad \qquad
  + \etac^{i_2}\, f_{a_1 \bar{a}_2}(x_1,x_2, y^2)
  + \etac^{i_1}\, \etac^{i_2}\,
            f_{\bar{a}_1 \bar{a}_2}(x_1,x_2, y^2) \Bigr] \,.
\end{align}
Here $i_1$ and $i_2$ refer to the currents in the matrix elements on the l.h.s.\ of \eqref{eq:t2-mat-els}.  An analogous relation holds between $I^t$ and the lowest moment of $f^t$.  The relations \eqref{eq:t2-mat-els} extend the well-known connection between the Mellin moments of PDFs and the matrix elements of a single local current to the case of two partons.

In analogy to the case of PDFs, the matrix element \eqref{eq:dpd-def} defining a DPD has support for both positive and negative $x_1$ and $x_2$, with positive $x_i$ corresponding to a parton $a_i$ and negative $x_i$ to its antiparton $\bar{a}_i$.  On the r.h.s.\ of \eqref{eq:mellin-mom-def}, we have limited the integration region to positive momentum fractions.  Note that if $a_1$ and $a_2$ are quarks and if $i=V$ or $T$ (but not $A$), then the quark-antiquark distributions on the r.h.s.\ enter with a minus sign.  This is of special importance for distributions in a pion, whose valence Fock state consists of a quark and an antiquark.
Relations analogous to \eqref{eq:t2-mat-els} exist for higher Mellin moments in $x_1$ and $x_2$ and involve local currents with covariant derivatives \cite{Diehl:2011yj}, as is the case for PDFs.

Contrary to $\Gamma_{\delta q}^{j}$ in \eqref{eq:quark-proj}, the tensor current $J_{q,T}^{\mu\nu}$ in \eqref{eq:local-ops} is defined without $\gamma_5$.  As a consequence, the vector indices $k_1$ and $k_2$ in \eqref{eq:t2-mat-els} do \emph{not} give the transverse quark spin direction but the transverse quark spin direction rotated by $+90^\circ$ in the $x-y$ plane.  This follows from the relation $i \sigma^{j+} \gamma_5 = \epsilon^{j k} \sigma^{k+}$.

The relations \eqref{eq:t2-mat-els} still refer to Minkowski space, because they involve plus-components.  To make contact with matrix elements evaluated in Euclidean space, we decompose the matrix elements \eqref{eq:mat-els}  in terms of basis tensors and of Lorentz invariant functions $A$, $B$, $C$, $D$, $E$ that depend on $y^2 = y^\mu y_\mu$ and $py = p^\mu y_\mu$.
We write
\begin{align}
  \label{eq:tensor-decomp}
\tfrac{1}{2} \bigl[ M^{\mu\nu}_{q_1 q_2, V V}(p,y)
                  + M^{\nu\mu}_{q_1 q_2, V V}(p,y) \bigr]
 & = t_{V V,A}^{\mu\nu}\, A_{q_1 q_2}^{}
   + t_{V V,B}^{\mu\nu}\, m^2\ms B_{q_1 q_2}^{}
   + t_{V V,C}^{\mu\nu}\, m^4\ms C_{q_1 q_2}^{}
\nonumber \\[0.1em]
 & \quad
   + t_{V V,D}^{\mu\nu}\, m^2\ms D_{q_1 q_2}^{} \,,
\nonumber \\[0.1em]
\operatorname{T} M^{\mu\nu\rho}_{q_1 q_2, T V}(p,y)
 & = u_{T V,A}^{\mu\nu\rho}\, m\ms A_{\delta q_1 q_2}^{}
   + u_{T V,B}^{\mu\nu\rho}\,  m^3\ms B_{\delta q_1 q_2}^{} \,,
\nonumber \\[0.3em]
\tfrac{1}{2} \, \bigl[ M^{\mu\nu\rho\sigma}_{q_1 q_2, T T}(p,y)
     + M^{\rho\sigma\mu\nu}_{q_1 q_2, T T}(p,y) \bigr]
 &= u_{T T,A}^{\mu\nu\rho\sigma}\, A_{\delta q_1 \delta q_2}^{}
    + u_{T T,B}^{\mu\nu\rho\sigma}\, m^2\ms B_{\delta q_1 \delta q_2}^{}
    + u_{T T,C}^{\mu\nu\rho\sigma}\, m^2\ms C_{\delta q_1 \delta q_2}^{}
\nonumber \\[0.3em]
 & \quad
    + u_{T T,D}^{\mu\nu\rho\sigma}\, m^4\ms D_{\delta q_1 \delta q_2}^{}
    + u_{T T,E}^{\mu\nu\rho\sigma}\, m^2\ms E_{\delta q_1 \delta q_2}^{} \,.
\end{align}
For the operator combination $T V$, we subtract trace terms according to
\begin{align}
  \label{eq:trace-subtractions}
\operatorname{T} u^{\mu\nu\rho} &= u^{\mu\nu\rho}
  + \tfrac{1}{3}\ms
      \bigl( g^{\mu\rho} u^{\nu\alpha}{}_{\!\alpha}
           - g^{\nu\rho} u^{\mu\alpha}{}_{\!\alpha} \bigr) \,,
\end{align}
where it is understood that $u^{\mu\nu\rho}$ is antisymmetric in $\mu$ and $\nu$.
The decomposition for $M_{q_1 q_2, A A}$ has the same form as the one for $M_{q_1 q_2, V V}$, involving the same basis tensors but different invariant functions $A_{\Delta q_1 \Delta q_2}$, \ldots, $D_{\Delta q_1 \Delta q_2}$.  The decomposition for $M_{q_1 q_2, V T}$ is like the one for $M_{q_1 q_2, T V}$ with an appropriate change in the role of the Lorentz indices.  In the following, we will not discuss the combination $V T$ any further, because it can be traded for $T V$ using the relation \eqref{eq:exchange-J}.
The basis tensors are chosen as
\begin{align}
  \label{basis-tensors}
t_{V V,A}^{\mu\nu} &= 2 p^\mu p^\nu - \half\ms g^{\mu\nu} p^2 \,,
\nonumber \\
t_{V V,B}^{\mu\nu} &= p^\mu y^\nu + p^\nu y^\mu - \half\ms g^{\mu\nu} py \,,
\nonumber \\
t_{V V,C}^{\mu\nu} &= 2 y^\mu y^\nu - \half\ms g^{\mu\nu} y^2 \,,
\nonumber \\
t_{V V,D}^{\mu\nu} &= g^{\mu\nu} \,,
\nonumber \\[0.3em]
u_{T V,A}^{\mu\nu\rho} &= 2 (y^\mu p^\nu - p^\mu y^\nu)\ms p^\rho
      + \tfrac{2}{3}\ms (g^{\mu\rho} y^\nu - g^{\nu\rho} y^\mu)\ms p^2
      - \tfrac{2}{3}\ms (g^{\mu\rho} p^\nu - g^{\nu\rho} p^\mu)\ms py \,,
\nonumber \\
u_{T V,B}^{\mu\nu\rho} &= 2 (y^\mu p^\nu - p^\mu y^\nu)\ms y^\rho
      + \tfrac{2}{3}\ms  (g^{\mu\rho} y^\nu - g^{\nu\rho} y^\mu)\ms py
      - \tfrac{2}{3}\ms  (g^{\mu\rho} p^\nu - g^{\nu\rho} p^\mu)\ms y^2 \,,
\nonumber \\[0.3em]
u_{T T,A}^{\mu\nu\rho\sigma} &= - 2\ms \bigl( g^{\mu\rho} p^\nu p^\sigma
    - g^{\mu\sigma} p^\nu p^\rho)
    + \tfrac{1}{2}\ms
      (g^{\mu\rho} g^{\nu\sigma} - g^{\mu\sigma} g^{\nu\rho})\ms p^2
    - \{ \mu\leftrightarrow\nu \} \,,
\nonumber \\
u_{T T,B}^{\mu\nu\rho\sigma} &= - y^2\, u_{T T,A}^{\mu\nu\rho\sigma}
    - 4\ms (y^\mu p^\nu - p^\mu y^\nu)\ms
           (y^\rho p^\sigma - p^\rho y^\sigma)
    + \tfrac{2}{3}\ms
      (g^{\mu\rho} g^{\nu\sigma} - g^{\mu\sigma} g^{\nu\rho})\ms
      \bigl[ p^2 y^2 - (py)^2 \bigr] \,,
\nonumber \\
u_{T T,C}^{\mu\nu\rho\sigma} &= - (g^{\mu\rho} p^\nu y^\sigma
    - g^{\mu\sigma} p^\nu y^\rho + g^{\mu\rho} y^\nu p^\sigma
    - g^{\mu\sigma} y^\nu p^\rho)
    + \tfrac{1}{2}\ms
      (g^{\mu\rho} g^{\nu\sigma} - g^{\mu\sigma} g^{\nu\rho})\ms py
    - \{ \mu\leftrightarrow\nu \} \,,
\nonumber \\
u_{T T,D}^{\mu\nu\rho\sigma} &= - 2\ms \bigl( g^{\mu\rho} y^\nu y^\sigma
    - g^{\mu\sigma} y^\nu y^\rho)
    +  \tfrac{1}{2}\ms
      (g^{\mu\rho} g^{\nu\sigma} - g^{\mu\sigma} g^{\nu\rho})\ms y^2
    - \{ \mu\leftrightarrow\nu \} \,,
\nonumber \\
u_{T T,E}^{\mu\nu\rho\sigma} &=  g^{\mu\rho} g^{\nu\sigma}
          - g^{\mu\sigma} g^{\nu\rho} \,.
\end{align}
The tensor components related to twist-two matrix elements can be identified from the l.h.s.\ of \eqref{eq:t2-mat-els}, taking into account that  $y^+ = 0$ and  $\tvec{p} = \tvec{0}$ in that equation.  For the basis tensors, a nonzero plus-component requires the vector $p$ on the r.h.s.\ of \eqref{basis-tensors}, whilst a nonzero transverse component requires the vector $y$ or the metric tensor.  One thus finds that the invariant functions corresponding to operators of twist two are $A_{q_1 q_2}$, $A_{\Delta q_1 \Delta q_2}$, $A_{\delta q_1 q_2}$, $A_{\delta q_1 \delta q_2}$ and $B_{\delta q_1 \delta q_2}$.  We will call them ``twist-two functions'' in the remainder of this work.   All of them are even functions of $py$ due to the symmetry relation \eqref{eq:time-reversal-constr}.

One can project out the invariant functions by multiplying the matrix
elements with suitable linear combinations of basis tensors.  For the twist-two functions, the relevant projections read
\begin{align}
  \label{eq:projector-meth}
A_{q_1 q_2} &= \frac{1}{8 N^2}\, \Bigr\{
  3 (y^2)^2\, t_{V V,A}^{\mu\nu} - 6 y^2 py\; t_{V V,B}^{\mu\nu}
           + \bigl[ p^2 y^2 + 2 (py)^2 \bigr]\ms t_{V V,C}^{\mu\nu}
  \Bigr\} \bigl[ M_{q_1 q_2, V V} \bigr]_{\mu\nu} \,,
\nonumber \\[0.2em]
m \ms A_{\delta q_1 q_2} &= \frac{3}{16 N^2}\, \Bigr\{
    y^2\, u_{T V,A}^{\mu\nu\rho} - py\, u_{T V,B}^{\mu\nu\rho}
   \Bigr\} \operatorname{T} \bigl[ M_{q_1 q_2, T V} \bigr]_{\mu\nu\rho} \,,
\nonumber \\[0.2em]
A_{\delta q_1 \delta q_2} &= \frac{1}{64 N^2}\, \Bigr\{
    3 (y^2)^2\, u_{T T,A}^{\mu\nu\rho\sigma}
  - 6 y^2 py\; u_{T T,C}^{\mu\nu\rho\sigma}
   + \bigl[ p^2 y^2 + 2 (py)^2 \bigr]\ms u_{T T,D}^{\mu\nu\rho\sigma}
  \Bigr\} \, \bigl[ M_{q_1 q_2, T T} \bigr]_{\mu\nu\rho\sigma} \,,
\nonumber \\[0.2em]
m^2 B_{\delta q_1 \delta q_2} &= \frac{1}{64 N^2} \Bigr\{
     3 \ms u_{T T,B}^{\mu\nu\rho\sigma}
   + 6 \ms py\; u_{T T,C}^{\mu\nu\rho\sigma}
   - 3 p^2\, u_{T T,D}^{\mu\nu\rho\sigma}
  \Bigr\} \, \bigl[ M_{q_1 q_2, T T} \bigr]_{\mu\nu\rho\sigma}
\end{align}
with a normalisation factor
\begin{align}
N = p^2 y^2 - (py)^2 \,.
\end{align}
For spacelike $y^\mu$, which we are interested in, one has $N < 0$, so that the projections are always well defined.

Using \eqref{eq:t2-mat-els} and \eqref{eq:tensor-decomp}, one can derive the
relation between Mellin moments of DPDs and integrals of twist-two functions over $py$:
\begin{align}
  \label{eq:mellin-inv-fct}
I_{a_1 a_2}(y^2)
  &= \int_{-\infty}^{\infty} d(py)\, A_{a_1 a_2}(py, y^2) \,,
\nonumber \\
I^t_{\delta q_1 \delta q_2}(y^2)
  &= \int_{-\infty}^{\infty} d(py)\, B_{\delta q_1 \delta q_2}(py, y^2) \,,
\end{align}
where in the first line we have all combinations of $(a_1, a_2)$ that appear on the r.h.s.\ of \eqref{eq:t2-mat-els}.

The matrix elements \eqref{eq:mat-els} can be evaluated in Euclidean space-time at $y^0$, i.e.\ with the two current operators taken at equal Euclidean time.  This entails the important restriction
\begin{align}
\label{eq:euclid-restr}
(py)^2 = (\mvec{p} \mvec{y})^2 \le \mvec{p}^2 \, \mvec{y}^2 \,,
\end{align}
where $\mvec{v} = (v^1, v^2, v^3)$ denotes the spatial components of a four-vector $v^\mu$.   Since the range of accessible hadron momenta $\vec{p}$ in a lattice calculation is finite, the range of the variable $py$ is limited, and one cannot directly evaluate the integrals in \eqref{eq:mellin-inv-fct}.  In addition, one needs data for nonzero hadron momentum $\mvec{p}$ to access even a finite range in $py$.

We note that the restriction \eqref{eq:euclid-restr} also applies if one computes the Mellin moments of transverse-momentum dependent single parton distributions (TMDs) on the lattice \cite{Hagler:2009mb,Musch:2010ka,Yoon:2017qzo}.  In that case, $y^\mu$ is the distance between the quark and the antiquark field in the matrix elements that define the distributions.  The same holds for lattice studies of single parton distributions in $x$ space.  There has been an enormous amount of activity in this area in recent years; we can only cite a few papers here \cite{Ji:2013dva,Ji:2014gla,Radyushkin:2017cyf,Orginos:2017kos,Ji:2018hvs,Braun:2018brg,Alexandrou:2019lfo,Ebert:2019okf,Ji:2019ewn} and refer to the recent reviews \cite{Cichy:2018mum,Ji:2020ect} for an extended bibliography.


\subsection{Skewed double parton distributions}
\label{sec:skewed-dpds}

Together with the restriction \eqref{eq:euclid-restr}, the necessity to perform an integral over all $py$ in \eqref{eq:mellin-inv-fct} presents a significant complication for relating matrix elements calculated on a Euclidean lattice with the Mellin moments of DPDs.  This prompts us to extend the theoretical framework in such a way that we can discuss the physical meaning of the twist-two functions $A_{a_1 a_2}$ and $B_{\delta q_1 \delta q_2}$ at a given value of $py$.

\begin{figure}
\begin{center}
\includegraphics[width=0.48\textwidth]{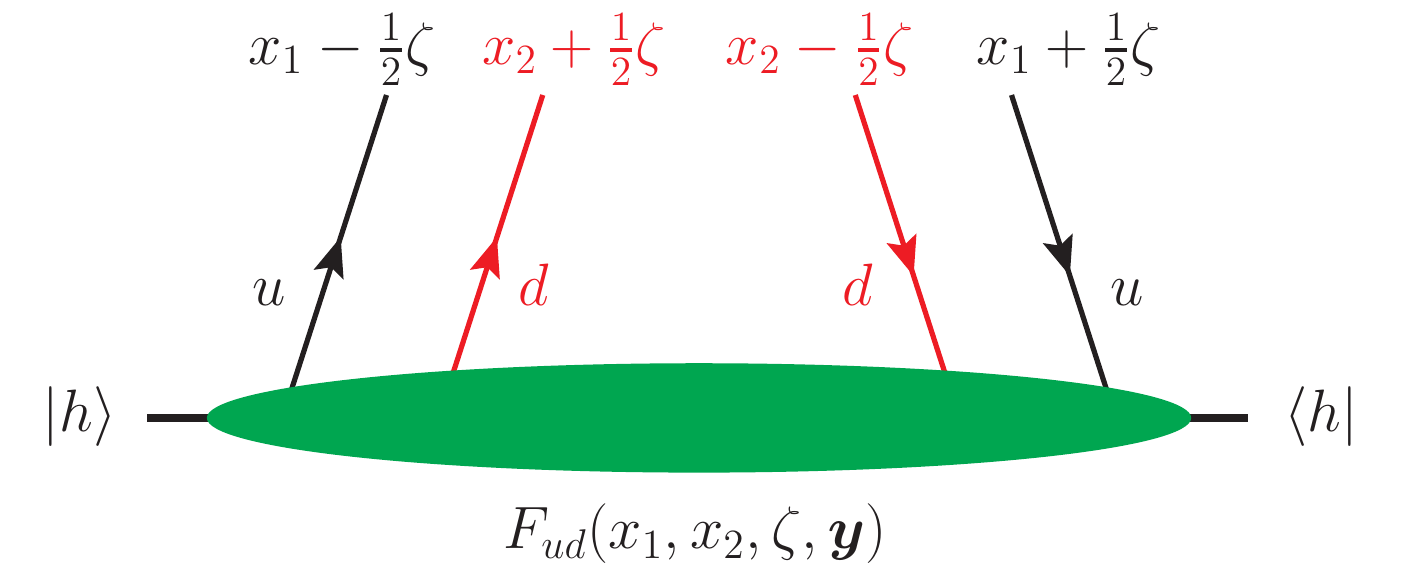}
\end{center}
\caption{\label{fig:zeta-distrib} Graphical representation of a skewed DPD for quark flavours $u$ and $d$ in the hadron $h$.  The configuration shown is for the case where all momentum fractions given at the top of the graph are positive.}
\end{figure}

To this end, we introduce \emph{skewed} double parton distributions\footnote{The term ``skewed'' refers to the parton momenta here, whilst the hadron momentum is the same in the bra and ket vector of \protect\eqref{eq:dpd-skew-def}.  This is different from ``skewed parton distributions'', now commonly called ``generalised parton distributions'', which involve two instead of four parton fields, such that there is an asymmetry both in the parton and in the hadron momenta.}
\begin{align}
\label{eq:dpd-skew-def}
F_{a_1 a_2}(x_1,x_2,\zeta, \tvec{y})
= 2p^+ \int dy^- e^{-i \zeta y^- p^+}
 &  \int \frac{dz^-_1}{2\pi}\, \frac{dz^-_2}{2\pi}\,
          e^{i\ms ( x_1^{} z_1^- + x_2^{} z_2^-)\ms p^+}
\nonumber \\
& \quad \times
    \bra{h(p)} \mathcal{O}_{a_1}(y,z_1)\, \mathcal{O}_{a_2}(0,z_2)
    \ket{h(p)} \,.
\end{align}
Compared with the definition \eqref{eq:dpd-def} of ordinary DPDs, we have an additional exponential $e^{- i \zeta p^+ y^-}$ here.  As a consequence, the partons created or annihilated by the fields $\bar{q}$ and $q$ in $\mathcal{O}_{a_1}$ and $\mathcal{O}_{a_2}$ have different longitudinal momentum fractions.
A sketch is given in figure~\ref{fig:zeta-distrib} for $(a_1, a_2) = (u, d)$ and the case where $x_1 - \half\zeta$, $x_1 + \half\zeta$, $x_2 - \half\zeta$ and $x_2 + \half\zeta$ are all positive.  If $x_1 - \half\zeta$ becomes negative, the $u$ quark in the wave function of $\ket{h}$ becomes an antiquark~$\bar{u}$ with momentum fraction $- x_1 + \half\zeta$ in the wave function of $\bra{h}$.  Corresponding statements hold for $x_1 + \half\zeta$, $x_2 - \half\zeta$ and $x_2 + \half\zeta$.

\begin{figure*}
\begin{center}
\includegraphics[width=0.99\textwidth]{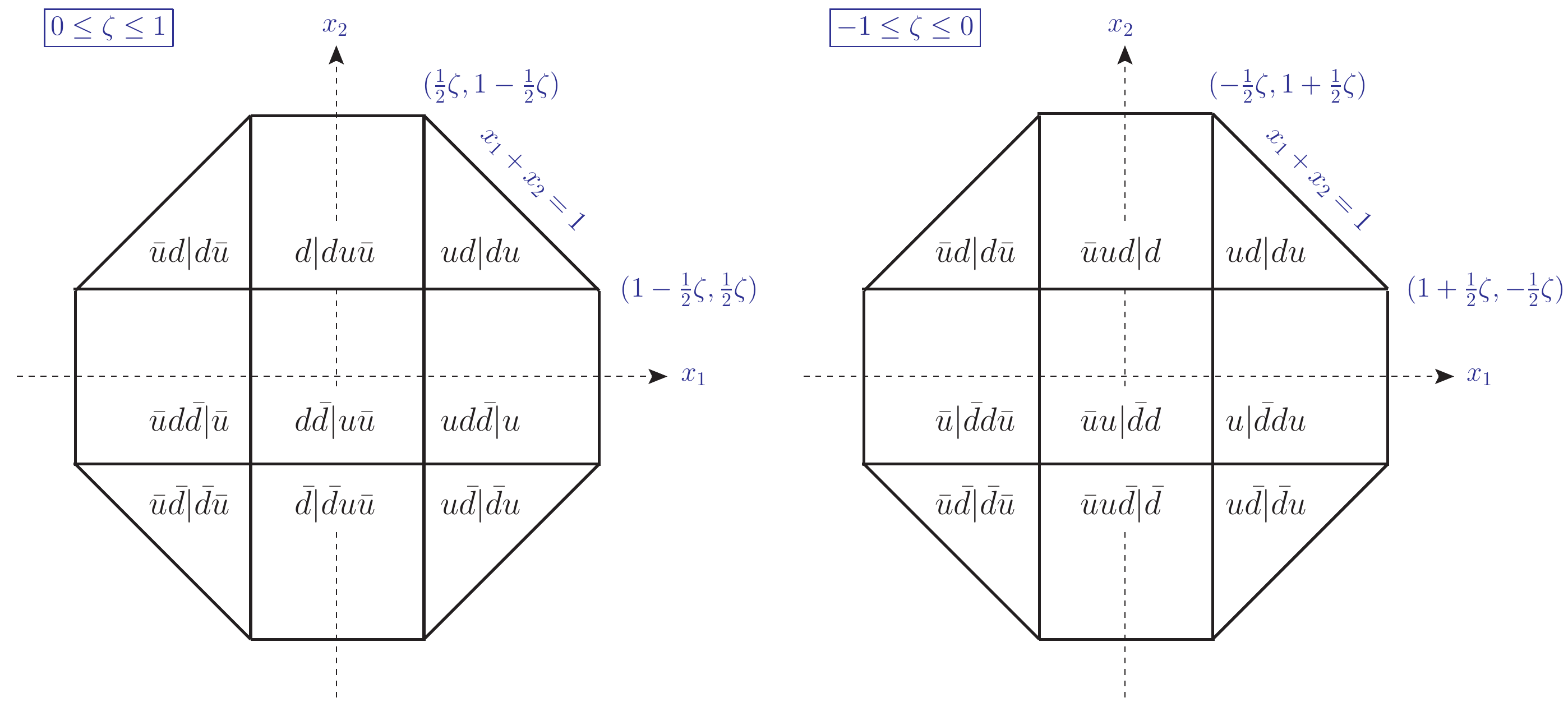}
\end{center}
\caption{\label{fig:dpd-support} Support region of the distribution
  $F_{u d}(x_1,x_2,\zeta, \tvec{y})$ in the momentum fraction arguments.  The notation $d \ms| d u \bar{u}$ means that one has one $d$ quark in the wave function of $\ket{h}$ and $d u \bar{u}$ in the wave function of $\bra{h}$.  In both panels, the triangle for the region $u d|d u$ has the corners $\bigl(\ms \half|\zeta|, \half|\zeta| \ms\bigr)$, $\bigl(\ms \half|\zeta|, 1-\half|\zeta| \ms\bigr)$ and $\bigl(\ms 1-\half|\zeta|, \half|\zeta| \ms\bigr)$.  Notice that the parton configuration in each of the four triangles is the same for positive and negative $\zeta$, whereas the configuration in each of the squares is different.}
\end{figure*}

For nonzero $\zeta$, the distributions \eqref{eq:dpd-skew-def} do not appear in cross sections for double parton scattering, but they may be regarded as a rather straightforward extension of the DPD concept.  Let us take a closer look at some of their properties.
The support region of the matrix element \eqref{eq:dpd-skew-def} in the momentum fraction arguments is the same as if all four parton fields were at the same transverse position. In that case, we would have a collinear twist-four distribution.  The support properties of these distributions were derived in \cite{Jaffe:1983hp}, and the argument given there does not depend on the transverse position arguments of the parton fields.  The result given in \cite{Jaffe:1983hp} is equivalent to the interpretation of $x_1 - \half\zeta$, $x_1 + \half\zeta$, $x_2 - \half\zeta$ and $x_2 + \half\zeta$ as positive or negative momentum fractions, as described in the previous paragraph.
For nonzero $\zeta$ there are hence different regions, in which one has
either 1, 2 or 3 partons in the wave function of $\ket{h}$.  With the
constraints that the partons in the wave function of $\ket{h}$ must carry
the same total longitudinal momentum as those in the wave function of $\bra{h}$, and that this cannot be larger than the longitudinal hadron momentum, one obtains the constraint
\begin{equation}
  \label{eq:zeta-range}
-1 \le \zeta \le 1
\end{equation}
and the support region for $(x_1, x_2)$ shown in figure~\ref{fig:dpd-support}.  For $\zeta=0$ this region becomes a square with corners $(0, \pm 1)$ and $(\pm 1, 0)$, whereas for $\zeta=\pm 1$ it becomes a square with corners $(\pm \half, \pm \half)$.

Using $PT$ symmetry, one finds that
\begin{align}
  \label{eq:reverse-zeta}
F_{a_1 a_2}(x_1,x_2,\zeta, \tvec{y})
&= \etapt^{i_1} \, \etapt^{i_2} \, F_{a_1 a_2}(x_1,x_2,-\zeta, -\tvec{y}) \,,
\end{align}
where $\etapt^i = +1$ for an unpolarised parton and $\etapt^i = -1$ for a polarised one.
The skewed DPDs can be decomposed in terms of scalar distributions as in \eqref{eq:invar-dpds}, with the distributions on both sides depending additionally on $\zeta$.  The symmetry property \eqref{eq:reverse-zeta} then implies
\begin{align}
  \label{eq:even-in-zeta}
f_{a_1 a_2}(x_1,x_2,\zeta, y^2)
&= f_{a_1 a_2}(x_1,x_2,-\zeta, y^2)
\end{align}
and an analogous relation for $f^t$.


\paragraph{Mellin moments.}

We define the lowest Mellin moments of skewed DPDs as
\begin{align}
  \label{eq:skewed-mellin-mom-def}
I_{a_1 a_2}(y^2, \zeta)
&= \int_{-1}^{1} dx_1^{} \int_{-1}^{1} dx_2^{} \;
   f_{a_1 a_2}(x_1,x_2,\zeta, y^2)
\end{align}
and likewise for $f^t$, where the integration region in $x_1, x_2$ follows from
figure~\ref{fig:dpd-support}.  The moments are nonzero for $\zeta$ in the interval $[-1,1]$.   The generalisation of \eqref{eq:mellin-inv-fct} to nonzero $\zeta$ reads
\begin{align}
\label{eq:skewed-mellin-inv-fct}
I_{a_1 a_2}(y^2, \zeta)
&= \int_{-\infty}^{\infty} d(py)\, e^{-i\zeta py}\, A_{a_1 a_2}(y^2, py) \,,
\end{align}
which can readily be inverted for the function $A_{a_1 a_2}(y^2, py)$.  In particular, one finds
\begin{align}
\label{eq:py-zero-fct}
A_{a_1 a_2}(y^2, py=0) = \frac{1}{\pi}
  \int_{0}^1 d\zeta \, I_{a_1 a_2}(y^2, \zeta) \,,
\end{align}
where we have used the symmetry relation \eqref{eq:even-in-zeta} to reduce the integration region to positive $\zeta$.  Rather than the Mellin moment of a DPD, a twist-two function at $py = 0$ is thus the average of the Mellin moment of a skewed DPD over the skewness parameter $\zeta$.

Quantities that characterise the $\zeta$ dependence of $I_{a_1 a_2}(y^2, \zeta)$  are the even moments in $\zeta$,
\begin{align}
  \label{eq:zeta-moments}
\langle \zeta^{2 m} \rangle_{a_1 a_2}^{}(y^2) &= \frac{%
    \int_{-1}^1 d\zeta \; \zeta^{2 m}\, I_{a_1 a_2}(y^2, \zeta)}{%
    \int_{-1}^1 d\zeta \, I_{a_1 a_2}(y^2, \zeta)}
 = \Biggl[\ms \frac{(-1)^m}{A_{a_1 a_2}(y^2, py)}
      \frac{\partial^{2 m} A_{a_1 a_2}(y^2, py)}{(\partial\ms py)^{2 m}}
   \ms\Biggr]_{py=0} \,.
\end{align}
Odd moments $\langle \zeta^{2 m + 1} \rangle$ are zero because of the symmetry \eqref{eq:even-in-zeta}.
To compute the moments $\langle \zeta^{2 m} \rangle$, one needs $A_{a_1 a_2}(y^2, py)$ in the vicinity of $py = 0$.  According to \eqref{eq:euclid-restr}, this can be evaluated from Euclidean data with nonzero hadron momentum $\mvec{p}$.

Relations analogous to \eqref{eq:skewed-mellin-inv-fct} to \eqref{eq:zeta-moments} can be written down for $I^t_{\delta q_1 \delta q_2}$ and $B_{\delta q_1 \delta q_2}^{}$ in the place of $I_{a_1 a_2}$ and $A_{a_1 a_2}$.


\subsection{Factorisation hypotheses}
\label{sec:fact-theory}

We now discuss how the factorisation hypothesis \eqref{eq:dpd-fact} for DPDs can be formulated at the level of Mellin moments and twist-two functions.  At this point, we specialise to the case where the hadron $h$ is a $\pi^+$.  This avoids complications due to the proton spin, which are discussed in \cite[\sect{4.3.1}]{Diehl:2011yj}.

Let us take the lowest Mellin moment in $x_1$ and $x_2$ of \eqref{eq:dpd-fact}.  The Mellin moment of an unpolarised impact parameter dependent parton distribution is
\begin{align}
\int_{-1}^1 dx\, f_q(x, \tvec{b}) &= \int \frac{d^2 \tvec{\Delta}}{(2\pi)^2}\;
   e^{-i \tvec{b} \tvec{\Delta}}\, F_{q, V}(-\tvec{\Delta}^2) \,,
\end{align}
where $F_{q, V}(t)$ is the form factor of the vector current
\begin{align}
\bra{\pi^+(p')} J_{q, V}^\mu(0) \ket{\pi^+(p)}
&= (p + p')^\mu \, F_{q, V}(t)
&
\text{with }
t &= (p - p')^2 \,.
\end{align}
We then obtain from \eqref{eq:dpd-fact}
\begin{align}
   \label{eq:mellin-fact}
I_{u d}(-\tvec{y}^2) & \overset{?}{=} \int \frac{d^2 \tvec{\Delta}}{(2\pi)^2}\;
   e^{-i \tvec{y} \tvec{\Delta}}\,
   F_{u,V}(- \tvec{\Delta}^2) \, F_{d,V}(- \tvec{\Delta}^2) \,.
\end{align}
We note that thanks to isospin invariance, one has $F_{u,V} = - F_{d,V}$.  As this is not essential in the present context, we will not use it here.

Since one cannot directly determine $I_{u d}(-\tvec{y}^2)$ from Euclidean correlation functions, one cannot directly test \eqref{eq:mellin-fact} with lattice data.  We therefore derive an analogous relation for the twist-two function $A_{u d}(y^2, py)$ at $py=0$.

\begin{figure}
\begin{center}
\subfigure[\label{fig:dpd-approx-ud}]{
\includegraphics[width=0.48\textwidth,trim=16 0 16 0,clip]{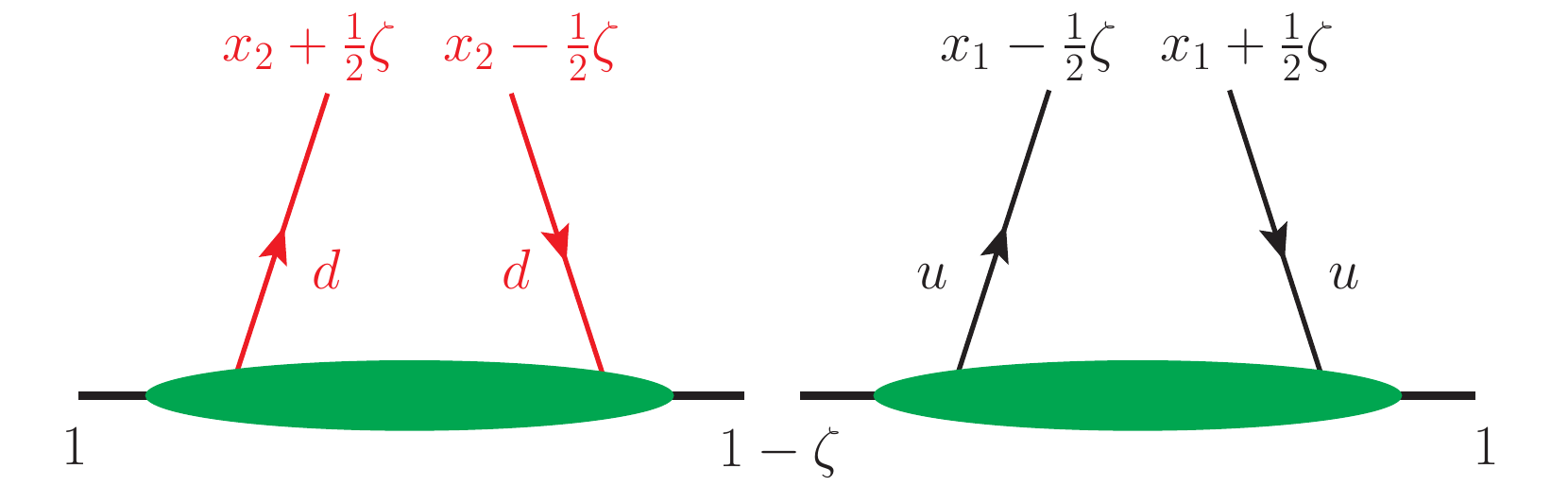}}
\hfill
\subfigure[\label{fig:dpd-approx-du}]{
\includegraphics[width=0.48\textwidth,trim=16 0 16 0,clip]{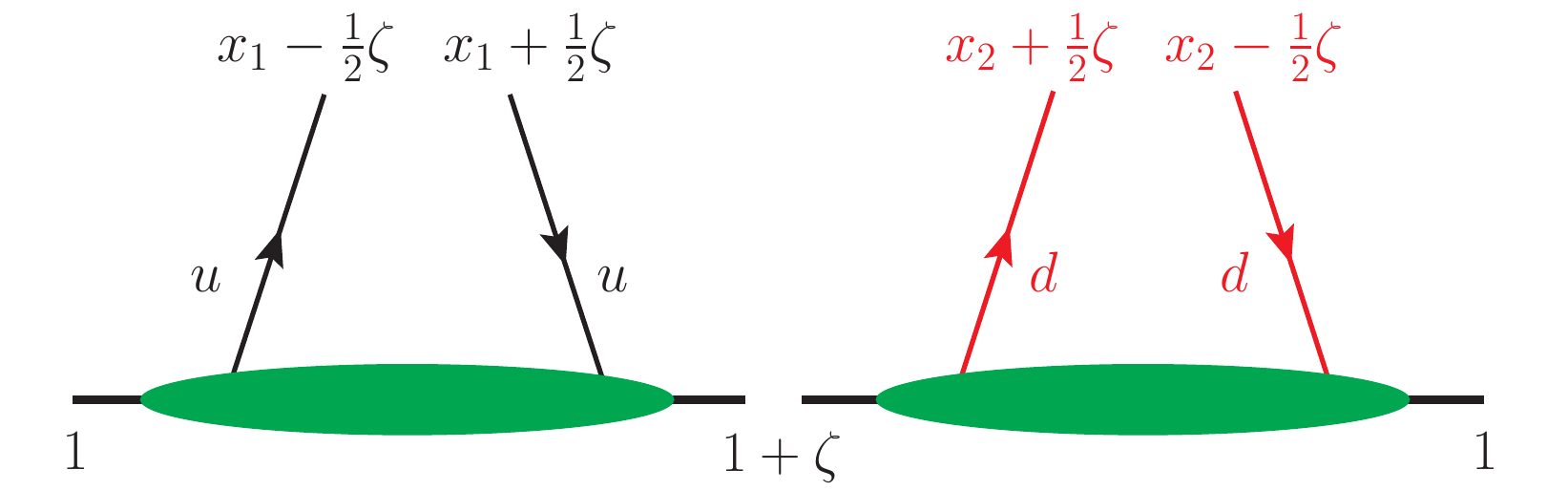}}
\end{center}
\caption{\label{fig:dpd-approx} (a): Pictorial representation of the r.h.s.\ of the factorisation hypothesis \protect\eqref{eq:dpd-skew-fact}.  This is obtained by inserting a full set of intermediate states between the operators in the matrix element $\bra{h} \mathcal{O}_u \ms\mathcal{O}_d \ket{h}$ and then retaining only the ground state.  (b): The representation obtained when inserting the full set of states after reordering the operators to $\bra{h} \mathcal{O}_d \ms\mathcal{O}_u \ket{h}$.  All momentum fractions refer to the hadron $h$ in the matrix element.  We use form (a) for $\zeta \ge 0$ and form~(b) for $\zeta < 0$.}
\end{figure}

We recall from \cite{Diehl:2011yj} that \eqref{eq:dpd-fact} can be obtained by inserting a complete set of intermediate states between the operators $\mathcal{O}_{a_1}(y,z_1)$ and $\mathcal{O}_{a_2}(0,z_2)$ in the DPD definition \eqref{eq:dpd-def} and then \emph{assuming} that the dominant term in this sum is the ground state.  Following exactly the same steps for the skewed DPD \eqref{eq:dpd-skew-def}, one obtains
\begin{align}
   \label{eq:dpd-skew-fact}
F_{u d}(x_1,x_2,\zeta, \tvec{y})
 \overset{?}{=}
 \int \frac{d^2 \tvec{\Delta}}{(2\pi)^2}\;
 e^{-i \tvec{y} \tvec{\Delta}}\;
    \frac{1}{1-\zeta} \;
 & H_u\biggl[ \frac{2 \ms x_1}{2-\zeta}, \frac{\zeta}{2-\zeta},
    t(\tvec{\Delta}^2, \zeta) \ms\biggr]\,
\nonumber \\[0.2em]
 \times & \, H_d\biggl[ \frac{2 \ms x_2}{2-\zeta}, \frac{\zeta}{2-\zeta},
    t(\tvec{\Delta}^2, \zeta) \ms\biggr]
\end{align}
with
\begin{align}
t(\tvec{\Delta}^2, \zeta)
&= - \frac{\zeta^2 m^2 + \tvec{\Delta}^2}{1 - \zeta} \,.
\end{align}
Here $H_q(x,\xi,t)$ is the generalised parton distribution (GPD) for
unpolarised quarks in a pion; its definition can be found e.g.\ in \cite[\sect{3.2}]{Diehl:2003ny}.  The momentum fraction arguments $x$ and $\xi$ of $H_q$ are defined in a symmetric way between the incoming and outgoing hadron and parton momenta, with $x$ referring to the sum of parton momenta and $\xi$ to their difference, and with momentum fractions normalised to the sum of hadron momenta in the bra and the ket state.  Both $x$ and $\xi$ are limited to the interval $[-1,1]$.  A pictorial representation of the GPDs that appear on the r.h.s.\ of \eqref{eq:dpd-skew-fact} is given in \fig{\ref{fig:dpd-approx-ud}}.

At this point, we must critically examine the support properties of the two sides of \eqref{eq:dpd-skew-fact} in $x_1$ and $x_2$.  The support of the l.h.s.\ is shown in \fig{\ref{fig:dpd-support}}, whereas the one of the r.h.s.\ is the square delineated by $-1+\half\zeta \le x_{1,2} \le 1-\half\zeta$ in the $(x_1,x_2)$ plane.  For $\zeta \ge 0$, this misses the kinematic constraint $|x_1| + |x_2| \le 1$ in $F_{u d}$, whereas for $\zeta < 0$ it is even larger.

In the matrix element \eqref{eq:dpd-skew-def}, the order of the two operators can be interchanged, because the respective fields are separate by spacelike distances.  In a schematic notation, we thus have $\bra{h} \mathcal{O}_{u}\ms \mathcal{O}_{d} \ket{h} = \bra{h} \mathcal{O}_{d}\ms \mathcal{O}_{u} \ket{h}$.  If we insert a set of intermediate states in the latter matrix element, we obtain \eqref{eq:dpd-skew-fact} with $\zeta$ replaced by $-\zeta$ on the r.h.s.  This is represented in \fig{\ref{fig:dpd-approx-du}}.  In that case, the mismatch between the support regions of the two sides is less bad for $\zeta \le 0$ than for $\zeta > 0$.  We therefore retain \eqref{eq:dpd-skew-fact} for $\zeta \ge 0$ and its analogue with $\zeta \to - \zeta$ on the r.h.s.\ for $\zeta < 0$.  This also satisfies the symmetry in $\zeta$ required by $\mathit{PT}$ invariance and stated in \eqref{eq:reverse-zeta}, which is violated if one uses \eqref{eq:dpd-skew-fact} for positive and negative~$\zeta$.

The mismatch of support properties just discussed also affects the case $\zeta = 0$ and is hence not special to the skewed kinematics we are considering here.  In fact, it is well known that the factorisation hypothesis \eqref{eq:dpd-fact} for DPDs violates the momentum conservation constraint $x_1 + x_2 \le 1$.  From a theoretical point of view, inserting a full set of intermediate states between the operators in the DPD definition \eqref{eq:dpd-def} or its skewed analogue \eqref{eq:dpd-skew-def} is of course a legitimate manipulation, but we see that the restriction of this set to the ground state leads to theoretical inconsistencies such as an incorrect support region or the loss of a symmetry required by $\mathit{PT}$ invariance.  How the sum over all states manages to restore the correct properties is difficult to understand in an intuitive manner.  We note that a similar observation was made in \cite{Jaffe:1983hp} when discussing the support properties of PDFs and of higher-twist distributions.

Integrating both sides of \eqref{eq:dpd-skew-fact} over their respective support regions in $x_1$ and $x_2$ and using the sum rule $\int dx\, H_q(x,\xi,t) = F_{q, V}(t)$, one obtains
\begin{align}
   \label{eq:I-zeta-fact}
I_{u d}(-\tvec{y}^2, \zeta)
& \overset{?}{=} \frac{(1-\half\zeta)^2}{1-\zeta}
   \int \frac{d^2 \tvec{\Delta}}{(2\pi)^2}\;
   e^{-i \tvec{y} \tvec{\Delta}}\,
   F_{u,V}\bigl( t(\tvec{\Delta}^2, \zeta) \bigr) \,
   F_{d,V}\bigl( t(\tvec{\Delta}^2, \zeta) \bigr) \,.
\end{align}
Using this for $\zeta \ge 0$ and inserting it into \eqref{eq:py-zero-fct}, we obtain
\begin{align}
  \label{eq:A-fact}
A_{u d}(-\tvec{y}^2, py=0)
& \overset{?}{=} \frac{1}{\pi} \int_0^1 d\zeta\,
   \frac{(1-\half\zeta)^2}{1-\zeta}
   \int \frac{d^2 \tvec{\Delta}}{(2\pi)^2}\;
   e^{-i \tvec{y} \tvec{\Delta}}\,
   F_{u,V}\bigl( t(\tvec{\Delta}^2, \zeta) \bigr) \,
   F_{d,V}\bigl( t(\tvec{\Delta}^2, \zeta) \bigr) \,.
\end{align}
We note that \eqref{eq:A-fact} is expressed in terms of a two-dimensional vector $\tvec{y}$.  This is different from the factorisation hypothesis we derived in \cite[\sect{5.3}]{Bali:2018nde}, which involved the zero-components of currents and a three-dimensional vector $\mvec{y}$.

Note that the two hypotheses \eqref{eq:A-fact} and \eqref{eq:mellin-fact} are based on the same assumption but are not equivalent to each other.  Both are special cases of \eqref{eq:I-zeta-fact}, obtained by either setting $\zeta=0$ or by integrating over $\zeta$ from $0$ to $1$.  The assumption that the ground state dominates the sum over intermediate states could be a better approximation in one or the other case.  Using our lattice results, we will investigate \eqref{eq:A-fact} in \sect{\ref{sec:fact-A}} and \eqref{eq:mellin-fact} in \sect{\ref{sec:fact-mellin}}.

\section{Lattice computation and lattice artefacts}
\label{sec:lattice}

We performed lattice simulations for the matrix elements \eqref{eq:mat-els} in a pion with the currents given in~\eqref{eq:local-ops}.  We set $y^0 = 0$, so that on the lattice the two currents are inserted at the same Euclidean time, but with a spatial separation $\mvec{y}$.  We generated data both for zero and nonzero pion momentum $\mvec{p}$.   The lattice techniques we employed are explained in detail in \cite[\sect{3}]{Bali:2018nde}.  In the following, we recall only the basic steps described in that work and then proceed to the specifics of our present analysis.


\subsection{Lattice graphs}
\label{sec:lattice-graphs}

\begin{figure*}
\begin{center}
\includegraphics[width=0.93\textwidth]{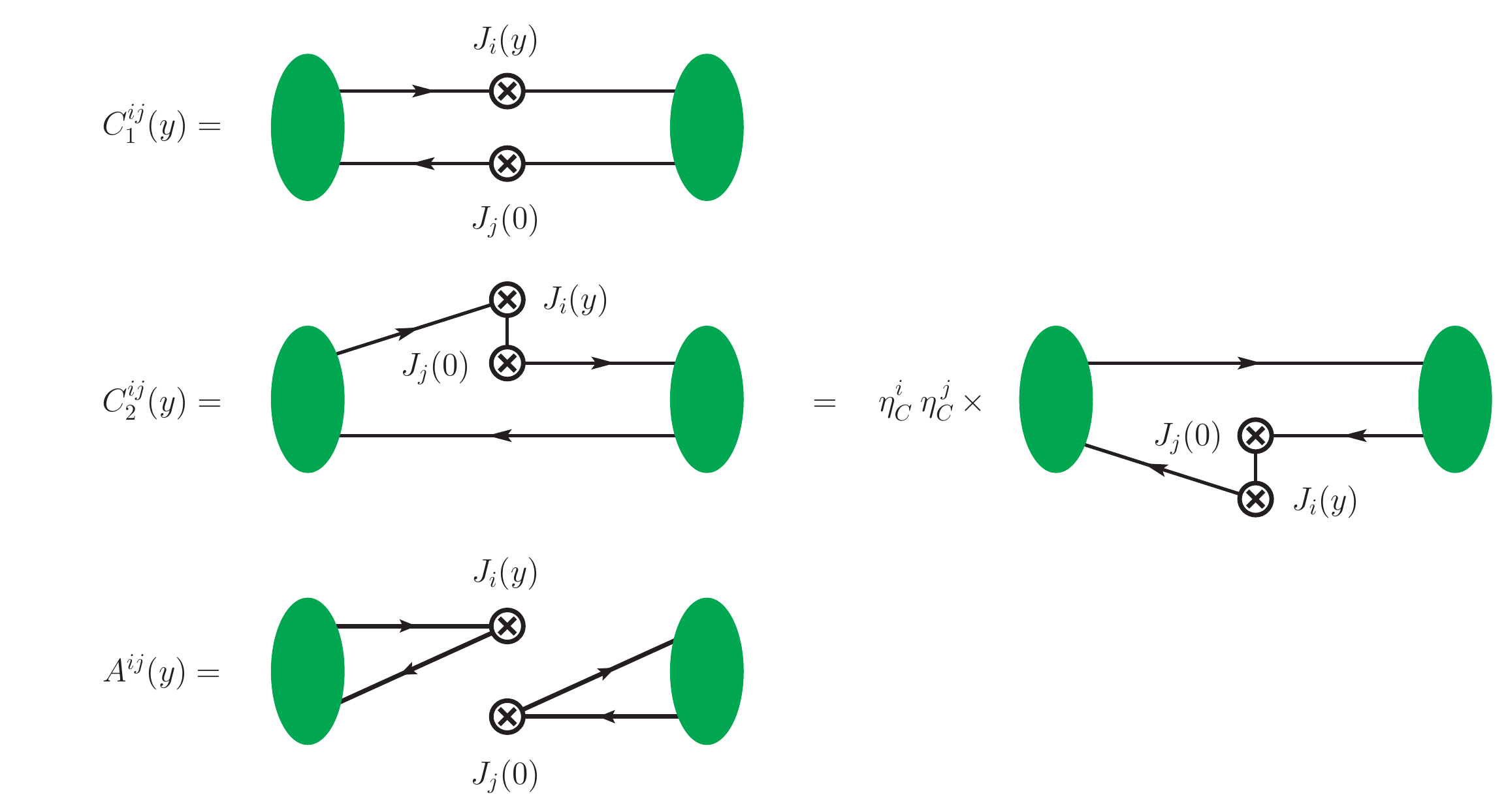} \\[1em]
\includegraphics[width=0.93\textwidth]{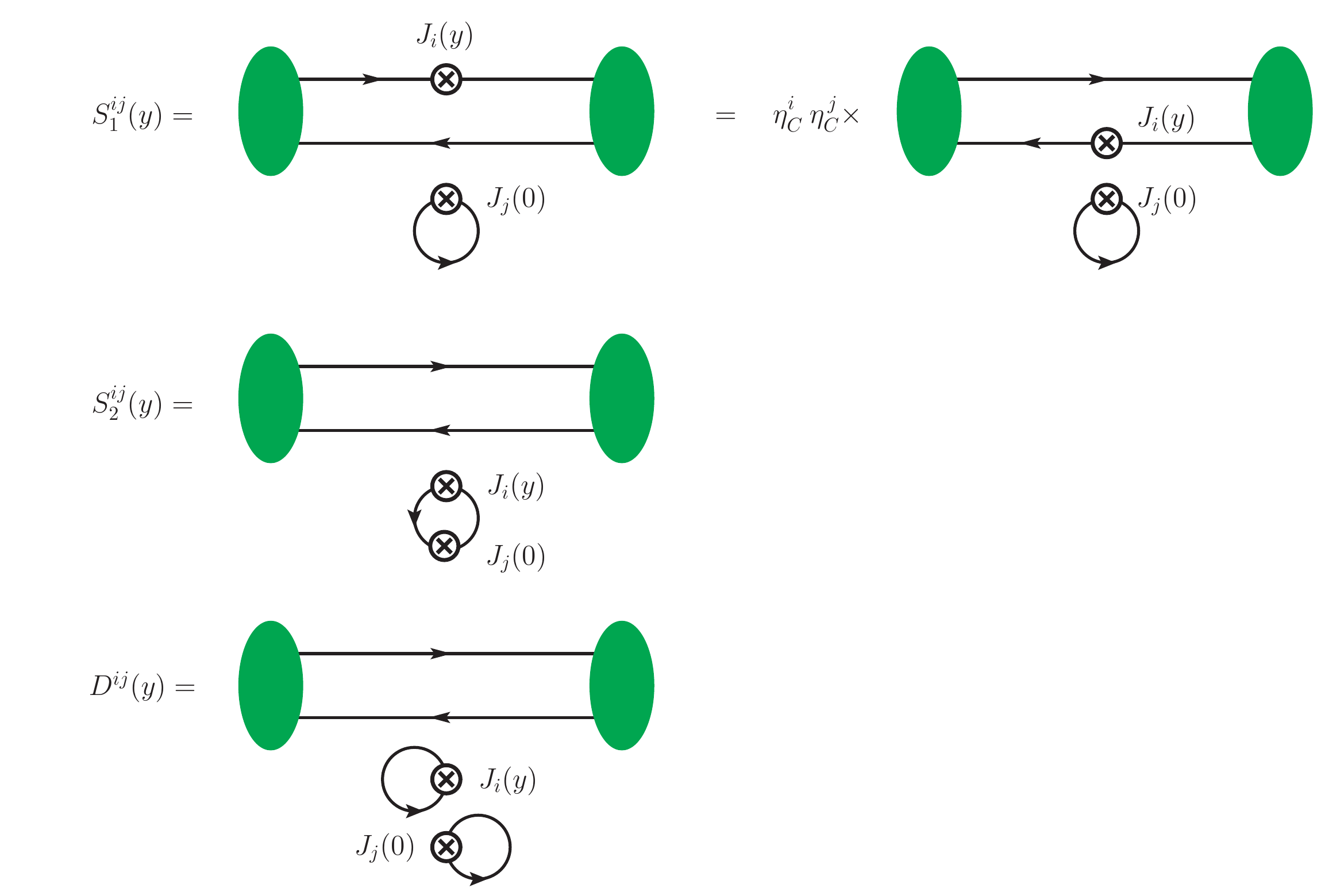}
\end{center}
\caption{\label{fig:contractions} Lattice graphs for the correlation functions used to extract the matrix elements \protect\eqref{eq:mat-els} in a pion.  The dependence on the pion momentum $\mvec{p}$ is not indicated for brevity.  $\etac^{i}$ denotes the charge conjugation parity of the current $J_i$ and is defined in \protect\eqref{eq:CPT-parity}.}
\end{figure*}

To evaluate the two-current matrix elements \eqref{eq:mat-els}, we compute the four-point correlation function of a pion source operator at Euclidean time $0$, a pion sink operator at Euclidean time $t$, and the two currents $J_i$ and $J_j$ at Euclidean time $\tau$.  The correlation function receives contributions from a large number of Wick contractions, which are shown in \fig{\ref{fig:contractions}}.  We will also refer to these contractions as ``lattice graphs'' or simply as ``graphs''.

The relation between pion matrix elements and lattice graphs depends on the product of $C$ parities of the currents.  Omitting Lorentz indices and the dependence on the pion momentum $p$, and using the shorthand notation
\begin{align}
\label{eq:contr-shorthand}
C_1^{} &= C_1^{i j}(y) \,,
&
C_2^{} &= \half \bigl[ C_2^{i j}(y) + C_2^{j i}(-y) \bigr] \,,
&
A &= \half \bigl[ A^{i j}(y) + A^{j i}(-y) \bigr] \,,
\nonumber \\[0.2em]
S_1^{} &= \half \bigl[ S_1^{i j}(y) + S_1^{j i}(-y) \bigr] \,,
&
S_2^{} &= S_2^{i j}(y) \,,
&
D &= D^{i j}(y)
\end{align}
for the graphs or their symmetrised combinations, we have
\begin{align}
\label{eq:phys-mat-els}
M_{u d, i j}(y) \ms\big|_{\pi^+}
&= C_1 + \bigl[ 2 S_1 + D \bigr] \,,
\nonumber \\
M_{u u, i j}(y) \ms\big|_{\pi^+}
&=  \bigl[ 2 C_2 + S_2 \bigr] + \bigl[ 2 S_1 + D \bigr] \,,
\nonumber \\
M_{u d, i j}(y) \ms\big|_{\pi^0}
&= \bigl[ 2 S_1 + D \bigr] - A \,,
\nonumber \\
M_{u u, i j}(y) \ms\big|_{\pi^0}
&=  C_1 + \bigl[ 2 S_1 + D \bigr] + \bigl[ 2 C_2 + S_2 \bigr] + A
\end{align}
for $\etac^{i \phantom{j}} \etac^{j} = +1$, which is satisfied for all combinations of currents considered in the present study.  We note that this is no longer the case if one includes operators with covariant derivatives (corresponding to higher Mellin moments).  One readily checks that \eqref{eq:phys-mat-els} satisfies the general symmetry relation \eqref{eq:val-sea-sum}, as it must.

To compute the different graphs on the lattice, we use a variety of techniques as detailed in \cite[\sect{3.3}]{Bali:2018nde}.  We make extensive use of stochastic sources, and for graph $C_2$ we use a hopping parameter expansion to reduce statistical noise for the propagation between the two currents.

For the disconnected graphs $S_2$ and $D$ we need to subtract vacuum contributions, namely the product of a two-point correlation function of the pion source and sink with a two-point correlation function of the two currents.  The latter corresponds to the vacuum expectation value $\bra{0} J_i(y) J_j(0) \ket{0}$.  The vacuum subtraction for the disconnected graph $S_1$ involves $\bra{0} J_i(y) \ket{0}$ or $\bra{0} J_j(0) \ket{0}$, which is zero because our currents carry Lorentz indices.

We anticipate that the doubly disconnected graph $D$ in general gives a good signal for the four-point correlation function, but that there is a near-perfect cancellation between this correlator and its vacuum subtraction term.  The result after subtraction is consistent with zero and has huge statistical uncertainties compared with those of any other graph.  We will hence not be able to report useful results for graph $D$.  Fortunately, we encounter no such problem for graph $S_2$.


\subsection{Lattice simulation and extraction of twist-two functions}
\label{sec:simulation}

We perform our simulations using the Wilson gauge action and $n_F = 2$ mass degenerate flavours of non-perturbatively improved Sheikholeslami-Wohlert (NPI Wilson-clover) fermions. The gauge configurations were generated by the RQCD and QCDSF collaborations.  We use two gauge ensembles, whose parameters are given in \tab{\ref{tab:lattice}}.  They have different spatial sizes, $L=32$ and $L=40$, which allows us to study finite volume effects in \sect{\ref{sec:vol-comp}}.  Despite having data for only a single lattice spacing, $a = 0.071 \fm$, we are also able to investigate discretisation effects, as discussed in \sect{\ref{sec:iso-boost}}.

\begin{table*}[!b]
\begin{center}
\renewcommand{\arraystretch}{1.2}
\begin{tabular}{c|cccccccccccc} \hline \hline
ensemble & $\beta$ &  $a \, [\fm]$ & $\kappa$ & $L^3 \times T$  &
$m_{\pi} \, [\mev]$ & $L m_{\pi}$ & $N_{\text{full}}$ & $N_{\text{used}}$ \\
\hline
IV & 5.29 & 0.071 & 0.13632 &   $32^3\times 64$  &  $294.6 \pm 1.4$ &  3.42   &
 2023 &  960 \\
 V & 5.29 & 0.071 & 0.13632 &   $40^3\times 64$  &  $288.8 \pm 1.1$ &  4.19   &
 2025 & 984 \\ \hline \hline
\end{tabular}
\end{center}
\caption{\label{tab:lattice} Details of the gauge ensembles used in this analysis.  $N_{\text{full}}$ is the total number of available gauge configurations, and $N_{\text{used}}$ is the number of configurations used in our simulations.  More detail can be found in \protect\cite{Bali:2014gha,Bali:2014nma}.}
\end{table*}

For the ensemble with $L = 40$, we performed simulations with different $\kappa$ values in the quark propagator:
\begin{align}
\label{eq:quark-masses}
& \text{light quarks:}   & \kappa &= 0.13632 \,,  & m_\pi &= 293 \mev \,,
\nonumber \\
& \text{strange:}        & \kappa &= 0.135616 \,, & m_\pi &= 691 \mev \,,
\nonumber \\
& \text{charm:}          & \kappa &= 0.125638 \,, & m_\pi &= 3018 \mev \,.
\end{align}
Here ``light quarks'' refers to the $\kappa$ value used for simulating the sea quarks, whereas the other two values correspond to the physical strange and charm quark masses, as determined in \cite{Bali:2016lvx} and \cite{Bali:2017pdv} by tuning the pseudoscalar ground state mass to $685.8 \mev$ in the first case and the spin-averaged $S$-wave charmonium mass to $3068.5 \mev$ in the second case.  Since our simulations are performed with an $n_F = 2$ fermion action, the strange and charm quarks are partially quenched.

The values of $m_\pi$ in \eqref{eq:quark-masses} are obtained from  exponential fits of the pion two-point function.  We quote them only for orientation and do not attempt to quantify their errors.  These masses are in reasonable agreement with the value in \tab{\ref{tab:lattice}} for light quarks, and with the mass of the pseudoscalar ground state quoted below \eqref{eq:quark-masses} for strange quarks.


\paragraph{Pion matrix elements.}
For all lattice graphs, we compute the correlation functions with zero three-momentum $\mvec{p}$ of the pion.  For the connected graphs $C_1$ and $C_2$, we additionally have data with finite pion momenta.  These data are restricted to the $L=40$ lattice and to light quarks.  The pion momenta that can be realised on the lattice are given by
\begin{align}
\label{eq:lattice-mom}
\mvec{p} &= \frac{2\pi}{L a} \, \mvec{P} \,,
\end{align}
where the components of $\mvec{P}$ are integers and $2\pi /(La) \approx 437 \mev$ in our case.  For simplicity we write $P = |\mvec{P}|$.
Graph $C_1$ is computed for all 18 nonzero momenta with $P^2 \le 2$, for 6 momenta with $P^2 = 3$, and for one momentum with $P^2 = 4$.  For $C_2$, we have results for all 6 momenta with $P = 1$.

The distance between the pion source and sink in the correlation functions is fixed to $t = 15 a \approx 1.07 \fm$ as a default.  To investigate the influence of excited states, we also calculate graphs $C_1$, $C_2$ and $S_1$ with $t = 32a$.  The matrix element \eqref{eq:mat-els} is extracted from the ratio between the four-point correlation function around $\tau = t/2$ and the pion two-point function.  For  graphs $C_1$ and $A$, we measure the $\tau$ dependence of the four-point function and fit to a plateau in the $\tau$ ranges specified in \cite[\eqn{(4.1)}]{Bali:2018nde}.  The quality of the corresponding plateaus is good for matrix elements that have a nonzero value within statistical uncertainties.  For the remaining graphs, we extract the matrix element from data at $\tau/a = 7$ and $8$ if $t/a = 15$.  For the $C_2$ and $S_1$ data with $t/a = 32$, we use $\tau/a = 16$.  A comparison of data with $t = 15a$ and $t = 32a$ is shown in section \ref{sec:t15-32}.

All lattice currents are converted to the $\overline{\text{MS}}$ scheme at the renormalisation scale
\begin{align}
\mu &= 2 \gev \,.
\end{align}
As described in \cite[\sect{3.4}]{Bali:2018nde}, this is done using a combination of non-perturbative and perturbative renormalisation and includes an estimate of the quark mass dependent order $a$ improvement term.


\paragraph{Invariant functions.}
From the matrix elements \eqref{eq:mat-els}, we determine the invariant functions for each individual value of $\mvec{y}$ and $\mvec{p}$.  This is done using a minimum $\chi^2$ fit of the data for all tensor components to the decomposition \eqref{eq:tensor-decomp}.  For invariant functions of twist two, we also use the projector method \eqref{eq:projector-meth}.  In both cases, the statistical error of an invariant function at given $\mvec{y}$ and $\mvec{p}$ is computed using the jackknife method.   To eliminate autocorrelations, we take the number of jackknife samples as $1/8$ times the number $N_{\text{used}}$ of gauge configurations given in \tab{\ref{tab:lattice}}.

For $P=0$, the twist-two functions extracted with one or the other method show excellent agreement with each other and have statistical uncertainties of almost the same size.  For $P>0$, the values obtained with the projection method have much larger statistical errors than those obtained with a fit and provide only a very weak cross check.  All data shown in the following are obtained by the fit method, both for $P=0$ and $P>0$.

In the remainder of this section, we investigate the extent to which our data are affected by lattice artefacts, largely following the corresponding studies in  \cite[\sect{4}]{Bali:2018nde}.  We only consider data with $py = 0$ here, because they have much smaller statistical errors than the data for $py \neq 0$.   We will return to the case of nonzero $py$ in \sect{\ref{sec:py-not-zero}}.

When discussing twist-two functions extracted from the correlation functions for particular lattice graphs, we will generically write $A_{q q}$, $A_{\Delta q \Delta q}$, \ldots, $B_{\delta q \delta q}$, without reference to specific quark flavors $q_1$ and $q_2$.  This is because the distinction between $u$ and $d$ quarks in a pion only appears when lattice graphs are combined as specified in \eqref{eq:phys-mat-els}.


\subsection{Isotropy and boost invariance}
\label{sec:iso-boost}

The decomposition \eqref{eq:tensor-decomp} of matrix elements in terms of basis tensors and functions of $y^2$ and $py$ assumes Lorentz invariance and thus requires both the continuum and the infinite-volume limit.  If our lattice simulations are sufficiently close to these limits, then the values of twist-two functions  extracted for individual points $\mvec{y}$ and $\mvec{p}$ with $\mvec{p} \mvec{y} = 0$ must not depend on the directions of $\mvec{y}$ or $\mvec{p}$ or on the size of $\mvec{p}$.

Let us test whether this is the case in our simulations for light quarks on the lattice with $L=40$.  We restrict our attention to graphs $C_1$ and $C_2$, for which statistical errors are small enough to reveal the effects of interest.
For the sake of legibility, we henceforth write $y = |\mvec{y}|$ for the length
of the spatial distance $\mvec{y}$ between the two currents.  We continue to use
$y^2$ and $py$ to denote the products $y^\mu y_\mu$ and $p^\mu y_\mu$ of
four-vectors in Minkowski space.  Since $y^\mu$ is always spacelike in our context, this implies that $y^2 < 0$.

\begin{figure*}
\begin{center}
\subfigure[$A_{q q}$, graph $C_1$]{
\includegraphics[height=14.6em,trim=0 0 0 17,clip]
{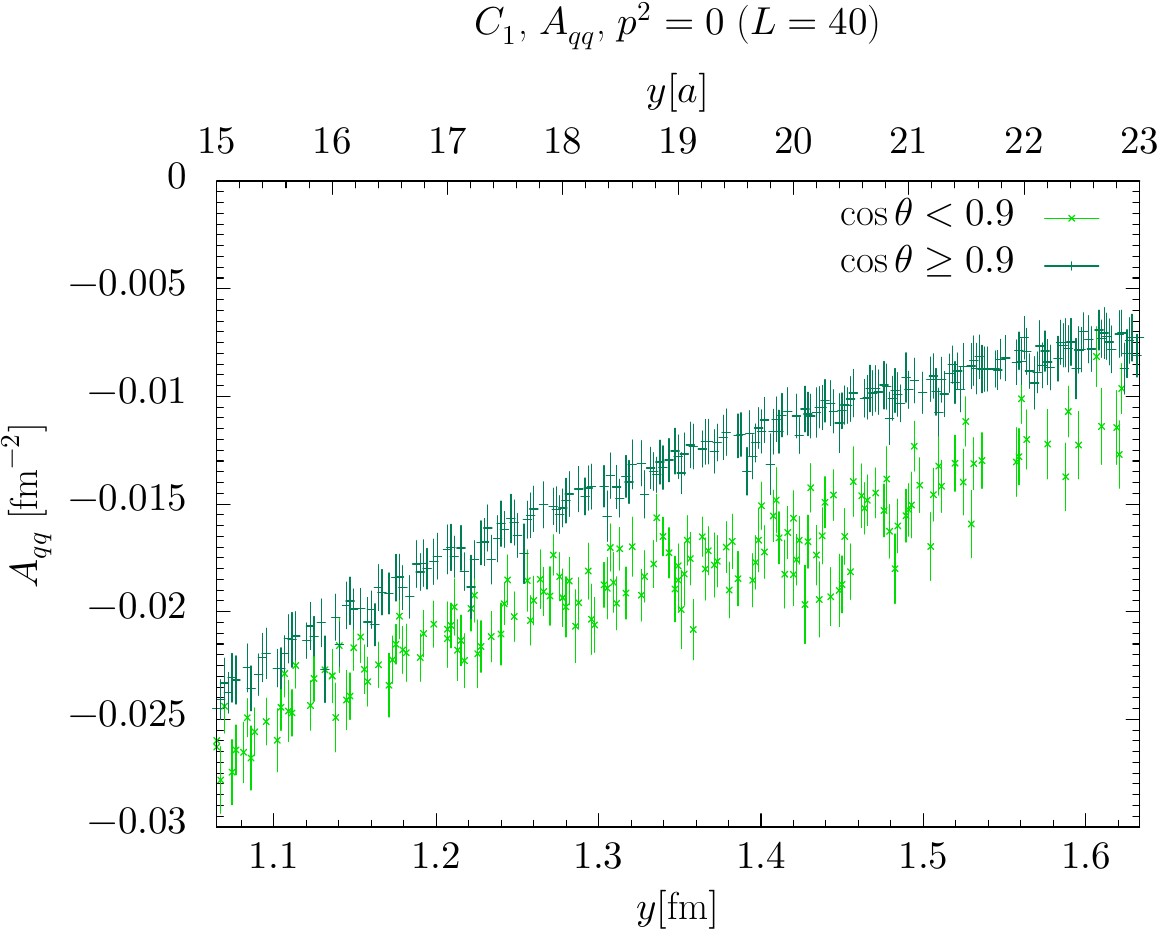}}
\hfill
\subfigure[$A_{\delta q \ms q}$, graph $C_1$]{
\includegraphics[height=14.6em,trim=0 0 0 17,clip]
{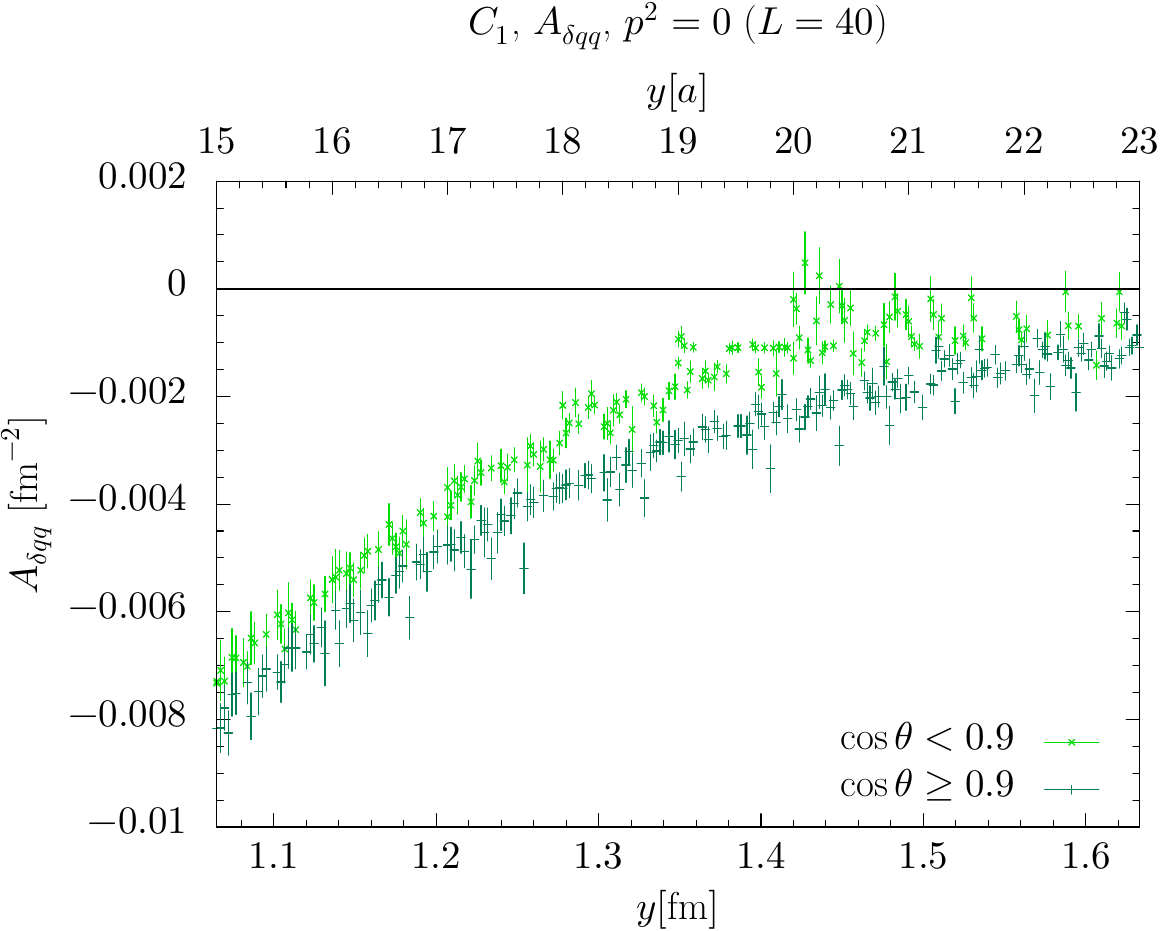}}
\caption{\label{fig:C1-aniso} Examples for the anisotropy in graph $C_1$ at large $y = |\mvec{y}|$.  The data shown are for $L=40$, zero pion momentum, and light quarks.
The results in this and all the following figures are given in the $\overline{\mathrm{MS}}$ scheme at the scale $\mu = 2 \gev$.  The error bars shown in the plots are statistical and obtained with the jackknife method.}
\end{center}
\end{figure*}

At large $y$ of order $L a/2$, we see a clear anisotropy with a saw-tooth pattern in all twist-two functions that have sufficiently small errors.  Examples are shown in \fig{\ref{fig:C1-aniso}}.  This pattern is expected on a lattice with periodic boundary conditions and can be understood in terms of ``mirror images''.  The same effect has been seen and discussed in previous lattice studies of two-current correlators \cite{Burkardt:1994pw,Alexandrou:2008ru}, including our study in \cite{Bali:2018nde} that employed the same lattice data as the present work.
As shown in \cite{Burkardt:1994pw}, the effect of mirror charges at a given distance $y$ is smallest for points $\mvec{y}$ close to one of the space diagonals, i.e.\ the lines given by $\mvec{z} = (z^1, z^2, z^3)$ with $|z^1| = |z^2| = |z^3|$.  To quantify this, we define $\theta(\mvec{y})$ as the angle between $\mvec{y}$ and the space diagonal in the same octant as $\mvec{y}$.  In \cite[\sect{4.2}]{Bali:2018nde}, we found that a cut
\begin{align}
\label{cos-cut}
\cos\theta(\mvec{y}) & \ge 0.9
\end{align}
on the data efficiently removes the effect of mirror charges at large $y$, whilst keeping sufficient statistics.

\begin{figure*}
\begin{center}
\subfigure[$A_{q q}$, graph $C_1$\label{fig:C1-aniso-A_VV}]{
\includegraphics[height=14.6em,trim=0 0 0 17,clip]
{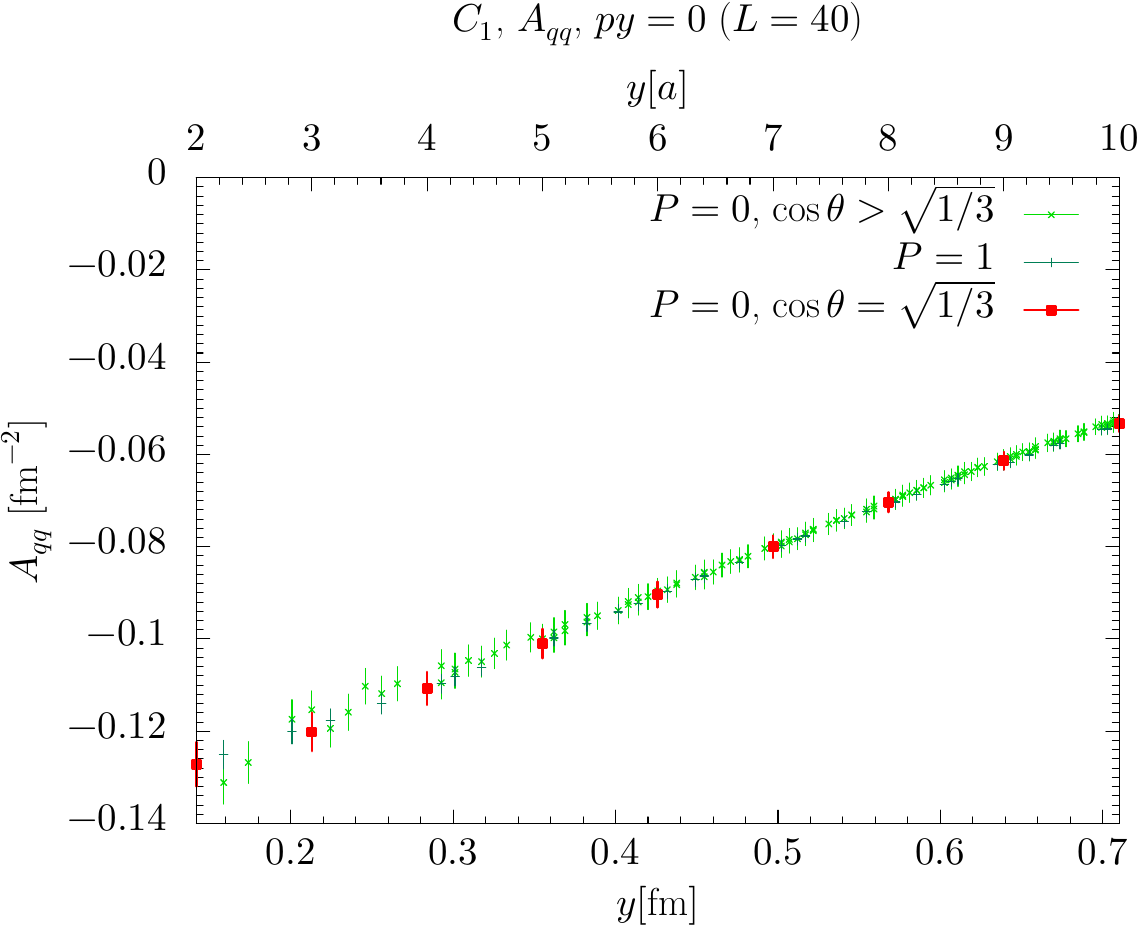}}
\hfill
\subfigure[$A_{\Delta q \Delta q}$, graph $C_1$]{
\includegraphics[height=14.6em,trim=0 0 0 17,clip]
{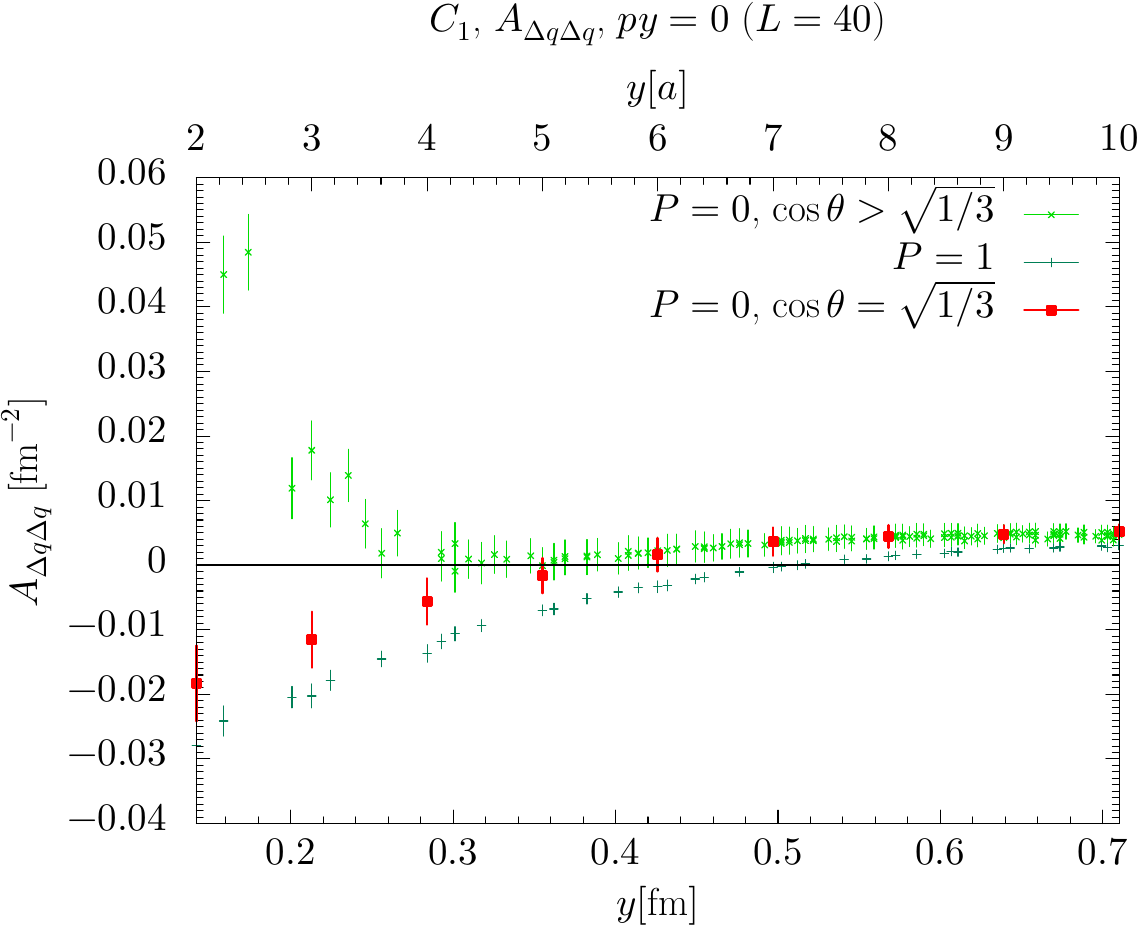}}
\\
\subfigure[$A_{\delta q \ms q}$, graph $C_1$]{
\includegraphics[height=14.6em,trim=0 0 0 17,clip]
{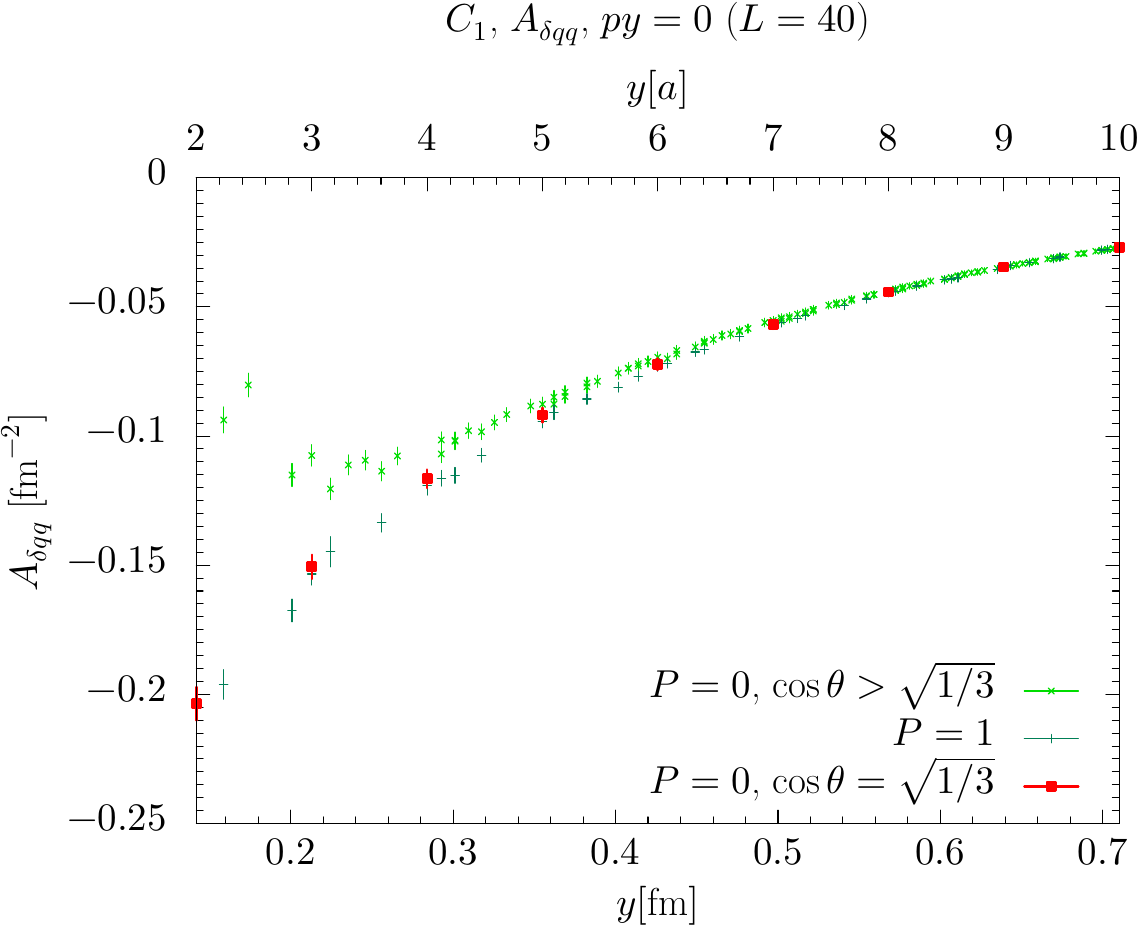}}
\hfill
\subfigure[$B_{\delta q \delta q}$, graph $C_1$]{
\includegraphics[height=14.6em,trim=0 0 0 17,clip]
{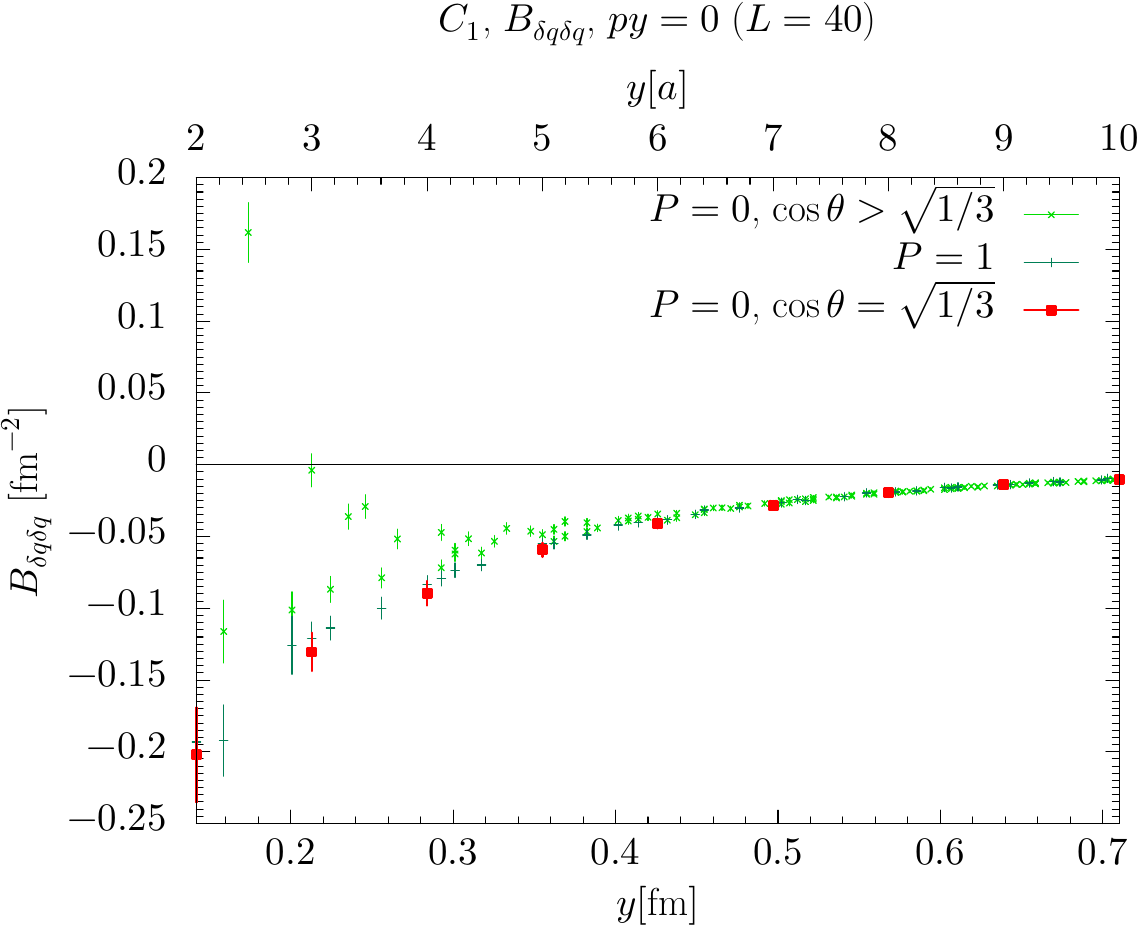}}
\caption{\label{fig:C1-pcomp} Twist-two functions at small $y$ for graph $C_1$, with scaled pion momenta $P = 0$ and $P = 1$ as defined below \protect\eqref{eq:lattice-mom}.  All points have $py = 0$ and are for $L=40$ and light quarks.  The data for $P = \sqrt{2}$, $P = \sqrt{3}$ and $P = 2$ agree with those for $P = 1$ within errors but are not shown for the sake of clarity.
Data with $\cos\theta = \sqrt{1/3}$ correspond to $\mvec{y}$ on a coordinate axis.}
\end{center}
\end{figure*}

A different type of anisotropy in the $C_1$ data is observed at small $y$ and shown in \fig{\ref{fig:C1-pcomp}}.  For $A_{\Delta q \Delta q}$, $A_{\delta q \ms q}$ and $B_{\delta q \delta q}$, the data with zero pion momentum exhibit a clear discrepancy between points $\mvec{y}$ on a coordinate axis (i.e.\ with two components being zero) and all other points.  This discrepancy is very strong for $y$ below $5a$ and ceases to be visible above $7a$.  The data for $A_{\delta q \delta q}$ (not shown in the figure) have larger errors and show only a weak anisotropy for $y < 4a$.  Only the function $A_{q q}$ is not affected by this phenomenon, for which we have no explanation.

By contrast, we find that the $C_1$ data with nonzero pion momentum and $py = 0$ are isotropic in $\mvec{y}$.  For nonzero momenta, we can hence average all data with the same values of $y$ and $P$, which greatly decreases statistical errors.  We find good agreement between the $P>0$ data and the $P=0$ data with $\mvec{y}$ on a coordinate axis for all twist-two functions except $A_{\Delta q \Delta q}$, where the agreement is only approximate.

\begin{figure*}
\begin{center}
\subfigure[$|A_{q q}|$, graph $C_2$]{
\includegraphics[height=14.6em,trim=0 0 0 17,clip]
{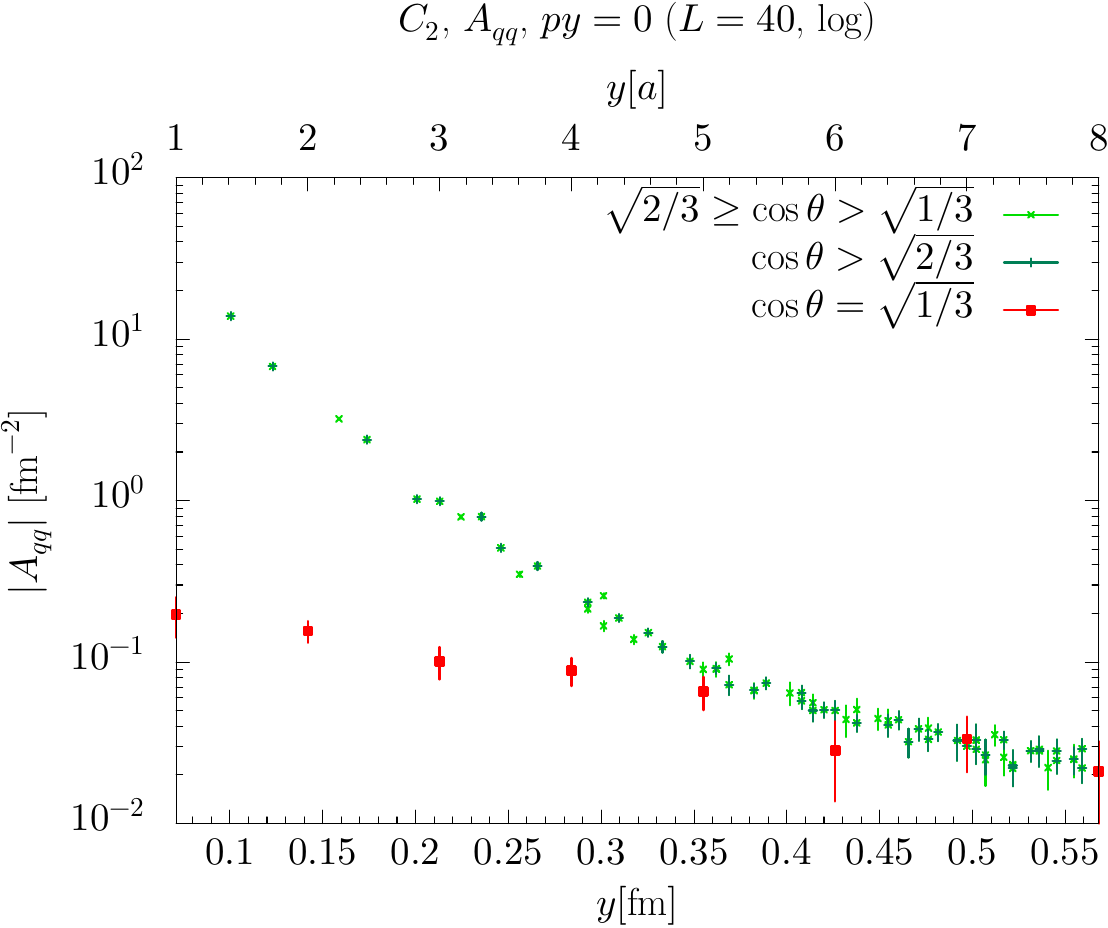}}
\hfill
\subfigure[$|A_{\Delta q \Delta q}|$, graph $C_2$]{
\includegraphics[height=14.6em,trim=0 0 0 17,clip]
{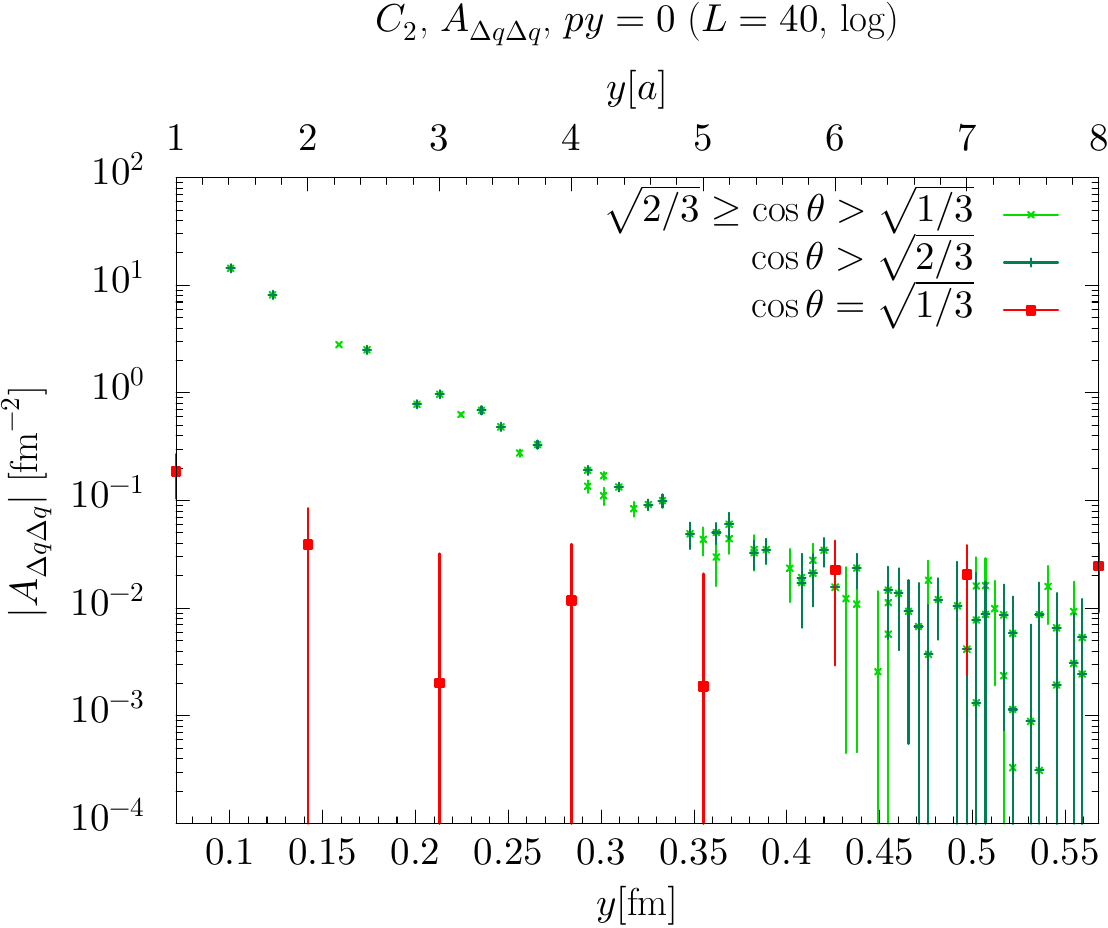}}
\\
\subfigure[$|A_{\delta q \ms q}|$, graph $C_2$]{
\includegraphics[height=14.6em,trim=0 0 0 17,clip]
{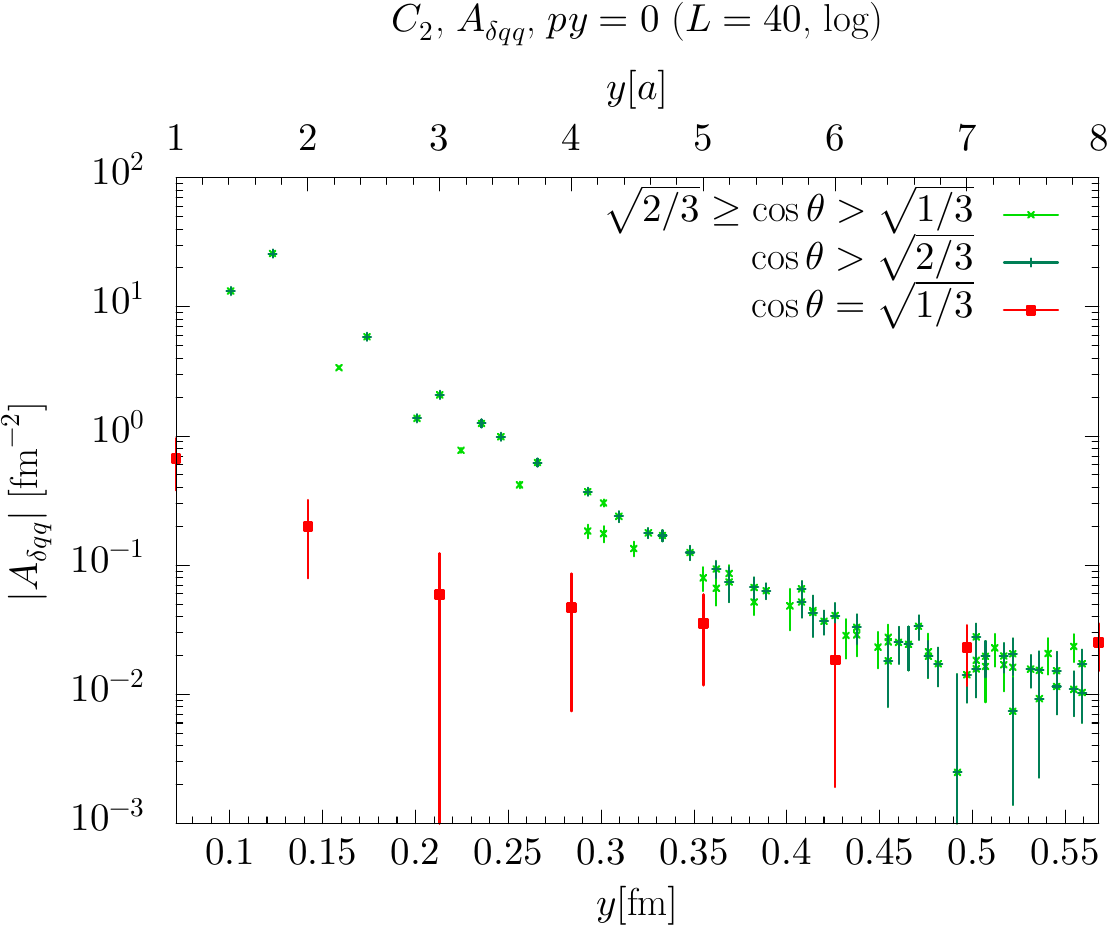}}
\hfill
\subfigure[$|B_{\delta q \delta q}|$, graph $C_2$]{
\includegraphics[height=14.6em,trim=0 0 0 17,clip]
{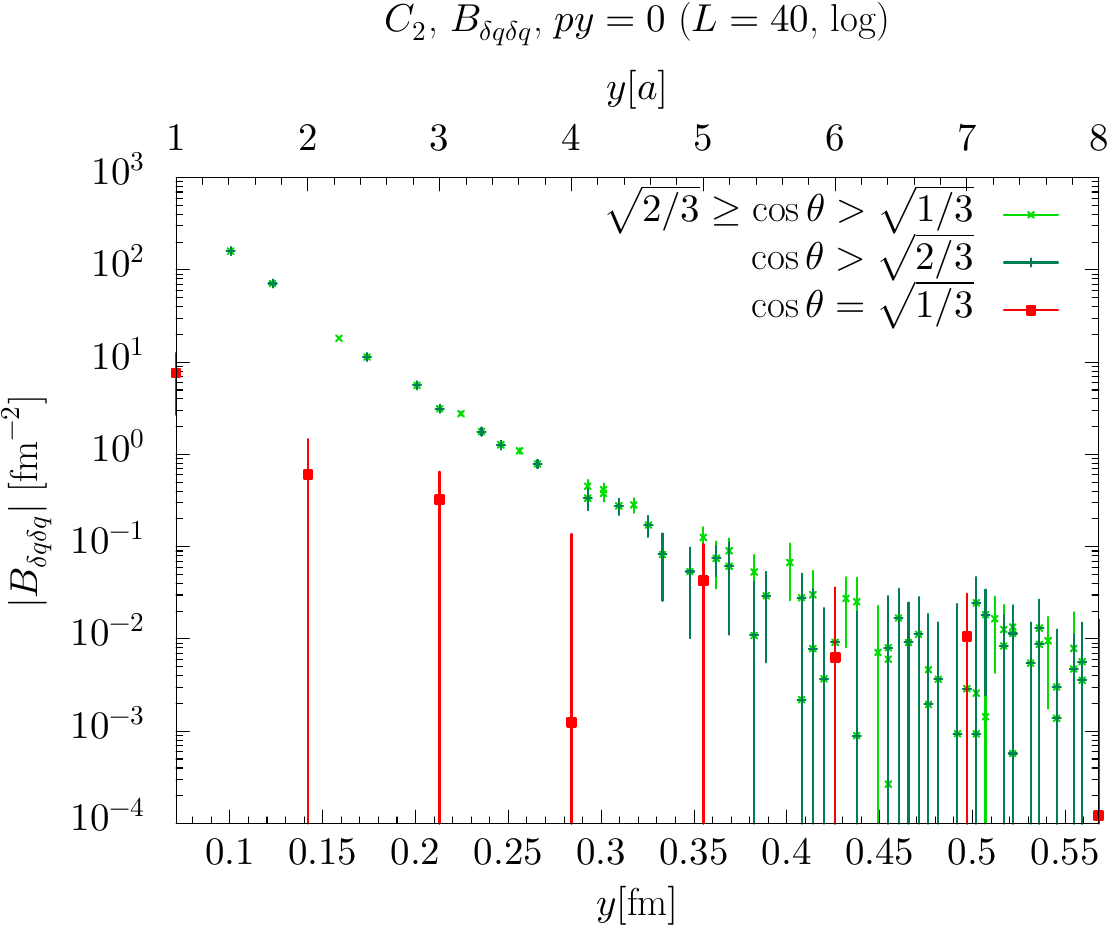}}
\caption{\label{fig:C2-aniso} Twist-two functions at small $y$ for graph $C_2$.  All points are for $L=40$, zero pion momentum, and light quarks.  For $A_{\delta q \delta q}$ (not shown), one finds a clear anisotropy at $y < 4a$, whilst at larger $y$ the statistical errors are too large for drawing firm conclusions.}
\end{center}
\end{figure*}

We now turn our attention to graph $C_2$ at small $y$.  Here, we find a very strong anisotropy in the $P=0$ data.  This is shown in \fig{\ref{fig:C2-aniso}}, where we distinguish points $\mvec{y}$ on the coordinate axes, which have $\cos\theta = \sqrt{1/3}$, points with $\sqrt{1/3} < \cos\theta \le \sqrt{2/3}$, and points with $\sqrt{2/3} < \cos\theta$.  We note that points in a coordinate plane, i.e.\ with at least one component of $\mvec{y}$ equal to zero, have $\sqrt{1/3} \le \cos\theta \le \sqrt{2/3}$.  In all channels, we see a clear discrepancy between the points $\mvec{y}$ on a coordinate axis and all other points.  In addition, there is a significant mismatch between points with $\cos\theta$ above or below $\sqrt{2/3}$ in several channels, most strongly so in $A_{\delta q \ms q}$.  We recall that a strong anisotropy for $C_2$ at small $y$ was also seen for the correlation functions in our study \cite{Bali:2018nde}.  In \sect{4.2} of that work, we argued that this reflects an anisotropy in the lattice propagator between the two currents, and that points selected by the cut \eqref{cos-cut} should give the most reliable results according to the analysis in \cite{Cichy:2012is}.

\begin{figure*}
\begin{center}
\subfigure[$A_{q q}$, graph $C_2$]{
\includegraphics[height=14.6em,trim=0 0 0 17,clip]
{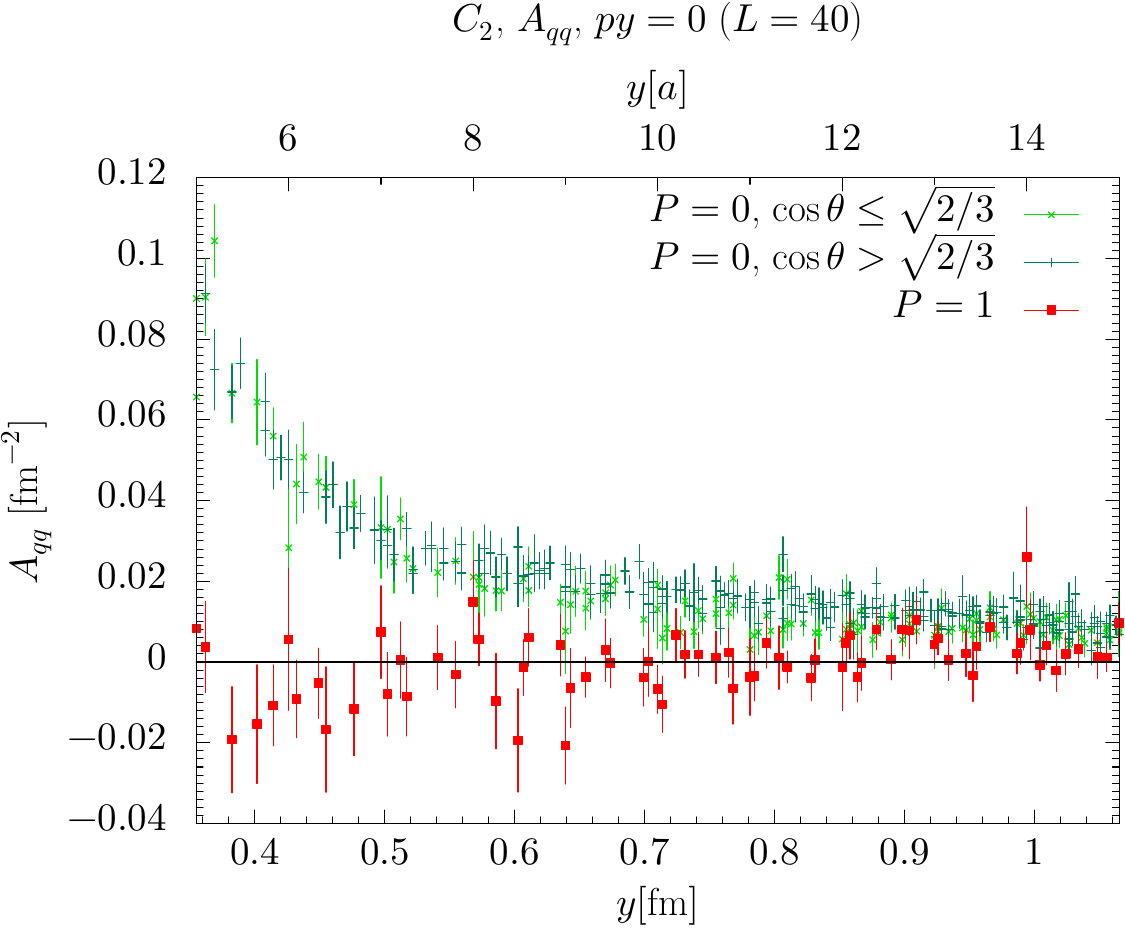}}
\hfill
\subfigure[$A_{\delta q \ms q}$, graph $C_2$]{
\includegraphics[height=14.6em,trim=0 0 0 17,clip]
{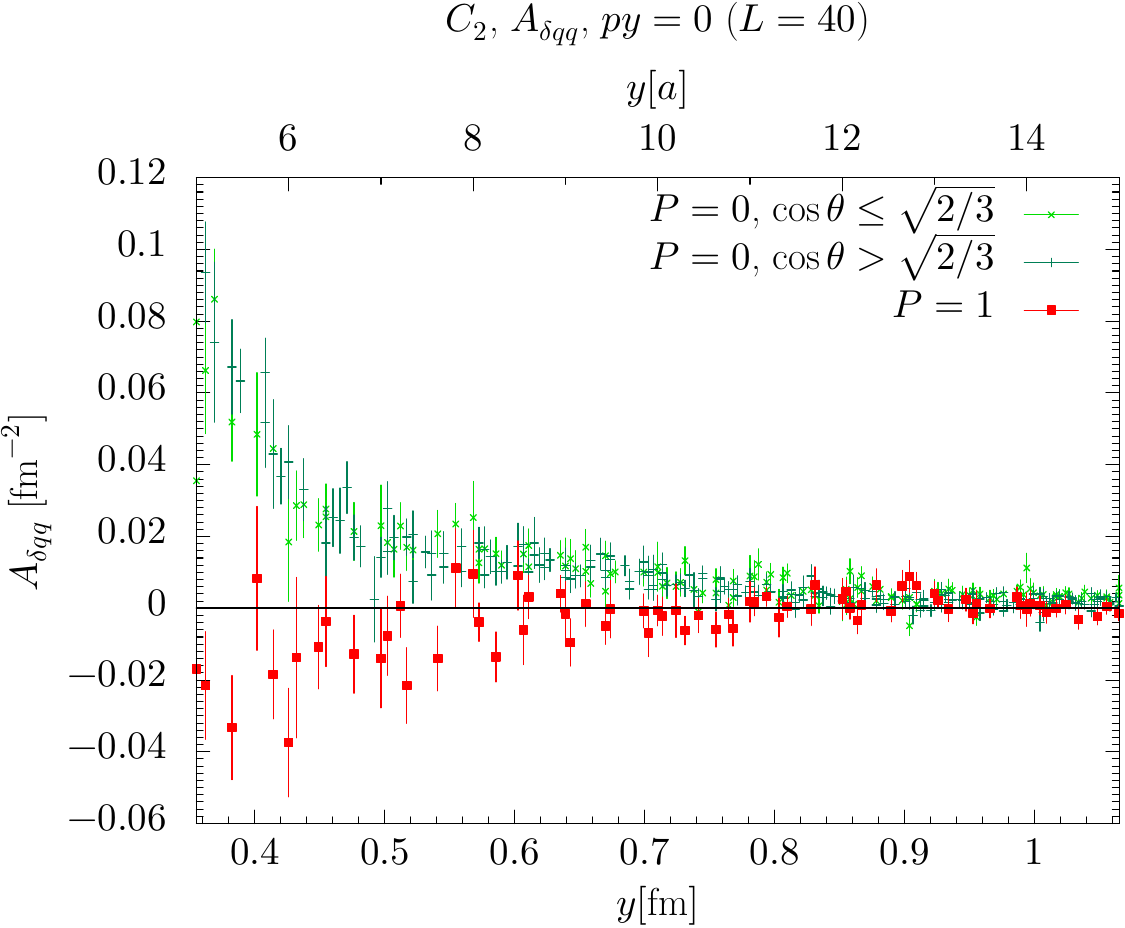}}
\caption{\label{fig:C2-pcomp} Twist-two functions at intermediate $y$ for graph $C_2$.  All points are for $py = 0$, $L=40$ and light quarks.}
\end{center}
\end{figure*}

We also have $P=1$ data for $C_2$, which we can compare with those for $P=0$.  As seen in \fig{\ref{fig:C2-pcomp}}, for $A_{q q}$ and $A_{\delta q \ms q}$ the data at $P=1$ are inconsistent with those at $P=0$, regardless of the value of $\cos\theta$ in the latter.  Since for $P=1$ the condition $py = 0$ requires $\mvec{y}$ to lie in a coordinate plane, we can in fact not select points satisfying the cut \eqref{cos-cut} in this case.  We therefore discard our data with nonzero $P$ for $C_2$.  Testing boost invariance of twist-two functions at $py = 0$ in the presence of the cut \eqref{cos-cut} would require data with at least $P=\sqrt{2}$, which we do not have for~$C_2$.

At $P=0$ and small $y$, we are now in a difficult situation.  Points with large $\cos\theta$ are preferred for $C_2$, while for $C_1$ the points with the smallest possible value $\cos\theta = \sqrt{1/3}$ seem to be more reliable, given that they agree with the $P>0$ data.  Applying different cuts in $\cos\theta$ to the data for $C_1$ and $C_2$ would prevent us from taking linear combinations of those graphs at a given $\mvec{y}$.  However, we regard combining data point by point in $\mvec{y}$ as highly desirable for a transparent and consistent treatment of statistical correlations in the jackknife analysis.

\begin{figure*}
\begin{center}
\subfigure[$A_{q q}$, graph $C_1$]{
\includegraphics[height=14.6em,trim=0 0 0 17,clip]
{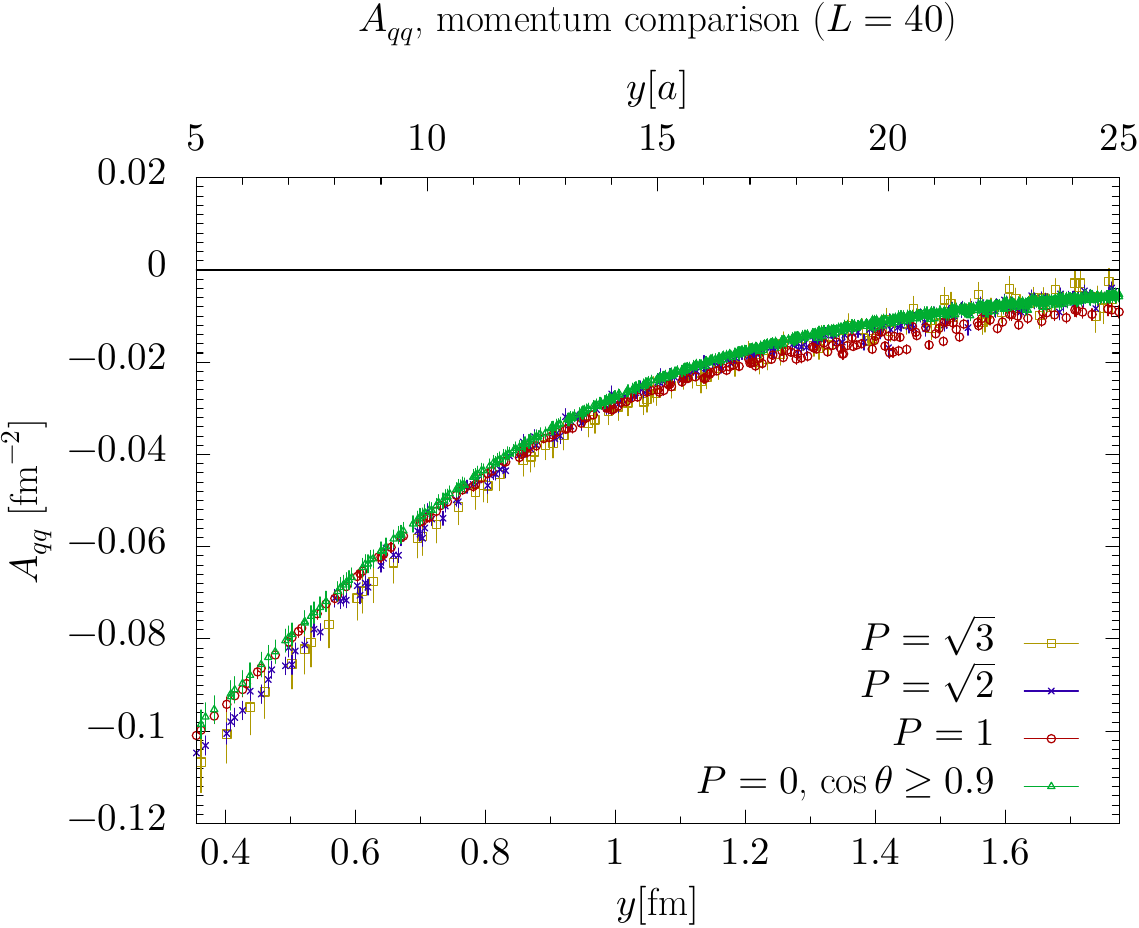}}
\hfill
\subfigure[$A_{\Delta q \Delta q}$, graph $C_1$\label{fig:C1-mom-A_AA}]{
\includegraphics[height=14.6em,trim=0 0 0 17,clip]
{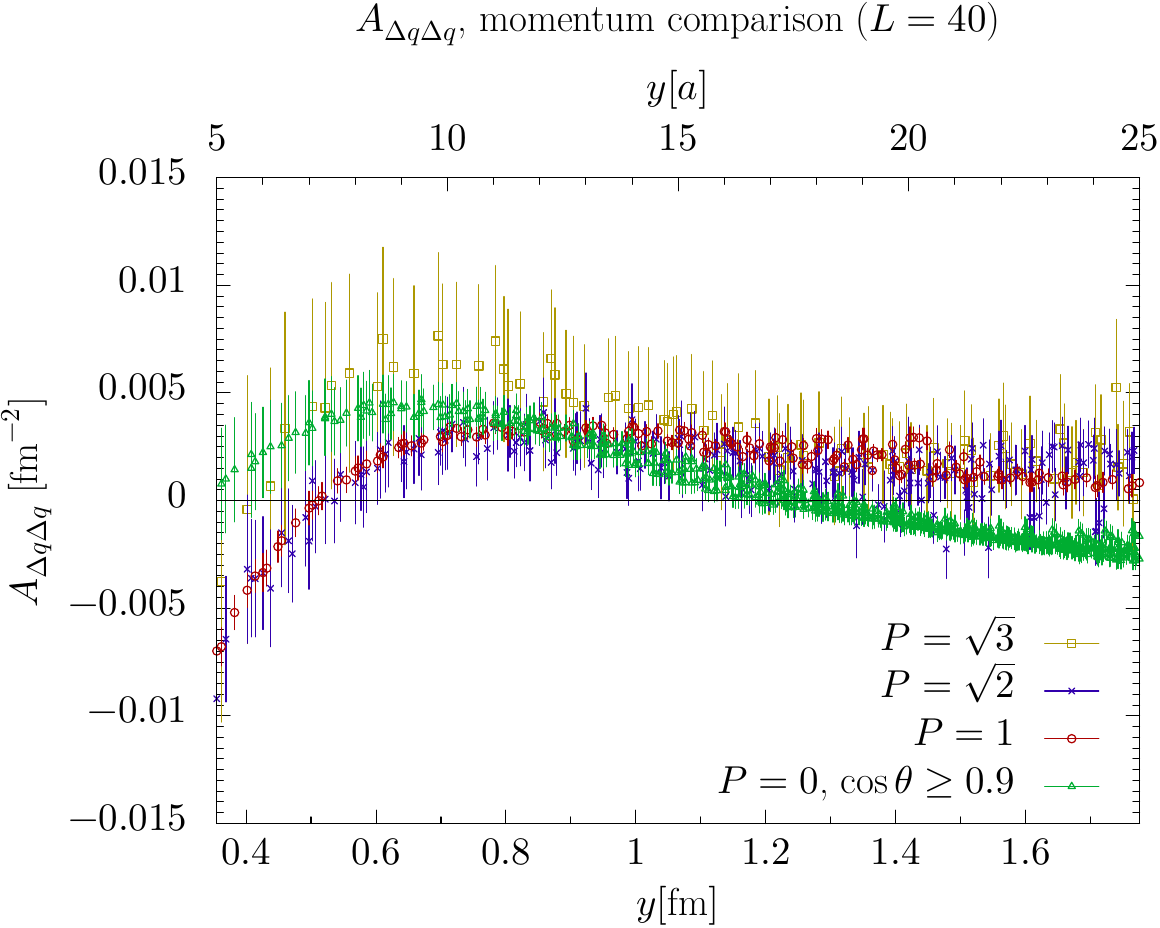}}
\\
\subfigure[$A_{\delta q \ms q}$, graph $C_1$]{
\includegraphics[height=14.6em,trim=0 0 0 17,clip]
{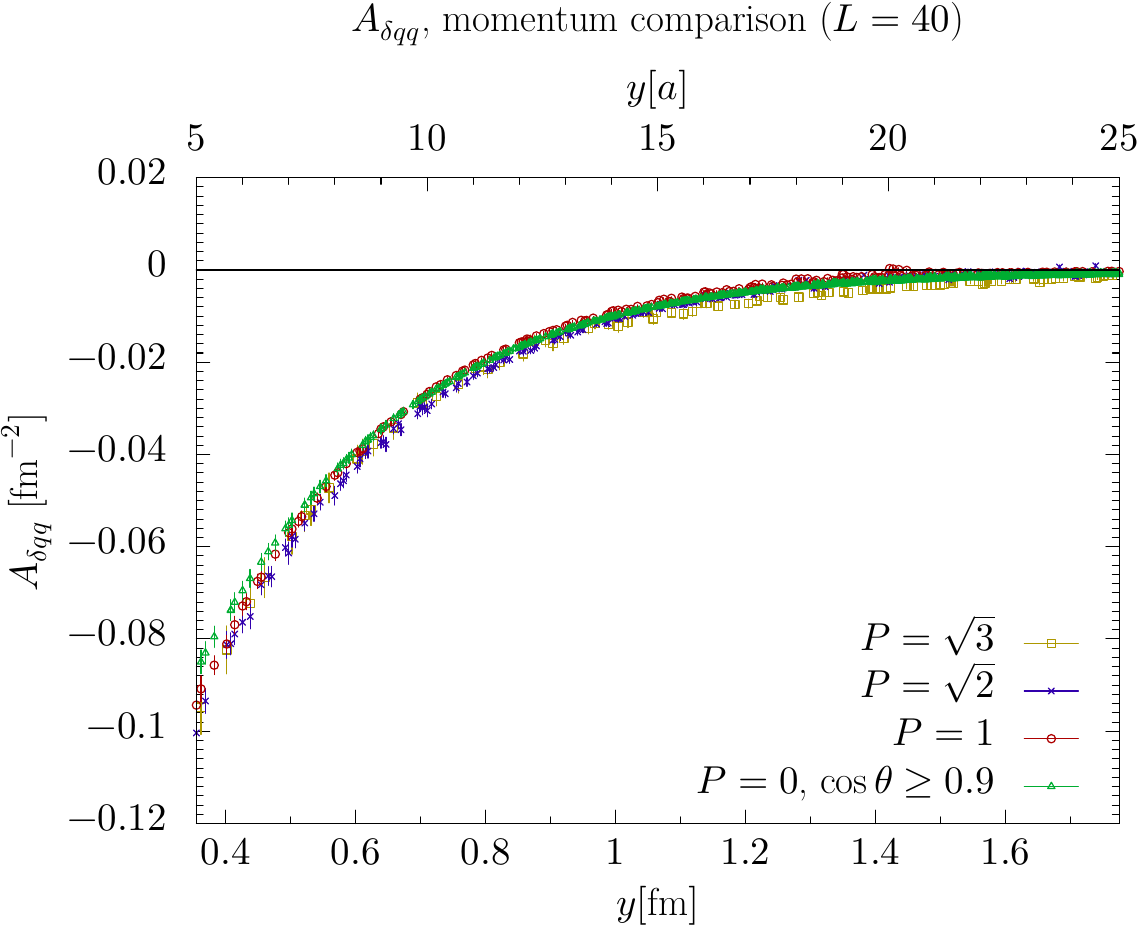}}
\hfill
\subfigure[$A_{\delta q \delta q}$, graph $C_1$]{
\includegraphics[height=14.6em,trim=0 0 0 17,clip]
{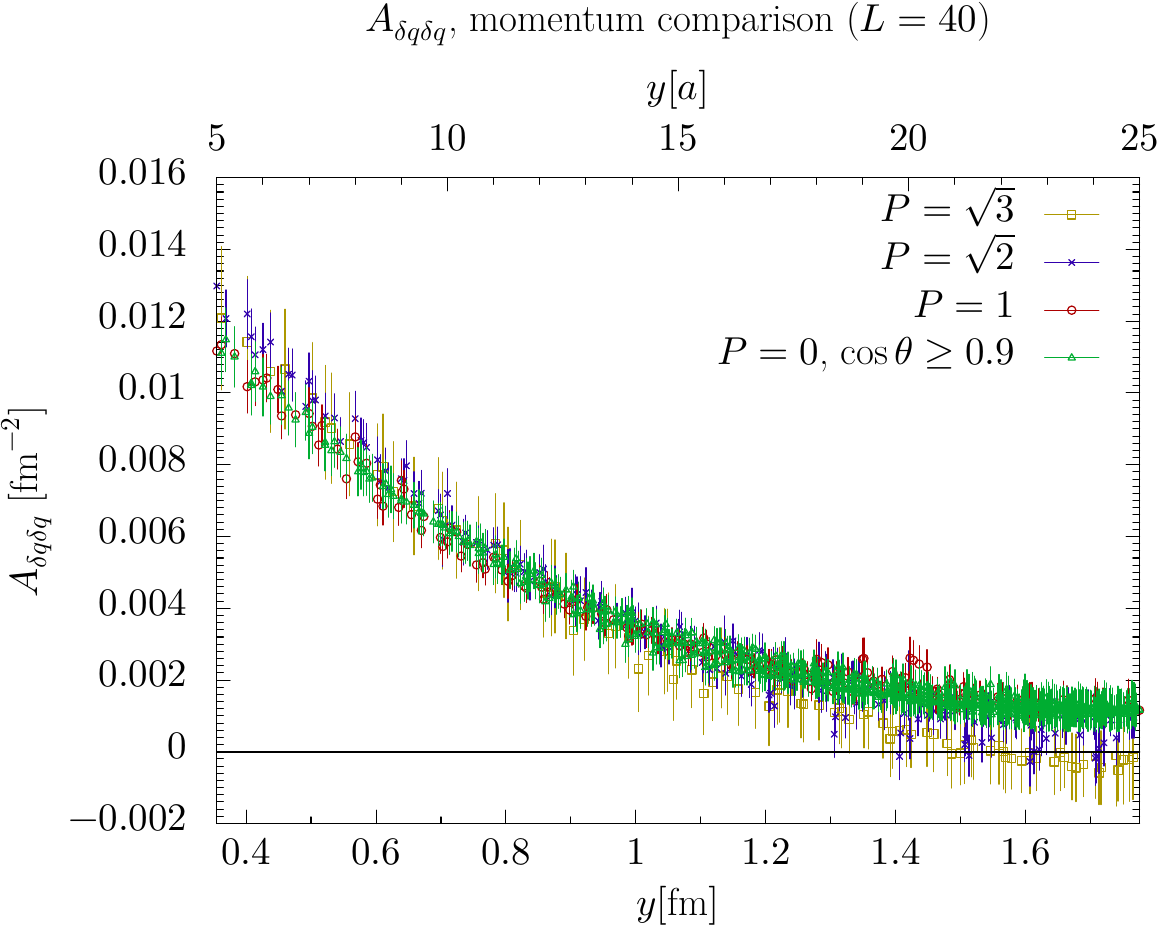}}
\caption{\label{fig:C1-mom} Comparison of twist-two functions for graph $C_1$ at different values of the scaled pion momentum $P$.   All points are for $py = 0$, $L=40$ and light quarks.  For $B_{\delta q \delta q}$ (not shown), one finds good agreement between all points, with a similar quality as for $A_{q q}$.}
\end{center}
\end{figure*}

To avoid this problem, we choose to discard points with $y < 5a$ in our further analysis, and to apply the cut in \eqref{cos-cut} to the $P=0$ data for all lattice graphs.  After this cut, data points with equal values of $y$ are averaged also for $P=0$.  We thus avoid the regions where the anisotropy for $C_1$ and $C_2$ seen at $P=0$ is most severe.  For $C_1$, a small discrepancy between the data with $P=0$ and $P>0$ is still visible up to about $y \sim 8a$, but we consider this to be at an acceptable level.
The result of this procedure is shown for graph $C_1$ in \fig{\ref{fig:C1-mom}}. The agreement between the data for different pion momenta is quite good, except for the function~$A_{\Delta q \Delta q}$.

As an exception to the selection just described, we will in \sect{\ref{sec:fact-A}} use the $C_1$ data for $A_{q q}$ down to $y = 3 a$, given that in this particular channel there are no indications of anisotropy or a pion momentum dependence, as can be seen in \fig{\ref{fig:C1-aniso-A_VV}}.


\subsection{Excited state contributions}
\label{sec:t15-32}

As specified in \sect{\ref{sec:simulation}}, we have a limited set of data with a separation of $t = 32a$ between pion source and sink.  Comparing this with our results for $t = 15a$ allows us to assess the relevance of excited state contributions in our extraction of the pion matrix elements \eqref{eq:mat-els}.

On our lattice with size $L=32$, we have $t = 32a$ data for graphs $S_1$ and $C_2$.  Unfortunately, these graphs give a statistical signal consistent with zero for all twist-two functions and for both source-sink separations.  We hence limit the following discussion to graph $C_1$ on our $L=40$ lattice.

\begin{figure*}[!ht]
\begin{center}
\subfigure[$A_{q q}$, graph $C_1$, light quarks\label{fig:t32-comp-A_VV}]{
\includegraphics[height=14.6em,trim=0 0 0 17,clip]
{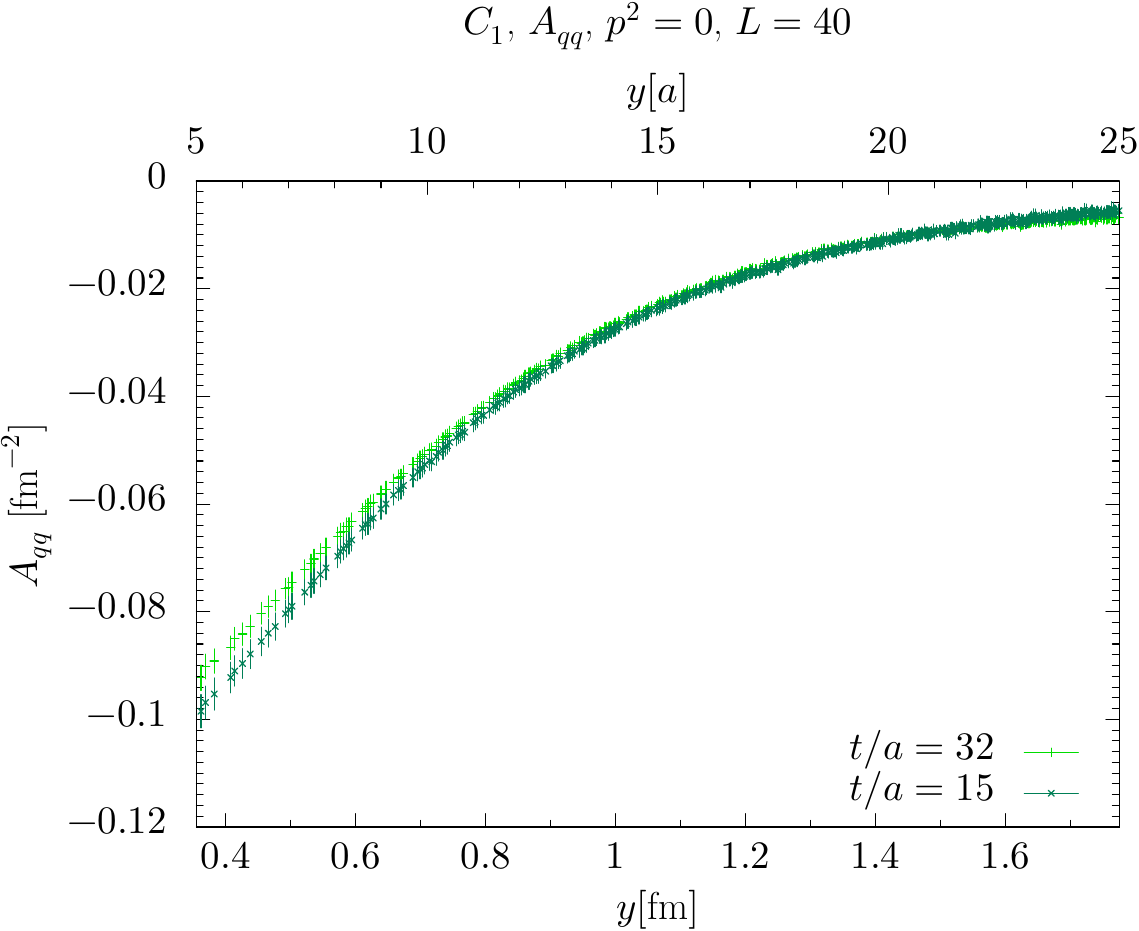}}
\hfill
\subfigure[$B_{\delta q \delta q}$, graph $C_1$, light quarks]{
\includegraphics[height=14.6em,trim=0 0 0 17,clip]
{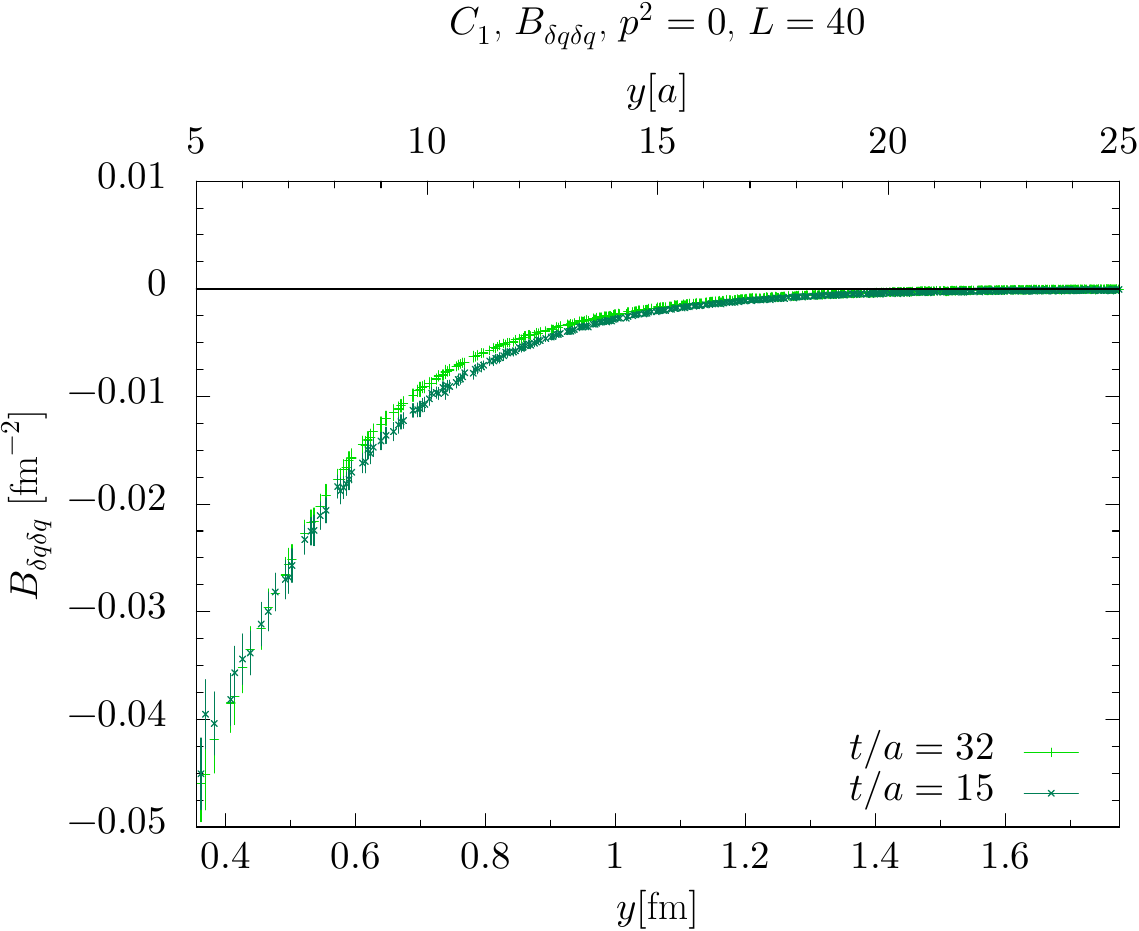}}
\\
\subfigure[$A_{\Delta q \Delta q}$, graph $C_1$, strange quarks\label{fig:t32-comp-A_AA}]{
\includegraphics[height=14.6em,trim=0 0 0 17,clip]
{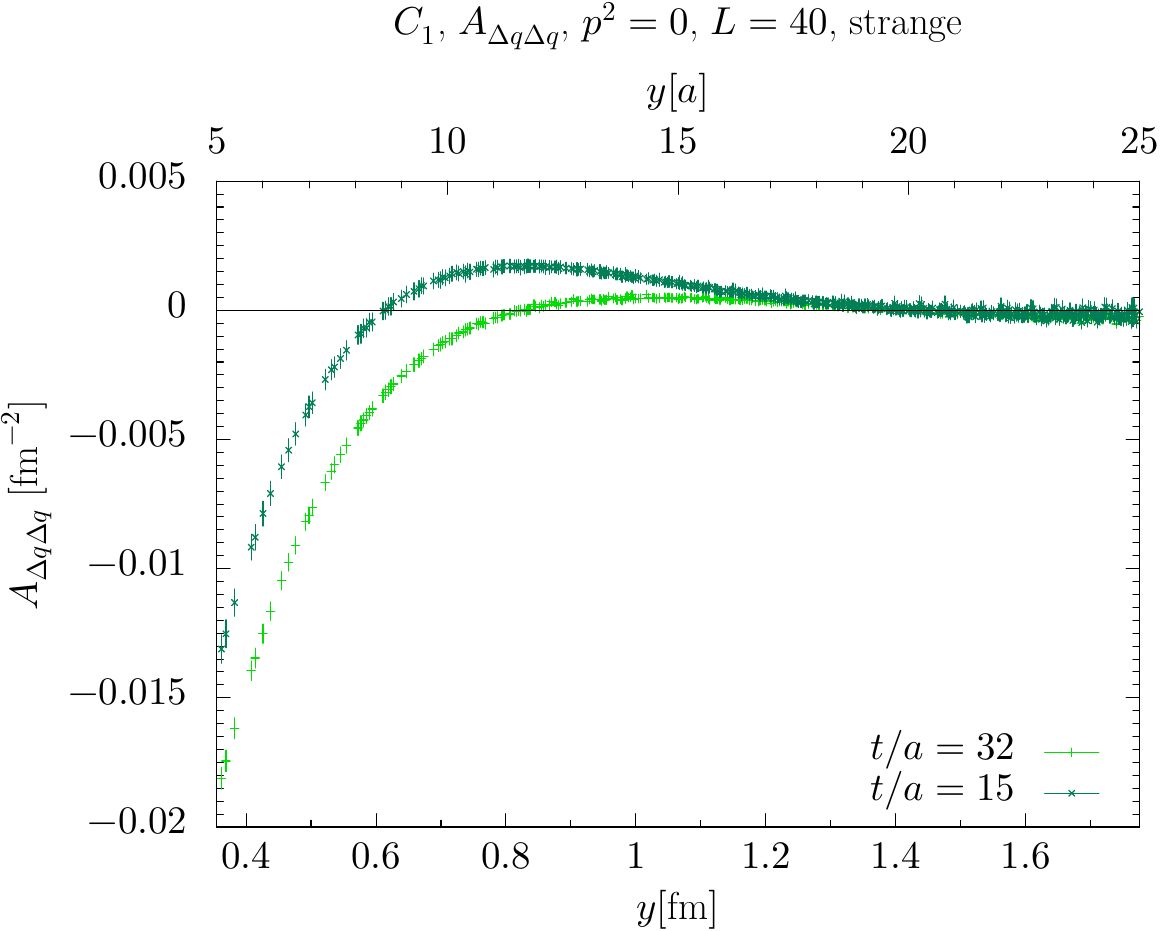}}
\hfill
\subfigure[$B_{\delta q \delta q}$, graph $C_1$, strange quarks]{
\includegraphics[height=14.6em,trim=0 0 0 17,clip]
{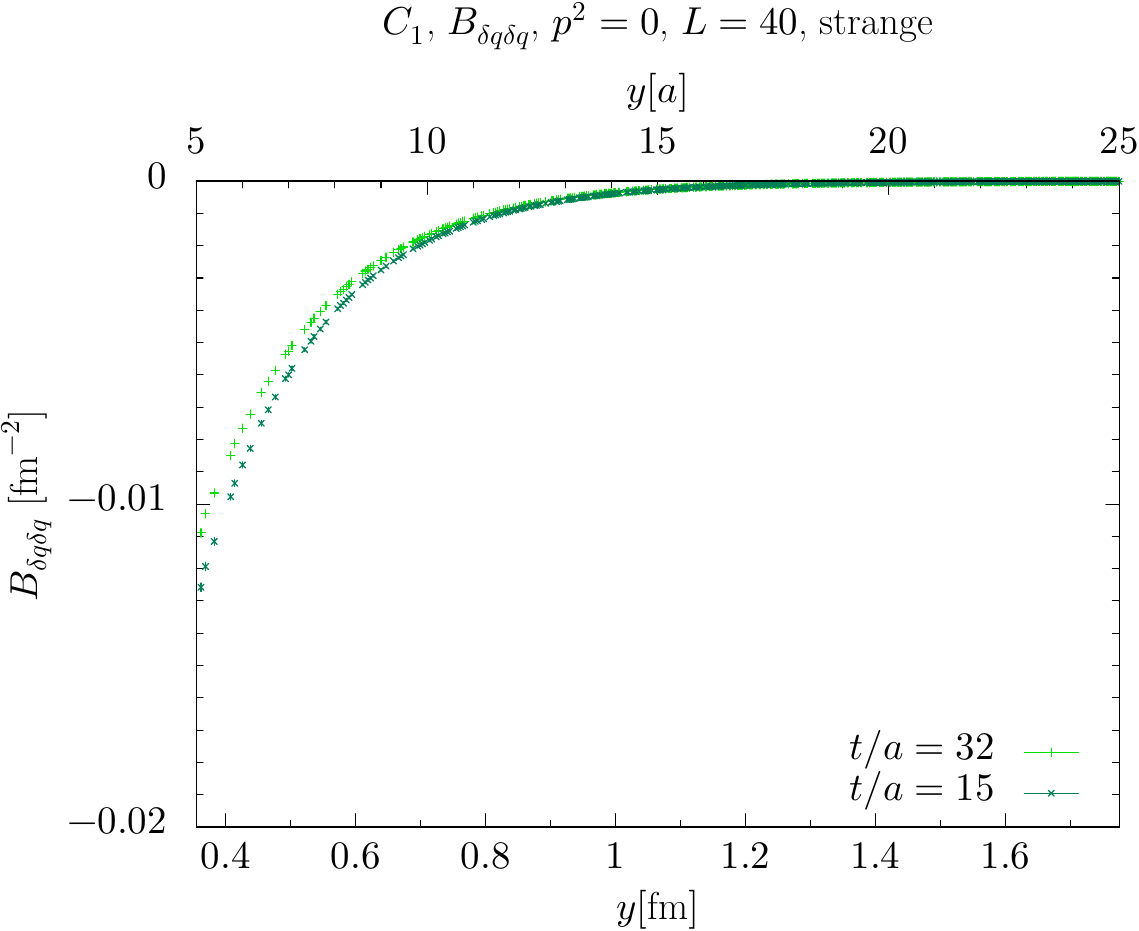}}
\caption{\label{fig:t32-comp} Comparison of twist-two functions extracted for graph $C_1$ with source-sink separations $t/a = 15$ or~$32$.  All data are for $L=40$, zero pion momentum, and subject to the cut \protect\eqref{cos-cut}.}
\end{center}
\end{figure*}

In general, we find that the results for the two source-sink separation agree
reasonably well for light quarks, as illustrated in the upper panels of \fig{\ref{fig:t32-comp}}.  For strange quarks, the data have smaller statistical errors and we can clearly see discrepancies between $t = 15 a$ and $32 a$, as shown in the lower panels of the figure.  Except for the case of $A_{\Delta q \Delta q}$, these discrepancies are, however, small when compared with the size of the twist-two functions.

In our data for charm quarks, the statistical signal and the agreement between the two source-sink separations is excellent for all twist-two functions, and even better than the one in \fig{\ref{fig:t32-comp-A_VV}}.  With the exception mentioned above, we thus find no indication for a sizeable contamination from excited states in our results.

\subsection{Volume dependence}
\label{sec:vol-comp}

\begin{figure*}
\begin{center}
\subfigure[$A_{q q}$, graph $C_1$\label{fig:vol-A_VV-C1}]{
\includegraphics[height=14.6em,trim=0 0 0 17,clip]
{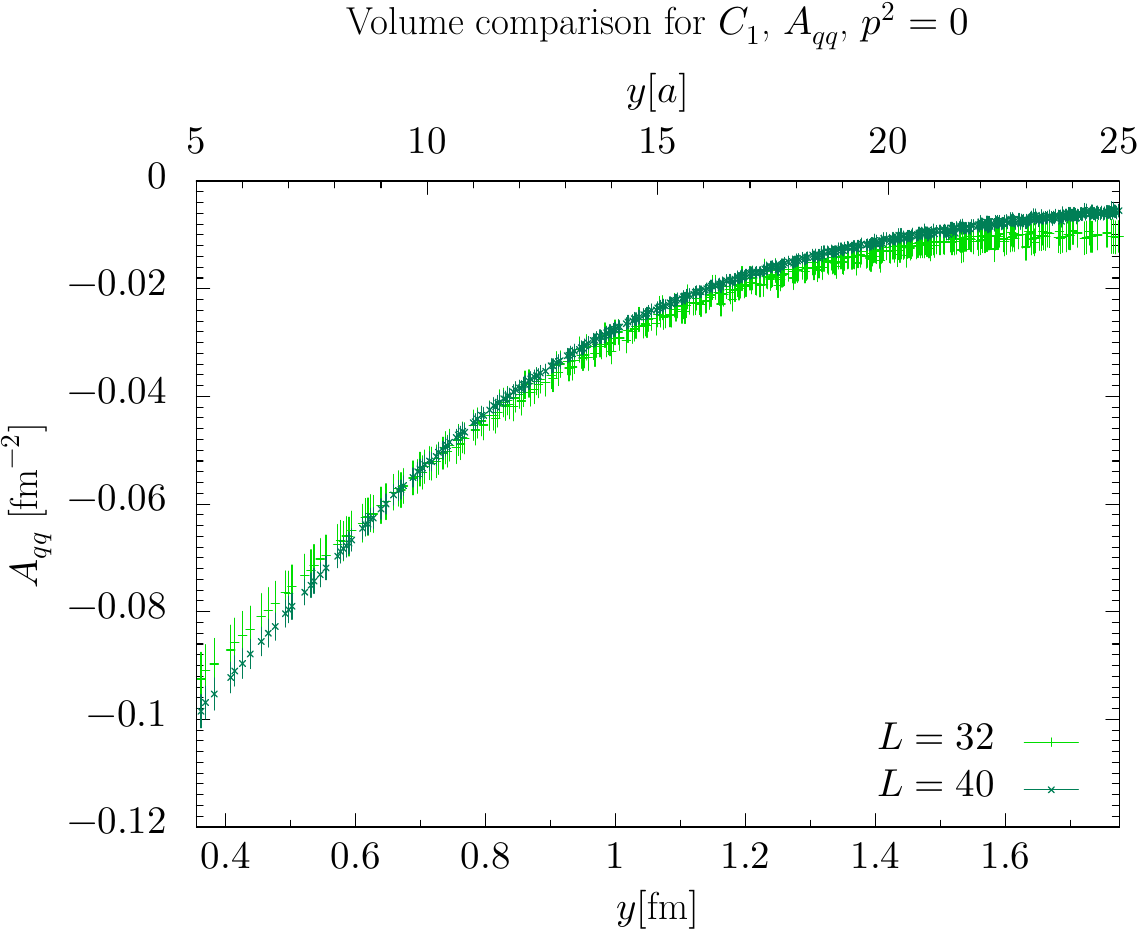}}
\hfill
\subfigure[$A_{\delta q \ms q}$, graph $C_1$\label{fig:vol-A_VT-C1}]{
\includegraphics[height=14.6em,trim=0 0 0 17,clip]
{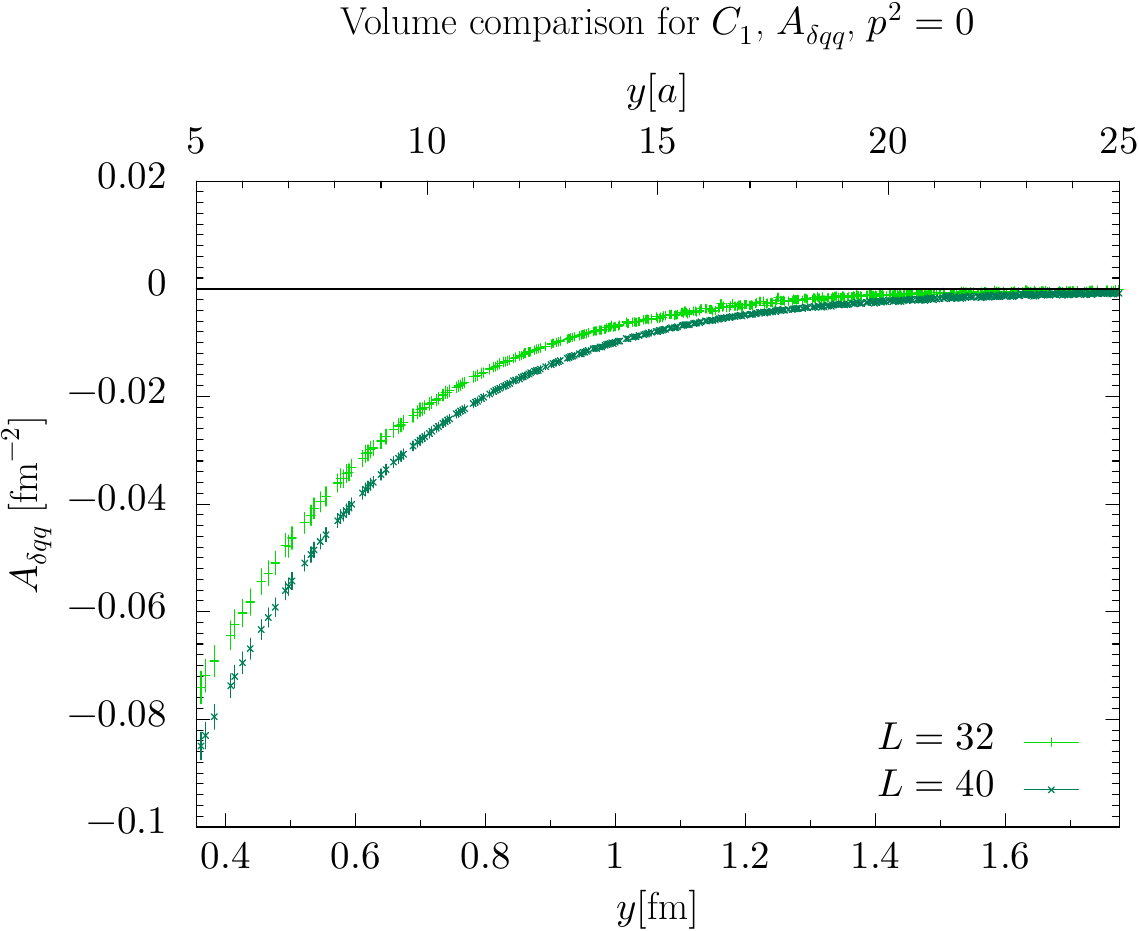}}
\\
\subfigure[$B_{\delta q \delta q}$, graph $C_1$\label{fig:vol-B_TT-C1}]{
\includegraphics[height=14.6em,trim=0 0 0 17,clip]
{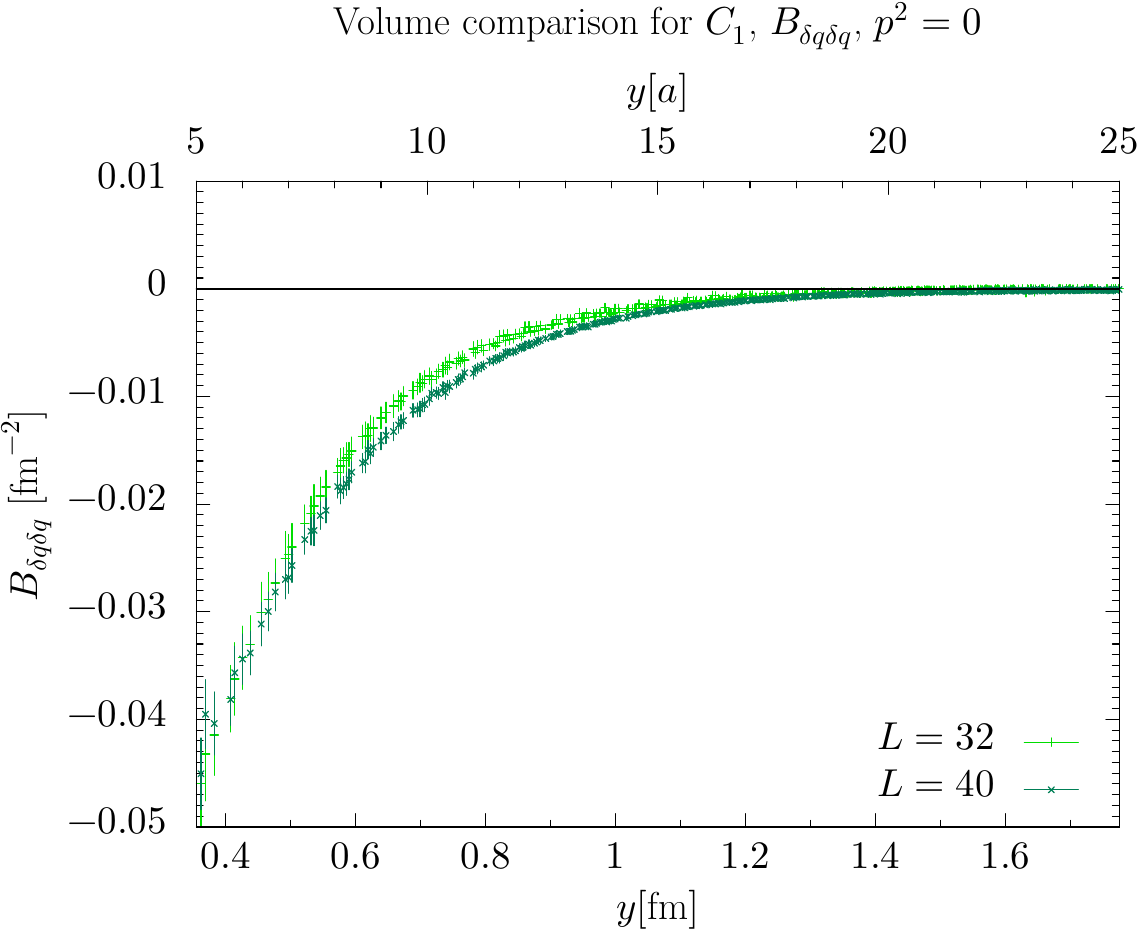}}
\hfill
\subfigure[$A_{\delta q \ms q}$, graph $C_2$\label{fig:vol-A_VT-C2}]{
\includegraphics[height=14.6em,trim=0 0 0 17,clip]
{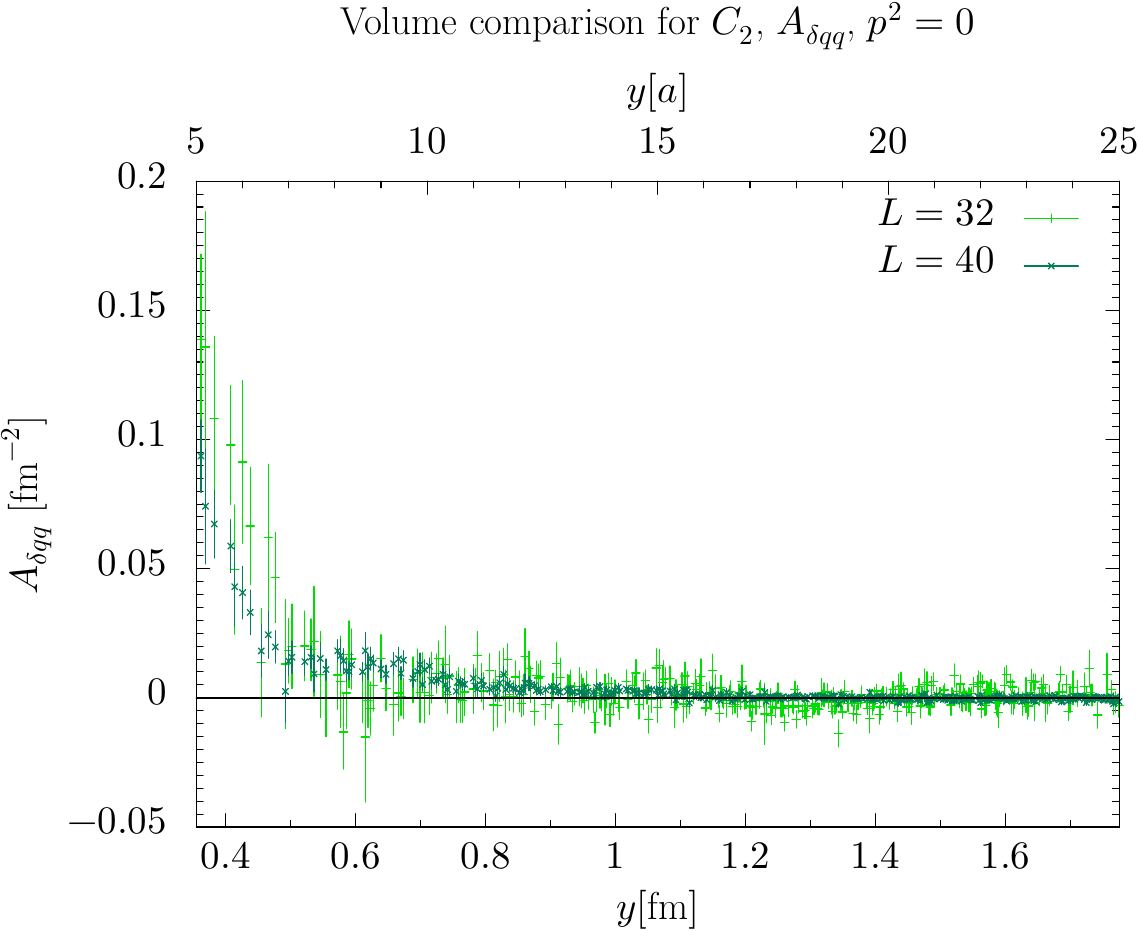}}
\\
\subfigure[$B_{\delta q \delta q}$, graph $A$\label{fig:vol-B_TT-A}]{
\includegraphics[height=14.6em,trim=0 0 0 17,clip]
{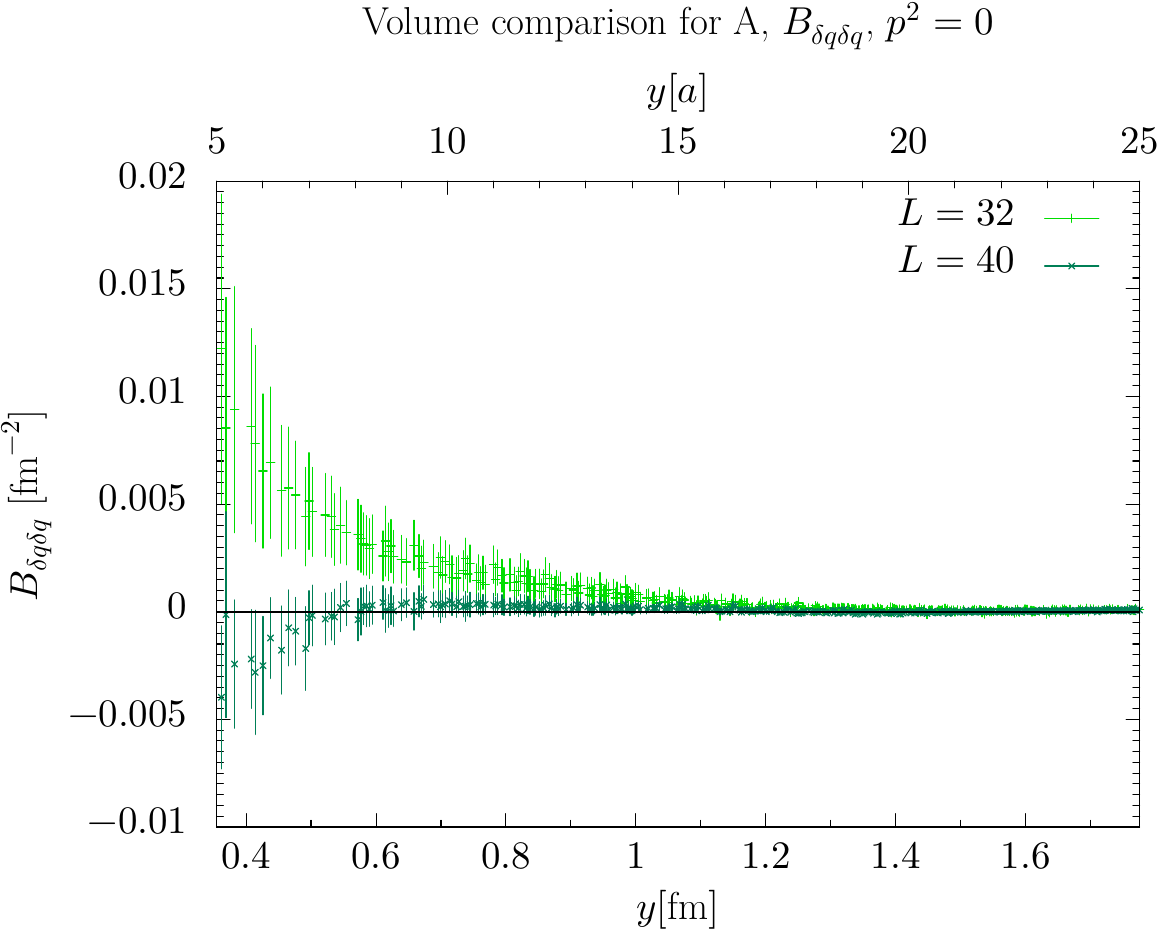}}
\hfill
\subfigure[$A_{\delta q \ms q}$, graph $S_2$\label{fig:vol-A_VT-S2}]{
\includegraphics[height=14.6em,trim=0 0 0 17,clip]
{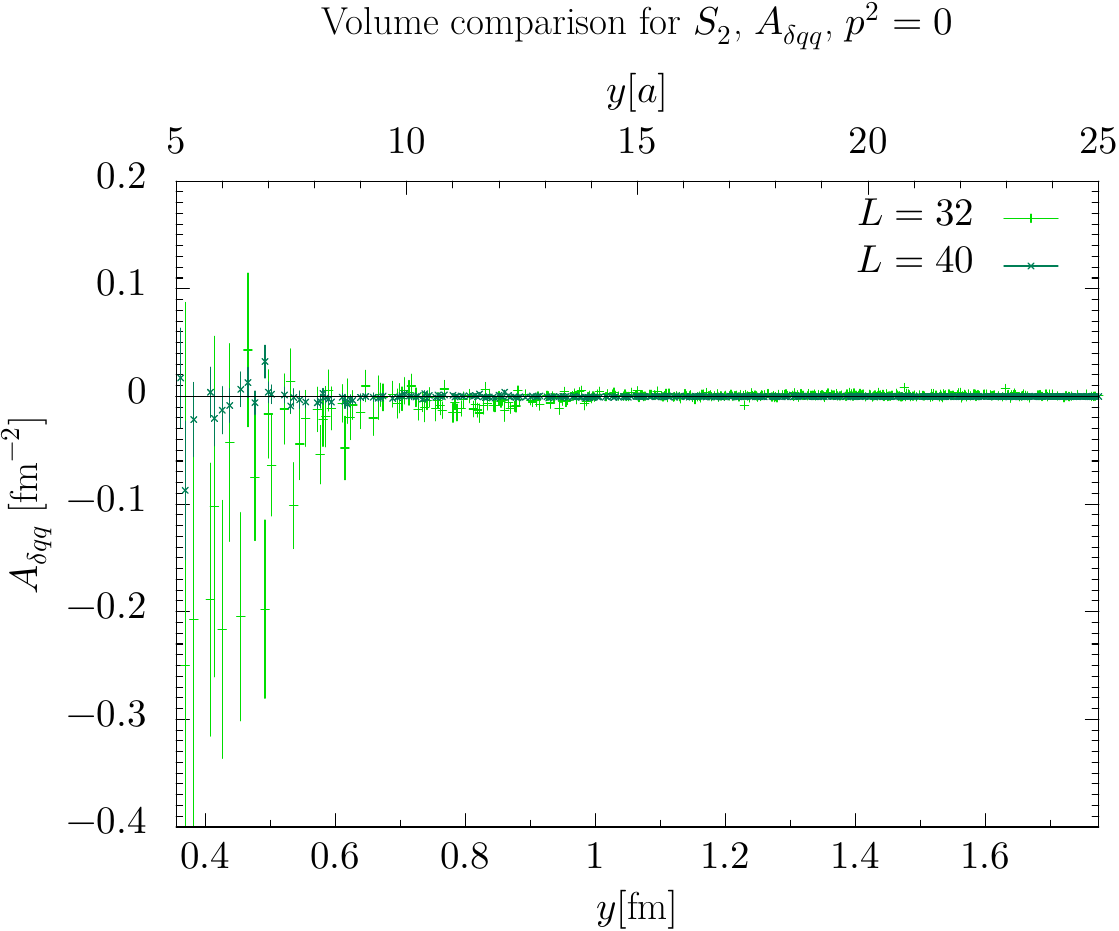}}
\caption{\label{fig:vol-comp} Comparison of data for the two different lattice sizes in our study.  All points are for zero pion momentum, light quarks, and subject to the cut \protect\eqref{cos-cut}.}
\end{center}
\end{figure*}

Let us finally compare our simulations for light quarks on the lattices with $L=40$ and $L=32$.  In general, the data for the smaller lattice have larger jackknife errors.  This is to  be expected from the parameters that determine the statistical averaging in our simulations.  Details for these are given in \tab{2} of \cite{Bali:2018nde}.

For twist-two functions with a small relative error, we typically find a weak volume dependence compared with the size of the functions themselves,
as shown in panels (a) to (c) of \fig{\ref{fig:vol-comp}}.  In the case of panel (b), this dependence is, however, statistically significant.
For functions that have large relative errors, the volume dependence appears to be more pronounced in some cases, especially at low $y$.  An example is \fig{\ref{fig:vol-B_TT-A}}.  One may take this as a general warning against over-interpreting statistically weak signals in our simulations.

\section{Results for zero pion momentum}
\label{sec:results}

In this section, we present our results for the twist-two functions at $py = 0$.  All data shown in the following are for zero pion momentum and have been extracted from the lattice with $L=40$ with our standard source-sink separation $t = 15a$.  The data selection described at the end of \sect{\ref{sec:iso-boost}} removes regions in which we see strong lattice artefacts in the form of broken rotational or boost symmetry.

As we explained in \sect{\ref{sec:skewed-dpds}}, twist-two functions at $py = 0$ are not directly related to the Mellin moments of DPDs.  Instead, they are Mellin moments of skewed DPDs, integrated over the skewness parameter $\zeta$.  As seen in \fig{\ref{fig:dpd-support}}, these moments receive contributions from parton configurations that are different from those in a DPD at $\zeta=0$.  When interpreting the results of the present section, we will \emph{assume} that these configurations are not dominant, and that the qualitative features of invariant functions at $py = 0$ are the same as for Mellin moments of DPDs at $\zeta=0$.  The results presented in \sect{\ref{sec:mellin-moments}} will lend support to this assumption.

\begin{figure*}[!b]
\begin{center}
\includegraphics[width=0.99\textwidth]{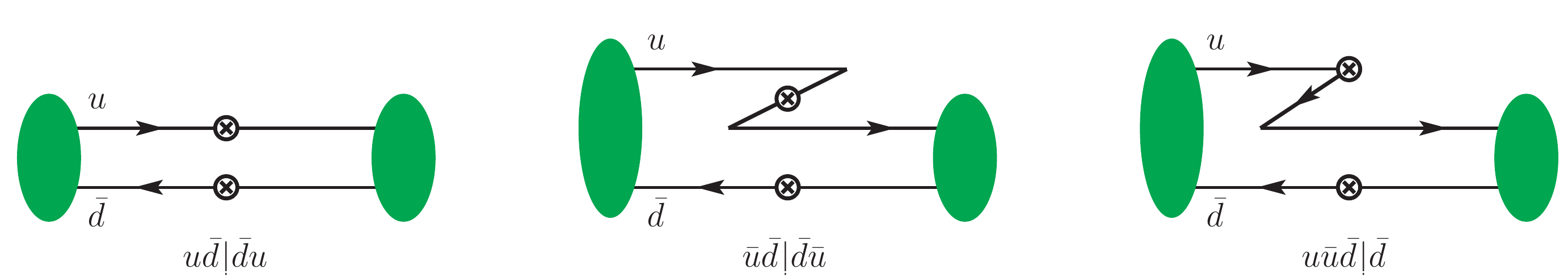}
\end{center}
\caption{\label{fig:parton-graphs-C1} Examples for the partonic regimes of graph $C_1$ in a $\pi^+$.  The notation $u \bar{u} \bar{d} \ms| \bar{d}$ is the same as in \fig{\protect\ref{fig:dpd-support}}, i.e.\ the partons to the left of the vertical bar belong to the pion on the left of the graph.}
\end{figure*}

\begin{figure*}
\begin{center}
\includegraphics[width=0.99\textwidth]{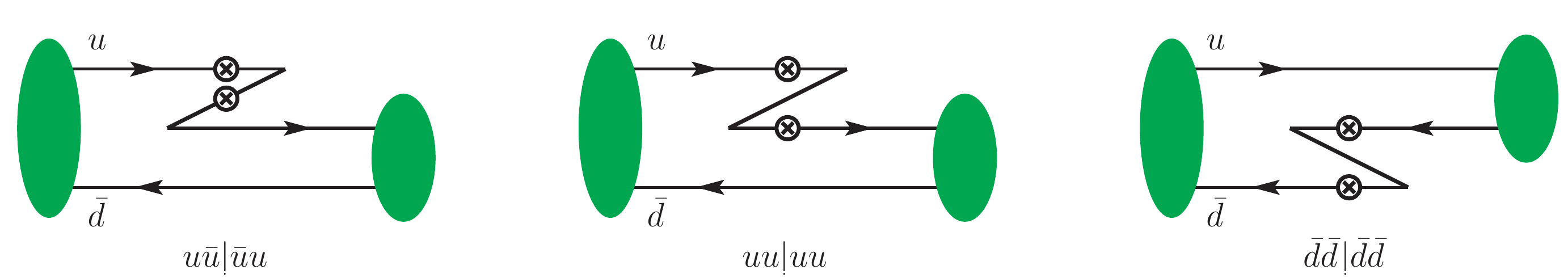}
\\
\includegraphics[width=0.64\textwidth]{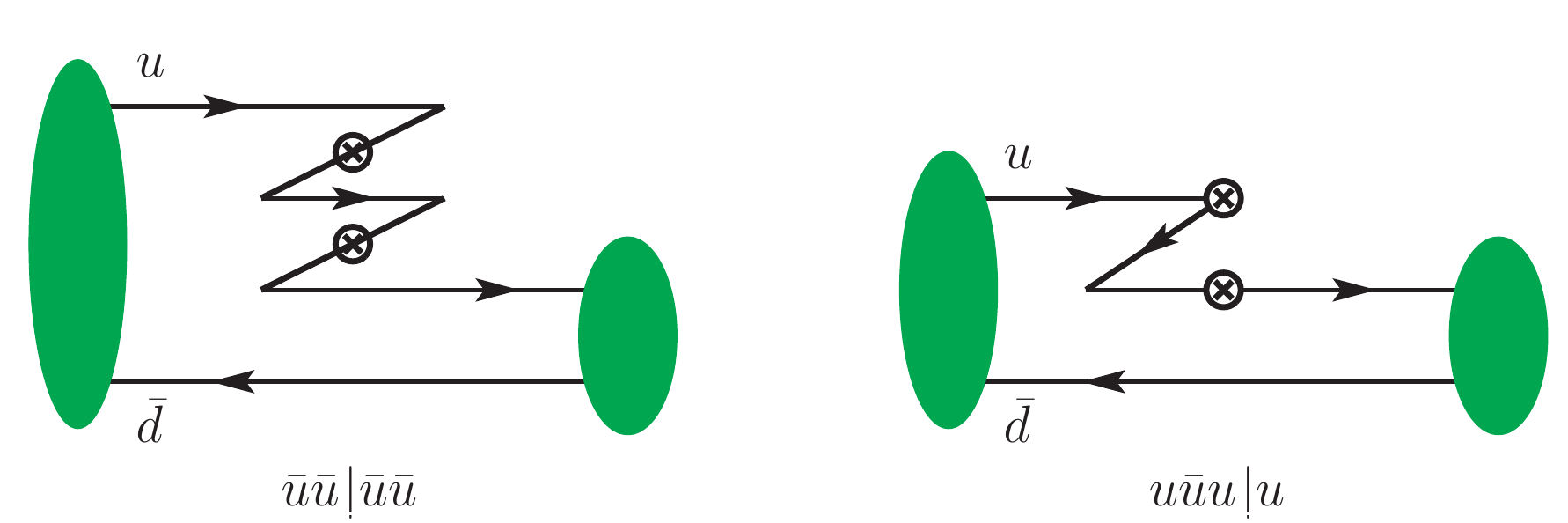}
\end{center}
\caption{\label{fig:parton-graphs-C2} As \fig{\protect\ref{fig:parton-graphs-C1}}, but for graph $C_2$.}
\end{figure*}

Notice that each of the lattice graphs in \fig{\ref{fig:contractions}} can contribute to each of the partonic regimes shown in \fig{\ref{fig:dpd-support}}.  Examples for different regimes of the connected graphs $C_1$ and $C_2$ are shown in \figs{\ref{fig:parton-graphs-C1}} and \ref{fig:parton-graphs-C2}.


\subsection{Comparison of graphs}
\label{sec:graph-comparision}

\begin{figure*}
\begin{center}
\subfigure[$A_{q q}$, light quarks\label{fig:light-graphs-1-A_VV}]{
\includegraphics[width=0.48\textwidth,trim=0 0 0 17,clip]
{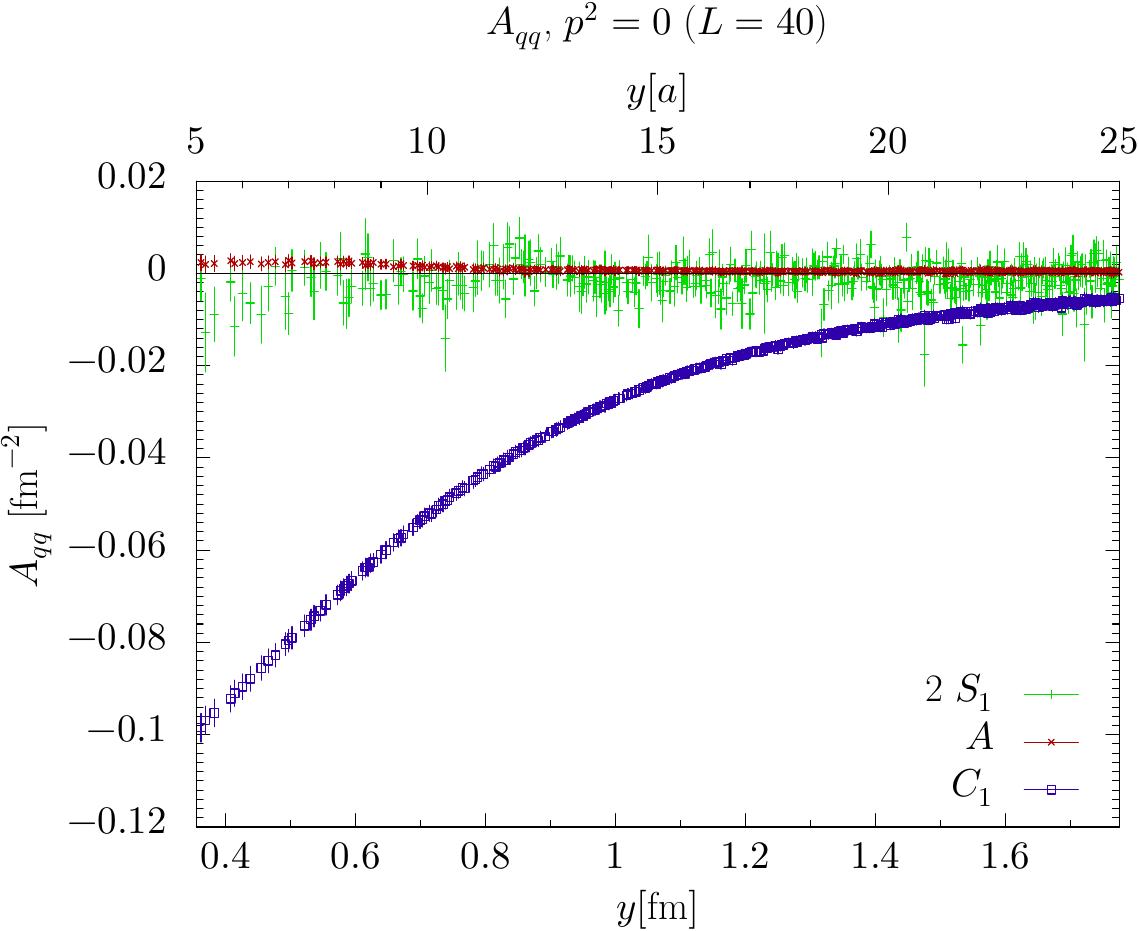}
\hfill
\includegraphics[width=0.48\textwidth,trim=0 0 0 17,clip]
{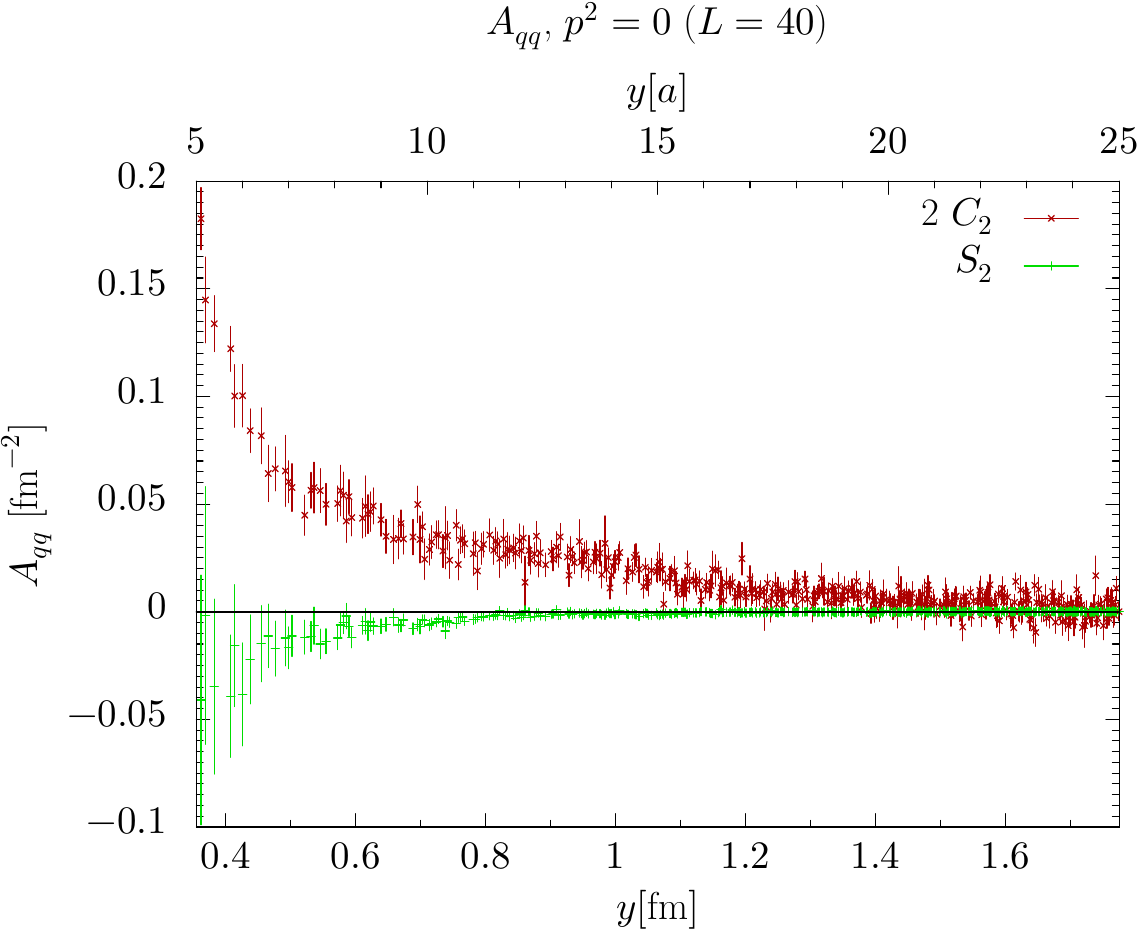}}
\\
\subfigure[$A_{\Delta q \Delta q}$, light quarks]{
\includegraphics[width=0.48\textwidth,trim=0 0 0 17,clip]
{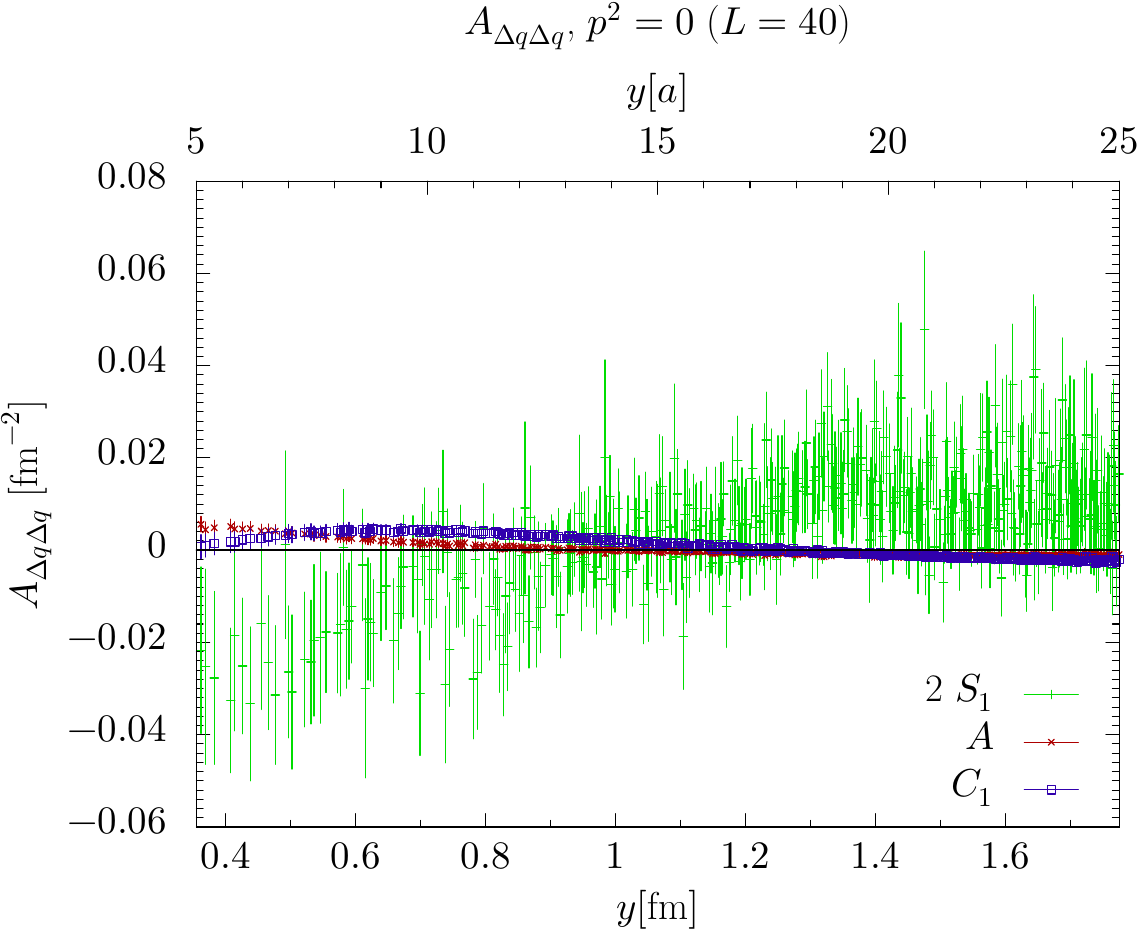}
\hfill
\includegraphics[width=0.48\textwidth,trim=0 0 0 17,clip]
{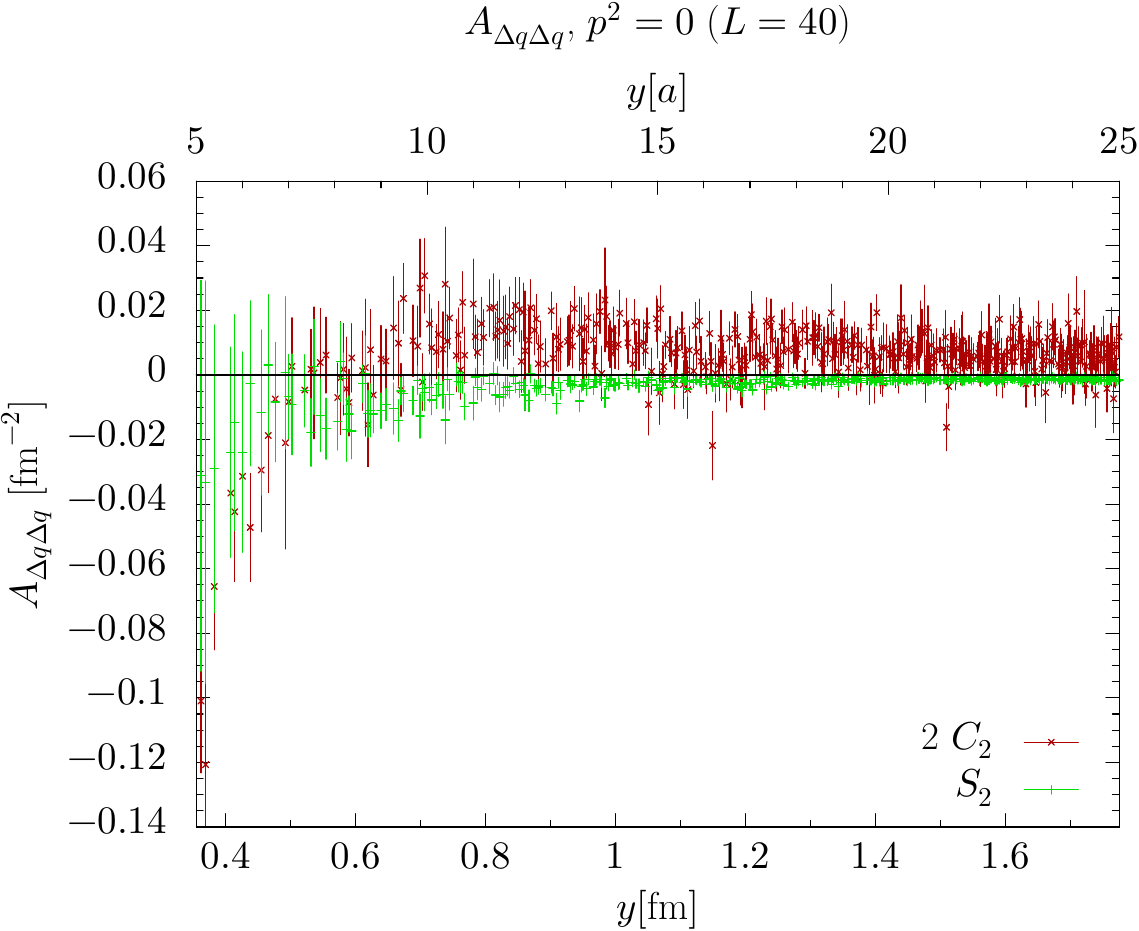}}
\\
\subfigure[$A_{\delta q \ms q}$, light quarks]{
\includegraphics[width=0.48\textwidth,trim=0 0 0 17,clip]
{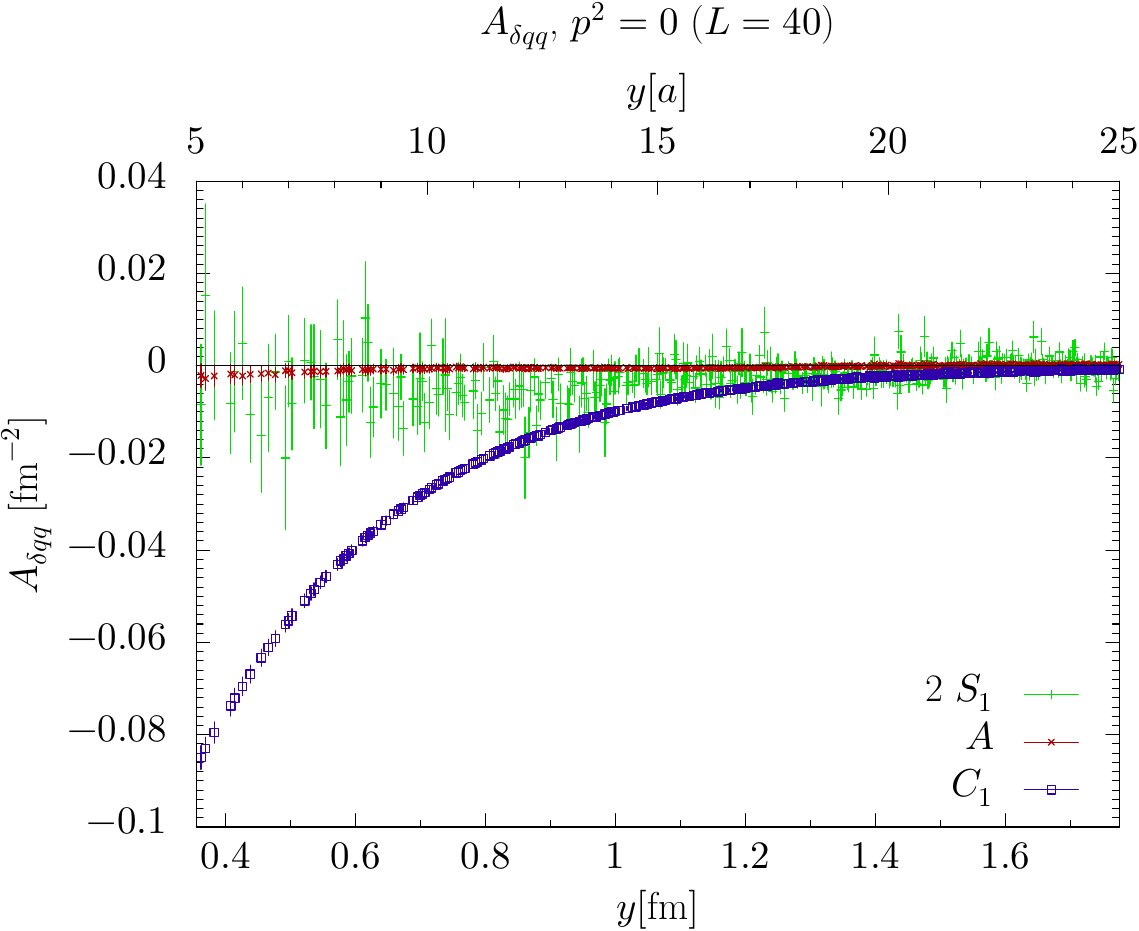}
\hfill
\includegraphics[width=0.48\textwidth,trim=0 0 0 17,clip]
{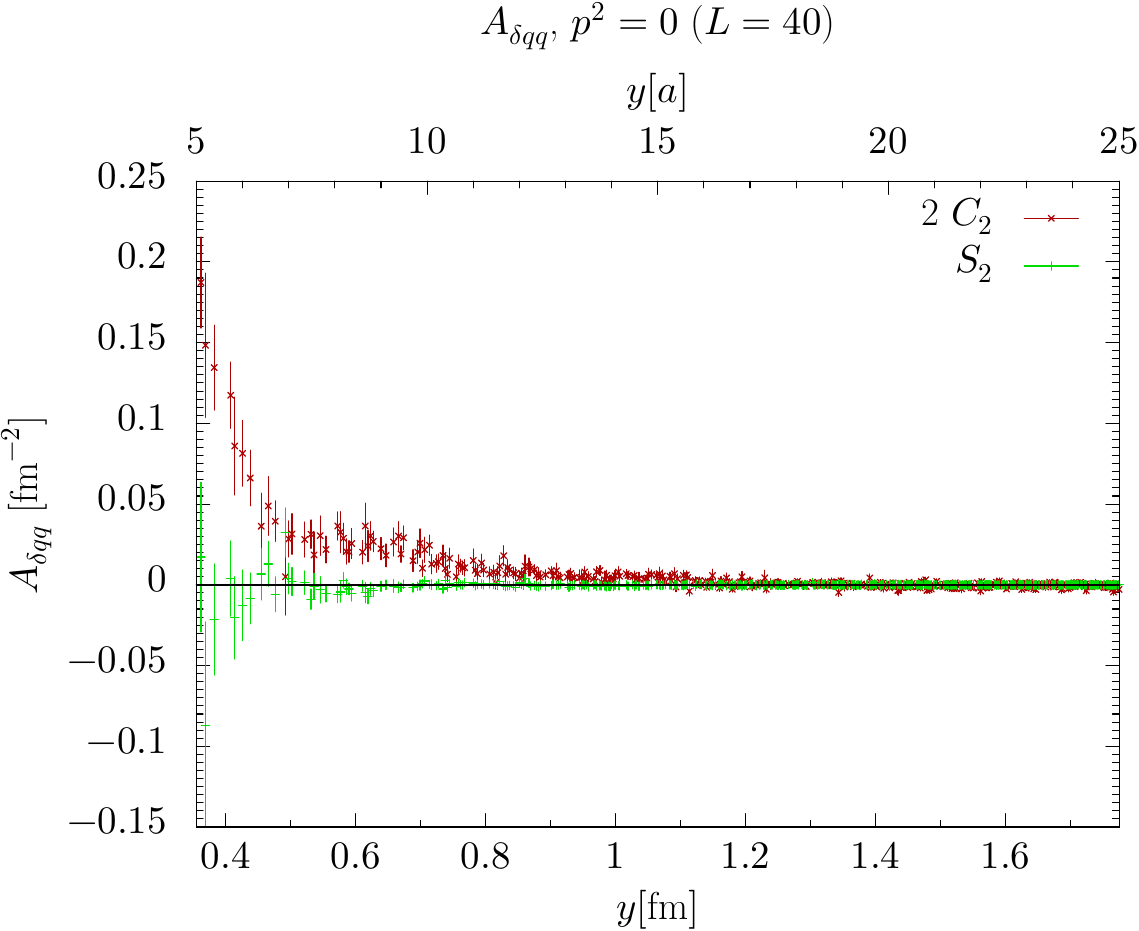}}
\caption{\label{fig:light-graphs-1} Contributions of the different lattice graphs to twist-two functions for light quarks.}
\end{center}
\end{figure*}

\begin{figure*}
\begin{center}
\subfigure[$A_{\delta q \delta q}$, light quarks]{
\includegraphics[width=0.48\textwidth,trim=0 0 0 17,clip]
{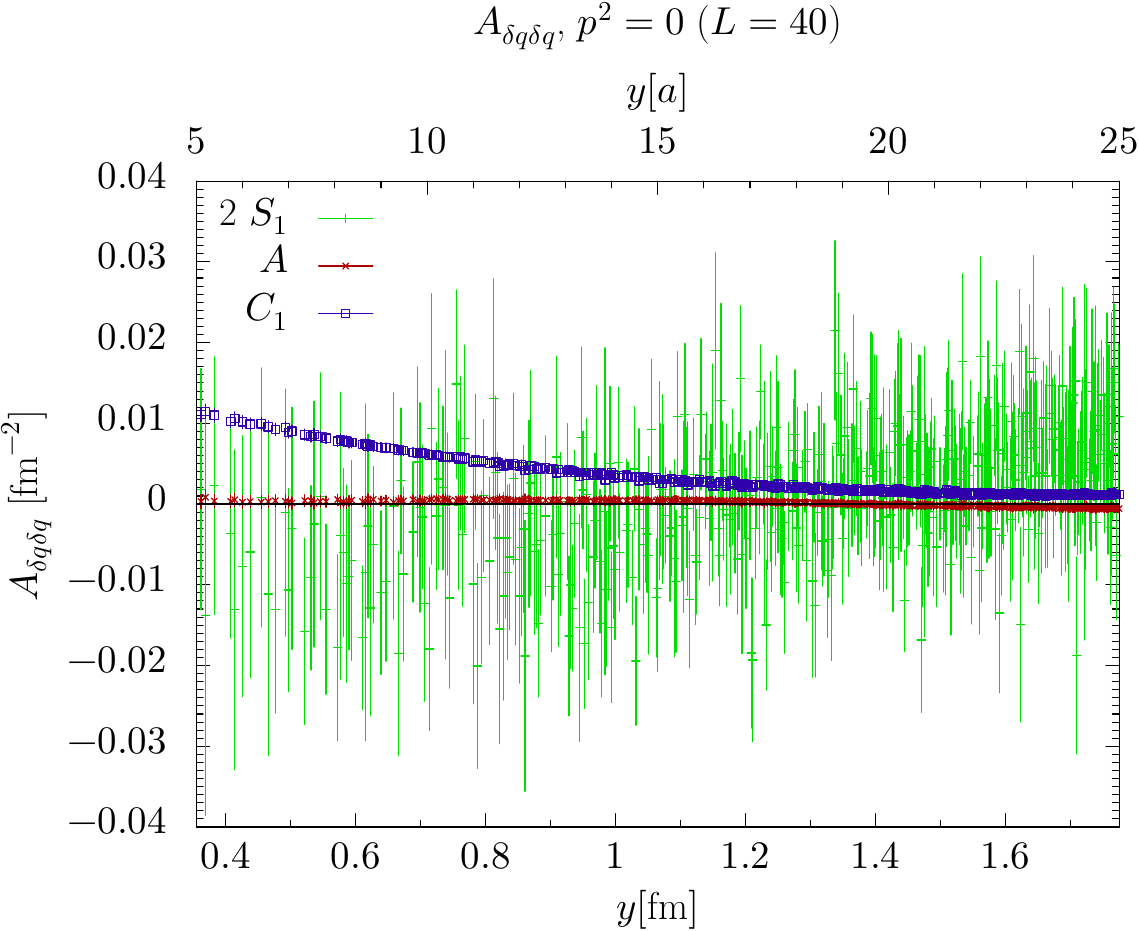}
\hfill
\includegraphics[width=0.48\textwidth,trim=0 0 0 17,clip]
{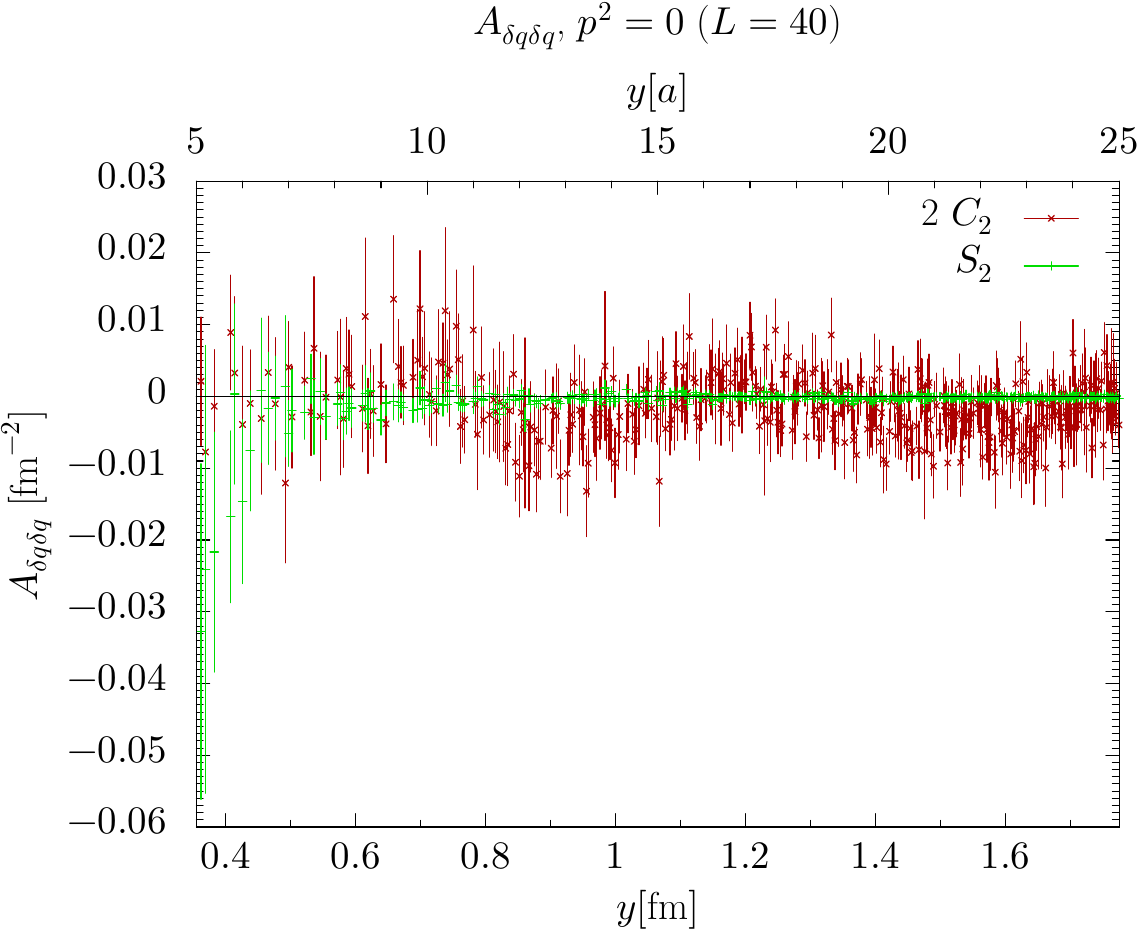}}
\\
\subfigure[$B_{\delta q \delta q}$, light quarks]{
\includegraphics[width=0.48\textwidth,trim=0 0 0 17,clip]
{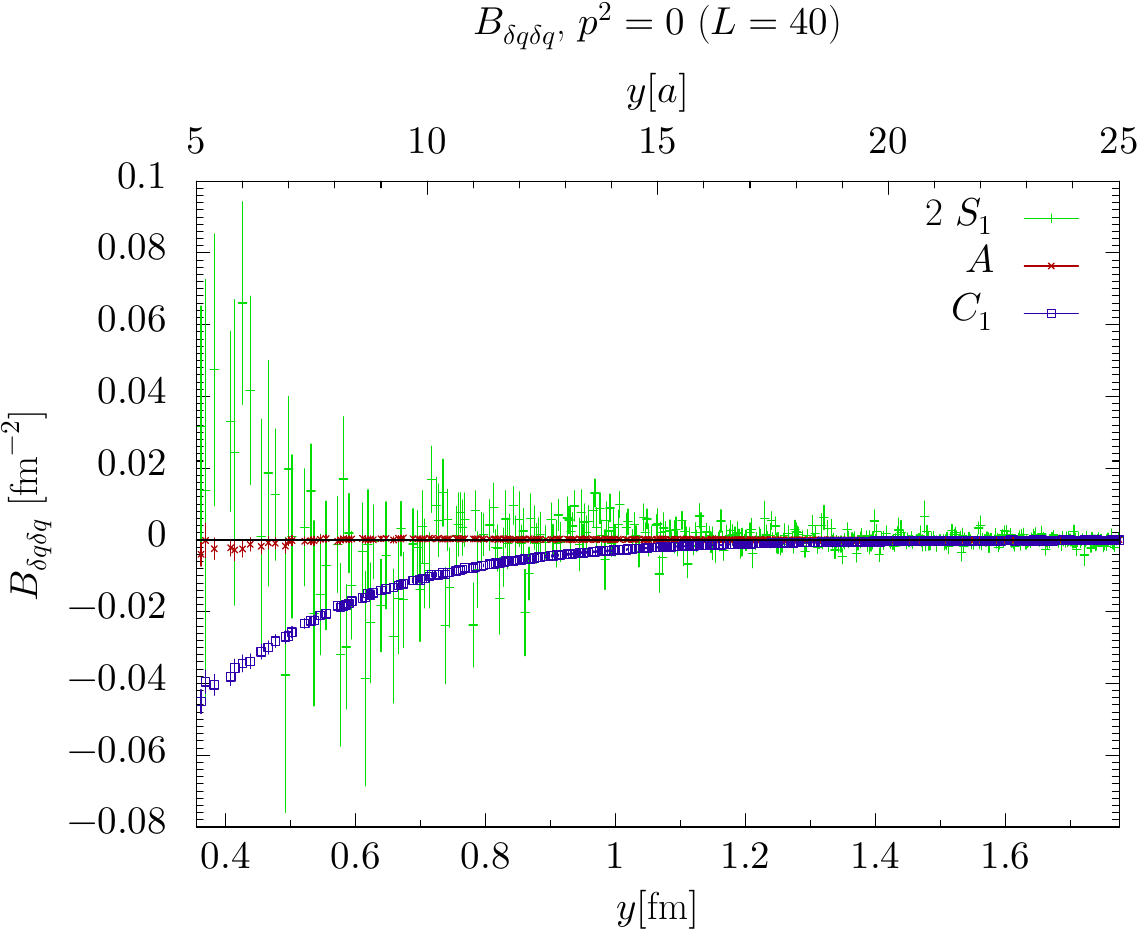}
\hfill
\includegraphics[width=0.48\textwidth,trim=0 0 0 17,clip]
{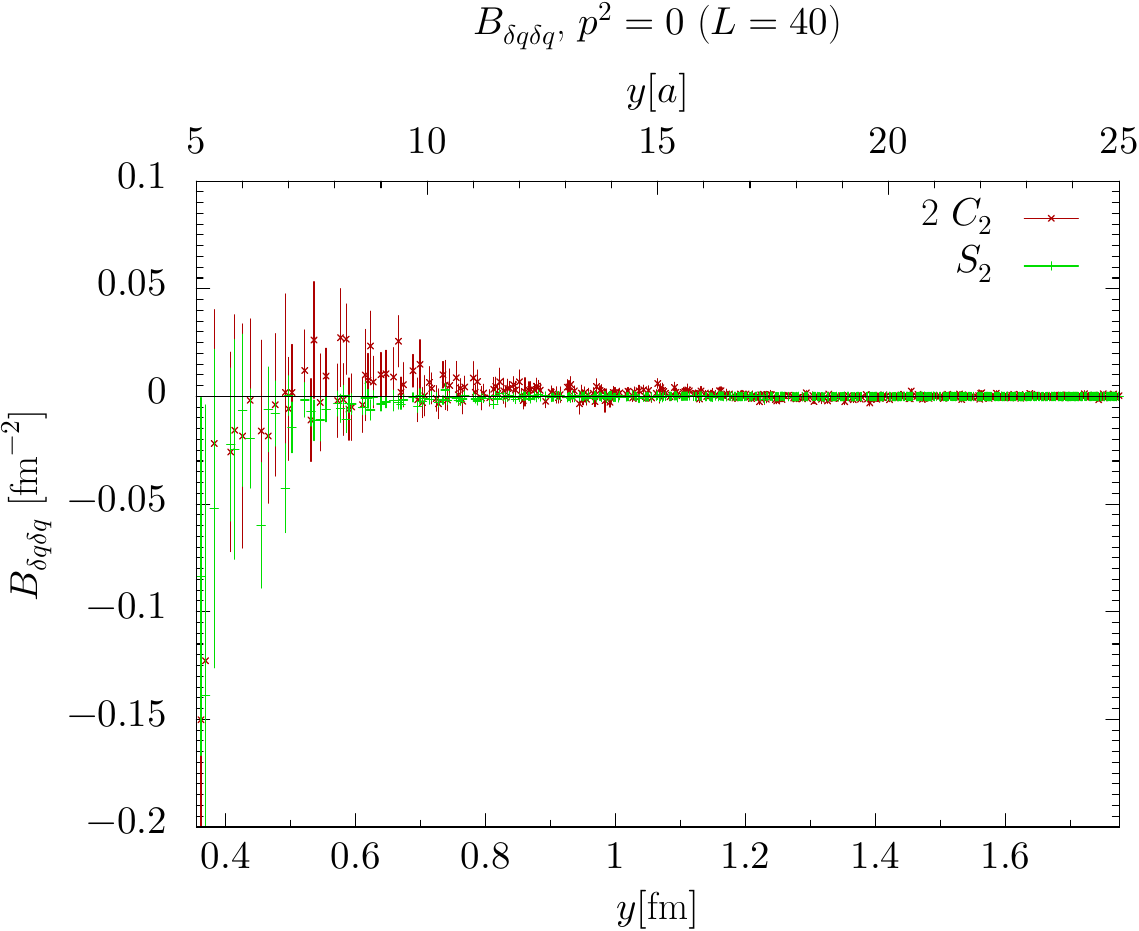}}
\caption{\label{fig:light-graphs-2} Continuation of \fig{\protect\ref{fig:light-graphs-1}}.}
\end{center}
\end{figure*}

In \figs{\ref{fig:light-graphs-1}} and \ref{fig:light-graphs-2}, we compare the contributions from different lattice graphs to the twist-two functions for light quarks.  The contributions from graphs $S_1$ and $C_2$ are multiplied with a factor $2$ in the figures, since they always appear with this weight in physical matrix elements according to \eqref{eq:phys-mat-els}.

For all twist-two functions except $A_{\Delta q \Delta q}$, graph $C_1$ gives a very clear signal, which is positive for $A_{\delta q \delta q}$ and negative for the other functions.  By comparison, the signal for the annihilation graph $A$ is smaller than the one for $C_1$ by an order of magnitude or more, except for $y > 20a$, where the statistical errors prevent us from making a clear statement.  The function $A_{\Delta q \Delta q}$ shows a different behaviour, with $C_1$ and $A$ being of similar size and much smaller than $C_1$ for all other twist-two functions.  We recall from \sect{\ref{sec:iso-boost}} that $A_{\Delta q \Delta q}$ is more strongly affected by lattice artefacts than the other channels, see \fig{\ref{fig:C1-mom-A_AA}}.  \rev{We nevertheless discuss this function here and in the following, because even with the large systematic uncertainties we have seen, its qualitative behaviour and overall size compared with the other polarisation channels are significant results of our simulations.}

A clear signal for the connected graph $C_2$ is only seen for $A_{q q}$ and $A_{\delta q \ms q}$, with a sign opposite to the one for graph $C_1$.  This signal is most important at small $y$.  For the graphs $S_1$ and $S_2$ with one disconnected fermion loop, the signal we obtain is rather noisy in all channels.  For graph $D$ with two disconnected fermion loops, the signal after vacuum subtraction is even more noisy and not shown.

\begin{figure*}
\begin{center}
\subfigure[$|A_{q q}|$, graphs $C_1$ and $A$]{
\includegraphics[width=0.48\textwidth,trim=0 0 0 17,clip]
{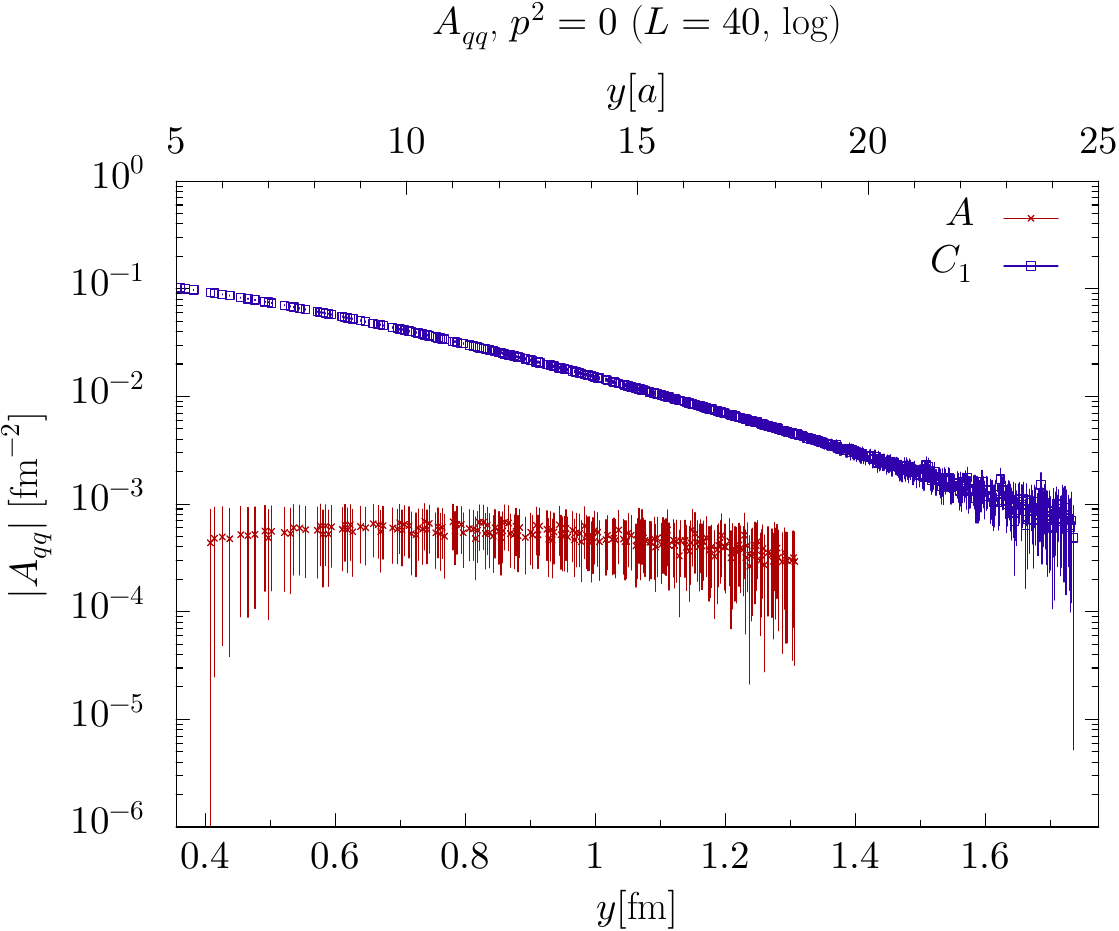}}
\hfill
\subfigure[$|A_{\delta q \delta q}|$, graphs $C_1$ and $A$]{
\includegraphics[width=0.48\textwidth,trim=0 0 0 17,clip]
{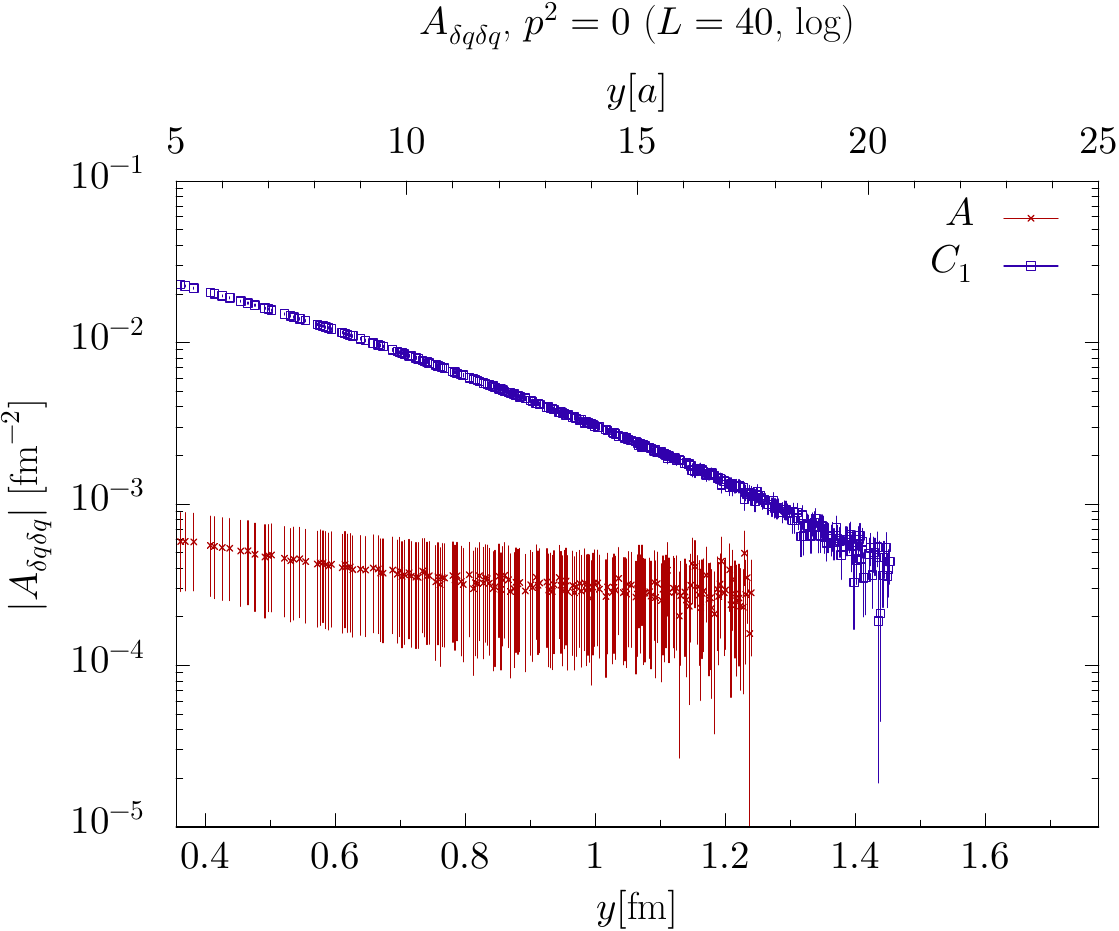}}
\caption{\label{fig:strange-graphs} Comparison of graphs $A$ and $C_1$ for strange quarks.  Here and in subsequent figures with a logarithmic scale, we stop showing data for individual quark masses at a value of $y$ beyond which the error bars become so large that they would obscure the plot.}
\end{center}
\end{figure*}

From our simulations with the strange quark mass, we have only data for graphs $C_1$ and $A$.  In all channels, we obtain an excellent signal for $C_1$, whereas for $A$ the statistical significance is typically not much larger than one standard deviation.  In the region $5a \le y \le 15a$, we find that $A$ is smaller than $C_1$ by one to two orders of magnitude, except for $A_{\Delta q \Delta q}$.  For $A_{q q}$ and $A_{\delta q \delta q}$, we see in \fig{\ref{fig:strange-graphs}} that the behaviour of $A$ is quite flat, unlike the one of $C_1$, so that at large $y$ the two graphs become more comparable in size.  As in the case of light quarks, the function $A_{\Delta q \Delta q}$ behaves differently, with graph $A$ being smaller than $C_1$ at $y \sim 5a$ and the data for both graphs having a zero crossing a bit below $y = 9a$.  Recall, however, that also for strange quarks we see stronger lattice artefacts in $A_{\Delta q \Delta q}$ than in other channels, as seen in \fig{\ref{fig:t32-comp-A_AA}}.

From our simulations with the charm quark mass, we have data for all graphs except $S_2$.  A clear nonzero signal is seen for $C_1$ and $C_2$ up to $y \sim 10a$ to $15a$, with $2 C_2$ being smaller than $C_1$ by at least one order of magnitude.
The signal for $A$ and $S_1$ is in general consistent with zero.  The only exception to this is $A_{\delta q \ms q}$.  For this function, we see a clear signal for $2 S_1$ at $y$ around $5a$, which is about 50 times smaller than the one for $C_1$.  We also see a weak $1\sigma$ signal for $A$, which we do not wish to over-interpret.

By and large, we find that for all quark masses the only graphs that give signals of appreciable size are $C_1$ and, in several cases, $C_2$.  We therefore take a closer look at these graphs in the next subsection.  The annihilation graph is negligible, except in the case of $A_{\Delta q \Delta q}$ for light or strange quarks, where the signal from graph $C_1$ is small by itself.  Disconnected graphs either have a negligibly small signal or large statistical errors.


\subsection{Results for connected graphs}
\label{sec:connected-graphs}

The contribution of graph $C_1$ to the twist-two function $A_{q q}$ for unpolarised partons is negative for all three quark masses in our study.  We recall from \eqref{eq:mellin-mom-def} that the regime with a quark and an antiquark in the pion contributes with a negative sign to the lowest Mellin moment of a DPD.  The same holds for the Mellin moment of a skewed DPD, and hence for $A_{q q}$ at $py=0$.  A negative sign of $A_{q q}$ is easily understood by the dominance of the valence $q\bar{q}$ Fock state, which is probed by graph $C_1$ as shown in the first panel of \fig{\ref{fig:parton-graphs-C1}}.

The situation is different for graph $C_2$, whose partonic representation always involves a higher Fock state of the pion.  The $Z$-graphs in \fig{\ref{fig:parton-graphs-C2}} probe the $q q$, $\bar{q}\bar{q}$ and $q\bar{q}$ regimes in a similar manner.  We find that for all quark masses, the contribution of $C_2$ to $A_{q q}$ is positive, which means that for a given distance $y$ this graph gives a larger probability for finding a $q q$ or $\bar{q}\bar{q}$ rather than a $q\bar{q}$ pair.

\begin{figure*}
\begin{center}
\subfigure[$- A_{q q}$, graph $C_1$]{
\includegraphics[width=0.48\textwidth,trim=0 0 0 17,clip]
{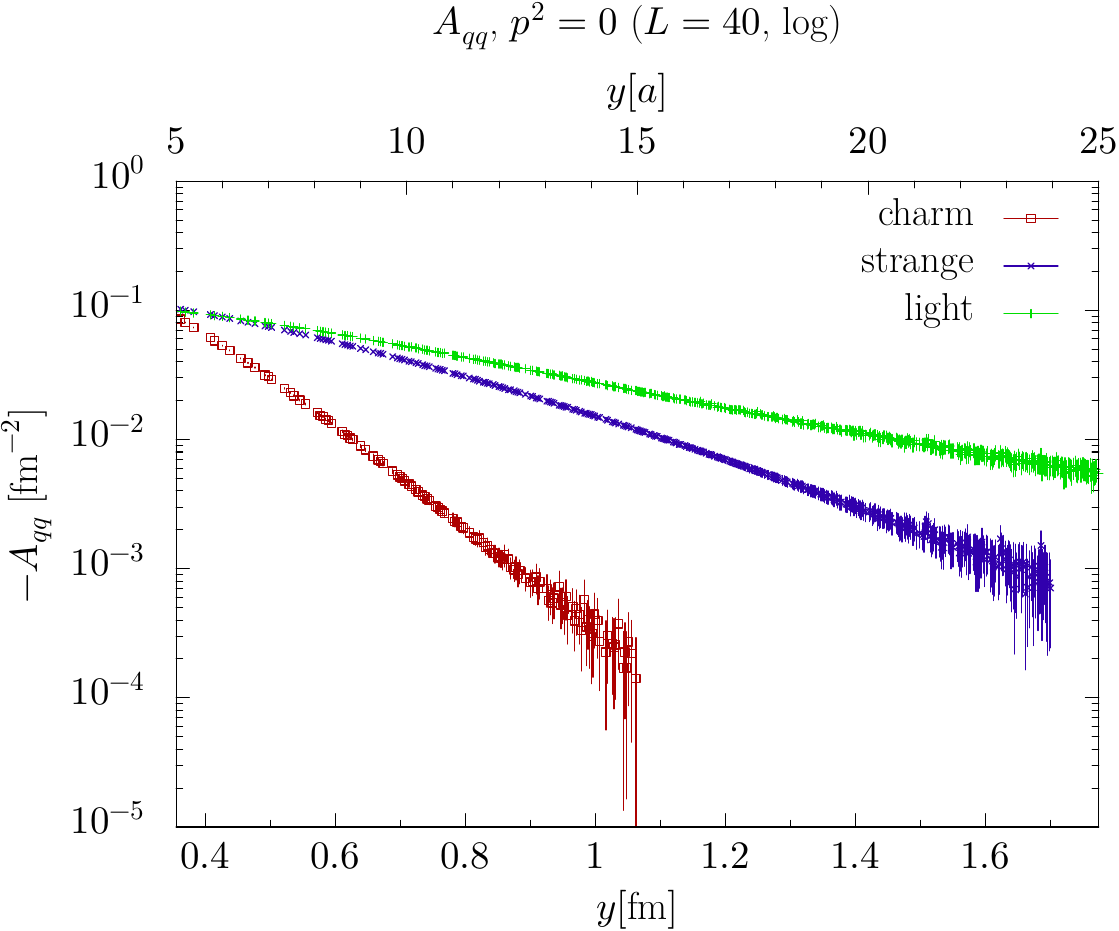}}
\hfill
\subfigure[$- m A_{\delta q \ms q}$, graph $C_1$]{
\includegraphics[width=0.48\textwidth,trim=0 0 0 17,clip]
{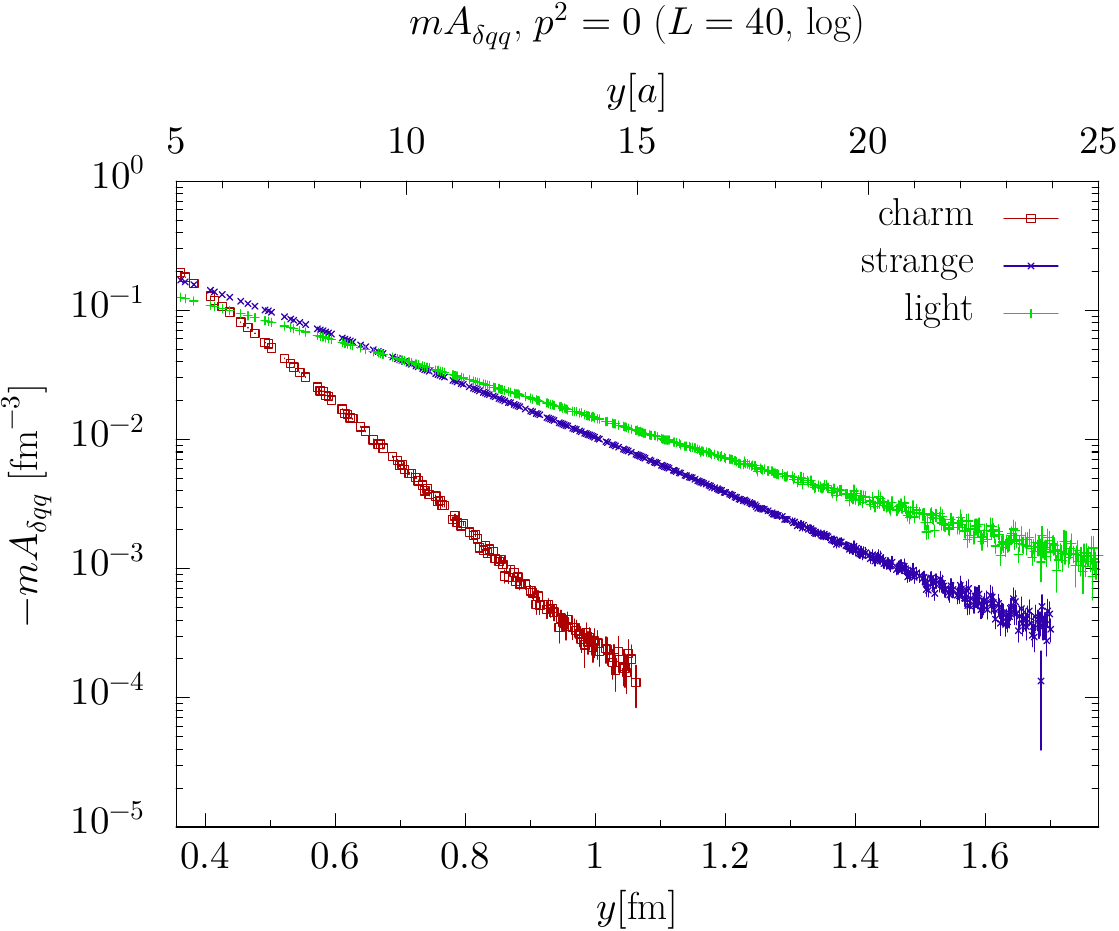}}
\\
\subfigure[$A_{\delta q \delta q}$, graph $C_1$]{
\includegraphics[width=0.48\textwidth,trim=0 0 0 17,clip]
{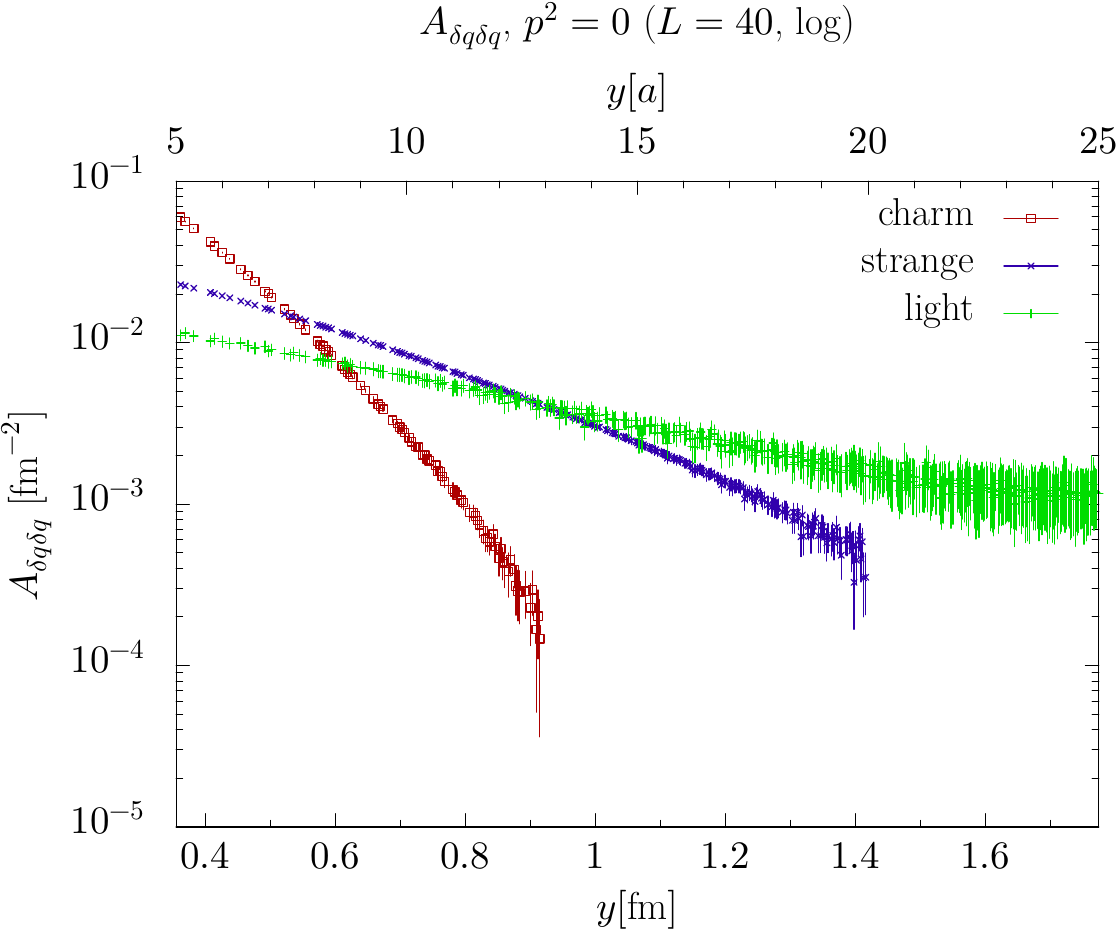}}
\hfill
\subfigure[$- m^2 B_{\delta q \delta q}$, graph $C_1$]{
\includegraphics[width=0.48\textwidth,trim=0 0 0 17,clip]
{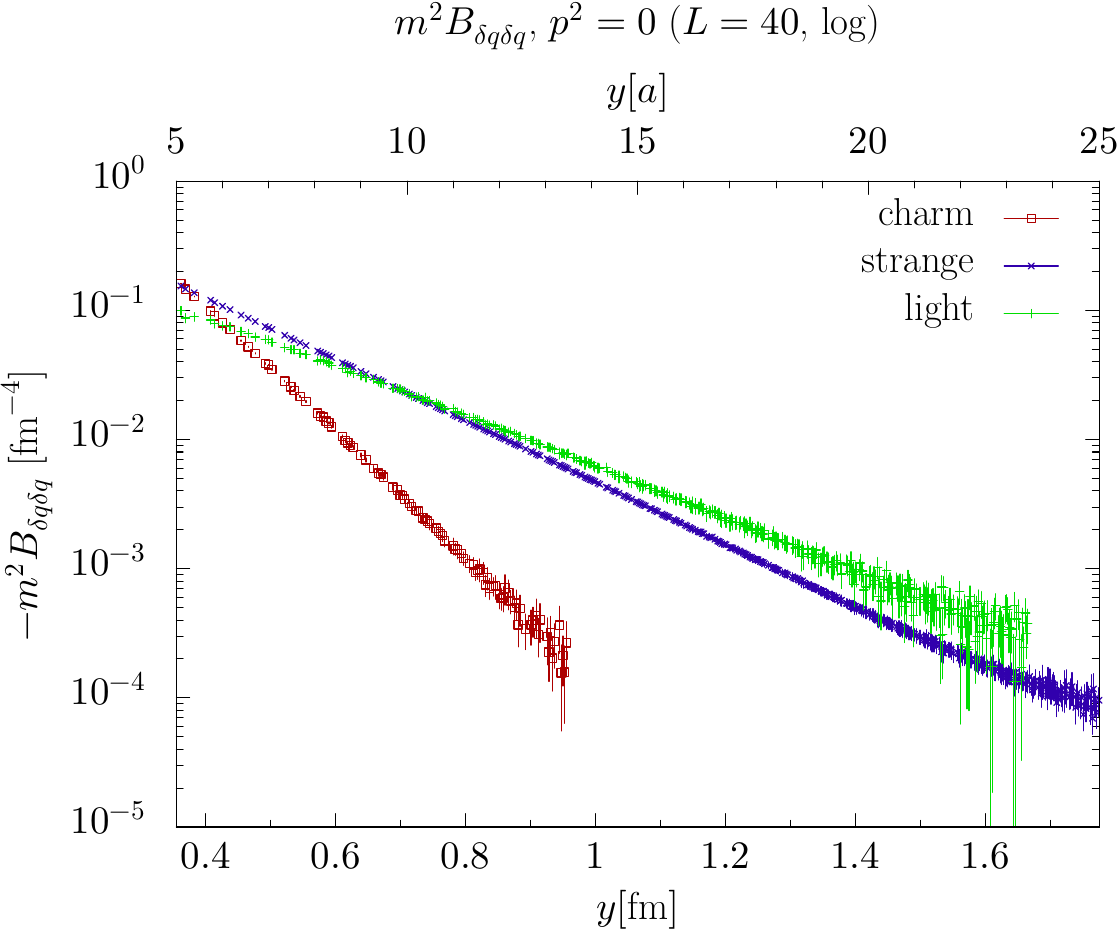}}
\caption{\label{fig:C1-mass-comp-log} Mass dependence of twist-two functions for graph $C_1$.}
\end{center}
%
\begin{center}
\subfigure[$A_{\Delta q \Delta q}$, graph $C_1$]{
\includegraphics[width=0.48\textwidth,trim=0 0 0 17,clip]
{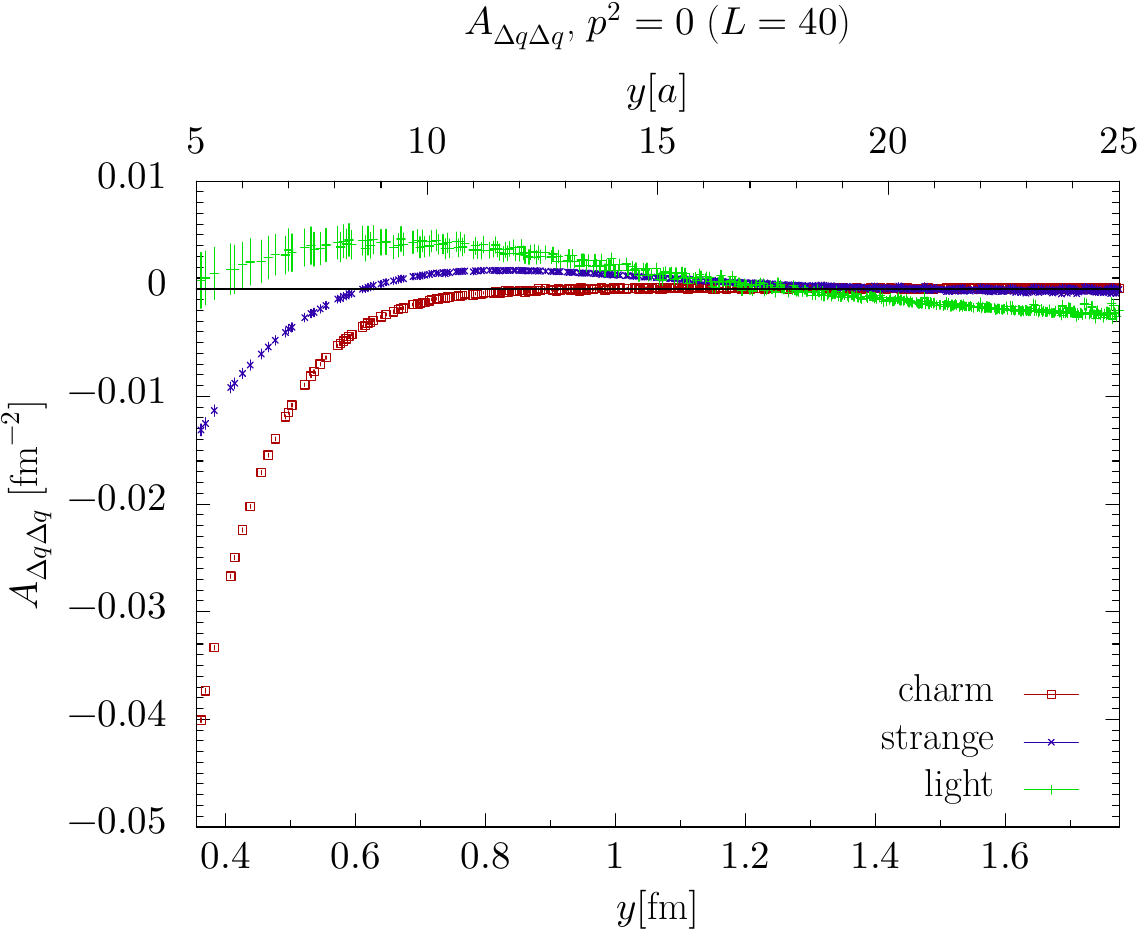}}
\caption{\label{fig:C1-mass-comp-lin} As \fig{\protect\ref{fig:C1-mass-comp-log}}, for $A_{\Delta q \Delta q}$ and with a linear instead of a logarithmic scale.}
\end{center}
\end{figure*}

Let us now take a closer look at the mass dependence of our results for graph $C_1$.  We multiply $A_{\delta q \ms q}$ and $B_{\delta q \delta q}$ with the power of the meson mass $m$ with which they appear in the decomposition \protect\eqref{eq:tensor-decomp} of two-current matrix elements.
We see in \fig{\ref{fig:C1-mass-comp-log}} that for all twist-two functions except $A_{\Delta q \Delta q}$, the decrease with $y$ becomes stronger with increasing quark mass, which simply reflects the decreasing size of the meson.  At $y \sim 5a$, the functions $A_{q q}$, $m A_{\delta q \ms q}$ and $m^2 B_{\delta q \delta q}$ are of comparable size for all quark masses, whereas $A_{\delta q \delta q}$ increases with the mass.  The behaviour of $A_{\Delta q \Delta q}$ for light and strange quarks is qualitatively different from the one of the other functions, as is evident from \fig{\ref{fig:C1-mass-comp-lin}}.  For charm quarks, $A_{\Delta q \Delta q}$ is approximately exponential in $y$, with a logarithmic slope similar to the one of $A_{q q}$.  A fit of the $y$ dependence of the twist-two functions for light quarks is presented in \sect{\ref{sec:py-fits}}.

\begin{figure*}
\begin{center}
\subfigure[$A_{q q}$, graph $C_2$]{
\includegraphics[width=0.48\textwidth,trim=0 0 0 17,clip]
{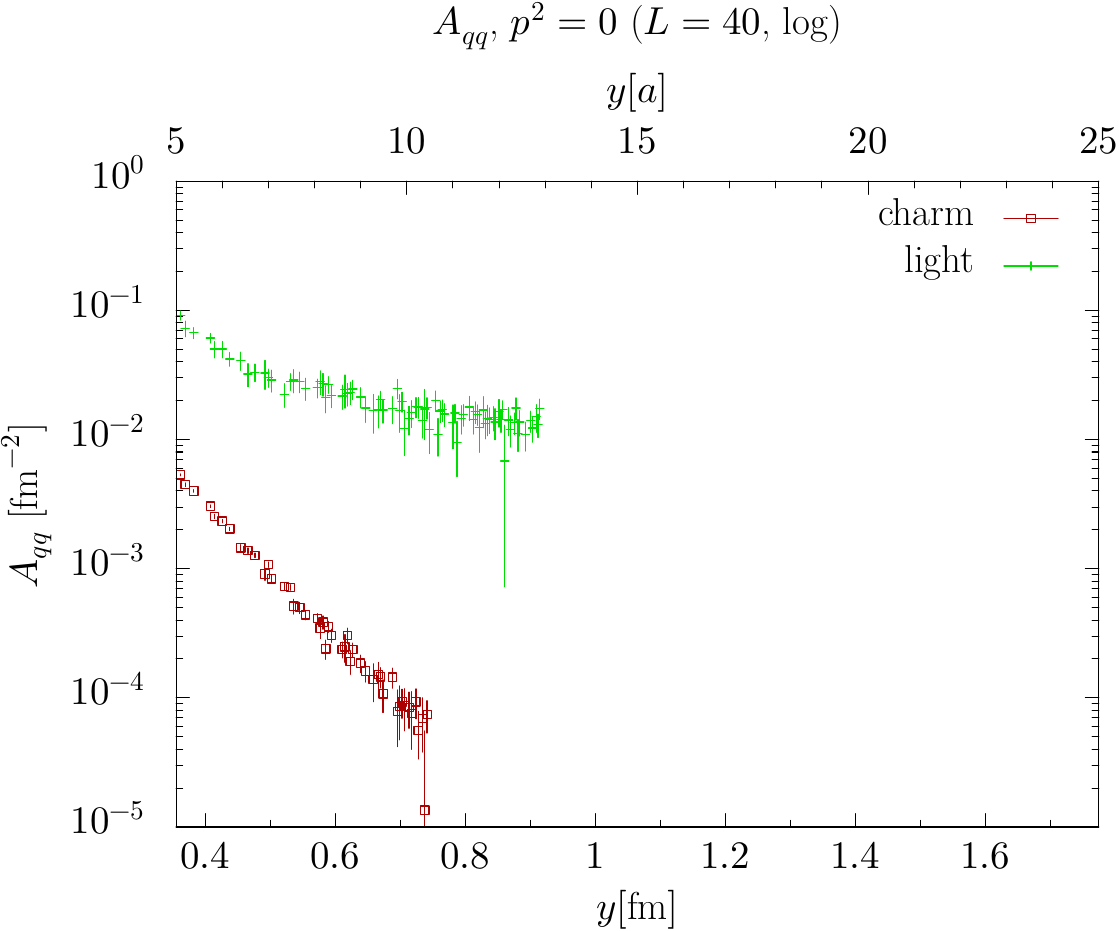}}
\hfill
\subfigure[$m A_{\delta q \ms q}$, graph $C_2$]{
\includegraphics[width=0.48\textwidth,trim=0 0 0 17,clip]
{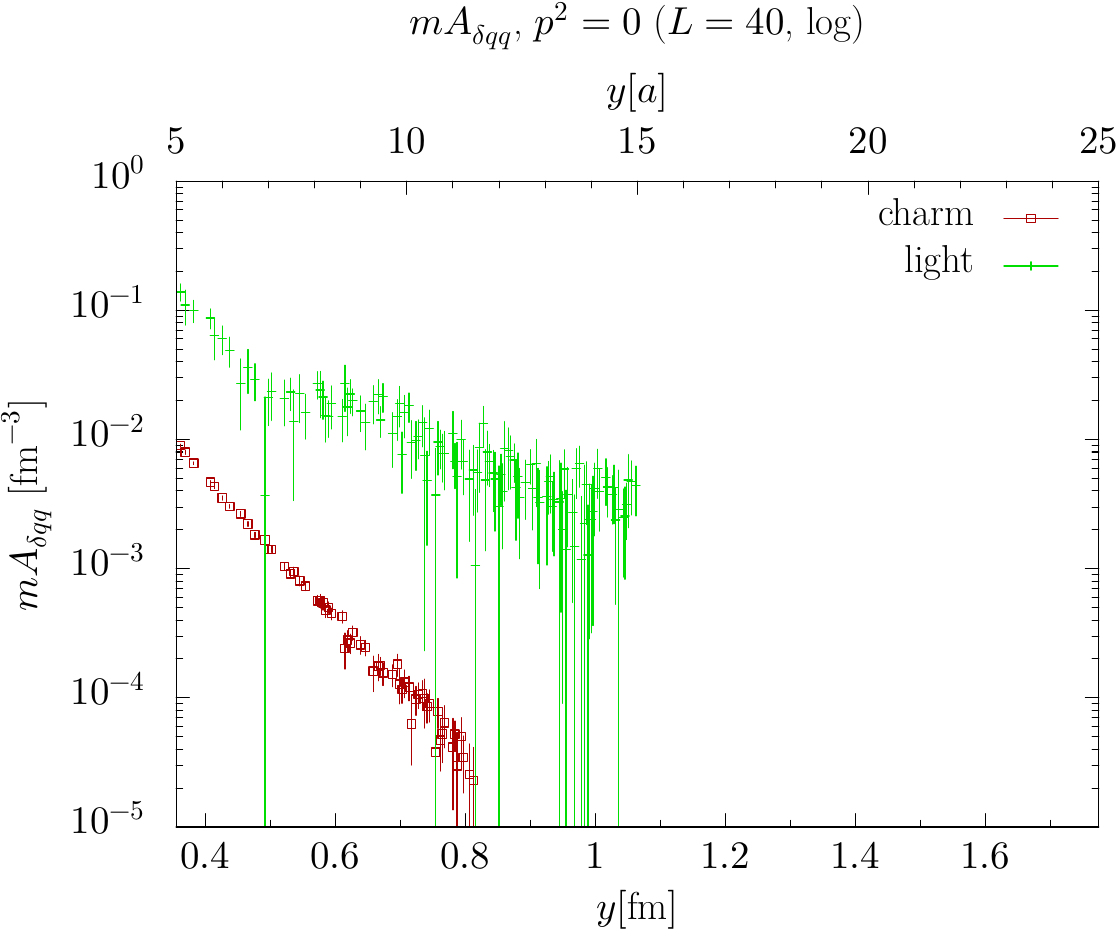}}
\caption{\label{fig:C2-mass-comp} Mass dependence of twist-two functions for graph $C_2$.}
\end{center}
\end{figure*}

We now discuss graph $C_2$, for which we have data with light quarks and with charm.  For the functions $A_{\Delta q \Delta q}$, $A_{\delta q \delta q}$ and $B_{\delta q \delta q}$, the light quark data is too noisy for a meaningful comparison with charm results, so that we focus on $A_{q q}$ and $A_{\delta q \ms q}$.  As is seen in \fig{\ref{fig:C2-mass-comp}}, the size of both functions is significantly smaller for charm quarks.  This is plausible: as discussed in the previous subsection, the partonic interpretation of graph $C_2$ always involves a Fock state with at least two quarks and two antiquarks in the meson, whereas for $C_1$ we have the regime shown in the first panel in \fig{\ref{fig:parton-graphs-C1}}, which involves only the quark-antiquark Fock state.
The $y$ dependence of $A_{q q}$ and $A_{\delta q \ms q}$ is also qualitatively different for the two masses: for charm we observe a clear and steep exponential falloff, whereas for light quarks, the logarithmic slope of both functions decreases around $y \sim 0.5 \fm$.

\FloatBarrier


\subsection{Polarisation effects}
\label{sec:polarisation}

A major aim of our study is to investigate the strength and pattern of spin correlations between two partons in a pion.  We spelled out the physical interpretation of polarised DPDs in \sect{\ref{sec:dps}}.  This interpretation extends to the corresponding twist-two functions at $py=0$, provided that these are dominated by partonic regimes associated with DPDs at $\zeta=0$.  Under this assumption, comparing $A_{\Delta q \Delta q}$ and $A_{\delta q \delta q}$ with $A_{q q}$ indicates whether two partons prefer to have their spins aligned or anti-aligned, with $A_{\Delta q \Delta q}$ referring to longitudinal and $A_{\delta q \delta q}$ to transverse polarisation.  We will refer to these as ``spin-spin correlations''.  Note that, according to \eqref{eq:mellin-mom-def}, a $q \bar{q}$ pair with aligned spins contributes with a negative sign to $A_{q q}$ and $A_{\delta q \delta q}$ and with a positive sign to $A_{\Delta q \Delta q}$, whereas a $q q$ pair with aligned spins contributes with a positive sign to all three functions.
The comparison of $m y A_{\delta q \ms q}$ and $m^2 |y^2| B_{\delta q \delta q}$ with $A_{q q}$  tells us about the strength of correlations between the transverse spin of one or both observed partons and the distance $\tvec{y}$ between these partons in the transverse plane.  We refer to this as ``spin-orbit correlations'' in the following.  The pre-factors $m y$ and $m^2 |y^2|$ in $m y A_{\delta q \ms q}$ and $m^2 |y^2| B_{\delta q \delta q}$ follow from the decompositions \eqref{eq:invar-dpds} and \eqref{eq:t2-mat-els}.

We note that the probability interpretation of polarised DPDs implies positivity constraints \cite{Diehl:2013mla} that extend the well-known Soffer bound for single parton distributions \cite{Soffer:1994ww}.  These bounds imply that  $|f_{\Delta q \Delta \bar{q}}|$, $|f_{\delta q \delta \bar{q}}|$, $|m y f_{\delta q \ms \bar{q}}|$ and $|m^2 y^2 f^t_{\delta q \delta \bar{q}}|$ are bounded by $f_{q \bar{q}}$.  Corresponding bounds do not hold for the lowest Mellin moments of DPDs because of the relative minus sign between quark and antiquark contributions in \eqref{eq:mellin-mom-def}.  They hold even less for the moments of skewed DPDs, which do not represent probabilities to start with.  Nevertheless, in a loose sense, the size of $A_{q q}$ sets a natural scale for the other twist-two functions (multiplied with $m y$ or $m^2 |y^2|$ as appropriate).

\begin{figure*}
\begin{center}
\subfigure[graph $C_1$, light quarks\label{fig:polar-C1-T}]{
\includegraphics[width=0.48\textwidth,trim=0 0 0 17,clip]
{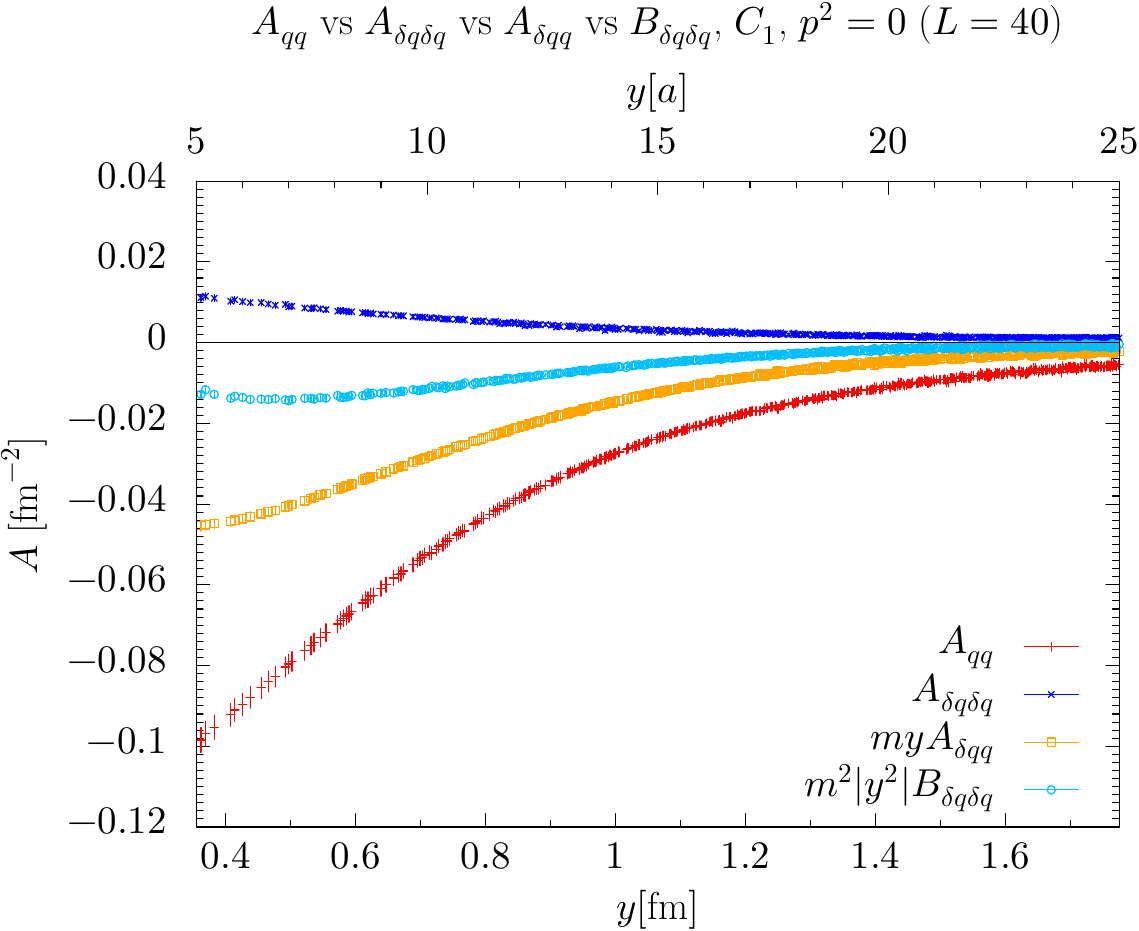}
\hfill
\includegraphics[width=0.48\textwidth,trim=0 0 0 17,clip]
{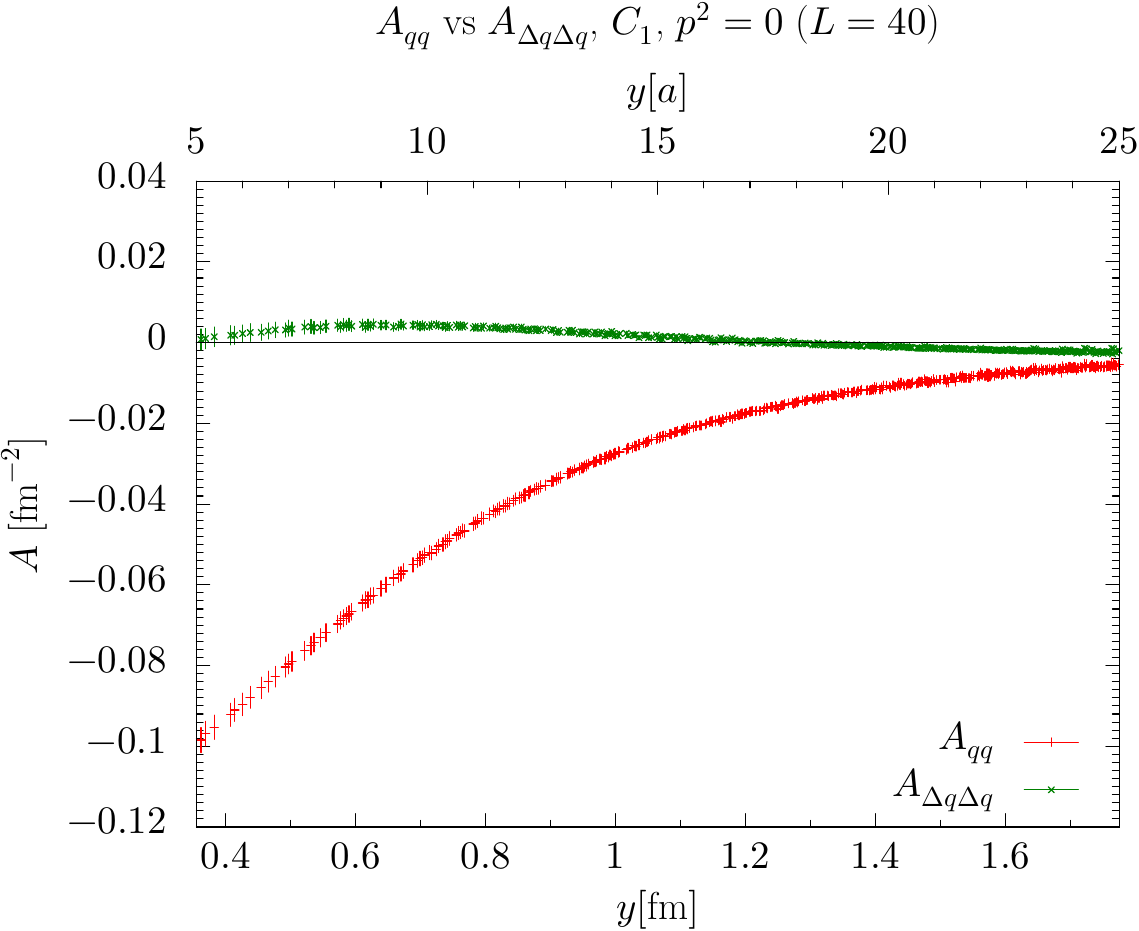}}
\\
\subfigure[graph $C_1$, strange quarks]{
\includegraphics[width=0.48\textwidth,trim=0 0 0 17,clip]
{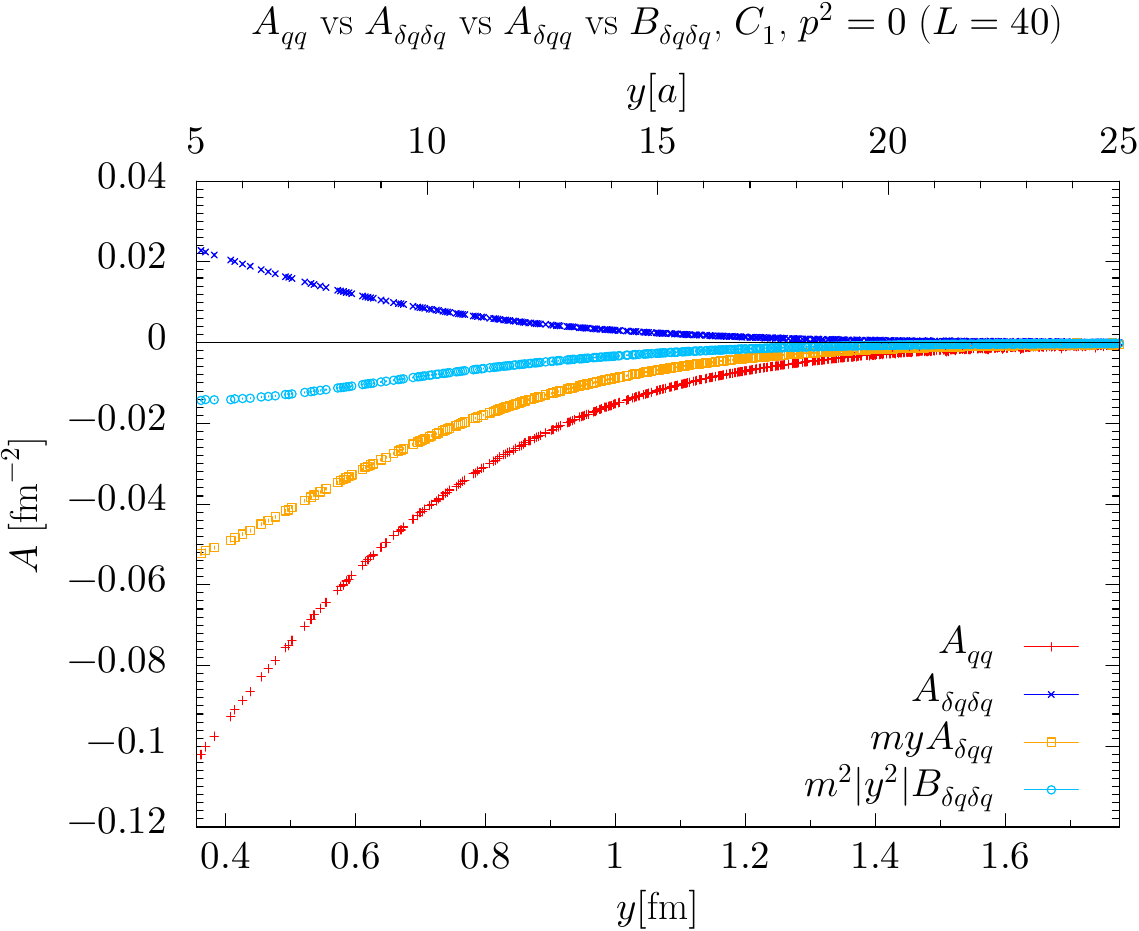}
\hfill
\includegraphics[width=0.48\textwidth,trim=0 0 0 17,clip]
{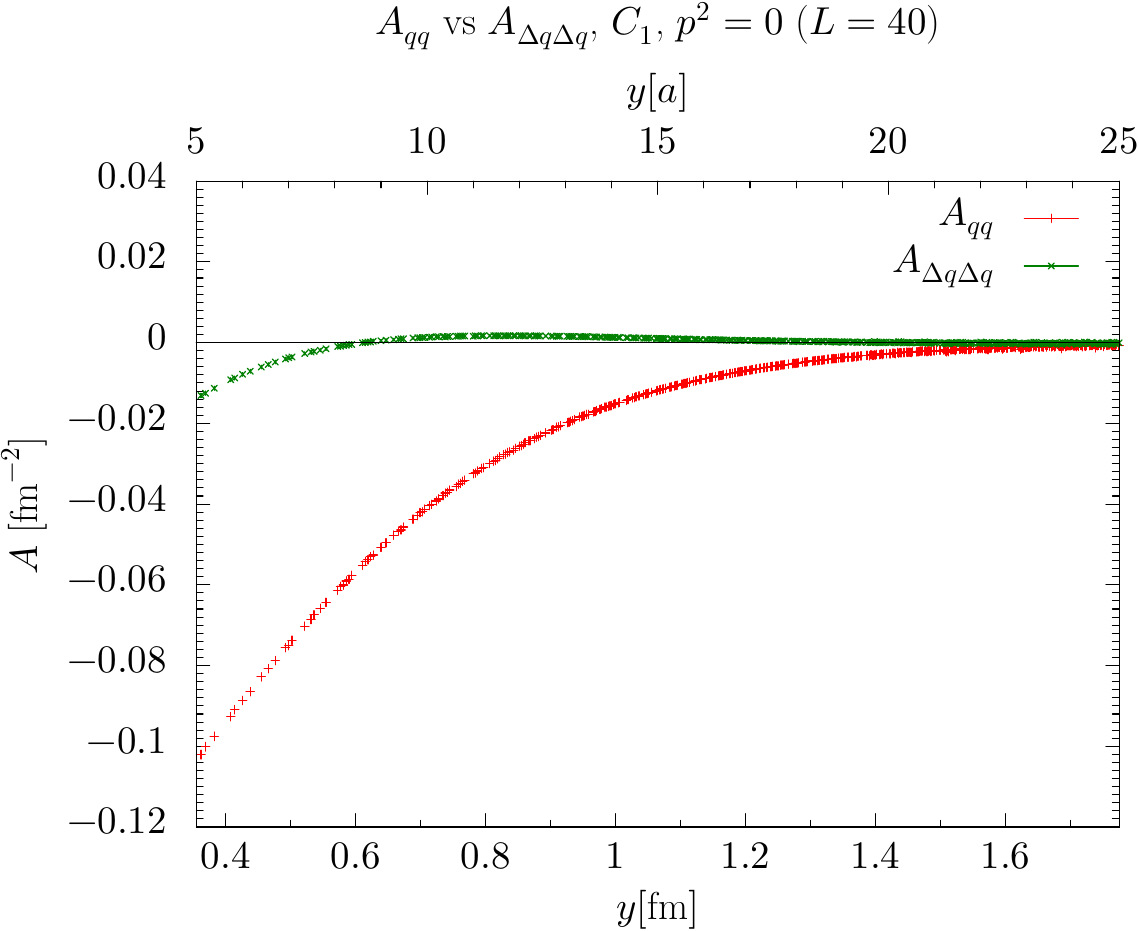}}
\\
\subfigure[graph $C_1$, charm quarks]{
\includegraphics[width=0.48\textwidth,trim=0 0 0 17,clip]
{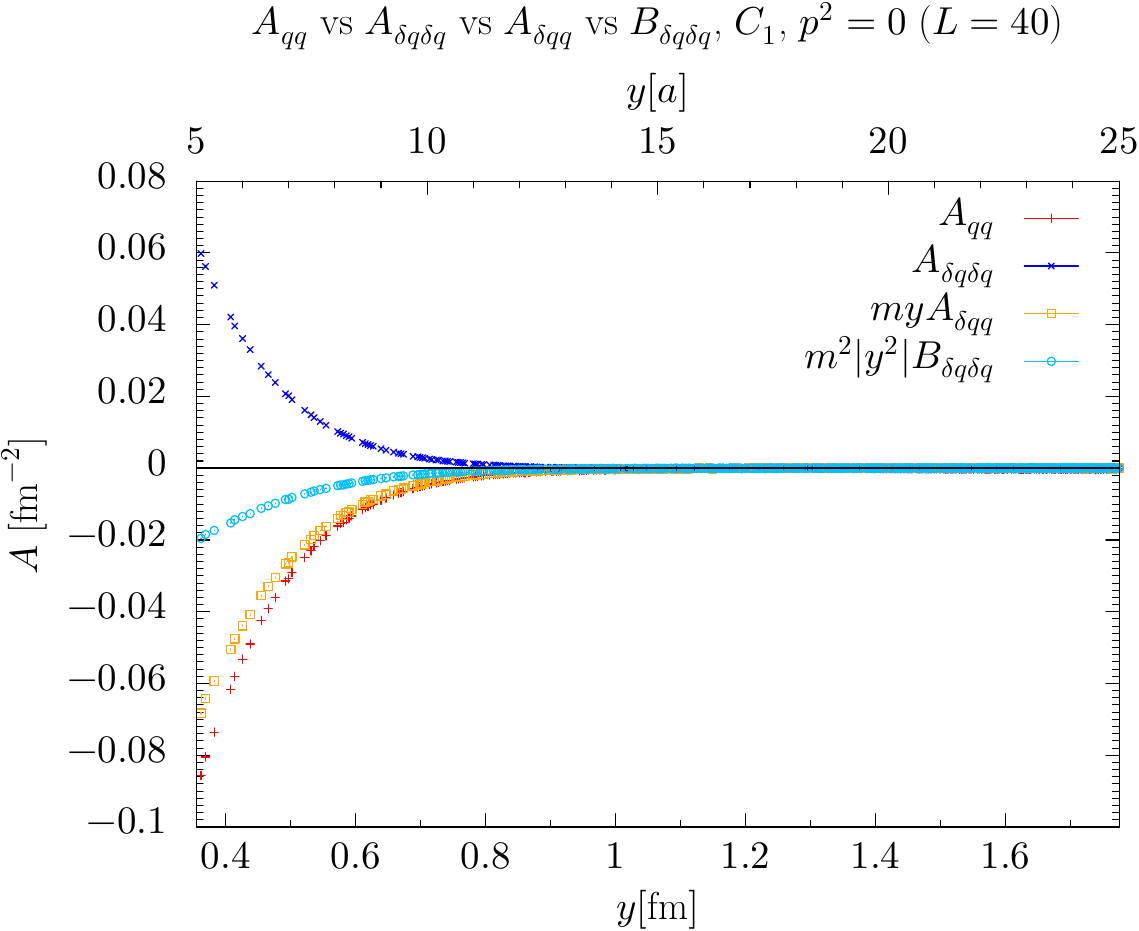}
\hfill
\includegraphics[width=0.48\textwidth,trim=0 0 0 17,clip]
{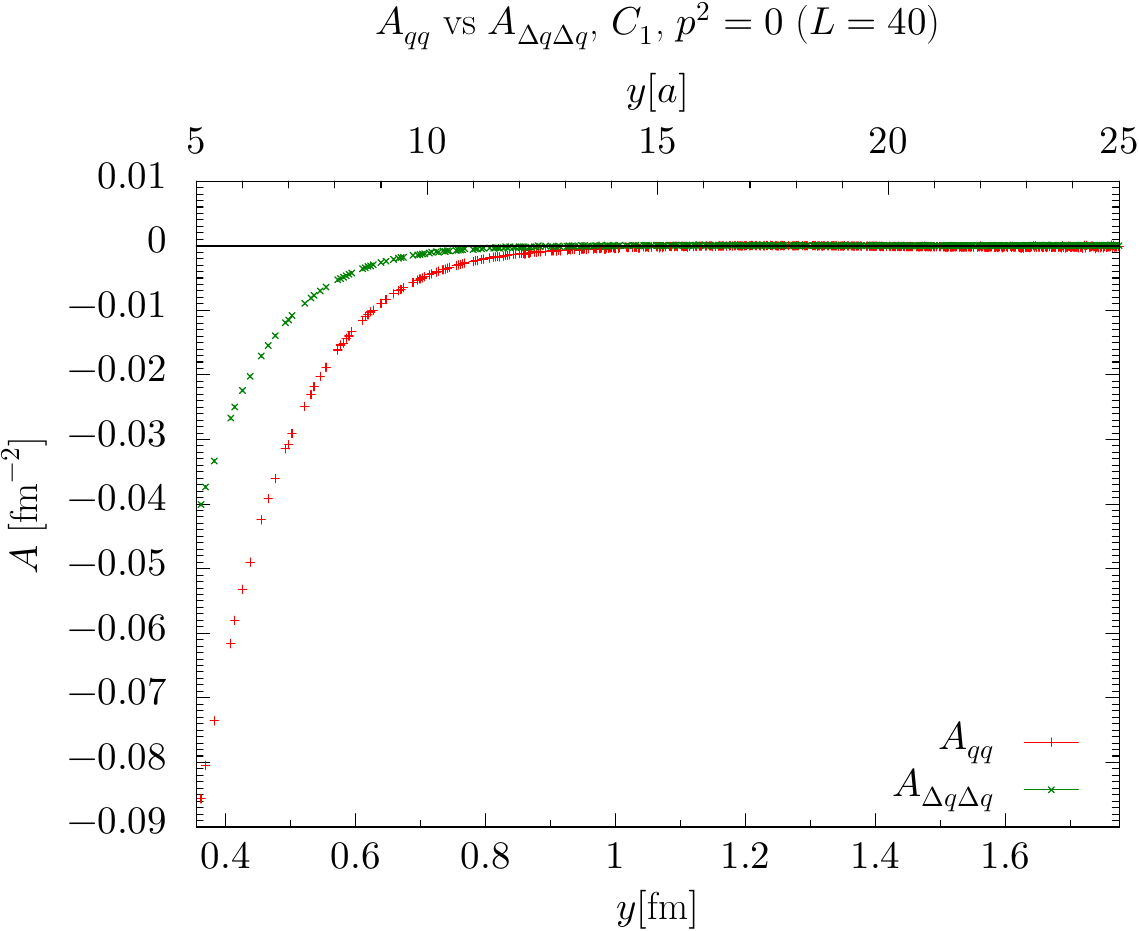}}
\caption{\label{fig:polar-C1} Effects of transverse (left) and longitudinal (right) polarisation for graph $C_1$.}
\end{center}
\end{figure*}

\rev{In the following, we consider polarisation effects separately for the connected graphs $C_1$ and $C_2$.  Their physical interpretation is rather different, as becomes clear from \figs{\ref{fig:parton-graphs-C1}} and \ref{fig:parton-graphs-C2} and our discussion in the previous subsection.  For graph $C_1$, polarisation effects reflect rather directly correlations between the quark and antiquark in the pion valence state, whereas for graph $C_2$ they are deeply connected with sea quark degrees of freedom.}

Starting our discussion with graph $C_1$, we see in the top panels of \fig{\ref{fig:polar-C1}} that by far the strongest polarisation effect seen for light quarks is the spin-orbit correlation for a single parton, followed by the spin-orbit correlation involving both partons.  Both the transverse and the longitudinal spin-spin correlations are very small.  This is completely different from the simple picture of a pion as a $q\bar{q}$ pair in an $S$-wave, for which one would obtain 100\% anti-alignment of both transverse and longitudinal spins.

All spin correlations increase considerably with the quark mass.  For charm quarks, $m y A_{\delta q \ms q}$ is almost as large as $A_{q q}$.  Spin-spin correlations are also important for charm: the spins of the quark and antiquark are anti-aligned by about 75\% for transverse and by about 50\% for longitudinal polarisation.  We note that this is still quite far away from the non-relativistic limit, in which transverse and longitudinal spin correlations become equal.

We note that the pion mass for our simulations with light quarks, $m_\pi \approx 295 \mev$, is quite a bit larger than the physical value.  A naive extrapolation of the polarisation patterns just described suggests that at the physical point the spin-orbit correlation for one polarised parton may be substantial, whilst correlations involving two quark spins might be even smaller than the ones we see for light quarks in the present study.

\begin{figure*}
\begin{center}
\subfigure[graph $C_2$, light quarks]{
\includegraphics[width=0.48\textwidth,trim=0 0 0 17,clip]
{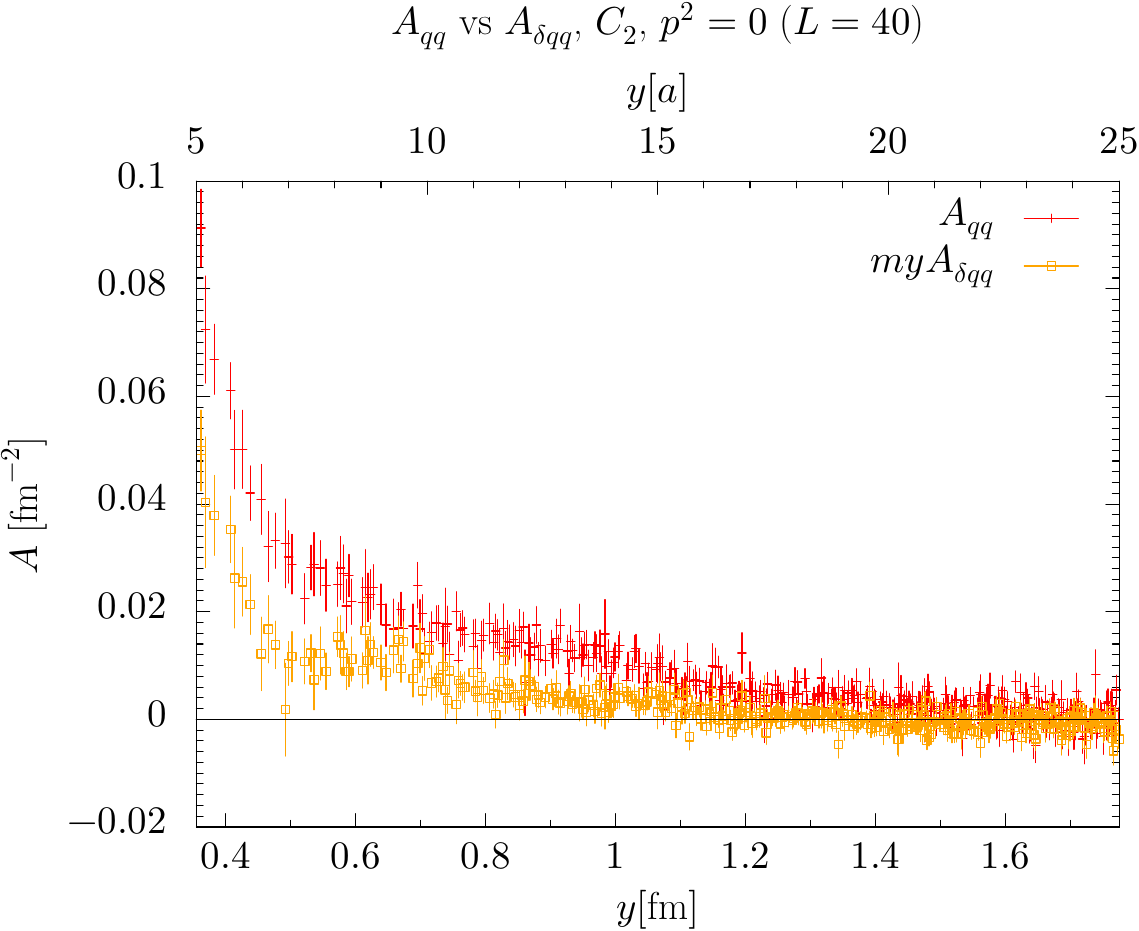}
\hfill
\includegraphics[width=0.48\textwidth,trim=0 0 0 17,clip]
{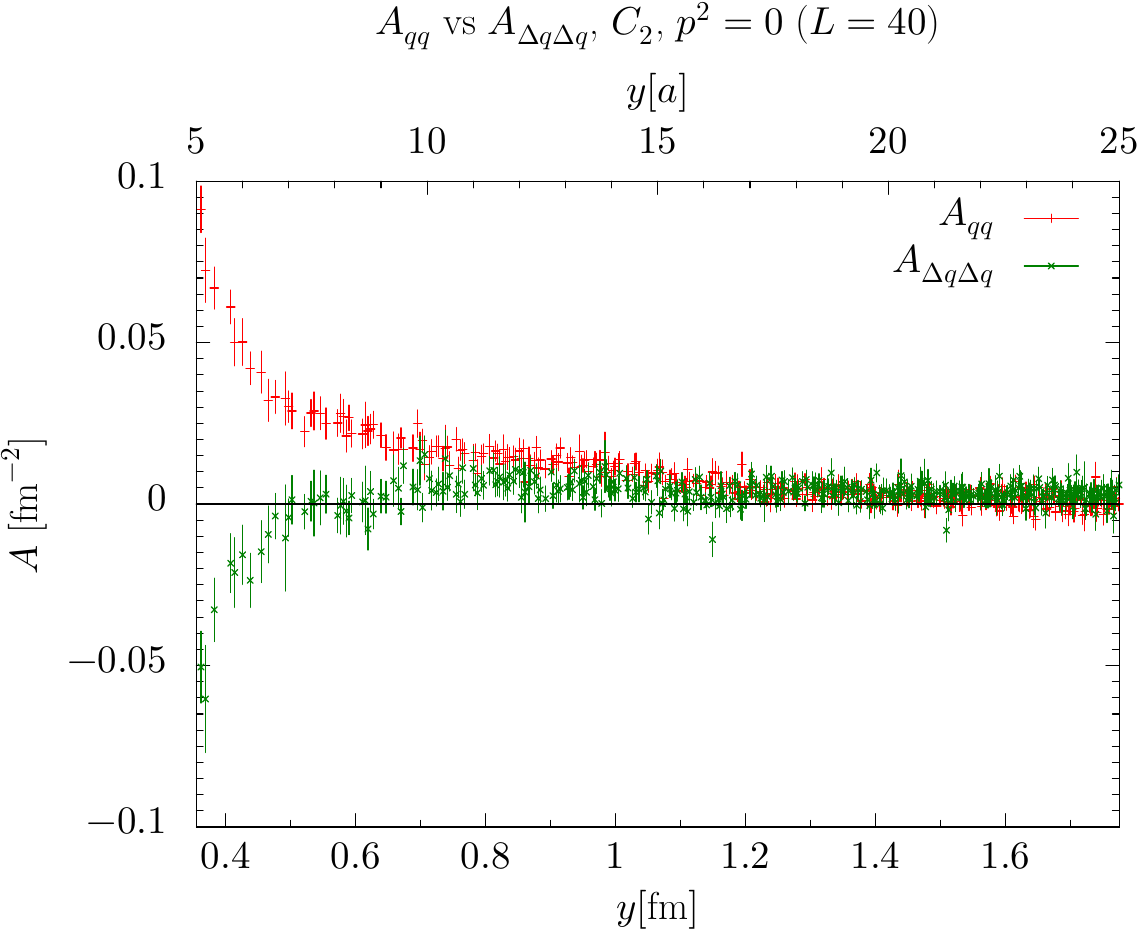}}
\\
\subfigure[graph $C_2$, charm quarks]{
\includegraphics[width=0.48\textwidth,trim=0 0 0 17,clip]
{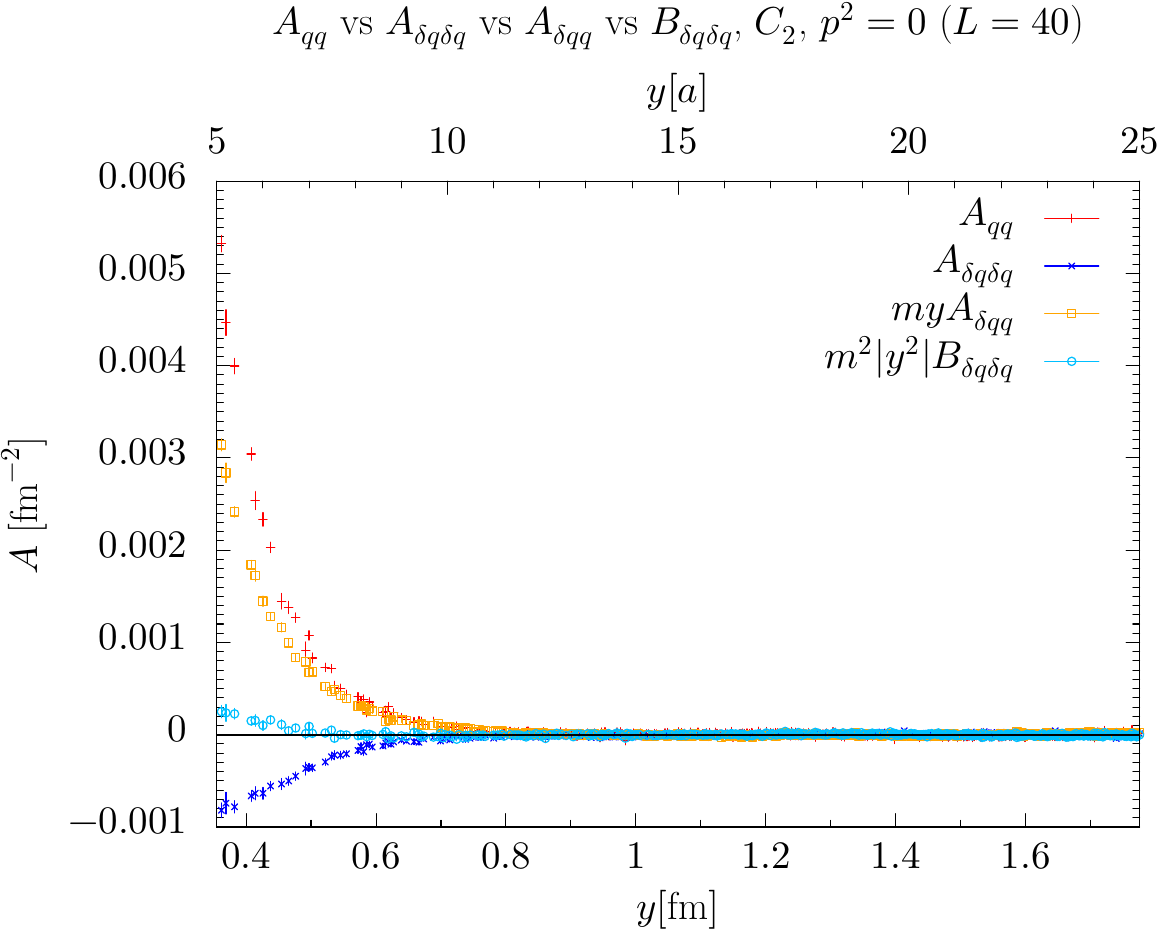}
\hfill
\includegraphics[width=0.48\textwidth,trim=0 0 0 17,clip]
{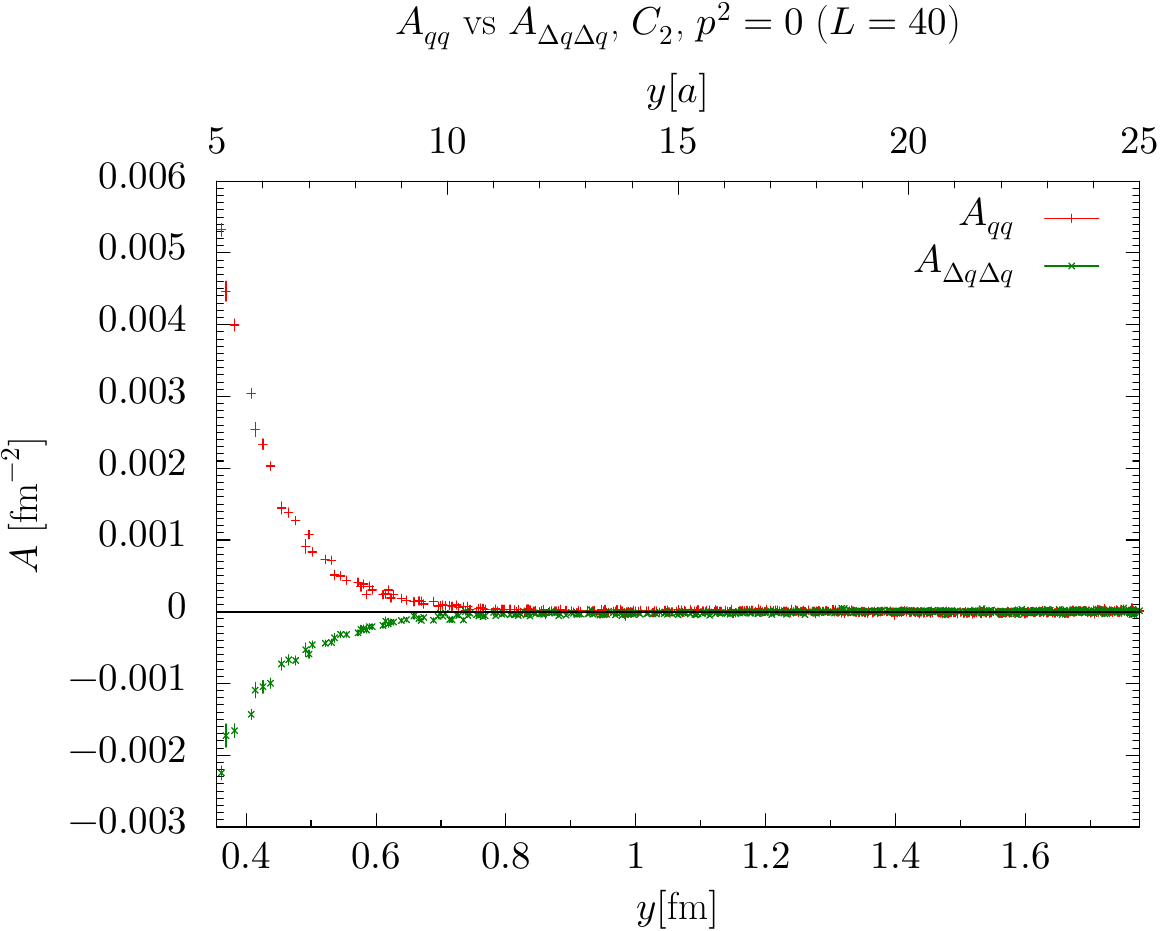}}
\caption{\label{fig:polar-C2} As \fig{\protect\ref{fig:polar-C1}}, but for graph $C_2$.  We have no strange quark results for this case.  The light quark data for $A_{\delta q \delta q}$ and $m^2 |y^2| B_{\delta q \delta q}$ is very noisy and not shown for the sake of clarity.}
\end{center}
\end{figure*}

We now turn to our results for graph $C_2$, which are shown in \fig{\ref{fig:polar-C2}}.  For light quarks, we see a substantial spin-orbit correlation of order 50\% for a single parton.  The spin-spin correlation for longitudinal polarisation is also of order 50\% for $y \sim 0.35 \fm$, but quickly decreases and is negligible already around $y \sim 0.5 \fm$.  For all other spin dependent correlations, the data for light quarks are too noisy to extract any physics.

With charm quarks, we have an excellent statistical signal for all twist-two functions.  We find that all spin correlations for graph $C_2$ are appreciable, apart from the one described by $m^2 |y^2| B_{\delta q \delta q}$. Notice that $A_{\Delta q \Delta q}$ has the same sign for $C_1$ and $C_2$, unlike all other twist-two functions.  If (as suggested by the sign of $A_{q q}$) the dominant parton configuration probed by the twist-two operators is a $c\bar{c}$ pair for graph $C_1$ and a $c c$ pair for graph $C_2$, then the longitudinal parton spins tend to be anti-aligned in both cases.



\subsection{Test of the factorisation hypothesis}
\label{sec:fact-A}

We now test the factorisation hypothesis for $A_{u d}(y^2, py=0)$ that we derived in \sect{\ref{sec:fact-theory}}.  We restrict ourselves to the contribution from the connected graph $C_1$.  Taking the full combination of graphs in the first line of \eqref{eq:phys-mat-els} is not an option because of the huge errors in our results for the doubly disconnected graph $D$.  By contrast, we see in \fig{\ref{fig:light-graphs-1-A_VV}} that $S_1$ is consistent with zero for $A_{q q}$ (although within errors much larger than those on $C_1$).  We find it plausible to expect that the contribution from $D$ is even smaller than the one of $S_1$, since $D$ has two disconnected fermion loops with one operator insertion.

The factorisation hypothesis \eqref{eq:A-fact} involves the vector form factor of the pion.  We have extracted this form factor from our lattice simulations, using the full number of 2025 gauge configurations available for our lattice with $L=40$.  As we consider only the connected contribution to the two-current correlation function, we restrict ourselves to the connected graph for the form factor as well.  We fit the form factor data to a power law
\begin{align}
\label{eq:ff-pole}
F_{u,V}(t) &= - F_{d,V}(t) = \bigl( 1 - t /M^2 \bigr)^{-p} \,.
\end{align}
We use two fit variants, which gives us a handle on the bias of the extrapolation to $-t > 1.15 \gev^2$, where we have no data.  Such an extrapolation bias is inevitable when we Fourier transform from momentum to position space, as is required in \eqref{eq:A-fact}.
In a monopole fit, we fix $p=1$ and obtain $M = 777(12) \mev$.  Leaving the power free, we obtain $p = 1.173(69)$ and $M = 872(16) \mev$.  Both fits give a very good description of our lattice data, as shown in \fig{14a} of \cite{Bali:2018nde}.

\begin{figure}
\begin{center}
\subfigure[\label{fig:comp-A}]{
\includegraphics[width=0.48\textwidth,trim=0 0 0 95,clip]
{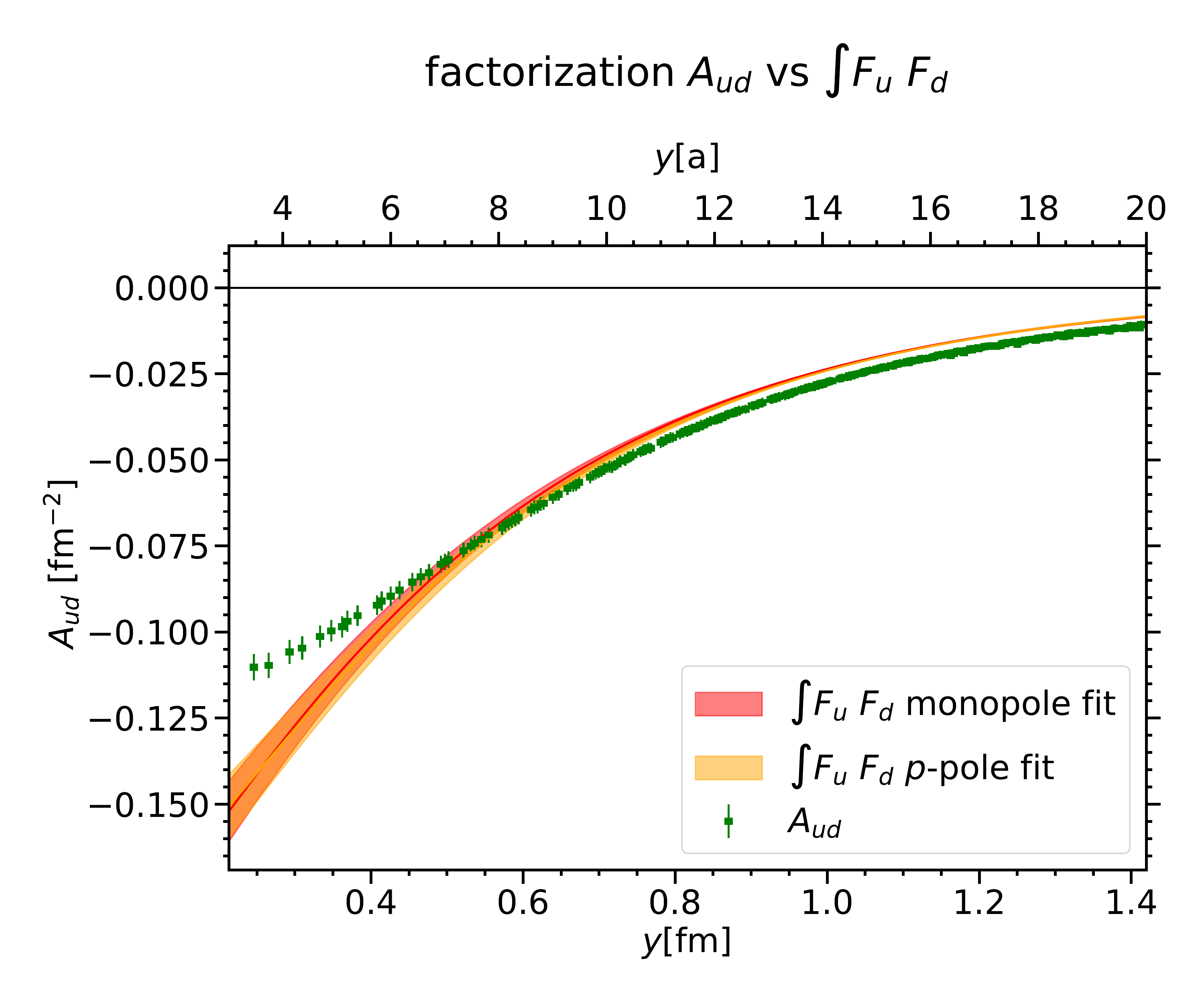}}
\hfill
\subfigure[\label{fig:rat-A}]{
\includegraphics[width=0.48\textwidth,trim=0 0 0 95,clip]
{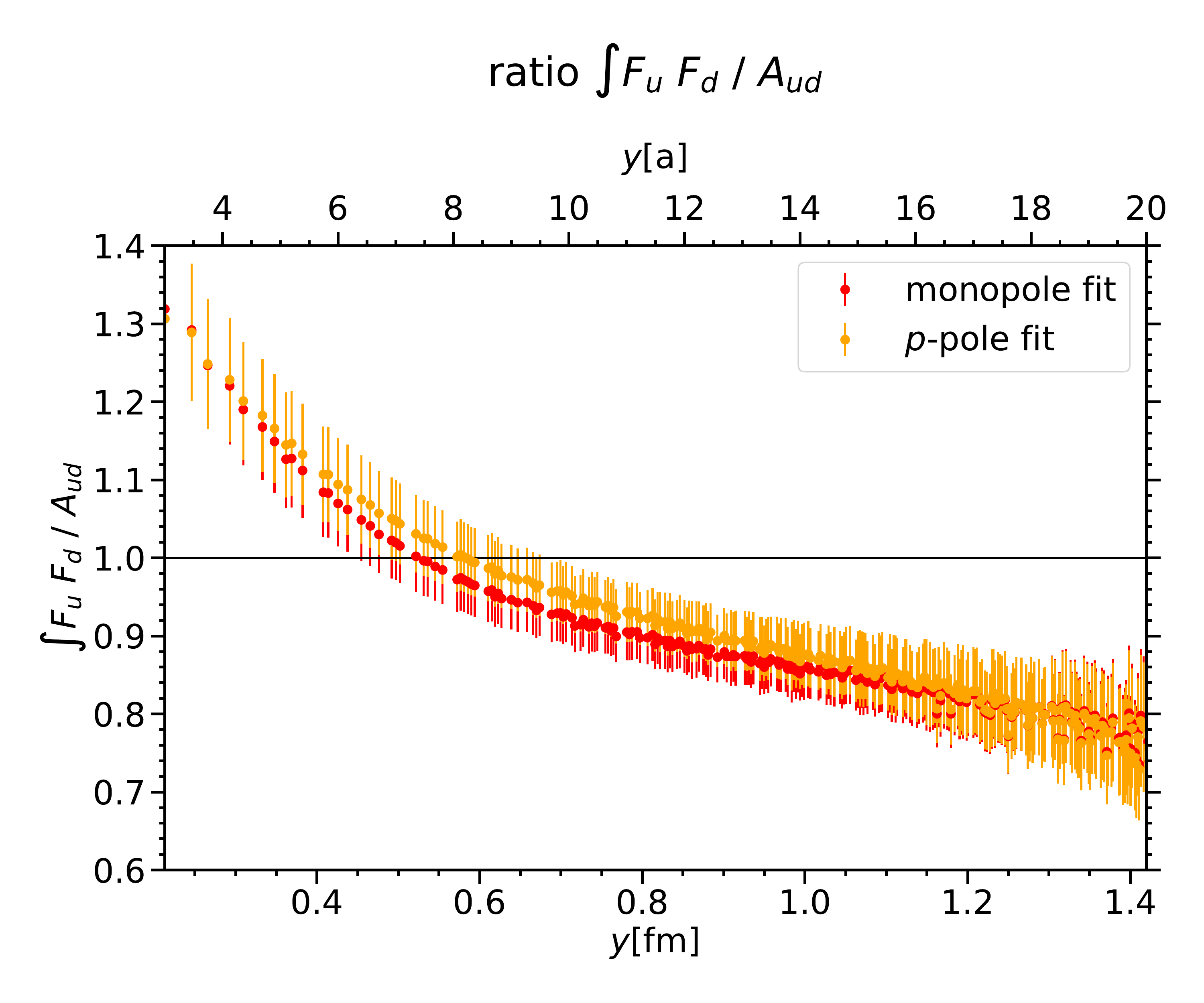}}
\caption{\label{fig:fact-test} Test of the factorisation hypothesis \protect\eqref{eq:A-fact} for the invariant function $A_{u d}$.
(a): data for $A_{u d}$ (restricted to the contribution from graph $C_1$) compared with the integral over form factors on the r.h.s.\ of \protect\eqref{eq:A-fact}.  The form factors are determined by a monopole or a $p$-pole fit.
(b): ratio of the form factor integral on the r.h.s.\ of \protect\eqref{eq:A-fact} to the data for $A_{u d}$.
}
\end{center}
\end{figure}

With the ansatz \eqref{eq:ff-pole}, the two-dimensional Fourier transform on the r.h.s.\ of \eqref{eq:A-fact} can be carried out analytically.  We compute the remaining integral over $\zeta$ numerically.  The results obtained with the two form factor fits agree very well for $y > 0.2 \fm$.  In panel (a) of \fig{\ref{fig:fact-test}} we compare the two sides of the factorisation hypothesis \eqref{eq:A-fact}, and in panel (b) we show the ratio of the r.h.s.\ to the l.h.s.\ of the equation.  We see a clear deviation from the factorised ansatz, which does however not exceed 30\% in the considered $y$ range.  One may thus say that the factorised ansatz provides a rough approximation of the two-current correlator.

\subsection{Physical matrix elements}
\label{sec:physical_combinations}

\begin{figure*}
\begin{center}
\subfigure[$A_{u d} \ms |_{\pi^+}$\label{phys-A_VV-ud}]{
\includegraphics[width=0.48\textwidth,trim=0 0 0 17,clip]
{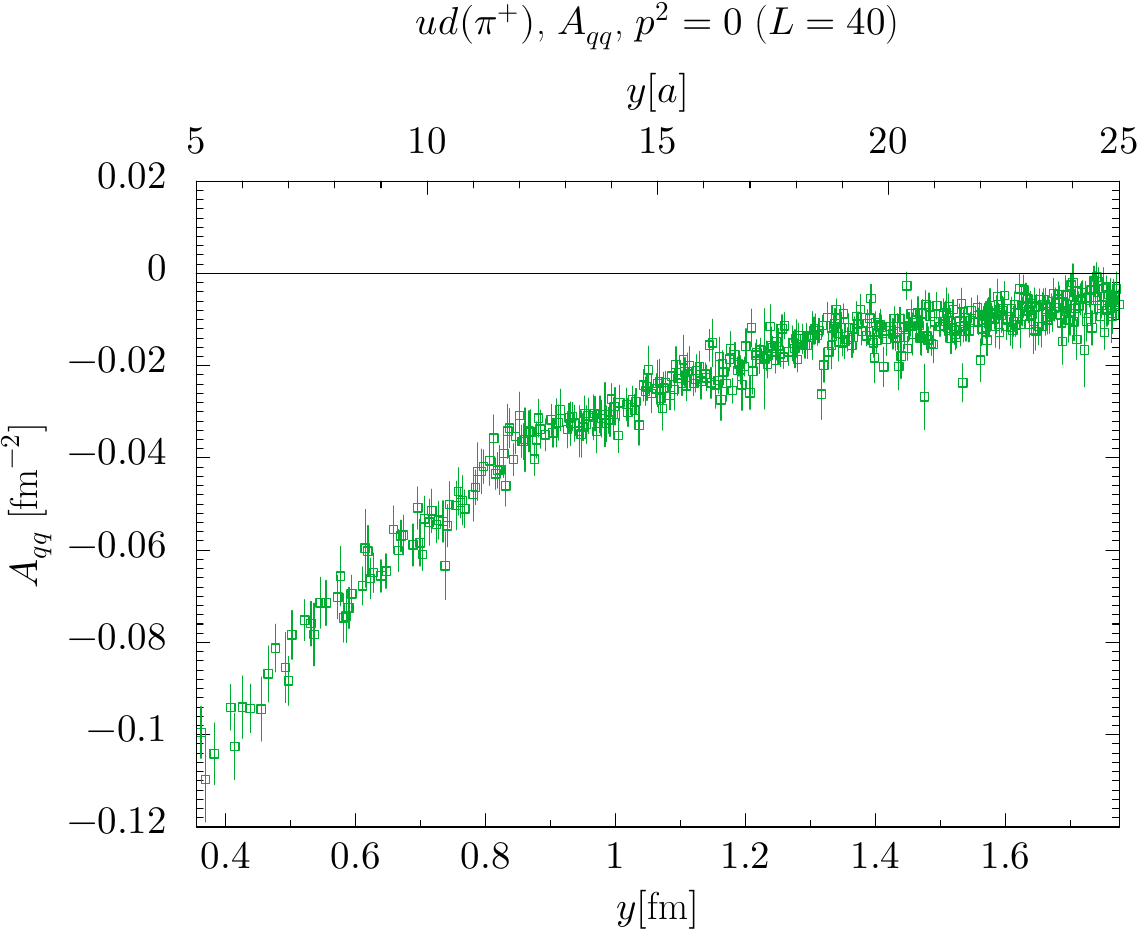}}
\hfill
\subfigure[$m y A_{\delta u \ms d} \ms |_{\pi^+}$]{
\includegraphics[width=0.48\textwidth,trim=0 0 0 17,clip]
{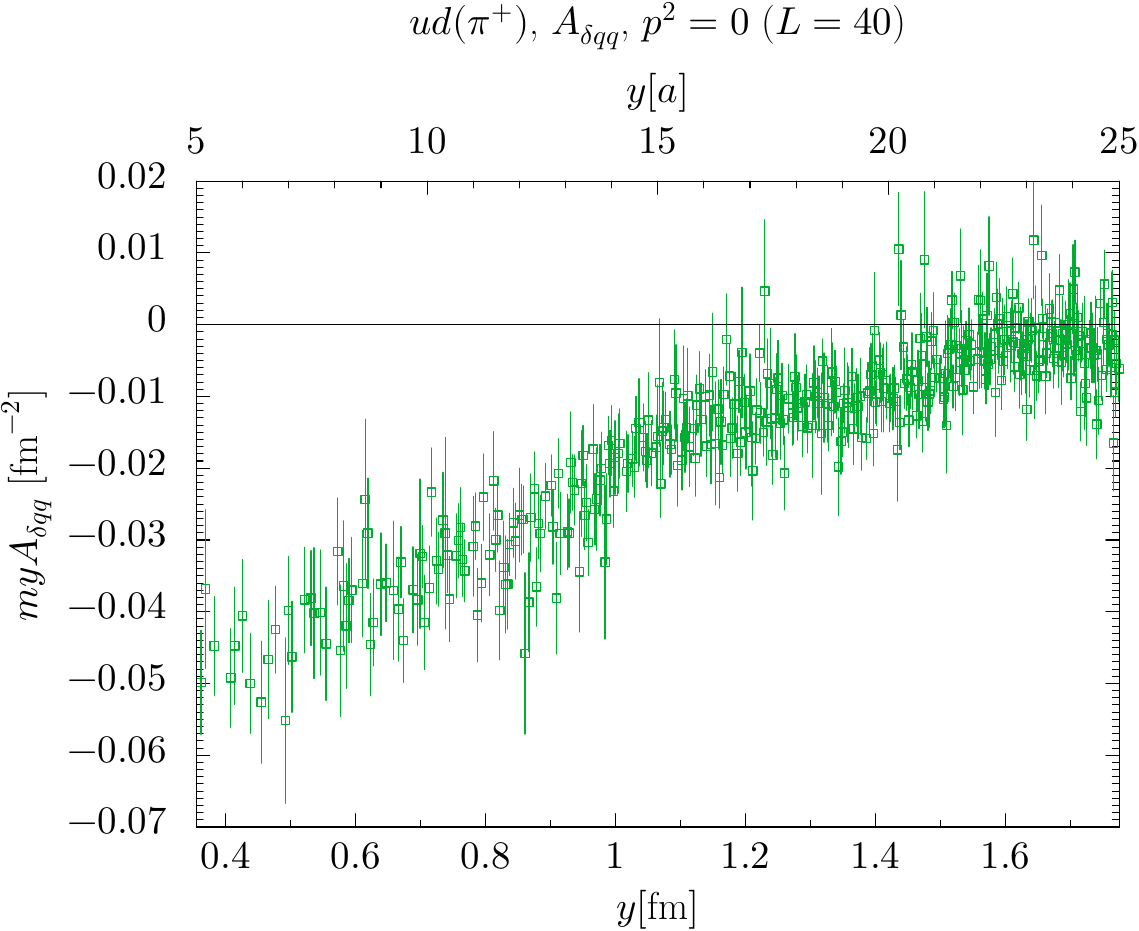}}
\\
\subfigure[$A_{u u} \ms |_{\pi^+}$\label{phys-A_VV-uu}]{
\includegraphics[width=0.48\textwidth,trim=0 0 0 17,clip]
{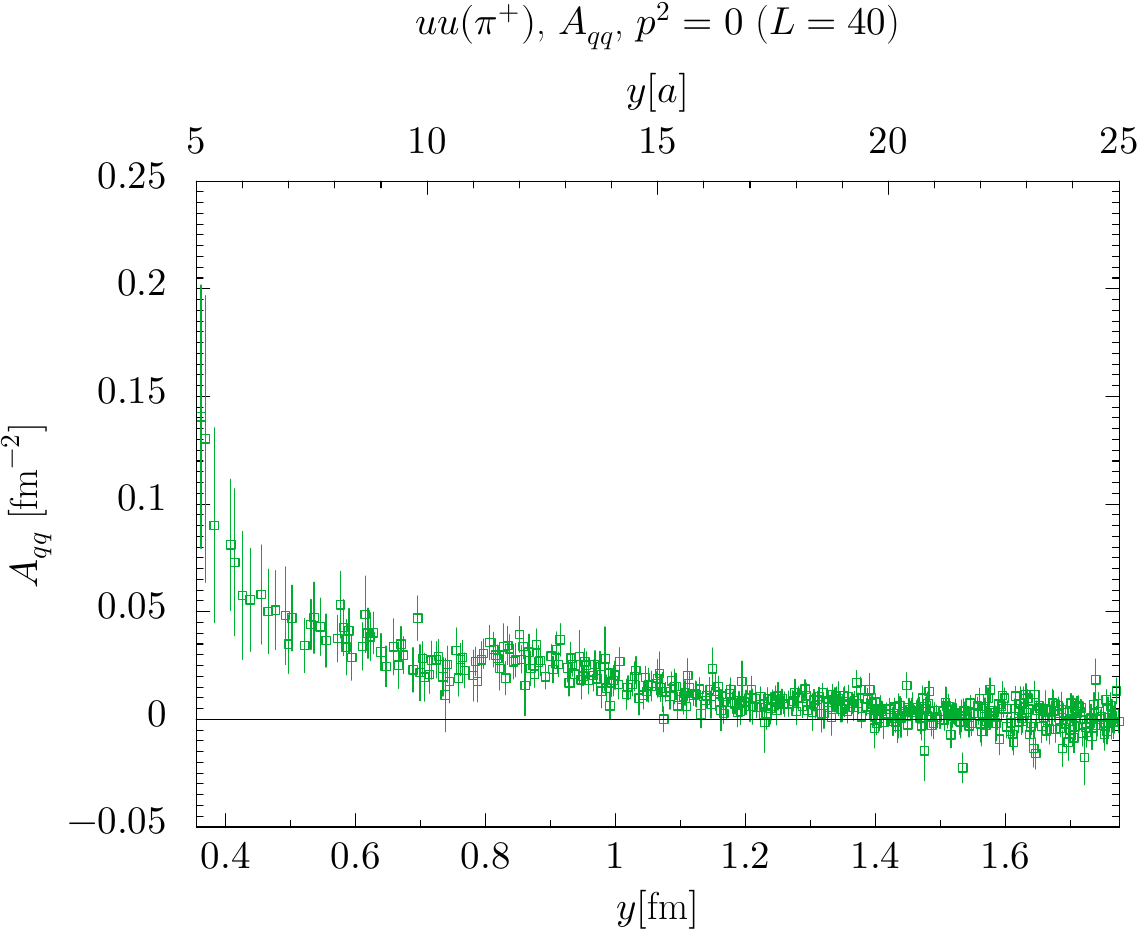}}
\hfill
\subfigure[$m y A_{\delta u \ms u} \ms |_{\pi^+}$]{
\includegraphics[width=0.48\textwidth,trim=0 0 0 17,clip]
{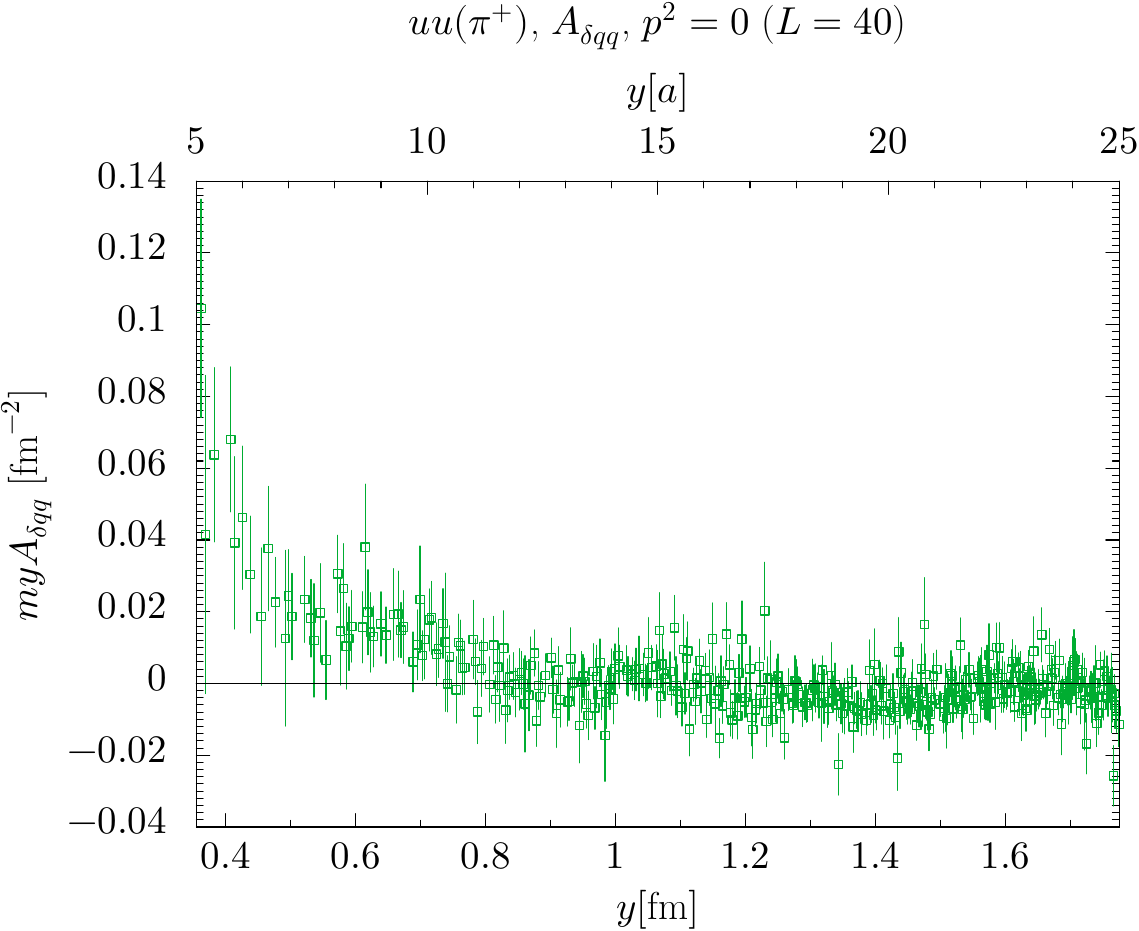}}
\\
\subfigure[$A_{u u} \ms |_{\pi^0}$]{
\includegraphics[width=0.48\textwidth,trim=0 0 0 17,clip]
{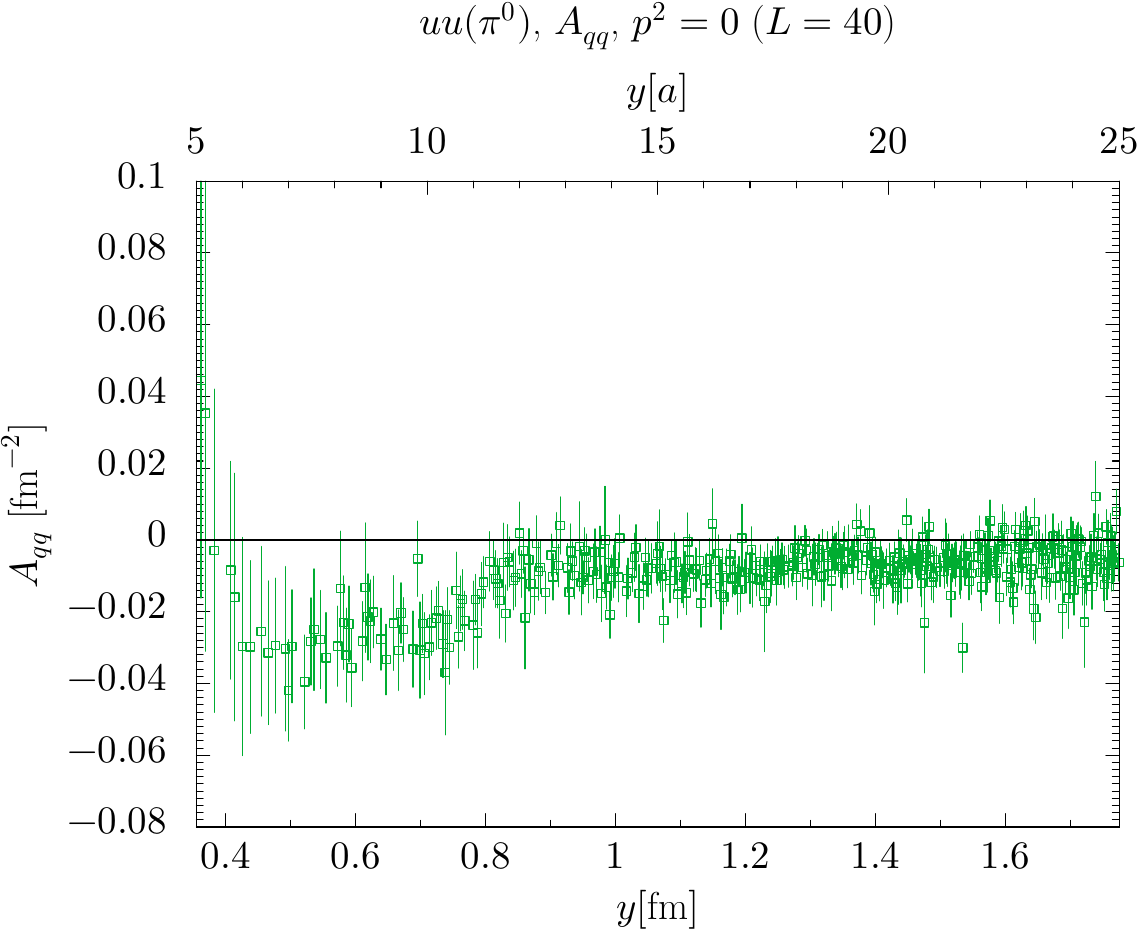}}
\hfill
\subfigure[$m y A_{\delta u \ms u} \ms |_{\pi^0}$]{
\includegraphics[width=0.48\textwidth,trim=0 0 0 17,clip]
{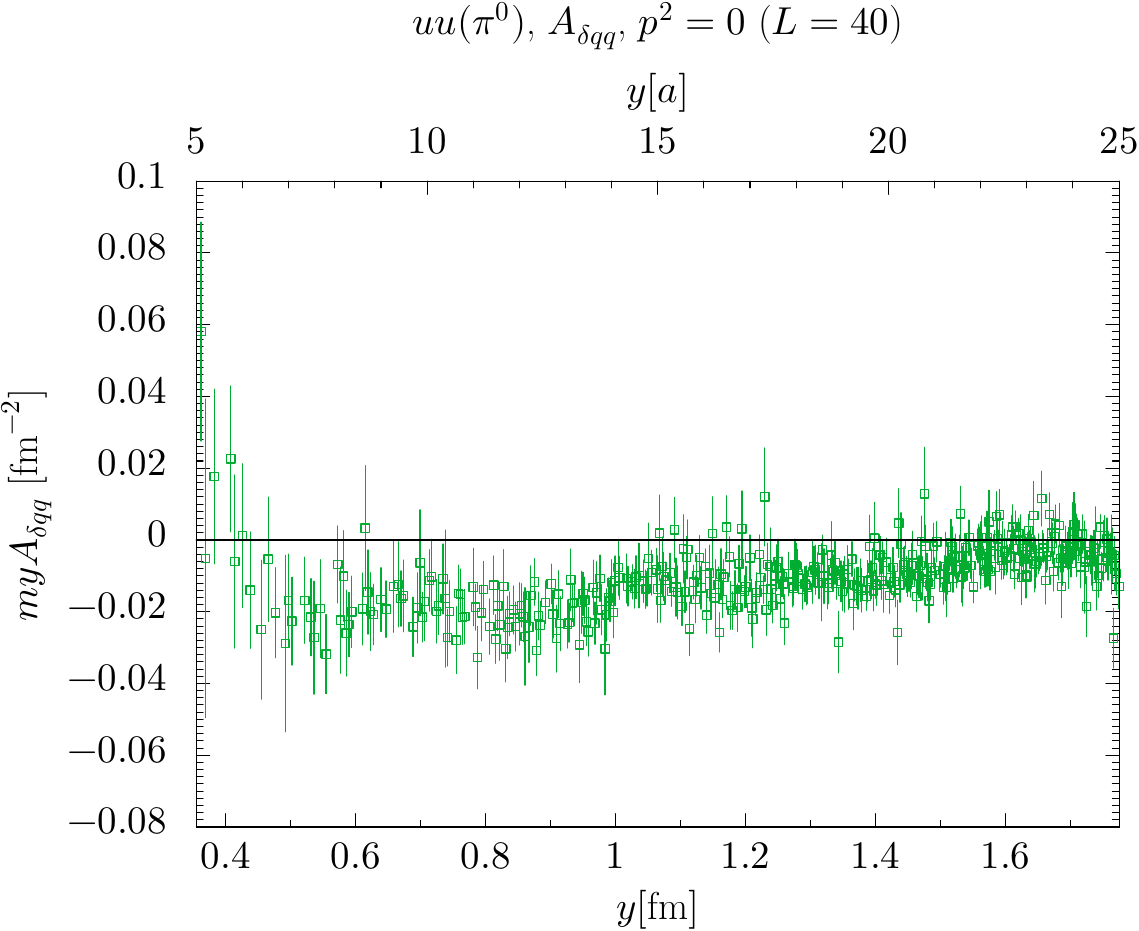}}
\caption{\label{fig:phys-A_VV-A_TV} Twist-two functions at $py = 0$ for the flavour combinations $u d$ or $u u$ in a $\pi^+$ or a $\pi^0$.  Lattice graphs are combined according to \protect\eqref{eq:phys-mat-els}, except for of graph $D$, which is affected by huge errors and hence omitted.  All results are for light quarks.}
\end{center}
\end{figure*}

\begin{figure*}
\begin{center}
\subfigure[$A_{\Delta u \Delta u} \ms |_{\pi^+}$]{
\includegraphics[width=0.48\textwidth,trim=0 0 0 17,clip]
{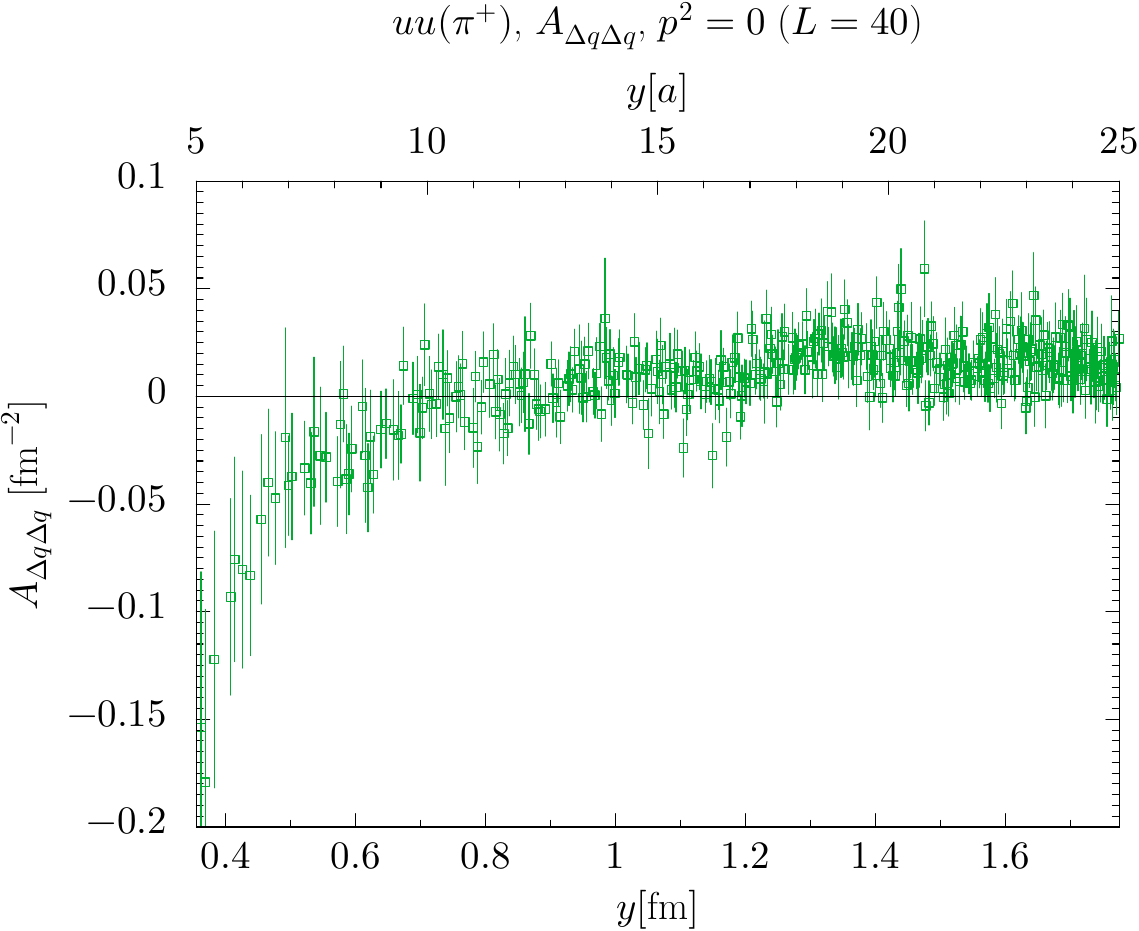}}
\hfill
\subfigure[$A_{\Delta u \Delta u} \ms |_{\pi^0}$]{
\includegraphics[width=0.48\textwidth,trim=0 0 0 17,clip]
{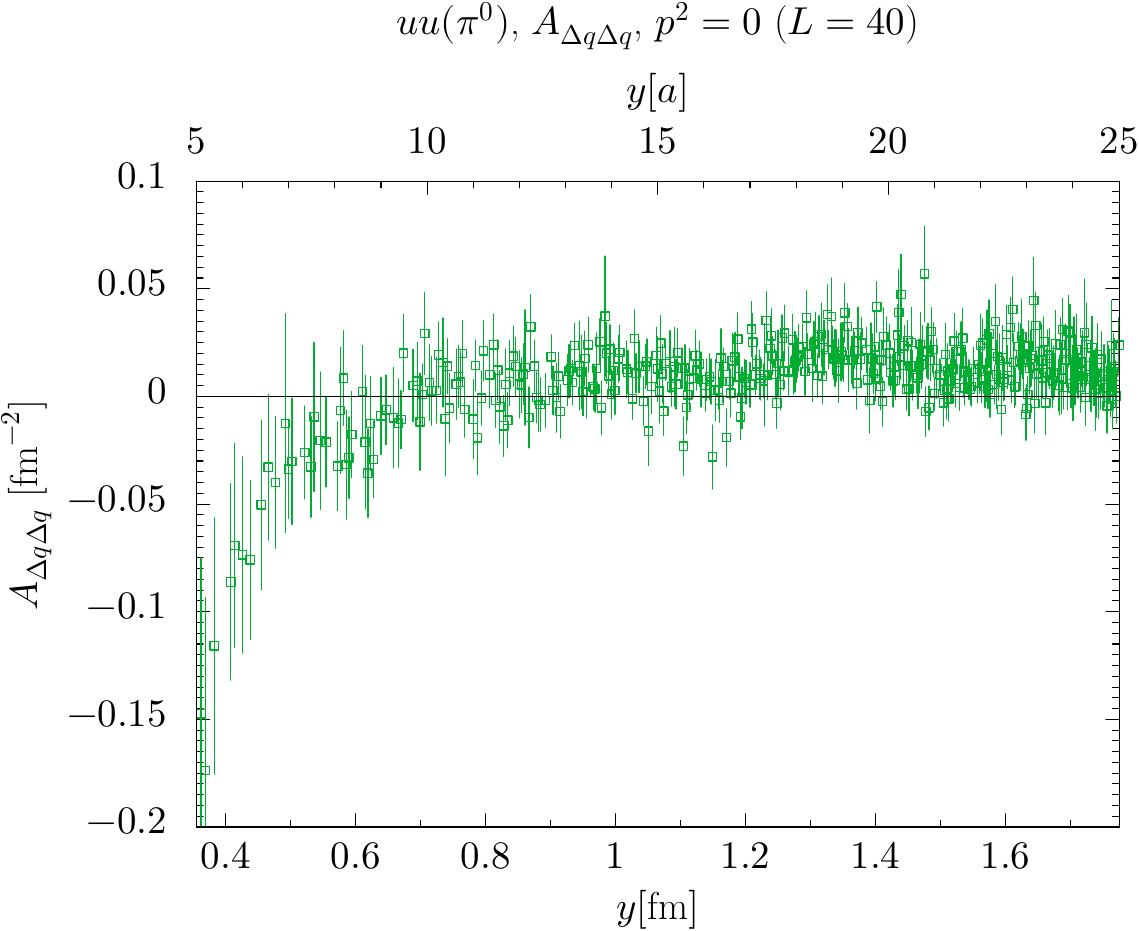}}
\caption{\label{fig:phys-A_AA} Continuation of \fig{\ref{fig:phys-A_VV-A_TV}}.}
\end{center}
\end{figure*}

We now investigate the combinations \eqref{eq:phys-mat-els} of lattice graphs that appear in the matrix elements of currents between charged or neutral pions.
We omit the doubly disconnected graph $D$ throughout, because its statistical errors are much larger than the signal for any other graph.  Since data for the full set of remaining graphs is only available for light quarks, we restrict our attention to this case.  The results are shown in \figs{\ref{fig:phys-A_VV-A_TV}} and \ref{fig:phys-A_AA} for the flavour combinations $u d$ and $u u$.  The combinations $d d$ and $d u$ can be obtained from the symmetry relations \eqref{eq:pion-symm}.

As can be expected from \figs{\ref{fig:light-graphs-1}} and \ref{fig:light-graphs-2}, the statistical errors of the physical combinations are significantly larger than those for the connected graphs alone.  Nevertheless, we see a clear negative signal for $A_{u d}$ in a $\pi^+$.  As discussed in \sect{\ref{sec:connected-graphs}}, this can be understood as a dominance of the valence Fock state $u \bar{d}$ over Fock states that contain $u d$, $\bar{u} \bar{d}$ or $\bar{u} d$.  The function $A_{u u}$ in a $\pi^+$ has a clear positive signal at small distances $y$.  This reflects the behaviour of graph $C_2$ and corresponds to a larger probability for finding two $u$ quarks rather than a $u \bar{u}$ pair at small separation $y$.  Remarkably, the signal at small $y$ is of comparable size for $A_{u u}$ and $A_{u d}$, which implies that Fock states containing sea quarks do play an important role in this region.
As for polarisation effects, a clear signal for $u d$ or $u u$ in a $\pi^+$ is only seen for $m y A_{\delta q \ms q}$ and shown in the right panels of \fig{\ref{fig:phys-A_VV-A_TV}}.  Comparing this with $A_{q q}$, we see that spin-orbit correlations are appreciable for both flavour combinations.

The flavour combination $u u$ in a $\pi^0$ involves the sum $C_1 + 2 C_2$.  We observe a very strong compensation between the two connected graphs, which results in a marginal signal for $A_{u u}$ and $m y A_{\delta u \ms u}$.
The twist-two functions for $u d$ in a $\pi^0$ receive no contribution from connected graphs at all.  Within errors, the corresponding results are zero for all combinations of currents, and we do not show them here.

Among all polarised twist-two functions other than $m y A_{\delta q \ms q}$, a marginally nonzero signal is only seen for the longitudinal spin correlation $A_{\Delta u \Delta u}$ in a $\pi^+$ or a $\pi^0$.  This is dominated by the contribution from $C_2$ in both cases and shown in \fig{\ref{fig:phys-A_AA}}.

\begin{figure*}
\begin{center}
\subfigure[$|A_{q q}|_{\pi^+}$ (linear scale)]{
\includegraphics[width=0.48\textwidth,trim=0 0 0 17,clip]
{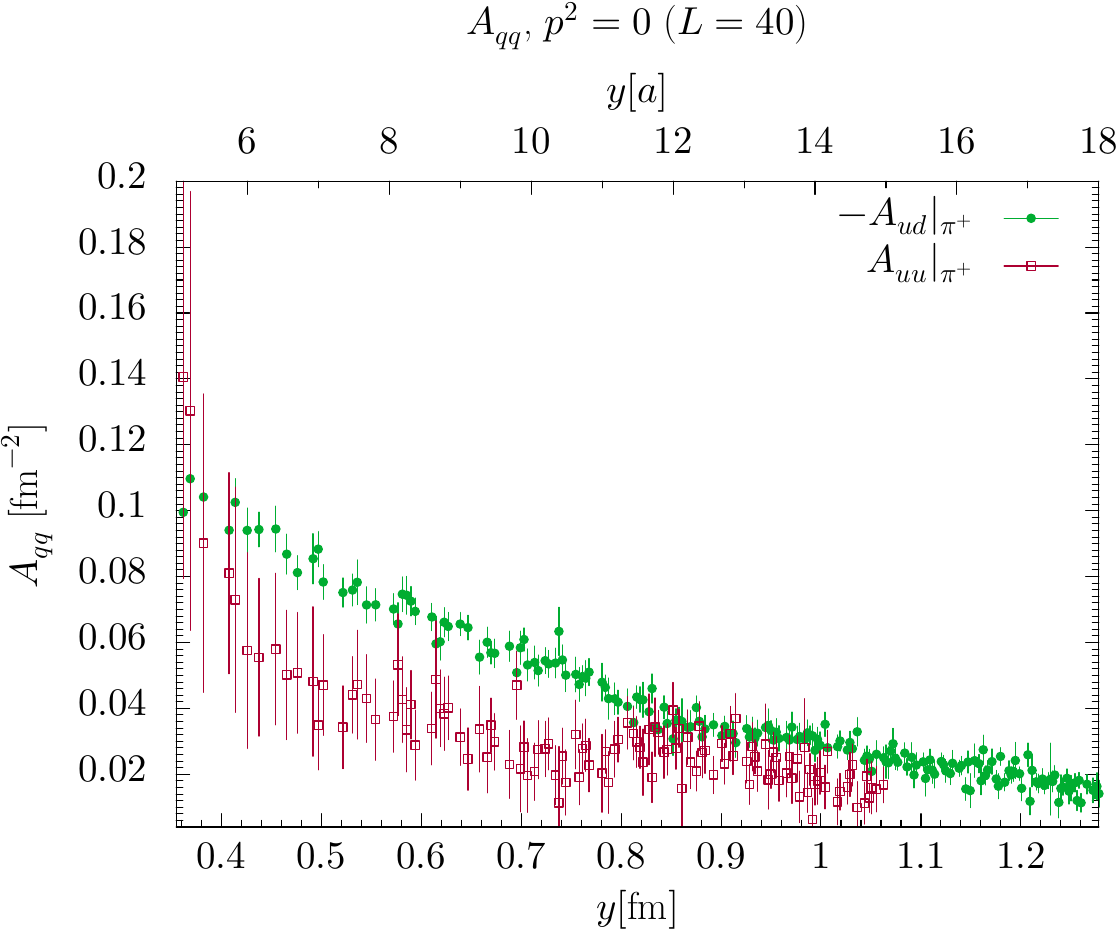}}
\hfill
\subfigure[$|A_{q q}|_{\pi^+}$ (logarithmic scale)]{
\includegraphics[width=0.48\textwidth,trim=0 0 0 17,clip]
{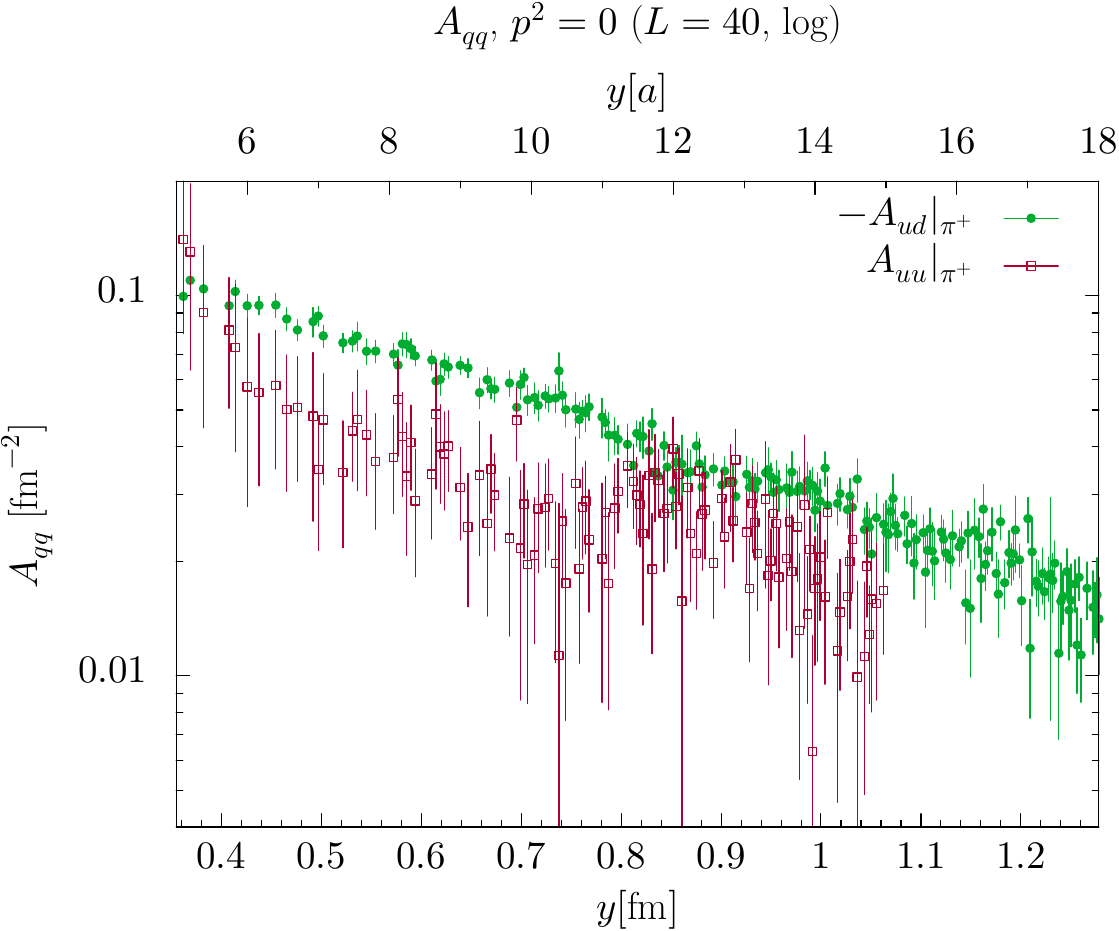}}
\caption{\label{fig:phys-A_ud_uu} Comparison of the twist-two functions $- A_{u d}$ and $A_{u u}$ in a $\pi^+$.  The data are the same as shown in \figs{\ref{phys-A_VV-ud}} and \ref{phys-A_VV-uu}.  Points with very large errors have been omitted for the sake of clarity.}
\end{center}
\end{figure*}

\rev{We recall that the assumption \eqref{eq:dpd-pocket} going into the ``pocket formula'' for double parton scattering implies that the unpolarised DPDs in a given hadron have the same $y$ dependence for all parton combinations.  If this were to hold also for skewed DPDs, the functions $A_{u d}$ and $A_{u u}$ in a $\pi^+$ should have the same $y$ dependence as well.  A comparison between these two functions is shown in figure~\ref{fig:phys-A_ud_uu}.  Although there is a hint for a different behaviour, especially at low $y$, the large errors in $A_{u u}$  prevent us from making a definitive statement.}

\FloatBarrier

\section{Results for nonzero pion momentum}
\label{sec:py-not-zero}

In this section, we use our data for nonzero pion momentum to study the $py$ dependence of the twist-two functions.  We restrict our study to graph $C_1$ for light quarks on the $L=40$ lattice: only in this case do we have simulations for a sufficient number of pion momenta.  Since graph $C_1$ dominates the twist-two matrix elements for $u d$ in a $\pi^+$, we will write $A_{u d}$, $A_{\delta u \ms d}$, \ldots for twist-two functions and $I_{u d}$, $I_{\delta u \ms d}$, \ldots for Mellin moments in what follows.


\subsection{Fit ansatz for the \texorpdfstring{$py$}{py} dependence}
\label{sec:py-fit-ansatz}

We start by proposing a functional ansatz for the twist-two functions, which is based on their relation \eqref{eq:py-zero-fct} with the Mellin moments of skewed DPDs.  We use this ansatz to fit the $py$ dependence of our lattice data.  This will allow for a model dependent extension of the twist-two functions  to all values of $py$, beyond the region \eqref{eq:euclid-restr} available on a Euclidean lattice.  This will in turn allow for a model dependent extraction of the Mellin moments of DPDs at zero skewness.
For ease of notation, we write $A(y^2, py)$ to denote any of the twist-two functions $A_{u d}$, \ldots, $A_{\delta u \delta d}$, $B_{\delta u \delta d}$.
Likewise, we write $I(y^2, py)$ for the Mellin moments $I_{u d}$, \ldots, $I_{\delta u \delta d}$, $I^t{}_{\!\delta u \delta d}$.

The basis of our ansatz is the assumption that, in its support region $-1 \le \zeta \le 1$, the skewed moment $I(y^2, \zeta)$ can be approximated by a polynomial in $\zeta$,
\begin{align}
   \label{eq:I-series}
I(y^2, \zeta) &= \pi \sum_{n=0}^N a_{n}(y^2)\, \zeta^{2n}
\end{align}
with some integer $N$, where we used the symmetry relation \eqref{eq:even-in-zeta} to restrict terms to even powers of $\zeta$.  We write $=$ instead of $\approx$ in the spirit of a fit ansatz, i.e.\ we do not claim that \eqref{eq:I-series} is exact.  \rev{We currently have no guidance from theory regarding the $\zeta$ dependence of $I(y^2, \zeta)$, and \eqref{eq:I-series} should be taken as a simple ansatz that is to be validated by data.  As there are no known constraints on the behaviour of skewed DPDs at the edge $\zeta = 1$ of their support region, we allow $I(y^2, \zeta)$ to be finite at that point.}

Inverting the Fourier transform in \eqref{eq:skewed-mellin-inv-fct}, we obtain
\begin{align}
   \label{eq:A-series}
A(y^2, py) &= \sum_{n=0}^N a_{n}(y^2)\, h_n(py) \,,
\end{align}
where we introduced the functions
\begin{align}
h_n(x) &= \frac{1}{2} \int_{-1}^1 d\zeta\, e^{i \zeta x}\, \zeta^{2 n} \,.
\end{align}
A crucial property of the ansatz \eqref{eq:A-series} is that its Fourier transformation \eqref{eq:skewed-mellin-inv-fct} has the correct support in $\zeta$.

Let us collect a few properties of the functions $h_n(x)$.  From their definition, one easily derives
\begin{align}
   \label{eq:h-props}
h_n(0) &= \frac{1}{1 + 2 n} \,,
&
\frac{d^2 h_n(x)}{d x^2} &= - h_{n+1}(x)
\end{align}
and thus obtains the Taylor series
\begin{align}
\label{eq:h-taylor}
h_n(x)
&= \sum_{m=0}^\infty \frac{(-1)^m}{1 + 2 n + 2 m}\; \frac{x^{2 m}}{(2 m)!} \;.
\end{align}
An explicit representation is given by
\begin{align}
h_n(x) = s_{n}(x) \sin x + c_{n}(x) \cos x
\end{align}
with rational functions
\begin{align}
s_{n}(x)
&= \sum_{m=0}^{n} \frac{(2 n)!}{(2 n - 2 m)!} \, \frac{(-1)^m}{x^{1 + 2 m}} \,,
&
c_{n}(x)
&= \sum_{m=0}^{n-1} \frac{(2 n)! }{(2 n - 2 m - 1)!} \,
   \frac{(-1)^m}{x^{2 + 2 m}} \,.
\end{align}
For $n=0$ and $n=1$, these functions read
\begin{align}
s_0(x) &= 1/x \,, & c_0(x) &= 0 \,,
&
s_1(x) &= (x^2 - 2) /x^3 \,, & c_1(x) &= 2 /x^2 \,.
\end{align}
In terms of the normalised quantities
\begin{align}
   \label{eq:normalised-A}
\widehat{A}(y^2, py)
   &= \frac{A(y^2, py)}{A(y^2, py=0)} \,,
&
\hat{a}_n(y^2)
   &= \frac{a_n(y^2)}{A(y^2, py=0)}
\end{align}
our ansatz \eqref{eq:A-series} reads
\begin{align}
   \label{eq:A-hat-series}
\widehat{A}(y^2, py) &= \sum_{n=0}^N \hat{a}_{n}(y^2)\, h_n(py) \,.
\end{align}
Using \eqref{eq:zeta-moments} and \eqref{eq:h-taylor}, we then obtain
\begin{align}
   \label{eq:zeta-series}
\langle \zeta^{2 m} \rangle (y^2)
&= \Biggl[\ms (-1)^{m} \,
      \frac{\partial^{2 m} \widehat{A}(y^2, py)}{(\partial\ms py)^{2 m}}
   \ms\Biggr]_{py=0}
 = \, \sum_{n=0}^N \frac{1}{1 + 2n + 2m} \; \hat{a}_n(y^2) \,.
\end{align}

Let us now describe our general fitting procedure.  In order to achieve stable fits, we first determine the $y^2$ dependence of $A(y^2, py=0)$.  This includes the information from data with zero pion momentum and has typically much smaller errors than the data for nonzero $py$.

In a second step, we fit the $y$ dependent coefficients $\hat{a}_{n}(y^2)$ in the ansatz \eqref{eq:A-hat-series}.  To make the degrees of freedom of this fit explicit, we consider the moments $\langle \zeta^{2 m} \rangle (y^2)$ for $m=0, \ldots, N$.  Inverting the relation \eqref{eq:zeta-series}, we obtain
\begin{align}
   \label{eq:a-hat-rel}
\hat{a}_n(y^2)
&= \sum_{m=0}^N ( T^{-1} )_{n m} \; \langle \zeta^{2 m} \rangle (y^2) \,,
\end{align}
where $T$ is the $(N+1) \times (N+1)$ matrix with elements
\begin{align}
   \label{eq:T-mat-def}
T_{m n} = (1 + 2 n + 2 m)^{-1} \,.
\end{align}
Since by definition $\langle \zeta^0 \rangle (y^2) = 1$, we can thus fit the $py$ dependence of the twist-two functions to \eqref{eq:A-hat-series} and \eqref{eq:a-hat-rel} with $N$ fit parameters $\langle \zeta^{2} \rangle$, \ldots, $\langle \zeta^{2 N} \rangle$ at each value of $y^2$.  We call this ``local fits'' in the following, where ``local'' means ``local in $y^2$''.

To obtain a parametrisation of both the $py$ and the $y^2$ dependence, we assume an expansion
\begin{align}
   \label{eq:zeta-y-fit}
\langle \zeta^{2 m} \rangle (y^2)
   &= \sum_{k=0}^{K} c_{m k} \, \sqrt{-y^2}^{\, k} \,.
\end{align}
This is referred to as our ``global fit''.  By virtue of \eqref{eq:A-hat-series} and \eqref{eq:a-hat-rel}, this corresponds to an expansion of $\widehat{A}(y^2, py)$ in powers of $\sqrt{-y^2}$.  The condition $\langle \zeta^0 \rangle (y^2) = 1$ implies $c_{0 \ms k} = \delta_{0 \ms k}$.


\subsection{Fitting the data}
\label{sec:py-fits}

We recall that we have data for $p = 0, 1, \sqrt{2}, \sqrt{3}$ and $2$ in units of $2\pi/(L a) \approx 437 \mev$.  For a given value of $y$, this allows for a maximum value $4 \pi y/(L a) \approx 6.28\, y /(20 a)$ for $|py|$.  We apply the cut \eqref{cos-cut} on the angle $\theta$ to the $p=0$ data, but not to the data with $p>0$.  We then average all data points with the same values of $py$ and $y^2$.

We find that the twist-two functions at $py = 0$ can be well described by a superposition of two exponentials,
\begin{align}
   \label{eq:y-dep-fit}
A(y^2, py=0) &= A_1 \, e^{- a_1 \ms (y - y_{\text{min}})}
              + A_2 \, e^{- a_2 \ms (y - y_{\text{min}})}
& \text{for } y_{\text{min}} \le y \le y_{\text{max}} \,,
\end{align}
with $y_{\text{min}} = 5a = 0.355 \fm$ and $y_{\text{max}} = 20a = 1.42 \fm$.  We do not include data with $y > y_{\text{max}}$, because they have large errors and are increasingly affected by finite size effects.   The resulting fit parameters are given in \tab{\ref{tab:y-dep-fit}}.  Let us emphasise that these fits are \emph{not} suitable for extrapolating the twist-two functions to values significantly below $y = y_{\text{min}}$.

\begin{table}[!b]
\begin{center}
\renewcommand{\arraystretch}{1.1}
\begin{tabular}{c|rrrr|c}
\hline \hline
function & $A_1 \, [\fm^{-2}]$ & $a_1 \, [\fm^{-1}]$ & $A_2 \, [\fm^{-2}]$ & $a_2 \, [\fm^{-1}]$ & $\chi^2 / \mathrm{dof}$\\
\hline
$A_{u d}$ & $-0.1163(39)$ & $2.150(68)$ & $0.0141(34)$ & $11.5 \pm 2.2$ & $0.95$ \\
\hline
$A_{\Delta u \Delta d}$ & $-0.0414(77)$ & $6.71(36)$ & $0.0326(74)$ & $4.21(52)$ & $0.94$ \\
\hline
$A_{\delta u \ms d}$ & $-0.1157(62)$ & $3.786(89)$ & $0.0222(68)$ & $6.38(30)$ & $1.76$ \\
\hline
$A_{\delta u \delta d}$ & $0.0133(24)$ & $2.11(28)$ & $-0.0018(23)$ & $7.8 \pm 7.4$ & $0.34$ \\
\hline
$B_{\delta u \delta d}$ & $-0.0491(59)$ & $4.50(20)$ & $-0.0084(64)$ & $9.4 \pm 2.0$ & $0.95$ \\
\hline \hline
\end{tabular}
\caption{\label{tab:y-dep-fit} Parameters for the fit \protect\eqref{eq:y-dep-fit} of twist-two functions at $py=0$ in the region $5a \le y \le 20a$.  Throughout this section, we consider the data for graph $C_1$ and light quarks on our lattice with $L=40$.}
\end{center}
\end{table}

We notice a relatively high value of $\chi^2 / \mathrm{dof}$ in the fit for $A_{\delta u \ms d}$.  This is due to some scatter in the data at high $y$, which comes from points with large $p$.  Repeating the fit with an upper limit $y \le 15a$, we find that $\chi^2 / \mathrm{dof}$ decreases from $1.76$ to $0.9$ for $A_{\delta u \ms d}$.  By comparison, the value of $\chi^2 / \mathrm{dof}$ in the fit for $A_{u d}$ decreases from $0.95$ to $0.6$ with the same reduction of the fitting range.

We then proceed and fit the $py$ dependence to \eqref{eq:A-hat-series} and \eqref{eq:a-hat-rel} locally in $y^2$.  To have enough data in these fits, we  introduce bins in $y$ and combine all points with $(n - 1/2) \ms a < y < (n + 1/2) \ms a$ for integer $n$ between $5$ and $20$.  In addition, we fit the combined $y^2$ and $py$ dependence of $\widehat{A}$ to \eqref{eq:A-hat-series}, \eqref{eq:a-hat-rel} and \eqref{eq:zeta-y-fit}.  We explored fits with different maximum values $N$ and $K$ in the sums and find that, given the fit range and the statistical quality of our data, an adequate choice is $N=1$ for local fits and $N=1$, $K=1$ for the global fit.  The parameters of the global fit are given in \tab{\ref{tab:py-fit}}.
If we take $N=2$ instead, the error bands of the fit results for $\widehat{A}$ increase significantly, whilst the decrease of $\chi^2/ \text{dof}$ is minor.  We hence conclude that we would over-fit the data by choosing $N=2$ or even higher values.

\begin{table}
\begin{center}
\renewcommand{\arraystretch}{1.25}
\begin{tabular}{c|cc|cccc|c}
\hline \hline
function & $c_{10}$ & $c_{11} \, [\fm^{-1}]$ & $\chi^2 / \mathrm{dof}$\\
\hline
$\widehat{A}_{u d}$ & $0.096(40)$ & $0.247(39)$ & $1.19$ \\
$\widehat{A}_{\Delta u \Delta d}$ & $-0.43(73)$ & $0.17(75)$ & $0.68$ \\
$\widehat{A}_{\delta u \ms d}$ & $0.102(50)$ & $0.111(47)$ & $1.37$ \\
$\widehat{A}_{\delta u \delta d}$ & $-0.05(14)$ & $0.31(12)$ & $0.80$ \\
$\widehat{B}_{\delta u \delta d}$ & $-0.023(90)$ & $0.242(92)$ & $0.99$ \\
\hline \hline
\end{tabular}
\caption{\label{tab:py-fit} Parameters of the fit of the combined $y^2$ and $py$ dependence of the normalised twist-two functions $\widehat{A}(y^2, py)$ to \protect\eqref{eq:A-hat-series}, \protect\eqref{eq:a-hat-rel} and \protect\eqref{eq:zeta-y-fit} with $N=K=1$.}
\end{center}
\end{table}

\begin{figure*}
\begin{center}
\subfigure[$\widehat{A}_{u d}$ at $y = 15 a$]{
\includegraphics[width=0.48\textwidth,trim=0 0 0 70,clip]
{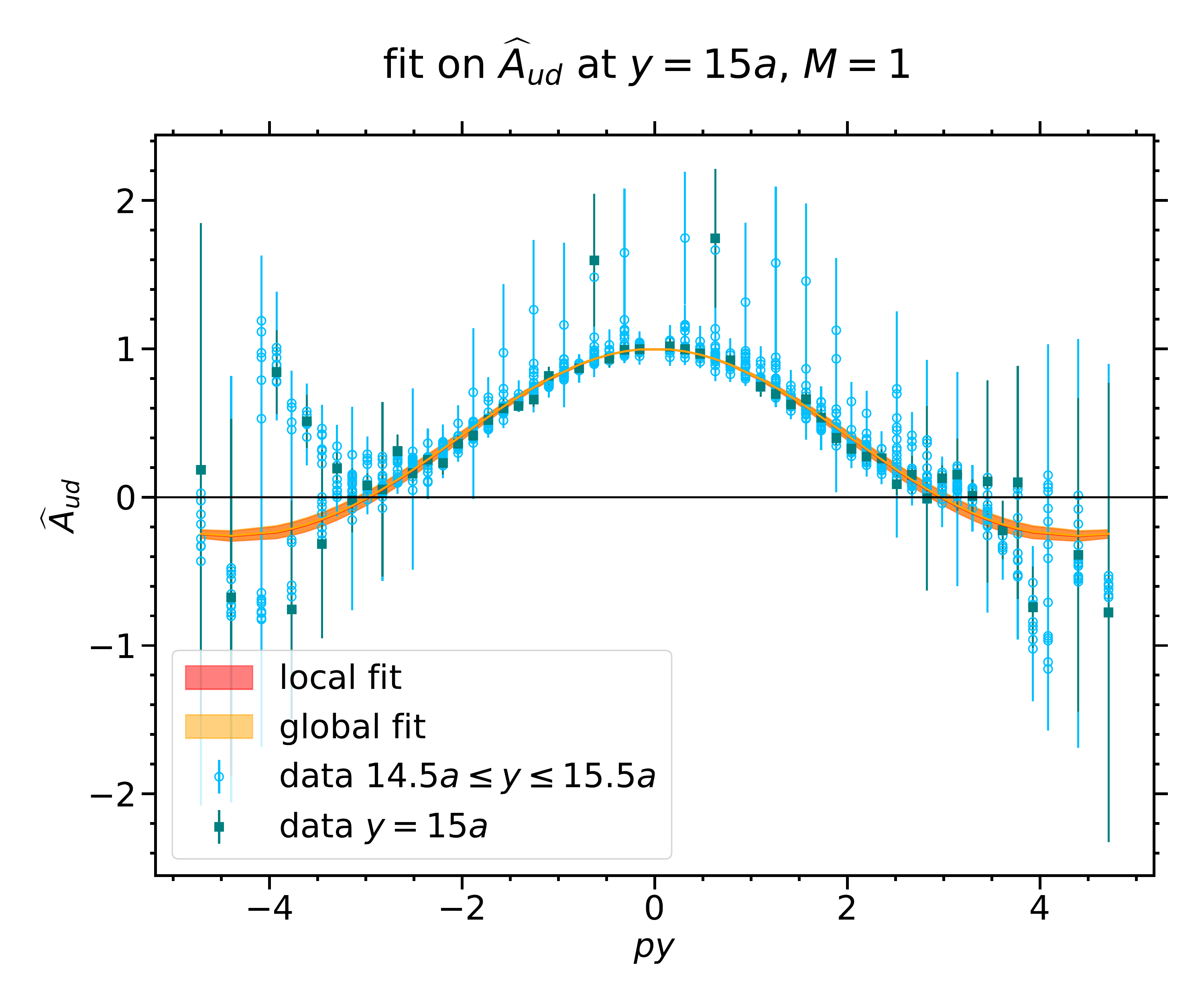}}
\hfill
\subfigure[$\widehat{A}_{\Delta u \Delta d}$ at $y = 15 a$\label{fig:py-fit-A_AA}]{
\includegraphics[width=0.48\textwidth,trim=0 0 0 70,clip]
{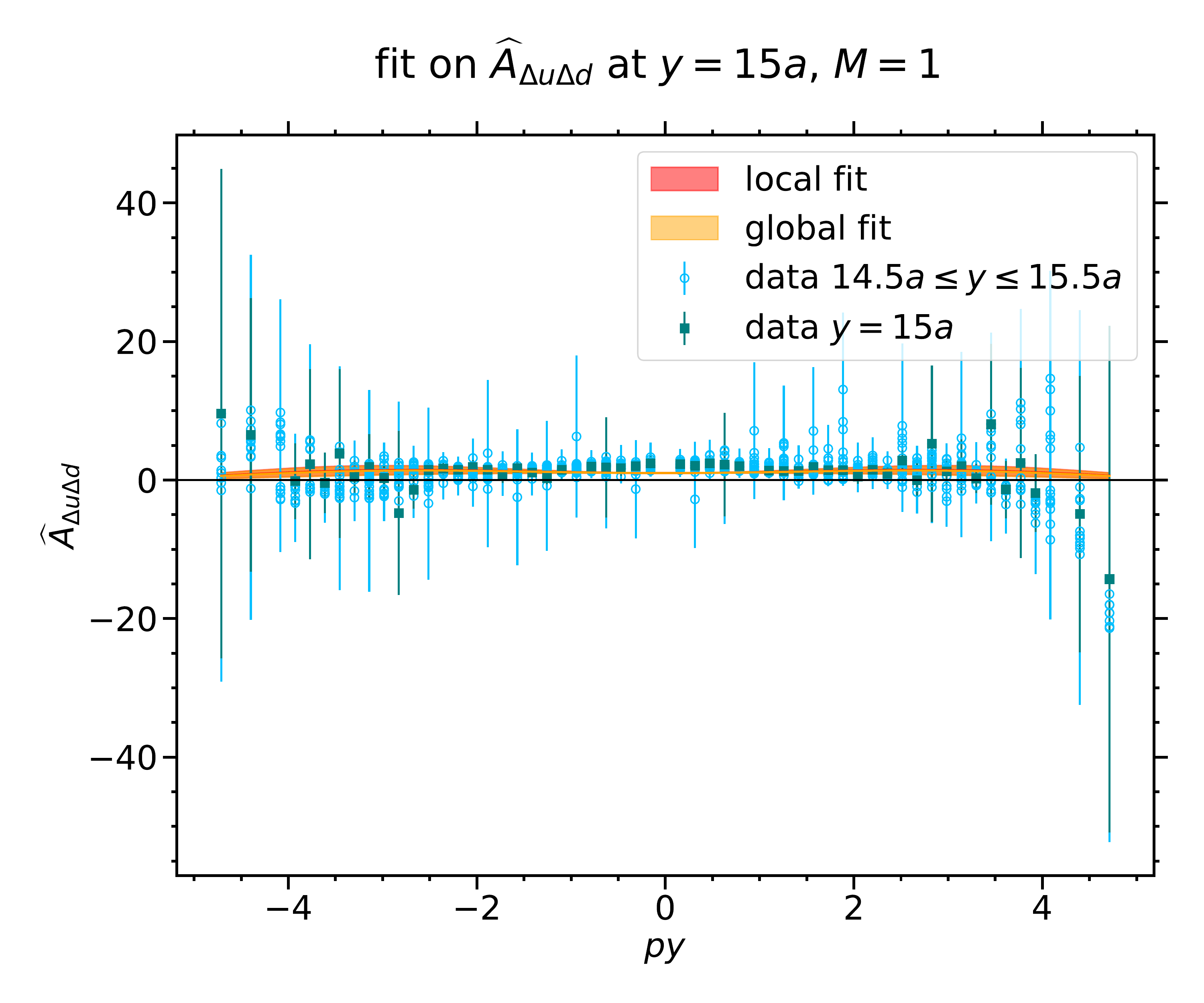}}
\\
\subfigure[$\widehat{A}_{\delta u \ms d}$ at $y = 15 a$]{
\includegraphics[width=0.48\textwidth,trim=0 0 0 70,clip]
{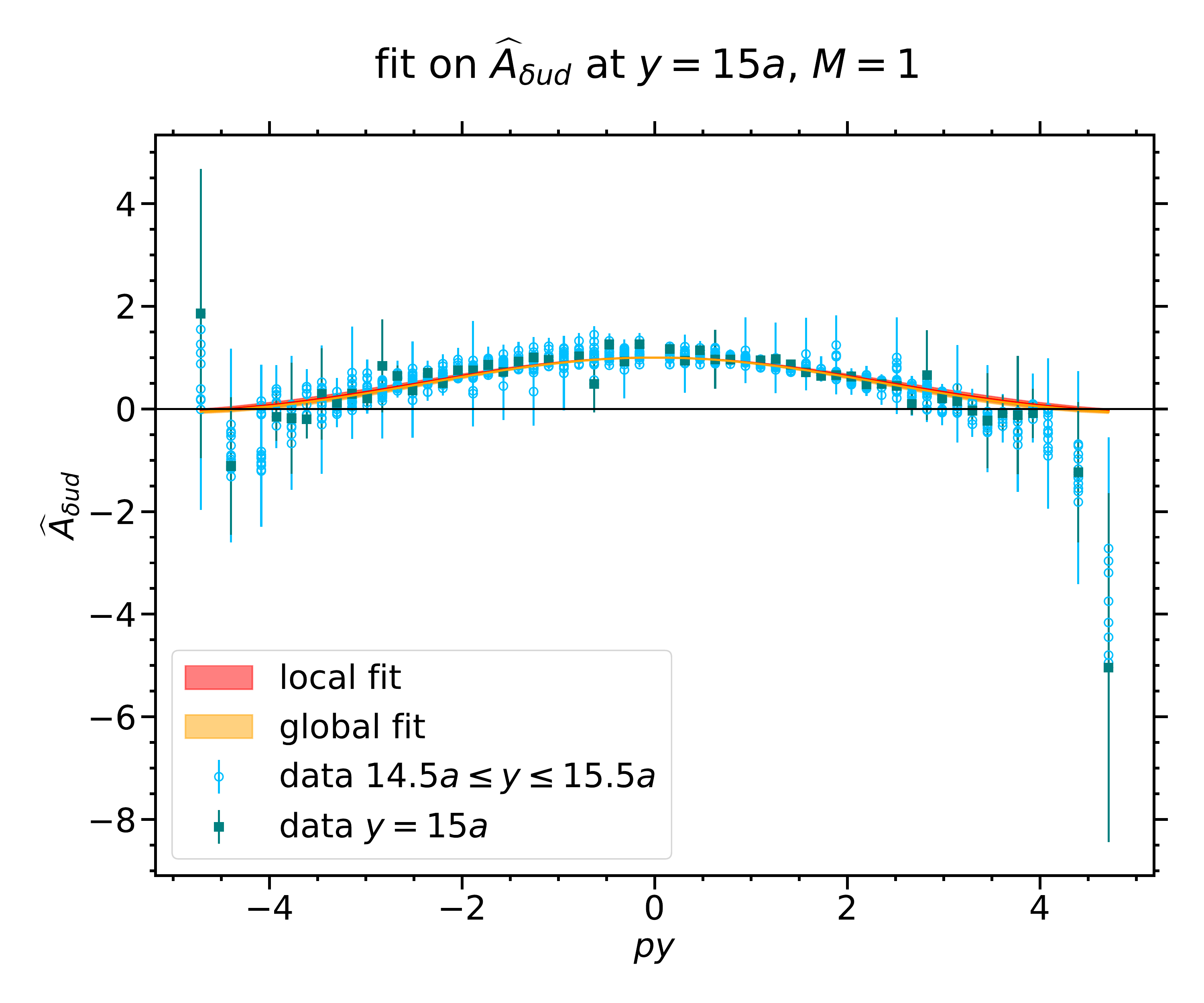}}
\hfill
\subfigure[$\widehat{B}_{\delta u \delta d}$ at $y = 15 a$]{
\includegraphics[width=0.48\textwidth,trim=0 0 0 70,clip]
{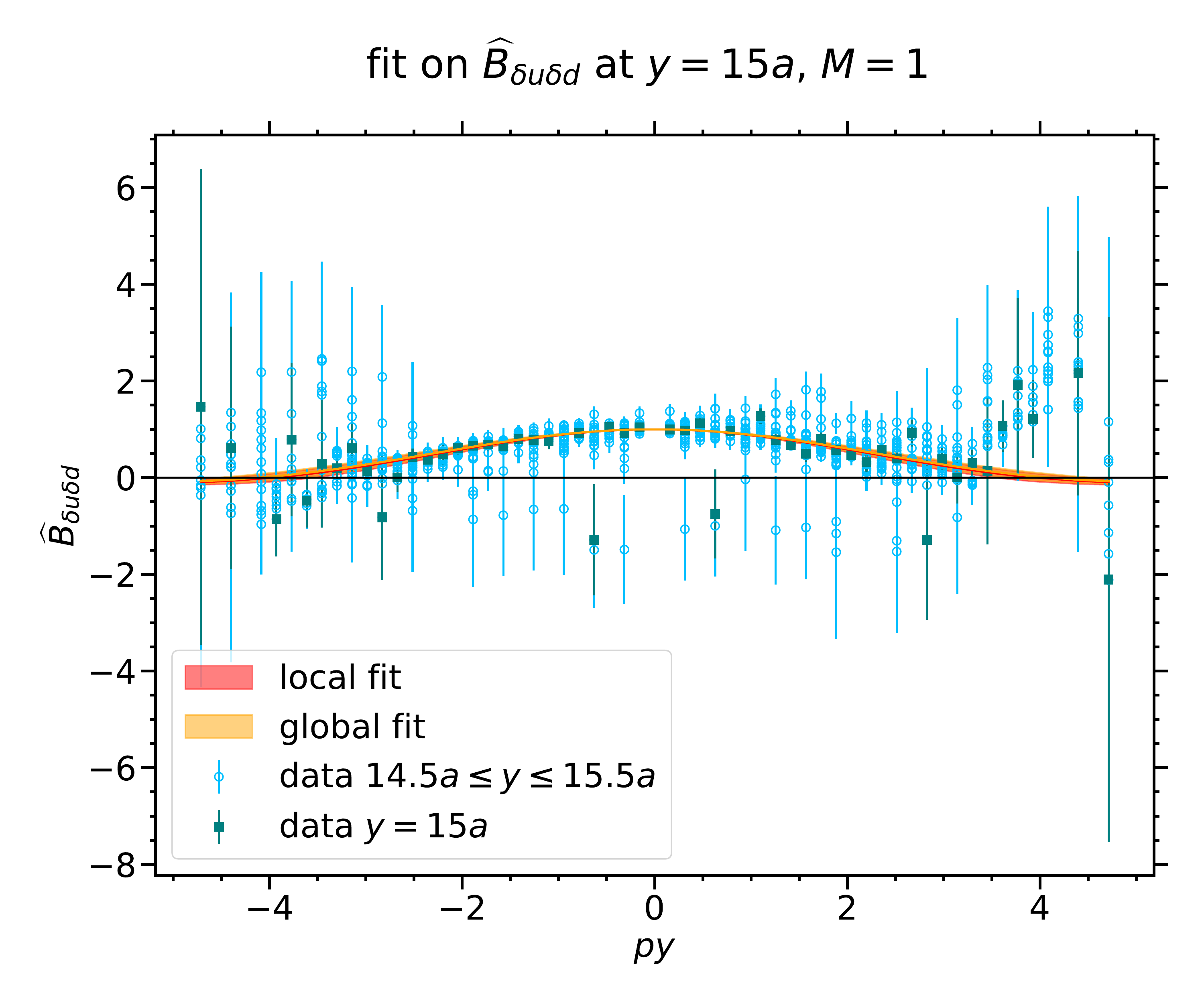}}
\caption{\label{fig:py-fit} Data and fits of the $py$ dependence of normalised invariant functions.  Dark points show data at $y = 15 a$.  Light points show data in a $y$ range of $a/2$ around $15 a$, which are included in the local fits.   The plot for $\widehat{A}_{\delta u \delta d}$ (not shown) is qualitatively similar to the one for $\widehat{B}_{\delta u \delta d}$.}
\end{center}
\end{figure*}

\begin{figure*}
\begin{center}
\subfigure[$\widehat{A}_{u d}$ at $y = 5 a$]{
\includegraphics[width=0.48\textwidth,trim=0 0 0 70,clip]
{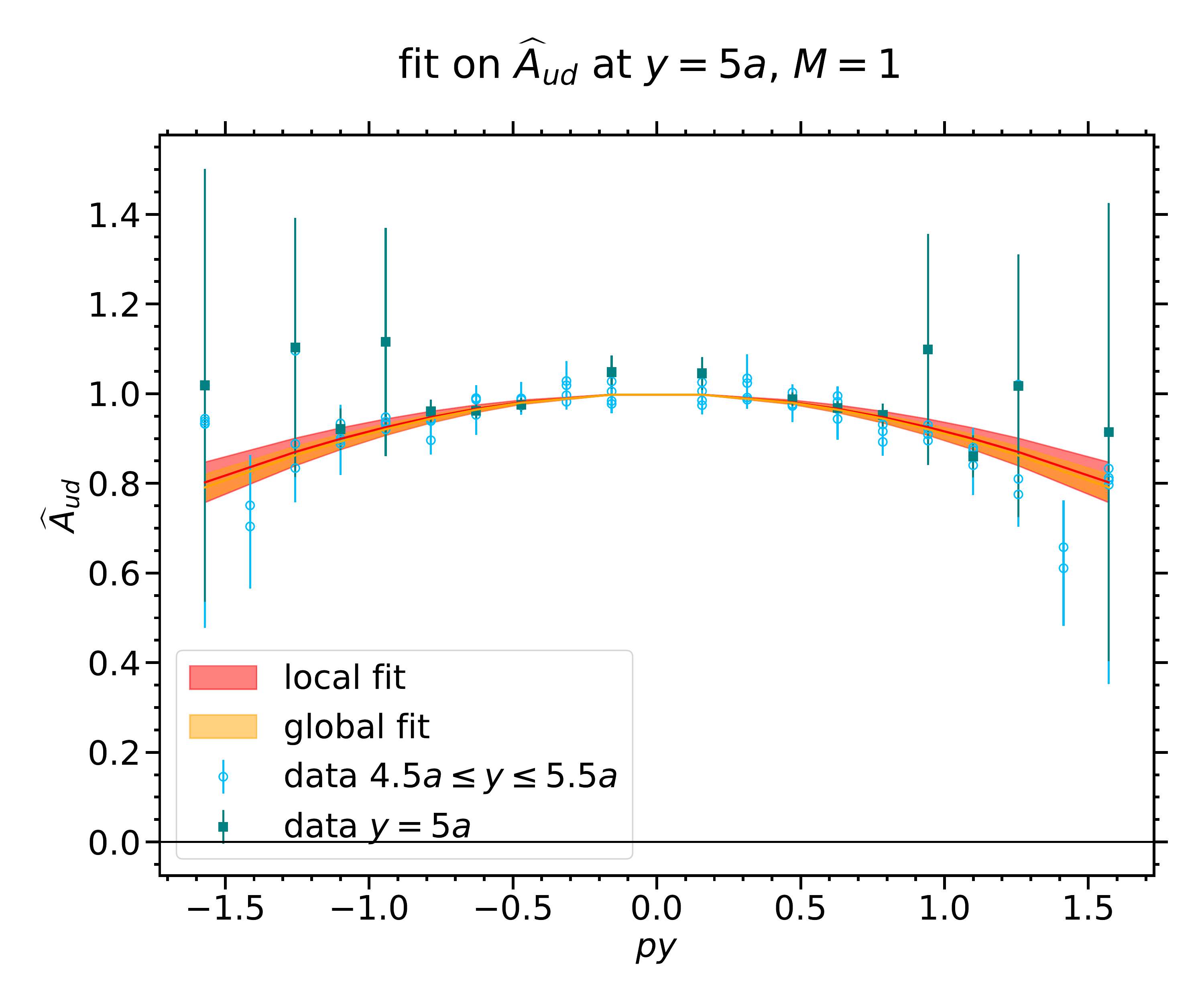}}
\hfill
\subfigure[$\widehat{A}_{u d}$ at $y = 10 a$]{
\includegraphics[width=0.48\textwidth,trim=0 0 0 70,clip]
{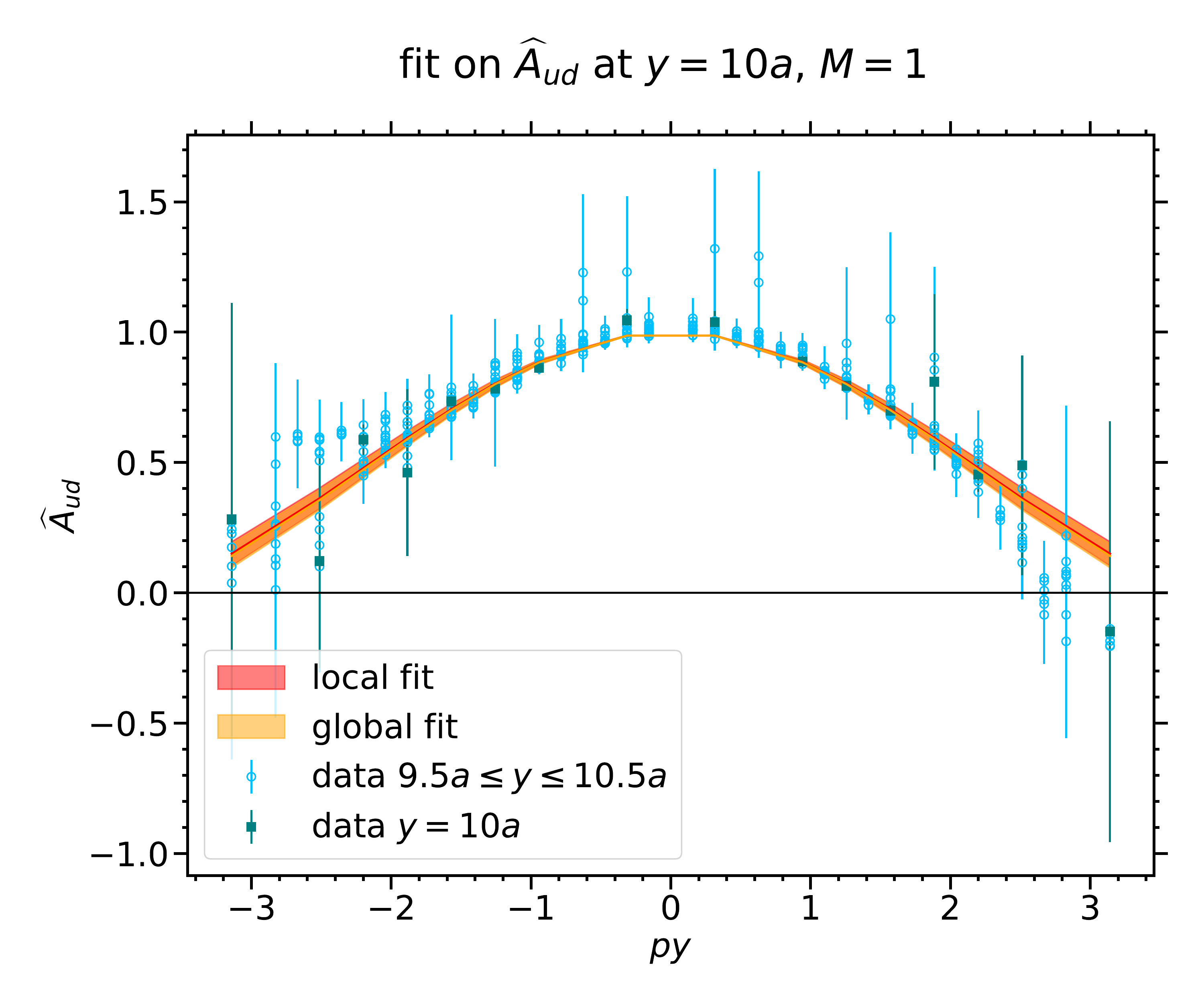}}
\caption{\label{fig:py-fit-A_VV} Data and fits of the $py$ dependence of $\widehat{A}_{u d}$ for different $y$.  The meaning of dark and light points is as in \fig{\protect\ref{fig:py-fit}}.}
\end{center}
\end{figure*}

We compare our data and fits in \fig{\ref{fig:py-fit}} for different functions at $y = 15a$ and in \fig{\ref{fig:py-fit-A_VV}} for $\widehat{A}_{u d}$ at $y = 5 a$ and $10 a$.  We find good agreement between the local and global fits.  Note that the twist-two functions are symmetric in $py$ due to $PT$ invariance, which is realised on the lattice.  A departure from this symmetry in the data must therefore be due to statistical fluctuations.  Many data points have admittedly large errors, which is a consequence of at least one of $y$ or $p$ being large.  Nevertheless, the fitted parameters for all functions except $\widehat{A}_{\Delta u \Delta d}$ are in general well determined, and the corresponding error bands of the fit results  are reasonably small.  As is seen in \fig{\ref{fig:py-fit-A_AA}}, the data for $\widehat{A}_{\Delta u \Delta d}$ are much too noisy for fitting the $py$ dependence, and we exclude this function from our further discussion.

In the data for $y = 15a$, we see an indication for zero crossings around $|py| = 4$ in several twist-two functions.  That this can be reproduced with a superposition of the two functions $h_0(py)$ and $h_1(py)$ gives us some confidence in our fit ansatz.

\begin{figure*}
\begin{center}
\subfigure[$I_{u d}$]{
\includegraphics[height=14.4em,trim=20 0 0 90,clip]
{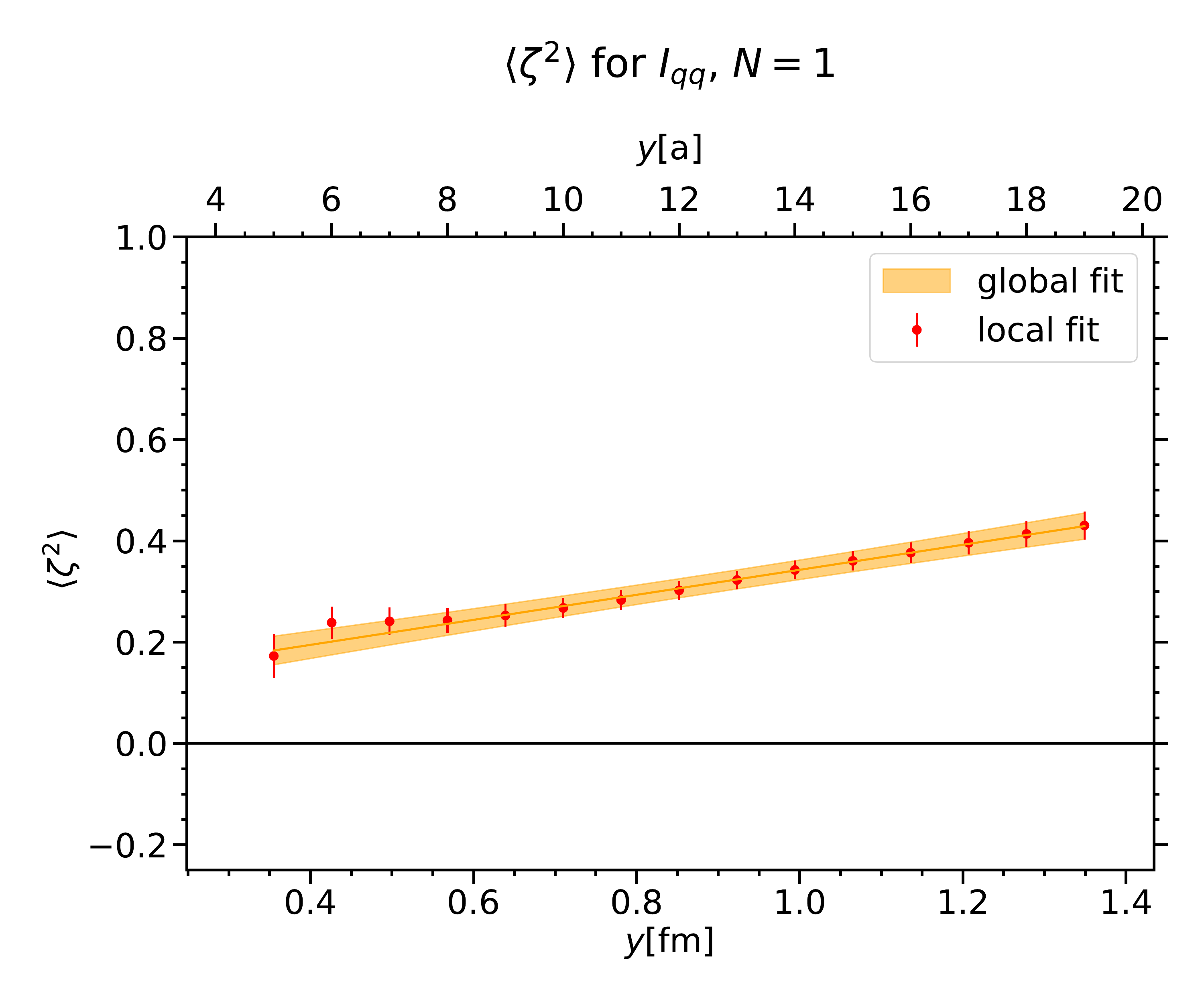}}
\hfill
\subfigure[$I_{\delta u \ms d}$]{
\includegraphics[height=14.4em,trim=20 0 0 90,clip]
{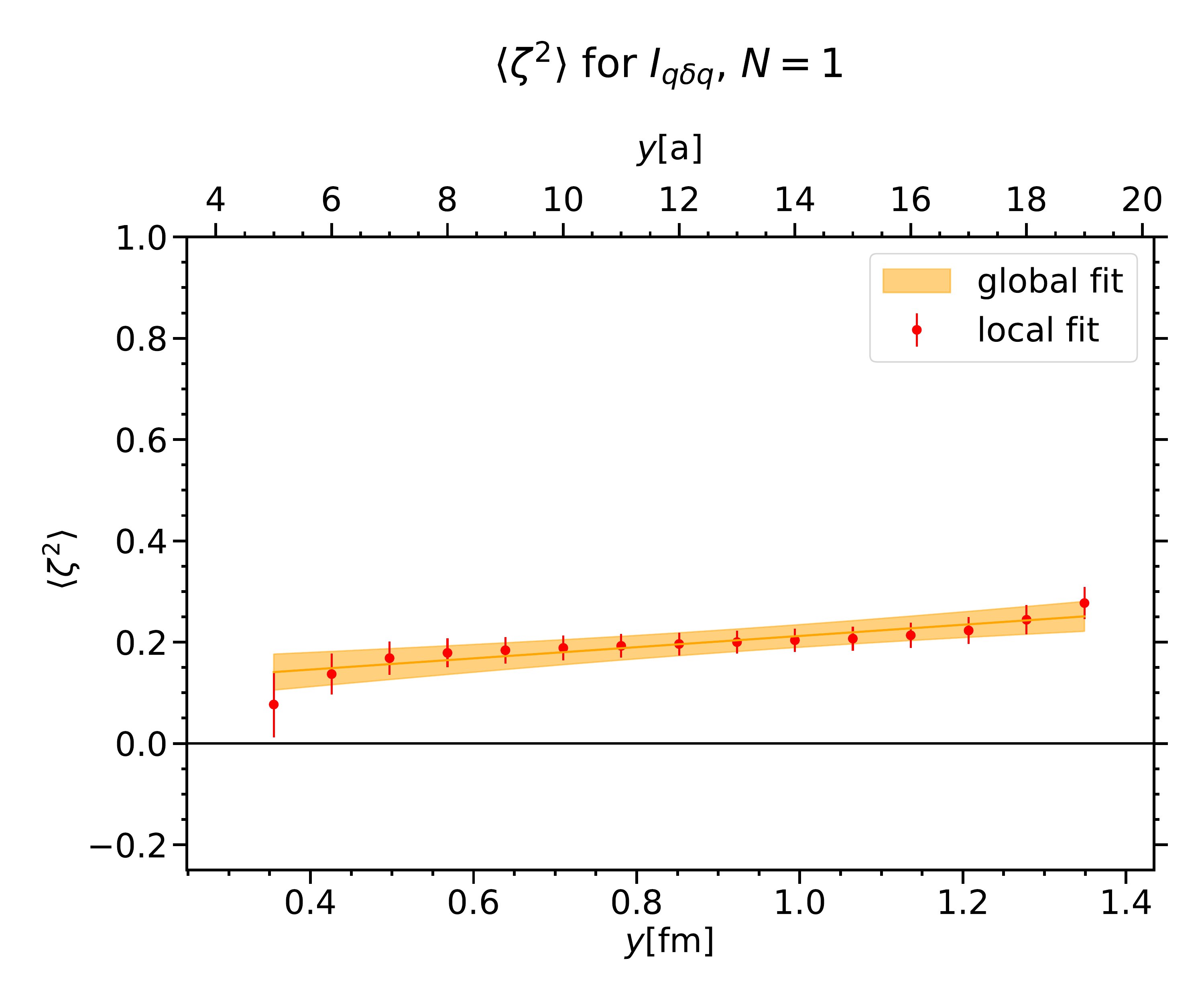}}
\\[0.8em]
\subfigure[$I^{}_{\delta u \delta d}$]{
\includegraphics[height=14.4em,trim=20 0 0 90,clip]
{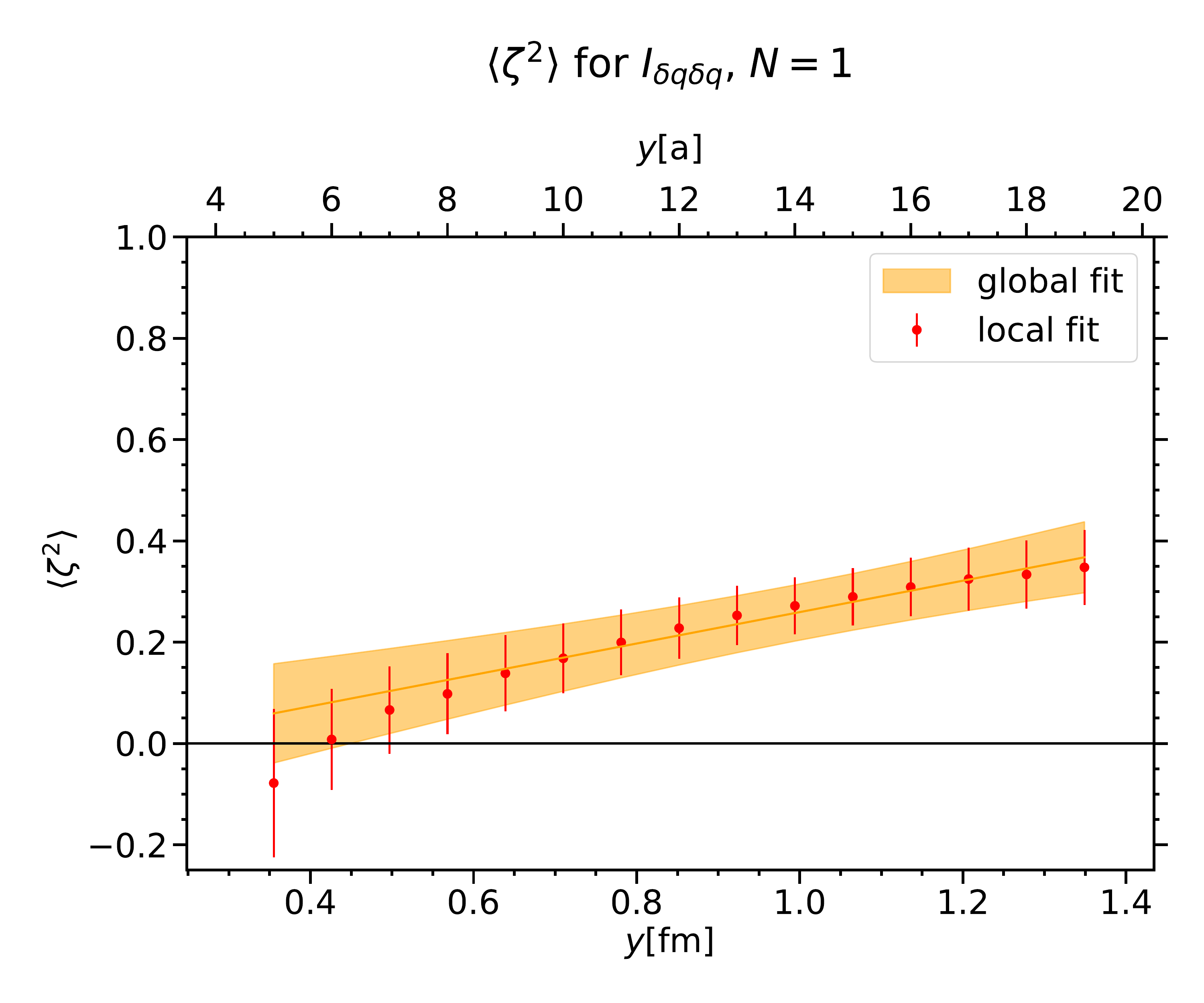}}
\hfill
\subfigure[$I^t_{\delta u \delta d}$]{
\includegraphics[height=14.4em,trim=20 0 0 90,clip]
{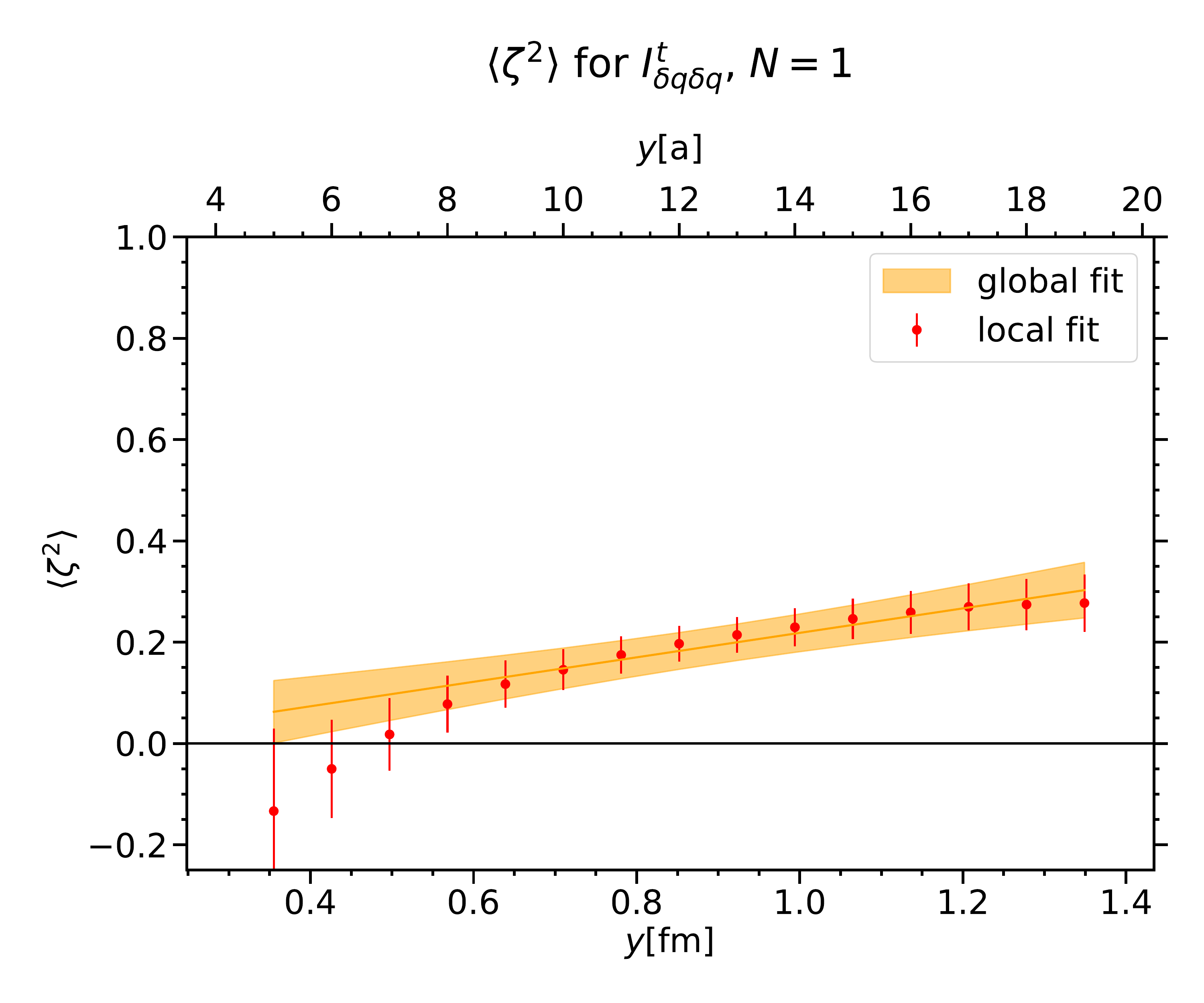}}
\caption{\label{fig:zeta} Values of the moment $\langle \zeta^2 \rangle(y^2)$ associated with $I(y^2, \zeta)$, extracted by local fits (data points) and the global fit (bands).}
\end{center}
\end{figure*}

Using our fits, we can compute the moment $\langle \zeta^2 \rangle(y^2)$ associated with $I(y^2, \zeta)$, which according to \eqref{eq:zeta-series} follows from the curvature of $\widehat{A}(y^2, py)$ at $py = 0$.  The results are shown in \fig{\ref{fig:zeta}}.  We find again good agreement between the local and global fits.  A clear $y$ dependence of $\langle \zeta^2 \rangle$ is observed, except for $I_{\delta u \ms d}$.  The values of $\langle \zeta^2 \rangle$ are not too large, especially for small $y$.  Their size does, however, imply that nonzero values of the skewness $\zeta$ must play some role in the integral representation $\pi A(y^2, py=0) = \int_0^1 \bs d\zeta\, I(y^2, \zeta)$.

\subsection{Mellin moments of DPDs}
\label{sec:mellin-moments}

We now use the global fit described in the last section to reconstruct the lowest Mellin moments of skewed DPDs.  Let us re-emphasise that such a reconstruction is necessarily dependent on the functional ansatz we have made, given the impossibility to constrain the full $py$ dependence of twist-two functions with lattice simulations.  We recall that the results for the spin correlation $\Delta u \Delta d$ are too noisy and hence omitted in the following.

\begin{figure*}
\begin{center}
\subfigure[$I_{u d}(y^2, \zeta)$]{
\includegraphics[width=0.48\textwidth,trim=20 0 0 85,clip]
{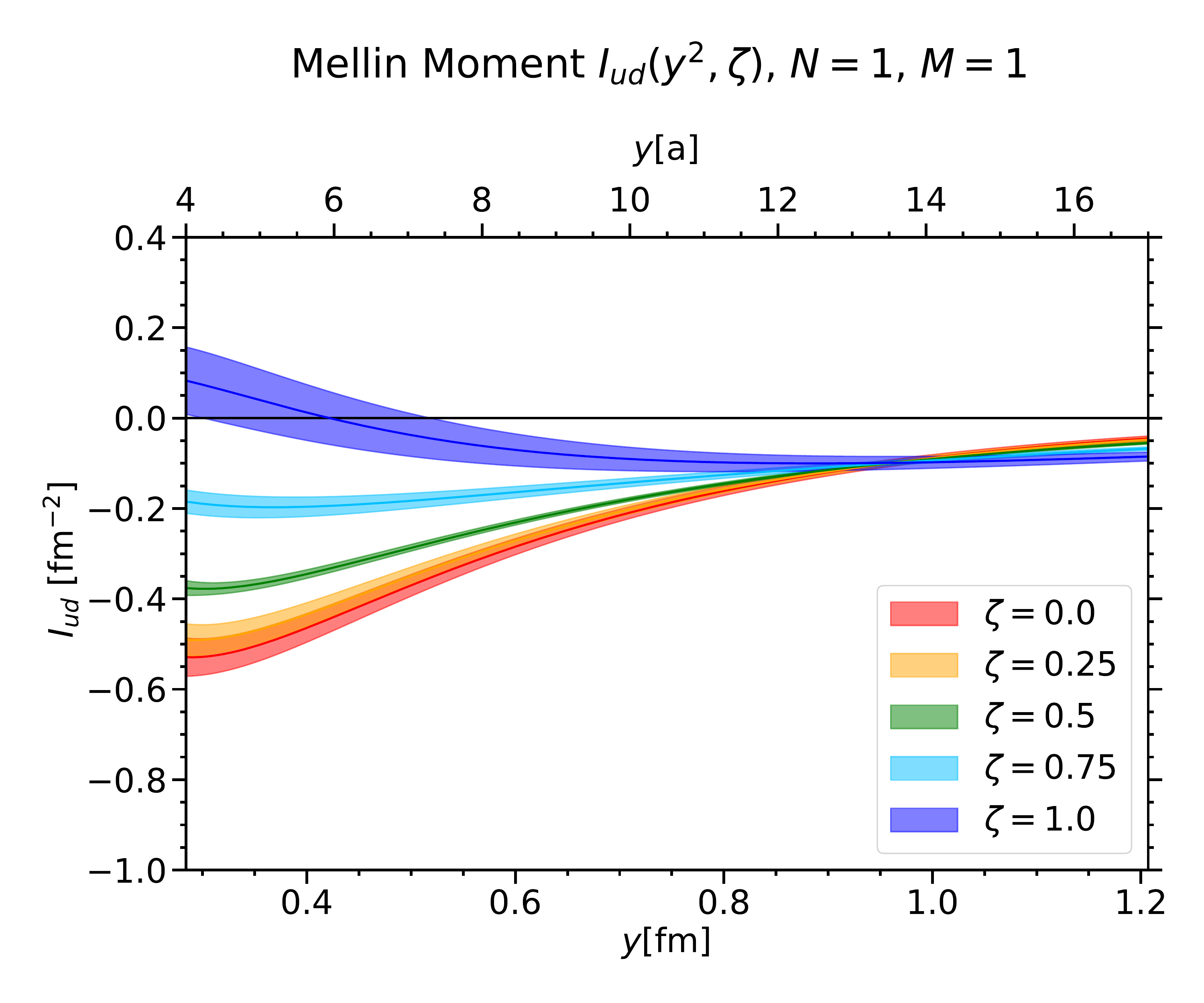}}
\hfill
\subfigure[$m y I_{\delta u \ms d}(y^2, \zeta)$]{
\includegraphics[width=0.48\textwidth,trim=20 0 0 85,clip]
{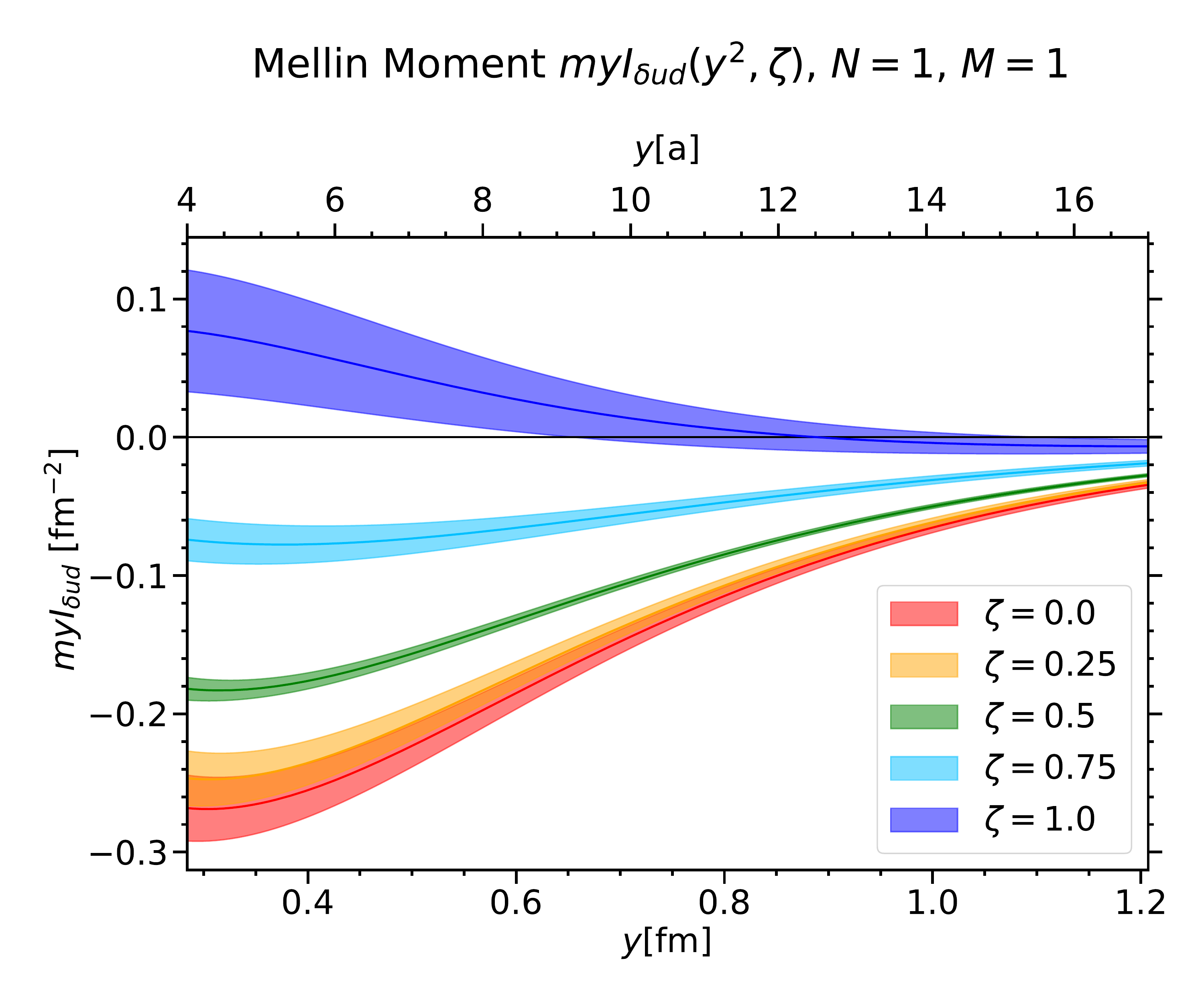}}
\\[0.8em]
\subfigure[$I_{\delta u \delta d}(y^2, \zeta)$]{
\includegraphics[width=0.48\textwidth,trim=20 0 0 85,clip]
{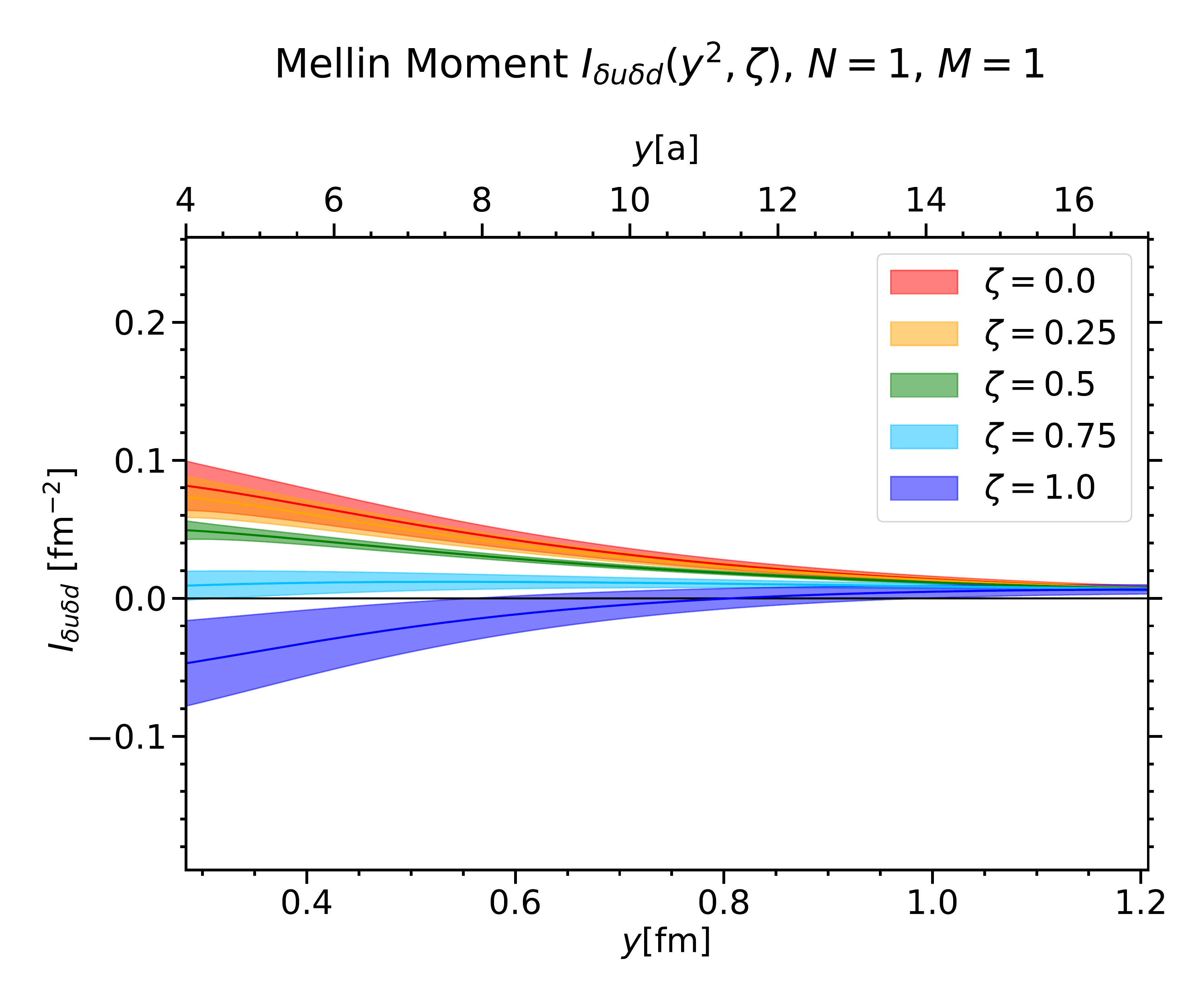}}
\hfill
\subfigure[$m^2 |y^2| I^t_{\delta u \delta d}(y^2, \zeta)$]{
\includegraphics[width=0.48\textwidth,trim=20 0 0 85,clip]
{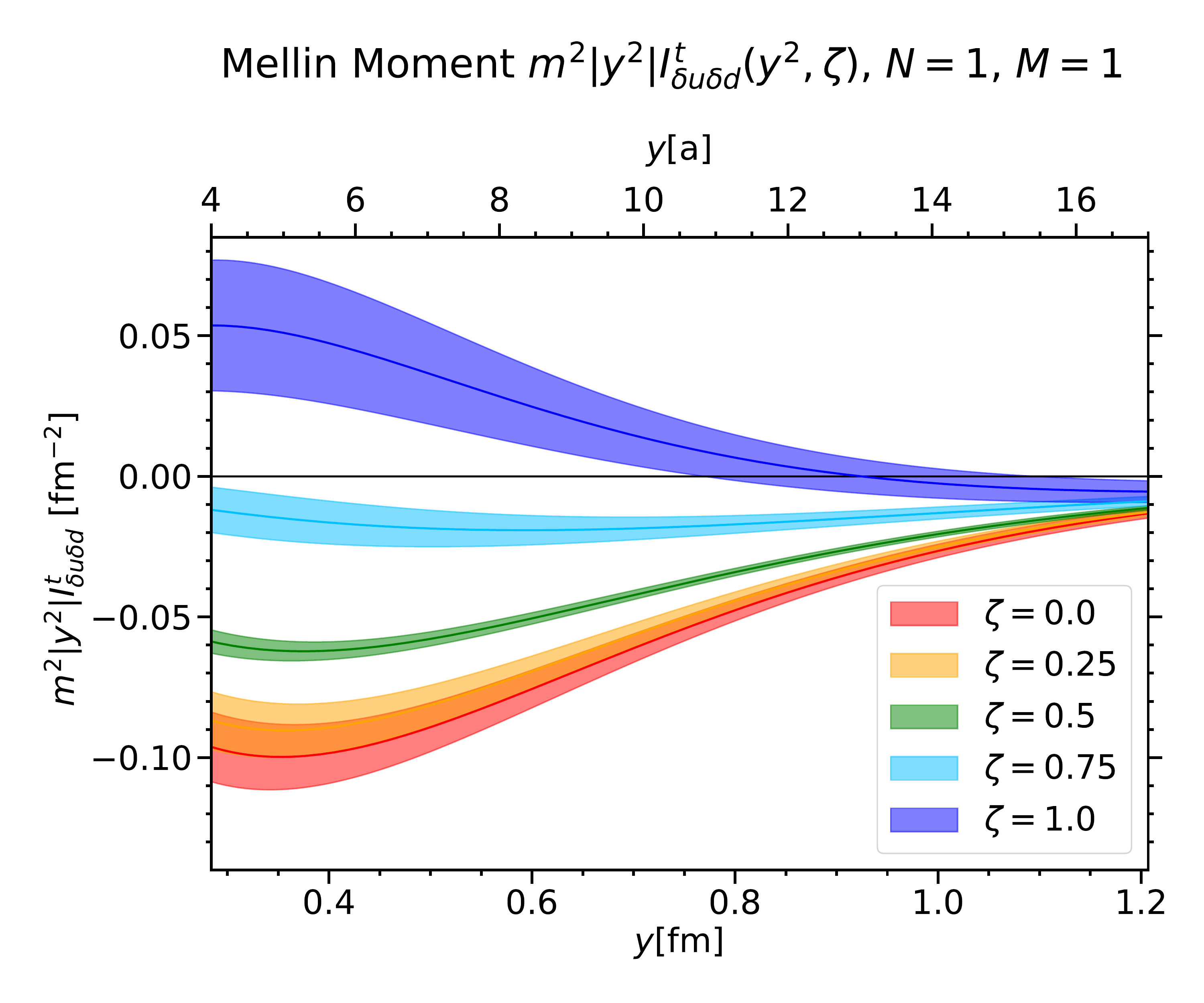}}
\caption{\label{fig:mellin-zeta} Mellin moments of skewed DPDs as a function of $y$, reconstructed from our global fit.}
\end{center}
\end{figure*}

\begin{figure*}
\begin{center}
\subfigure[$I_{u d}(y^2, \zeta) / I_{u d}(y^2, 0)$]{
\includegraphics[width=0.48\textwidth,trim=0 0 0 60,clip]
{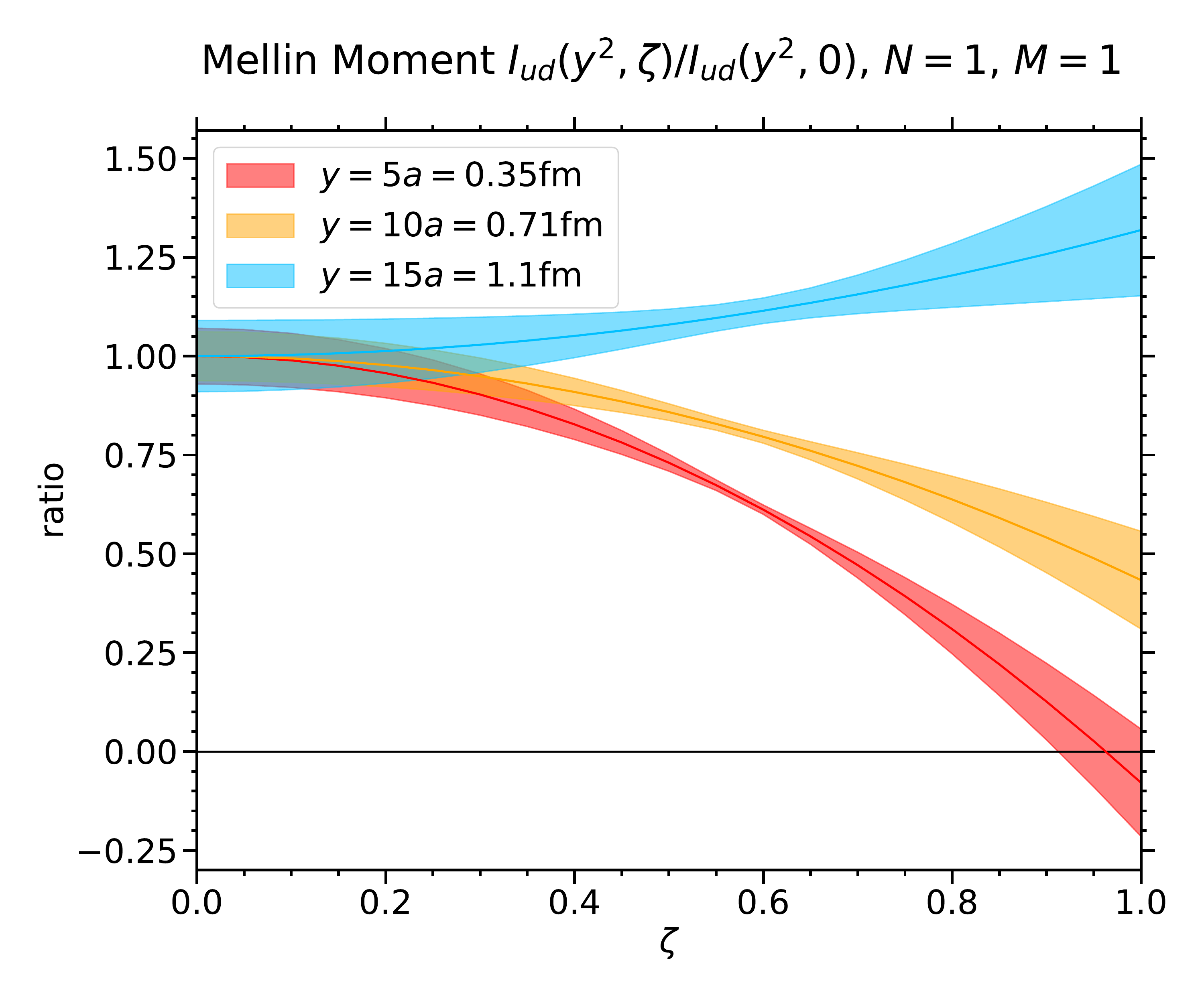}}
\hfill
\subfigure[$I_{\delta u \ms d}(y^2, \zeta) / I_{\delta u \ms d}(y^2, 0)$]{
\includegraphics[width=0.48\textwidth,trim=0 0 0 60,clip]
{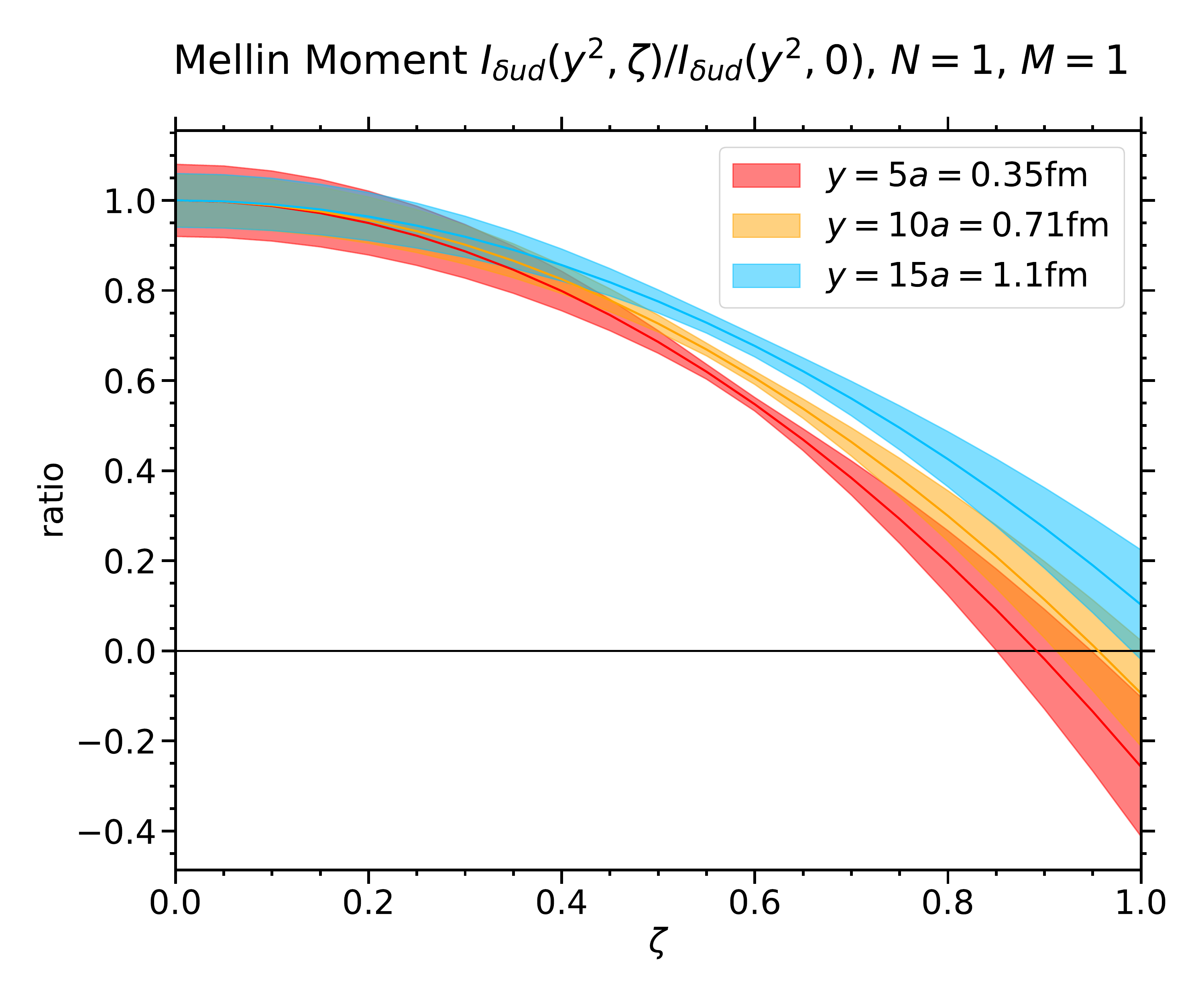}}
\caption{\label{fig:mellin-zeta-dep} Mellin moments of skewed DPDs as a function of $\zeta$, reconstructed from our global fit and normalised to their value at $\zeta=0$.}
\end{center}
\end{figure*}

We can easily derive the analytic form of the Mellin moments for our fits by inverting the $2 \times 2$ matrix $T_{m n}$ in \eqref{eq:T-mat-def}.  This gives
\begin{align}
\label{eq:fitted-mom}
I(y^2, \zeta) &= \frac{3 \pi}{4} \ms
  \biggl\{\ms 3 - 5 \ms \langle \zeta^2 \rangle(y^2)
    - 5 \ms \zeta^2 \ms \Bigl[ 1 - 3 \ms \langle \zeta^2 \rangle(y^2) \ms\Bigr]
  \biggr\} \, A(y^2, py=0) \,.
\end{align}
The values of $\langle \zeta^2 \rangle(y^2)$ for $y \le 20a$ are in the range between $0$ and $0.5$ for all twist-two functions.  The combination $3 - 5 \langle \zeta^2 \rangle$ in \eqref{eq:fitted-mom} is therefore always positive and varies between $3$ and $0.5$.  We can hence anticipate that the dependence of the Mellin moments $I(y^2, \zeta=0)$ on $y$ and on the polarisation indices should roughly follow the corresponding dependence of $A(y^2, py=0)$.  By contrast, the coefficient of $\zeta^2$ in \eqref{eq:fitted-mom} has a larger variation and can change sign as a function of $y$.
Our results for the $y$ and $\zeta$ dependence of the Mellin moments are visualised in \figs{\ref{fig:mellin-zeta}} and \ref{fig:mellin-zeta-dep}.  Compared with the data entering our fit, we have slightly extended the $y$ range from $5a$ down to $4a$.

\begin{figure}
\begin{center}
\subfigure[Mellin moments $I(y^2, \zeta=0)$\label{fig:mellin-pol-a}]{
\includegraphics[height=14.4em,trim=0 25 0 85,clip]
{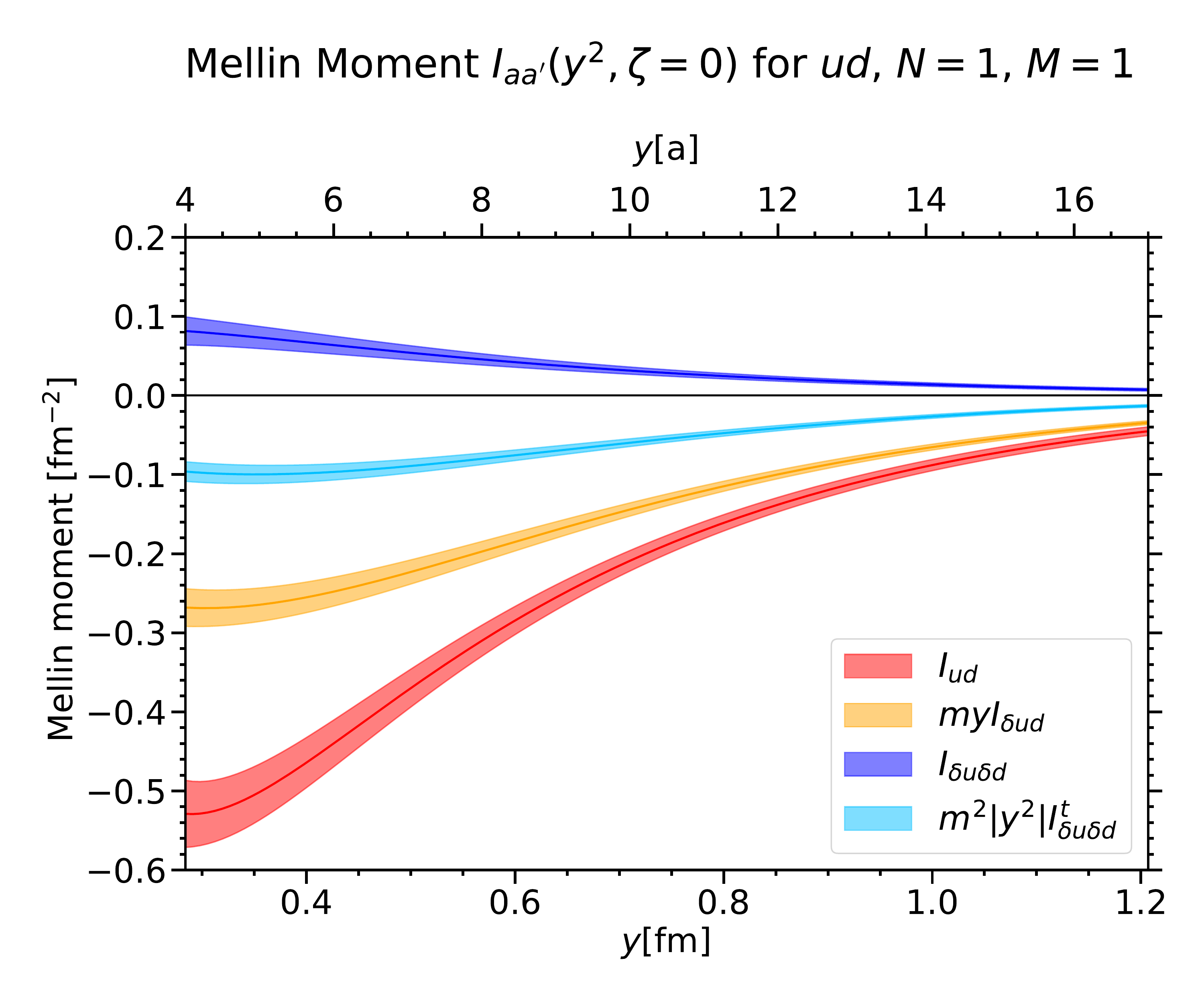}}
\hfill
\subfigure[twist-two functions $A(y^2, py=0)$]{
\includegraphics[height=14.4em,trim=0 0 0 17,clip]
{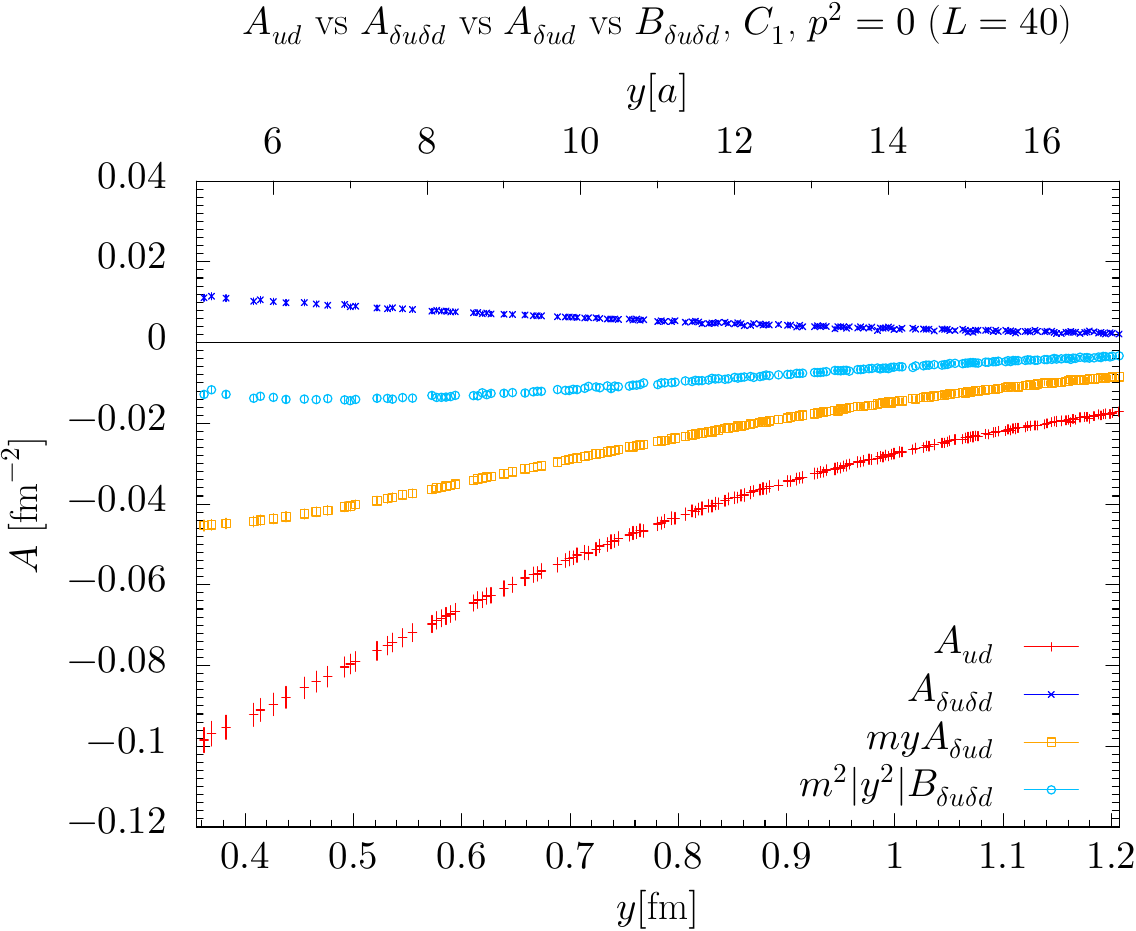}}
\caption{\label{fig:mellin-pol} (a): Mellin moments of DPDs for the flavour combination $u d$ in a $\pi^+$, reconstructed from our global fit.  (b): Lattice data for the corresponding twist-two functions at $py=0$.  This shows the same data as \fig{\protect\ref{fig:polar-C1-T}}, but is limited to $y \le 17a$ for ease of comparison.  Notice that the minimum $y$ in panels (a) and (b) is slightly different.}
\end{center}
\end{figure}

In the left panel of \fig{\ref{fig:mellin-pol}}, we show the Mellin moments at $\zeta=0$ for the different polarisation combinations.  Comparison with the data of the corresponding twist-two functions at $py=0$ shows the close similarity between the two quantities.  This corroborates the basic assumption of our discussion in \sect{\ref{sec:results}}, namely that the qualitative features of twist-two functions at $py=0$ are representative of the Mellin moments of ordinary DPDs.

With the caveats of choosing a functional ansatz and restricting ourselves to the connected graph $C_1$, we can in particular extend our discussion for light quarks in \sect{\ref{sec:polarisation}} to the Mellin moments of DPDs for the flavour combination $u d$ in a $\pi^+$: there is a substantial spin-orbit correlation for one transversely polarised quark or antiquark, whereas correlations involving transverse polarisation of both partons are rather small.  This is one of the main results of our work.

DPDs at $\zeta=0$ satisfy sum rules, which have been proposed in
\cite{Gaunt:2009re} and can be proven rigorously in QCD \cite{Gaunt:2012ths,Diehl:2018kgr}.  These sum rules express momentum and quark number conservation.  The quark number sum rule for the flavour combination $u d$ in a $\pi^+$ implies that
\begin{align}
   \label{eq:dpd-sr}
2\pi \int\limits_{y_{\text{cut}}}^\infty \!
   dy\, y \ms I_{u d}(y^2; \mu)
  &= -1 + \mathcal{O}(\Lambda^2 y_{\text{cut}}^2)
        + \mathcal{O}\bigl( \alpha_s^2(\mu) \bigr) \,,
\end{align}
where $\Lambda$ denotes a hadronic scale.  The necessity of a lower cutoff on the $y$ integral and the presence of an $\mathcal{O}(\alpha_s^2)$ term on the r.h.s.\ result from the singular behaviour of DPDs at perturbatively small distances $y$, as explained in \cite{Diehl:2020xyg}.  To avoid large logarithms in the $\mathcal{O}(\alpha_s^2)$ term, one should take $y_{\text{cut}} \sim 1/\mu$, and a standard choice is $y_{\text{cut}} = b_0 / \mu$, where $b_0 = 2 e^{-\gamma} \approx 1.12$ and $\gamma$ is the Euler-Mascheroni constant.  With the renormalisation scale $\mu = 2 \gev$ of our analysis, this gives $y_{\text{cut}} \approx 0.11 \fm \approx 1.56 \ms a$.  Extrapolating our global fit down to this value and evaluating the integral over $y$, we obtain
\begin{align}
   \label{eq:dpd-sr-result}
2\pi \int\limits_{b_0 /\mu}^\infty \!
   dy\, y \ms I_{u d}(y^2; \mu) = -0.915(78) \,.
\end{align}
This result is not too sensitive to the extrapolation in $y$: taking an upper integration boundary of $20a$, we obtain $-0.908(63)$, whilst raising the lower integration boundary by a factor $2$, we obtain $-0.885(72)$.  Note that with a larger $y_{\text{cut}}$, one expects a larger $\mathcal{O}(\Lambda^2 y_{\text{cut}}^2)$ term on the r.h.s.\ of \eqref{eq:dpd-sr}.  Given the presence of this power correction in the theory prediction, we find its agreement with our result \eqref{eq:dpd-sr-result} quite satisfactory.  We regard this as a strong cross check of our analysis, and in particular of the fit ansatz we have made \rev{in \eqref{eq:I-series}, \eqref{eq:zeta-y-fit} and \eqref{eq:y-dep-fit}}.


\subsection{Factorisation hypothesis for Mellin moments}
\label{sec:fact-mellin}

\begin{figure}[!th]
\begin{center}
\subfigure[\label{fig:comp-I}]{
\includegraphics[width=0.48\textwidth,trim=0 0 0 95,clip]
{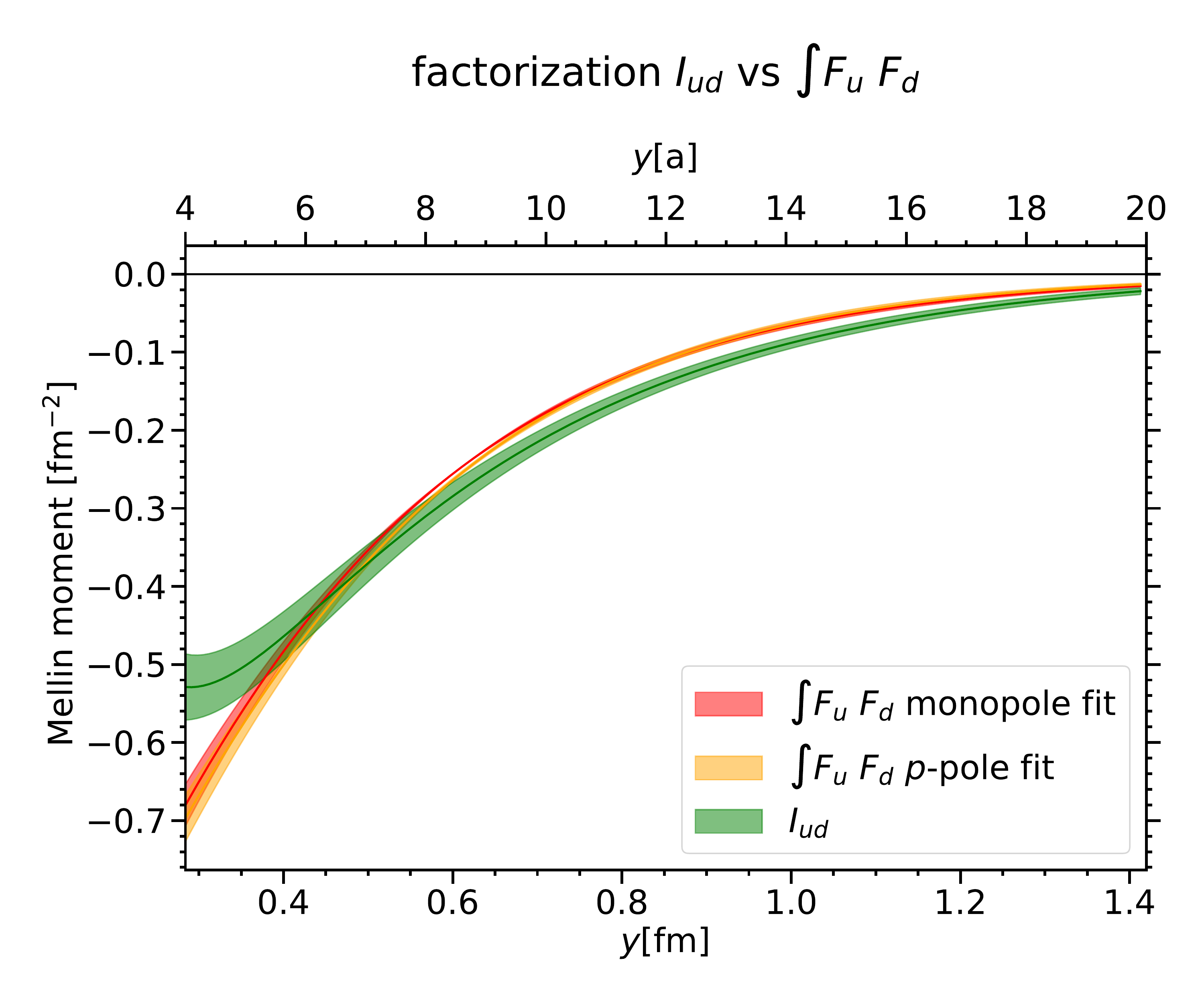}}
\hfill
\subfigure[\label{fig:rat-I}]{
\includegraphics[width=0.48\textwidth,trim=0 0 0 95,clip]
{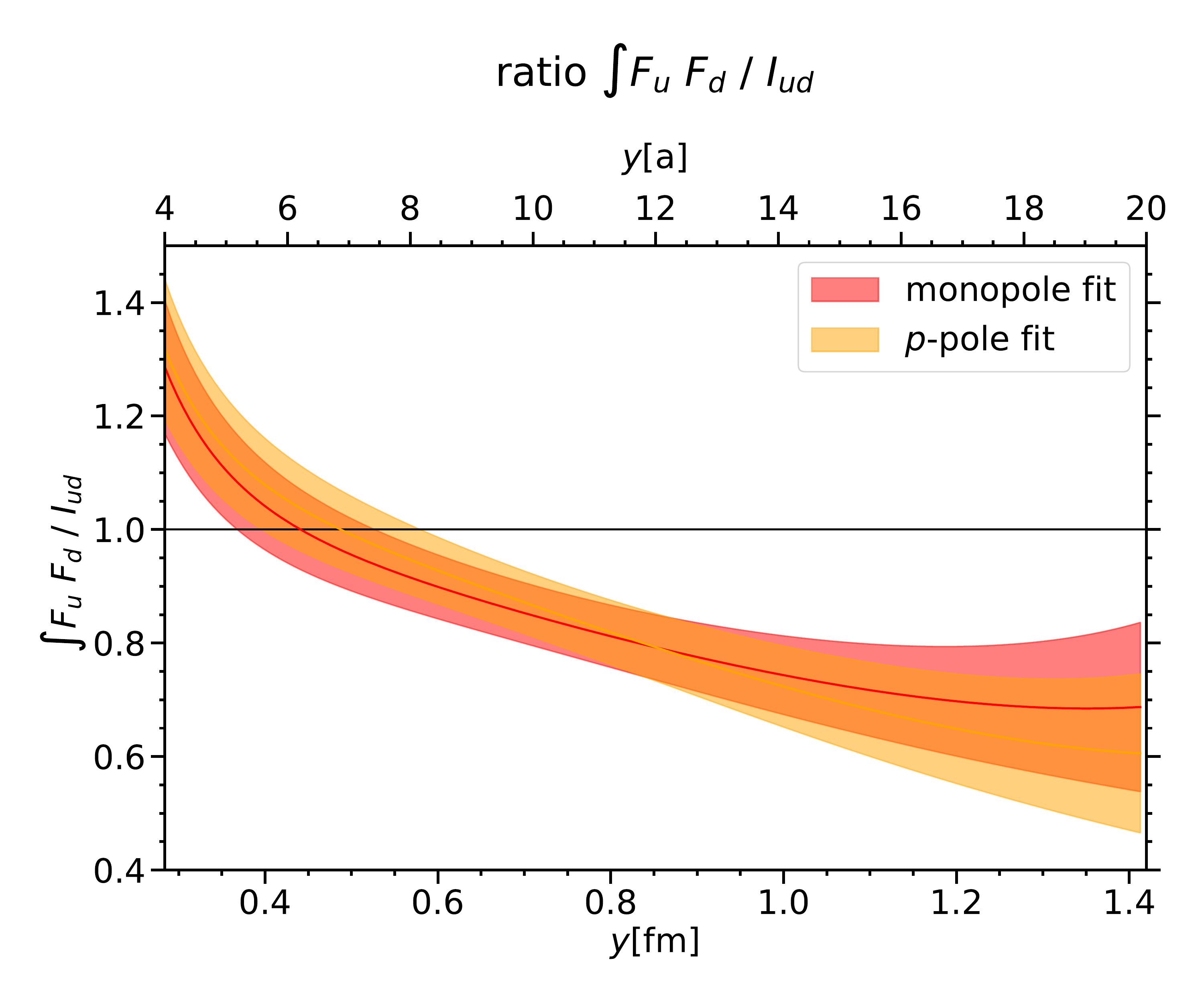}}
\caption{\label{fig:fact-mellin} Test of the factorisation hypothesis \protect\eqref{eq:mellin-fact} for the lowest Mellin moment $I_{u d}$ of the unpolarised DPD $F_{u d}$ at $\zeta=0$ in a $\pi^+$.
(a): Comparison of $I_{u d}$ determined by the global fit of \sect{\protect\ref{sec:py-fits}} with the integral over form factors on the r.h.s.\ of \protect\eqref{eq:mellin-fact}.  The form factors are determined by a monopole or a $p$-pole fit as described in \sect{\protect\ref{sec:fact-A}}.
(b): Ratio of the integral over form factors and the Mellin moment.  The factorisation hypothesis predicts this ratio to be $1$.}
\end{center}
\end{figure}

With the Mellin moments reconstructed from our global fit, we can also test the factorisation hypothesis \eqref{eq:mellin-fact}, which directly follows from the corresponding hypothesis \eqref{eq:dpd-fact} for DPDs.  To evaluate the r.h.s.\ of \eqref{eq:mellin-fact}, we use the same two fits for the vector form factor of the pion as we did in \sect{\ref{sec:fact-A}}.  The comparison of the left and right-hand sides of \eqref{eq:mellin-fact}, as well as their ratio is shown in \fig{\ref{fig:fact-mellin}}.  We see the same trend as we did in \fig{\ref{fig:fact-test}} for $A_{u d}$ at $py = 0$.  At small $y$, the result of the factorised ansatz is too large in absolute size, and at large $y$ it is too small.  The discrepancy at large $y$ is even somewhat stronger for the Mellin moment $I_{u d}$ than it is for $A_{u d}$.  We draw the same conclusion as we did in \sect{\ref{sec:fact-A}}: the factorised ansatz for the unpolarised $u d$ flavor combination in a $\pi^+$ can provide a rough approximation at the level of several $10\%$.  In the sense that the factorised ansatz represents the assumption that the $u$ and the $\bar{d}$ in a $\pi^+$ have independent spatial distributions, our result for $I_{u d}$ indicates that the two partons prefer to be farther apart than if they were uncorrelated.

\subsection{Comparison with quark model results}
\label{sec:quark-models}

As we mentioned in the introduction, there are a few calculations of pion DPDs within quark models \cite{Rinaldi:2018zng,Courtoy:2019cxq,Broniowski:2019rmu,Broniowski:2020jwk}.  Most of the results presented in these papers are for distributions differential in $x_1$ and $x_2$, which are not accessible in our lattice calculation.  However, reference \cite{Courtoy:2019cxq} also gives the lowest Mellin moment of polarised and unpolarised DPDs as a function of $y$, as we did in the present work.  Predictions for the lowest Mellin moment of the unpolarised DPD are also shown in \cite{Broniowski:2019rmu,Broniowski:2020jwk}.  They are quite similar to the ones in \cite{Courtoy:2019cxq}.

The results presented in \cite{Courtoy:2019cxq} are for the physical value of the pion mass.  The authors of that work have provided us with the numbers they obtain when taking $m_\pi = 300 \mev$ instead \cite{Courtoy:2020prc}.  This parameter setting was also used in \cite{Courtoy:2020tkd}, where two-current matrix elements in the pion computed in the same model were compared with the results of our lattice study~\cite{Bali:2018nde}.

\begin{figure}[!t]
\begin{center}
\subfigure[$-y I_{u d}$]{
\includegraphics[width=0.48\textwidth,trim=0 0 0 65,clip]
{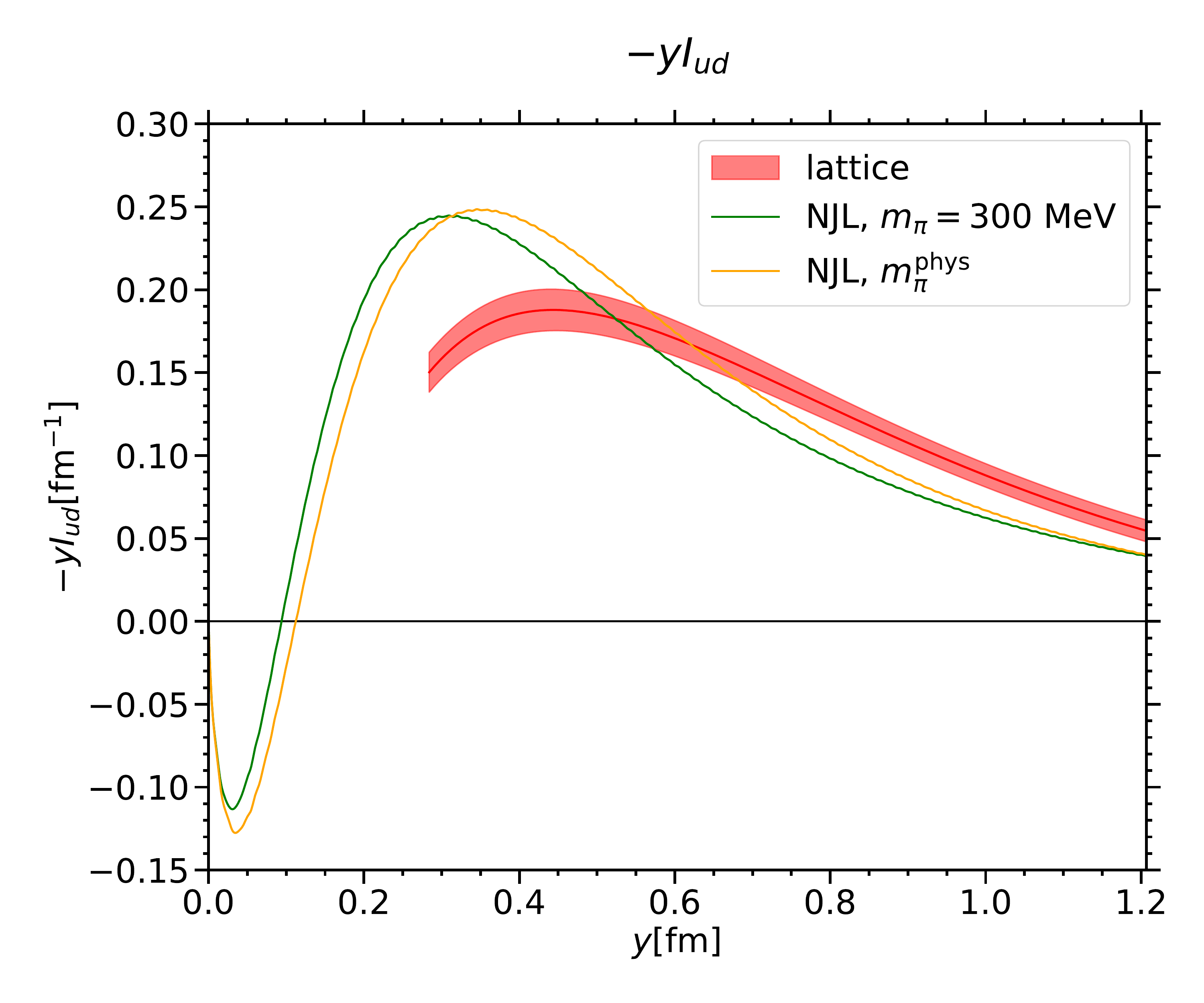}}
\hfill
\subfigure[$- m |y^2| I_{u \ms \delta}$]{
\includegraphics[width=0.48\textwidth,trim=0 0 0 65,clip]
{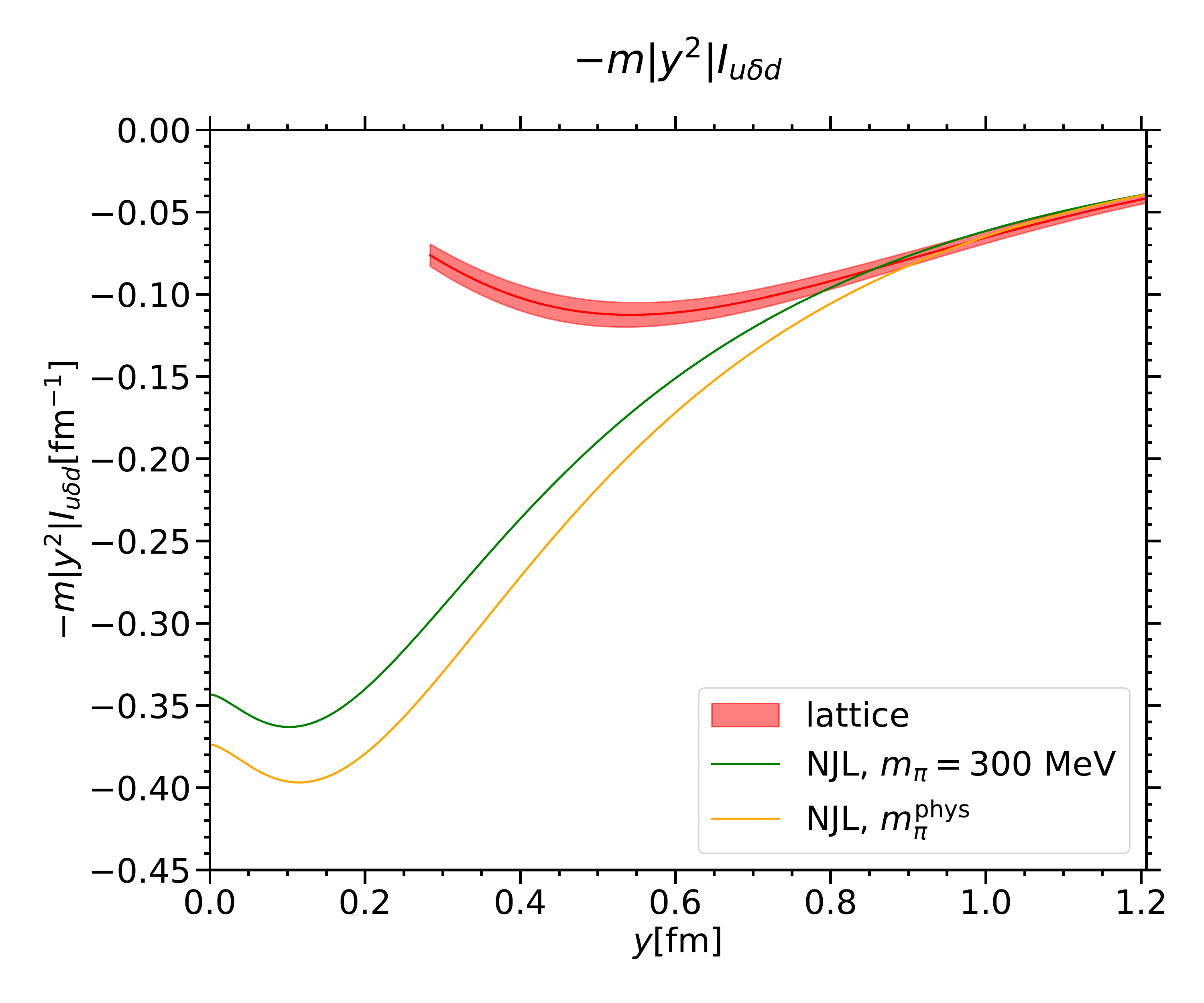}}
\\
\subfigure[$- y I_{\delta u \delta d}$]{
\includegraphics[width=0.48\textwidth,trim=0 0 0 65,clip]
{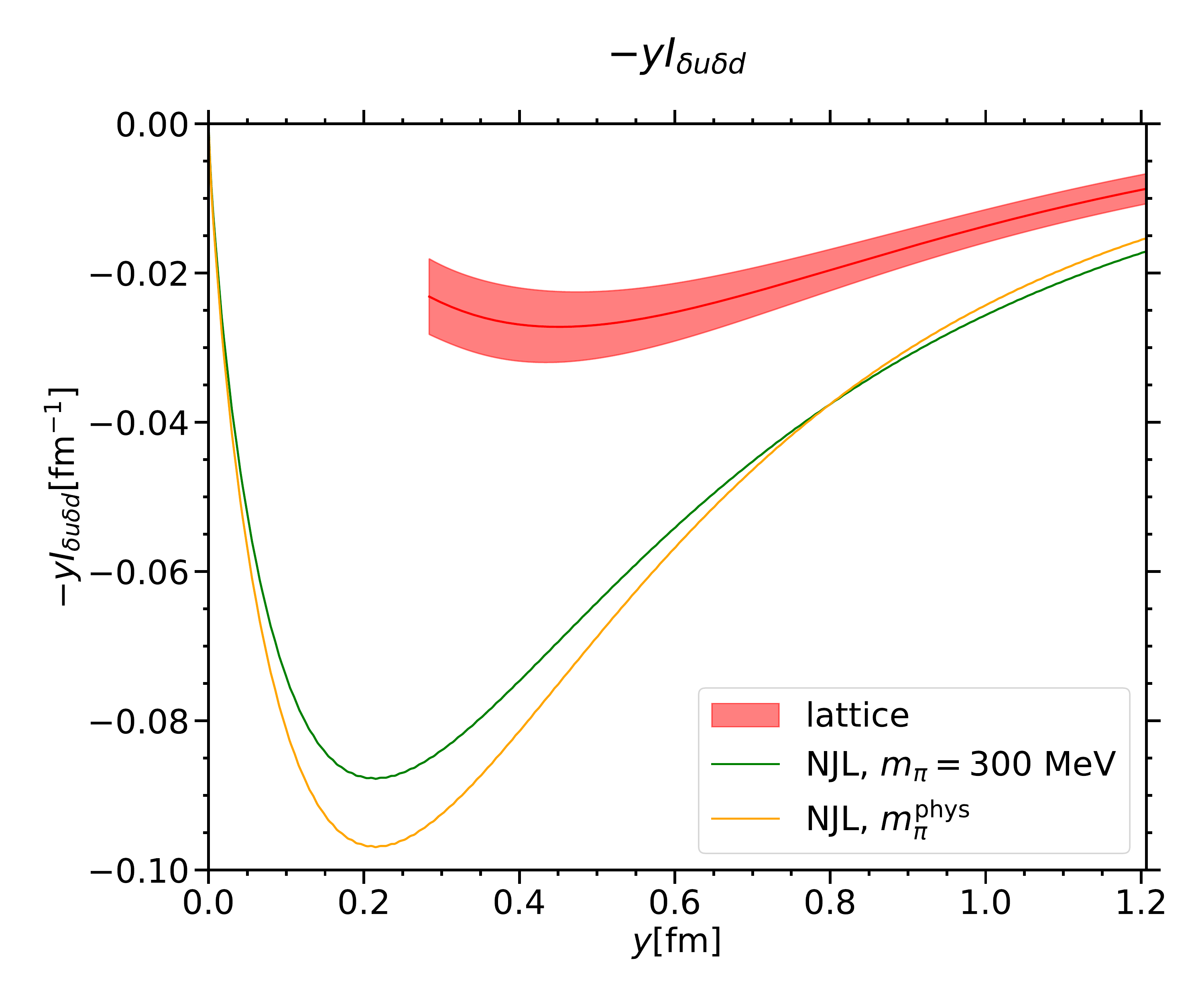}}
\hfill
\subfigure[$- m^2 |y|^3 I^t_{\delta u \delta d}$]{
\includegraphics[width=0.48\textwidth,trim=0 0 0 65,clip]
{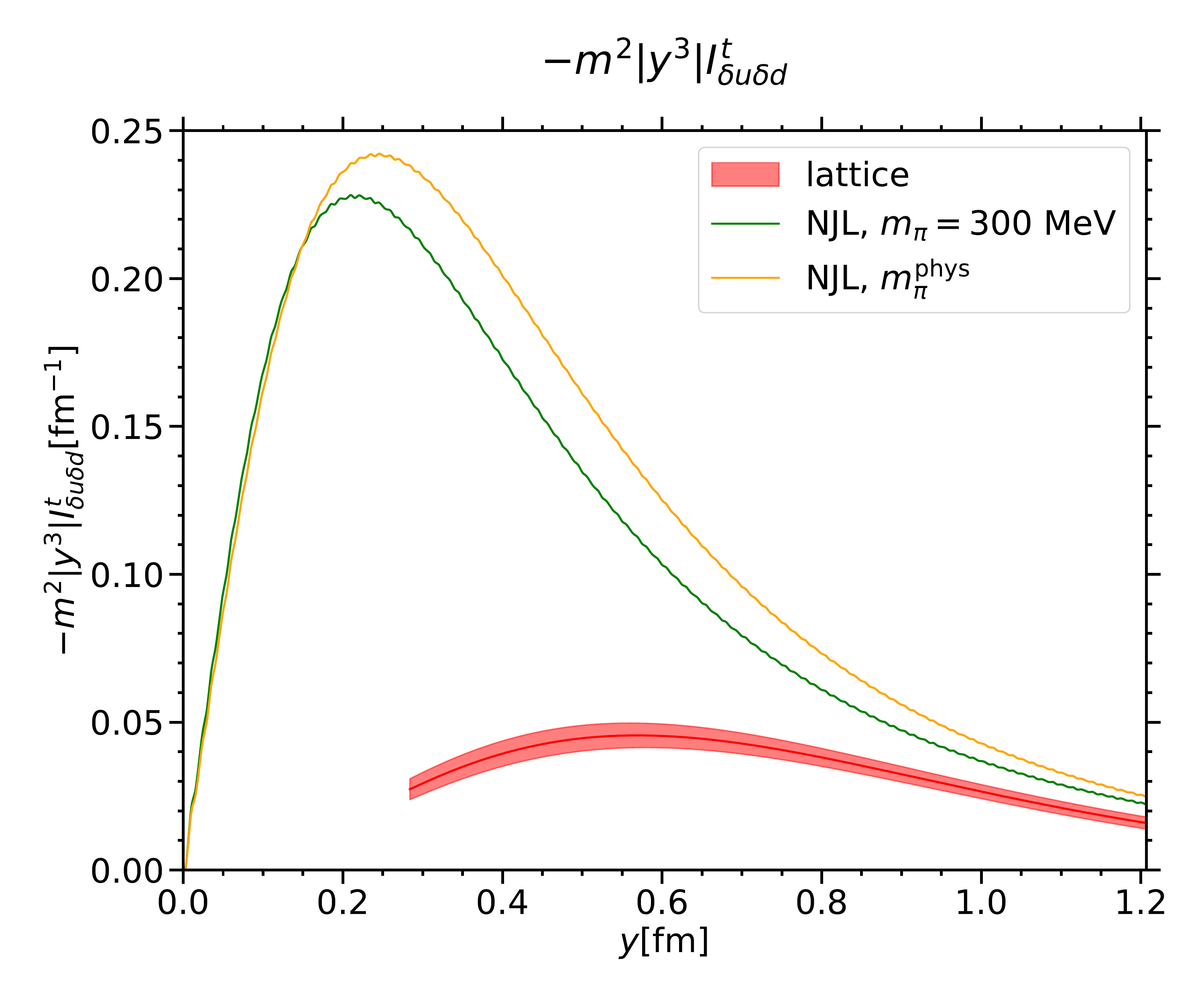}}
\caption{\label{fig:quark-model} Comparison of our results for Mellin moments $I(y,\zeta=0)$ of DPDs in a $\pi^+$ with those obtained with the Nambu Jona-Lasinio model in \protect\cite{Courtoy:2019cxq}.  The lattice results correspond to the ones in our \fig{\protect\ref{fig:mellin-pol-a}} but are multiplied by an additional factor $-y$.  The model curves for $m_\pi^{\text{phys}}$ are the same as the ones labelled ``PV'' in \fig{3} of \protect\cite{Courtoy:2019cxq}.  The lattice and model results refer to different factorisation scales, as specified in the text.}
\end{center}
\end{figure}

In \fig{\ref{fig:quark-model}}, we show the quark model results along with the Mellin moments reconstructed from our global fit.\footnote{%
   The notation for DPDs involving transverse polarisation in \cite{Courtoy:2019cxq} is related to our notation in \eqref{eq:invar-dpds} as $F^v_{u \ms \delta d} = m y f_{u \ms \delta d}$, $F^s_{\delta u \delta d} = f_{\delta u \delta d}$, and $F^t_{\delta u \delta d} = m^2 |y^2| f^t_{\delta u \delta d}$, where all functions depend on $x_1, x_2$, and $y$.}
Compared with our \fig{\protect\ref{fig:mellin-pol-a}}, the moments are multiplied with an additional factor $-y$ so as to correspond to the curves in \fig{3} of \cite{Courtoy:2020tkd}.  Notice that \cite{Courtoy:2020tkd} gives results for two different regulators of ultraviolet divergences.  We only show the ones obtained with Pauli-Villars regularisation here and note that the difference between the two regulators in \fig{3} of \cite{Courtoy:2020tkd} is quite noticeable for $I_{u \ms \delta d}$ at $y < 0.4 \fm$ and for $I^t_{\delta u \delta d}$ at $y > 0.2 \fm$.

The difference between the model results for $m_\pi = 300 \mev$ and $m_\pi = 140 \mev$ is quite small for the moments shown in \fig{\ref{fig:quark-model}}.  A larger mass dependence is found for $I_{\Delta u \Delta d}$, which we do not show because our lattice data in this channel is too noisy for a stable reconstruction of the Mellin moment.

The quark model curves in \fig{\ref{fig:quark-model}} refer to the typical renormalisation scale of the model, which in \cite{Courtoy:2008nf} was estimated to be $\mu = 290 \mev$ for evolution at LO and $\mu = 430 \mev$ for evolution at NLO in the strong coupling.  The Mellin moment $I_{u d}$ for unpolarised quarks involves only the vector current and is therefore scale independent, so that the model curves can be directly compared with our lattice values at $\mu = 2 \gev$.  We find that the results of the two approaches are remarkably close to each other, despite a visible difference around $y \sim 0.3 \fm$.  Let us note that the agreement between our result and the one shown for the Spectral Quark Model in \fig{6b} of \cite{Broniowski:2019rmu} is even better.

Evolution from the quark model scale to $\mu = 2 \gev$ will reduce the moment $I_{u \ms \delta d}$ by a factor $r$ and the moments $I^{}_{\delta u \delta d}$ and $I^t_{\delta u \delta d}$ by a factor $r^2$.  We refrain from estimating this factor here, because it involves evolution in a region where perturbation theory becomes quite unstable.  Despite this uncertainty, we can state that for all Mellin moments involving transverse quark polarisation, the two approaches agree in the sign and qualitative shape for $y > 0.3 \fm$.  However, it is also clear that no value of the evolution factor $r$ can bring the lattice and model results for all three moments into quantitative agreement.

\FloatBarrier

\section{Summary}
\label{sec:summary}

This paper presents the first lattice calculation that provides information about double parton distributions in a pion.  Our simulations are for a pion mass of $m_\pi \approx 300 \mev$, a lattice spacing of $a \approx 0.07 \fm$, and two lattice volumes with $L=32$ and $L=40$ points in the spatial lattice directions, respectively.  We also have results for the pseudoscalar ground state made of strange or of charm quarks at their physical masses, in a partially quenched setup.

We compute the pion matrix elements of the product of two local currents that are separated by a space-like distance.  From these tensor-valued matrix elements, we extract Lorentz invariant functions associated with the twist-two operators in the definition of DPDs.  In the continuum and infinite volume limits, these functions depend on the pion momentum $p^\mu$ and the distance $y^\mu$ between the currents only via the invariant products $py$ and $y^2$.  This allows us to detect discretisation and finite size effects in our data, and to devise cuts that minimise these artefacts.  In particular, most results reported here are limited to distances $y$ above $5a \approx 0.35 \fm$.  Comparing the data from our two lattice volumes, we find only mild differences in channels that have a good statistical signal.  Comparing results obtained with different source-sink separation, we find little evidence for contributions from excited states in our analysis.  The invariant twist-two function in the axial vector channel appears to be most strongly affected by several of the lattice artefacts.

Comparing the importance of different Wick contractions in the twist-two functions, we find that the connected graphs $C_1$ and $C_2$ in \fig{\ref{fig:contractions}} are the most important ones in almost all cases.  For light quarks, graph $C_2$ is as important as $C_1$ at small distances $y$ between the two partons, which indicates that Fock states containing sea quarks are important in that region.  As one would expect, this importance is strongly reduced for charm quarks, but it is still visible at a level below $10\%$.

We compute matrix elements for different combinations of the vector, axial vector and tensor currents, which respectively correspond to unpolarised partons and partons with longitudinal or transverse polarisation.  For light quarks, we find surprisingly small correlations between the longitudinal or transverse spins of the two partons.  By contrast, a large spin-orbit correlation is seen between the transverse component of $y$ and the transverse polarisation of one of the partons.  All spin correlations increase considerably with the quark mass, and for charm quarks we observe large spin-spin correlations for both longitudinal and transverse polarisation.

The invariant twist-two functions that we can determine on the lattice are not directly related to the Mellin moments of DPDs, but rather to the moments of what can be called ``skewed'' DPDs.  To compute the Mellin moments of ordinary DPDs from two-current matrix elements, one needs the dependence of the invariant functions on the variable $py$ on the full real axis.  This is inaccessible on a Euclidean lattice.  Fitting an ansatz for the $py$ dependence to our lattice data, we can however reconstruct the \rev{lowest} Mellin moments by extrapolating this ansatz to the full $py$ range.  We find that the moments obtained in this way have a behaviour very similar to the one of the twist-two functions at $py = 0$.  A valuable cross check of our procedure is the fact that the result for the unpolarised Mellin moment is in good agreement with the number sum rule that must be obeyed by the DPD for the flavour combination $u d$ in a $\pi^+$.
\rev{Comparing our results for the Mellin moments with those obtained in quark models, we find rather close agreement for unpolarised quarks.  For moments involving transverse quark polarisation, we observe qualitative agreement but quantitative differences.  We have not reconstructed the lowest Mellin moment for longitudinal quark polarisation, because we consider our lattice data in this channel to be too noisy for this purpose.}

A starting point of many phenomenological studies is the assumption that unpolarised DPDs can be ``factorised'' into the single-particle distributions of each parton, which would mean that the two partons are independent of each other.  We have formalised this assumption and tested it, both for the twist-two functions directly extracted from the lattice data and for the Mellin moments reconstructed by extrapolating a fit to these data.  In both cases, we find that the two-parton correlator deviates from the factorisation ansatz by a few $10\%$, and that the sign of the deviation depends on the transverse distance $y$.  More specifically, the two partons tend to be farther apart from each other than if they were independent of each other.

We see several directions into which the studies reported here should be extended.  First and foremost comes the extension from a pion to a nucleon, which is of direct relevance for double parton scattering in proton-proton collisions.  Work in this direction is underway.  On a longer time scale, one will want to have simulations with finer lattice spacing and smaller quark masses.  Data of sufficient quality for higher hadron momenta will extend the range in $py$ that can be probed and thus allow for a better controlled extrapolation in this variable.  Given the results obtained in the present work, we think that the efforts required for such studies will be rewarded with valuable physics insights.


\section*{Acknowledgements}

We gratefully acknowledge input from Sara Collins and Andr{\'e} Sternbeck, and discussions with Michael Engelhardt.  \rev{We thank A. Courtoy,  S. Noguera, and S. Scopetta for kindly providing us with the numbers of their results in \cite{Courtoy:2019cxq} and with the corresponding values for a pion mass close to the one in our simulations.}

We used a modified version of the Chroma~\cite{Edwards:2004sx} software package, along with the locally deflated domain decomposition solver implementation of openQCD~\cite{Luscher:2012av}. The gauge ensembles have been generated by the QCDSF and RQCD collaborations on the QPACE computer using BQCD~\cite{Nakamura:2010qh,Nakamura:2011cd}. The graphs  were produced with JaxoDraw~\cite{Binosi:2003yf,Binosi:2008ig}.
The simulations used for this work were performed with resources provided by the North-German Supercomputing Alliance (HLRN).  This work was supported by the Deutsche Forschungsgemeinschaft SFB/TRR~55, by BMBF Verbundprojekt O5P2018(ErUM-FSP T01), and by the European Union's Horizon 2020 research and innovation programme under grant agreement No.~824093 (STRONG-2020) and Marie Sk\l{}odowska-Curie grant agreement No.~813942 (EuroPLEx).


\FloatBarrier

\phantomsection
\addcontentsline{toc}{section}{References}

\bibliographystyle{JHEP}
\bibliography{twist-two}

\end{document}